\documentclass[11 pt]{article}
\pdfoutput=1 
\usepackage{jheppub} 
\usepackage{amsmath}
\usepackage{orcidlink}
\usepackage{lscape}
\usepackage{multicol}
\usepackage{booktabs}
\usepackage{float}
\usepackage{mathrsfs}  
\usepackage{longtable}
\usepackage{graphicx}
\usepackage{orcidlink}
\usepackage{xcolor}
\usepackage{color, colortbl}
\usepackage{shuffle}
\usepackage{empheq}
\usepackage{tabularx}
\usepackage{xcolor}
\definecolor{baikovbrown}{HTML}{6B3E26}
\usepackage{adjustbox}
\usepackage{bm}
\usepackage{physics}
\usepackage{caption}
\usepackage{subcaption,longtable,stmaryrd}
\usepackage{slashed}
\usepackage{bigints}
\usepackage{cancel}
\usepackage{tikz}
\usetikzlibrary{arrows.meta,calc}
\usepackage{enumitem}
\usepackage[customcolors,shade]{hf-tikz}
\usepackage{microtype}
\usepackage{changepage}
\usepackage{cleveref}
\usepackage[most]{tcolorbox}
\DeclareSymbolFont{usualmathcal}{OMS}{cmsy}{m}{n}
\DeclareMathAlphabet\mathbfcal{OMS}{cmsy}{b}{n}
\DeclareSymbolFontAlphabet{\mathcal}{usualmathcal}
\DeclareSymbolFont{rmlargesymbols}{OMX}{mdbch}{m}{n}
\DeclareMathSymbol{\rmintop}{\mathop}{rmlargesymbols}{82}
\DeclareMathSymbol{\rmointop}{\mathop}{rmlargesymbols}{72}

\definecolor{mygray}{gray}{0.5}

\newcommand{\ee}{\mathrm{e}}
\newcommand{\ii}{\mathrm{i}}
\newcommand{\MaxCut}{\operatorname{MaxCut}}
\newcommand{\res}{\textrm{Res}}
\allowdisplaybreaks

\title{\boldmath{Geometry of massive loops in gravitational scattering: Maximal cut, Riemann's bilinear relations and genus drop}}

\author[a]{Arpan Bhattacharyya\,\orcidlink{0000-0002-7933-6441
}}
\author[b]{and Saptaswa Ghosh\,\orcidlink{0000-0002-0237-7042} }
\affiliation[a]{Department of Physics, Indian Institute of Technology Gandhinagar,\\
Palaj, Gandhinagar 382055, India.}
\affiliation[b]{Centre for High Energy Physics, Indian Institute of Science,\\ Bangalore 560012, India.}
\emailAdd{abhattacharyya@iitgn.ac.in}
\emailAdd{saptaswag@iisc.ac.in}

\abstract{We study a representative two-loop family of worldline quantum field theory integrals relevant for classical gravitational scattering with a massive mediator and investigate the geometric structure of its maximal cut. We find that, unlike the massless case, the massive loops already give rise to an elliptic maximal cut at two loops, which degenerates into a nodal curve in the massless limit and
in the static limit. In the global Baikov representation, two successive integrations reduce the maximal cut to a period of an elliptic curve. We further establish five primary consequences of this study: (a) The integration-by-parts method generates five master integrals, which, under reflection, split into a three-dimensional elliptic block and a two-dimensional polylog block. We derive the differential equation in two ways, one by Laporta reduction and subsequently by intersection theory in two cocycle bases, and construct the rational map between them. (b) The period matrices in arbitrary dimension have been computed in terms of Appell and Beta functions. Furthermore, we derive twisted Riemann period relations that yield nine quadratic identities among them. (c) We study the global monodromy relation of the differential system that satisfies Riemann's bilinear relation. (d) We explicitly derive the function space of the iterated integrals in terms of elliptic multiple polylogs. (e) In impact-parameter space, the Fourier-transformed cut is a Bessel-twisted differential on the elliptic curve, which we show has extra involution symmetries. Thus, the parity of the differential further decides which quotient - rational or elliptic its periods live on. All results presented concern the maximal cut and, therefore, the homogeneous solutions, but they open an important pathway to tackling the full massive two-loop integral.}

\begin{document}
\maketitle
\flushbottom

\section{Introduction}
\label{sec:intro}
Precision calculations in both high-energy particle physics and gravitational wave (GW) physics remain increasingly challenging, and their advancement relies crucially on our ability to compute Feynman integrals efficiently. Although these two areas probe seemingly disparate physical scales, they share a common fundamental requirement: the need for accurate theoretical predictions to interpret experimental and observational data. In particular, recent detection of gravitational waves from binary-black-hole mergers \cite{LIGOScientific:2016aoc,LIGOScientific:2016sjg,LIGOScientific:2016vlm} has made this requirement particularly
important for the relativistic two-body problem. Hence, improving our understanding of the dynamics of compact objects is not only necessary to test general relativity (GR) but also necessary to probe whether any beyond-GR effects exist or not, which may leave minuscule effects in the detected signal \cite{LIGOScientific:2021sio,LIGOScientific:2026qni,LIGOScientific:2026oim}. In this context, methods originally developed for studying scattering in particle physics have become useful tools for studying classical gravitational dynamics.   

In physics, the scattering process is one of the best ways to extract information about the interaction between two objects. At subatomic scales, one uses scattering experiments to extract this information, but at the larger length scale suitable for classical gravitational physics, it can be related to the conservative and radiative dynamics of compact objects. Using techniques of effective field theories, one can make this connection more systematic by resorting to the description of the compact objects as a point particle (to the leading order in their size) moving on a worldline.  Then their interactions can be organized as a systematic expansion of Newton's constant ($G_N$) known as the post-Minkowskian (PM) expansion (retaining the dependency on the relative velocity). Scattering amplitude based methods \cite{Brandhuber:2022qbk,Brandhuber:2023hhy,Kosower:2018adc,Bjerrum-Bohr:2018xdl,Bjerrum-Bohr:2013bxa,Alessio:2024wmz, Bern:2019crd, Bern:2019nnu, Bern:2020buy,Bern:2024adl,Bern:2023ity,Bern:2022kto, Bern:2021yeh,Bern:2021dqo,Bern:2020gjj,Bern:2025zno, Brandhuber:2019qpg,AccettulliHuber:2019jqo,AccettulliHuber:2020oou,Bjerrum-Bohr:2021wwt,Bjerrum-Bohr:2021vuf,Bjerrum-Bohr:2020syg, Bjerrum-Bohr:2019kec, Cristofoli:2019neg,Bjerrum-Bohr:2016hpa,Chen:2024mmm,Alessio:2022kwv,Alessio:2023kgf, ashoke1, Gonzo:2024zxo,Aoude:2023vdk,Aoude:2023dui,Adamo:2024oxy,Adamo:2021rfq,Georgoudis:2023eke,Georgoudis:2024pdz,Akpinar:2024meg,Bini:2024rsy, DeAngelis:2023lvf, Cristofoli:2021vyo,Mougiakakos:2022sic,Bhattacharyya:2025ghl} and post-Minkowskian effective theories \cite{Kalin:2020mvi, Kalin:2020lmz,Dlapa:2024cje,Dlapa:2023hsl,Dlapa:2025biy} have developed this approach extensively. In recent times, worldline quantum field theory (WQFT) has made this connection especially direct. In this framework,  the classical quantities such as the impulse and emitted radiation are obtained from expectation values of worldline
and field observables, evaluated through Feynman diagrams \cite{Mogull:2020sak,Jakobsen:2021lvp,Jakobsen:2023ndj,Jakobsen:2023hig,Jakobsen:2022psy,Wang:2022ntx, Driesse:2024feo,Bhattacharyya:2024kxj}. The calculation of classical observables, therefore, leads naturally to the computation of several specialized families of loop integrals.

This program has now advanced to the point where the main challenges one often encounters are computing integrals or understanding their cut structures to devise efficient computational methods, rather than generating these Feynman diagrams. For GR, at 4PM, conservative dynamics require computations of elliptic integrals which appear at the 3-loop \cite{Dlapa:2023hsl,Jakobsen:2023ndj}. Furthermore, at 5PM order, the topologies of the four loop integrals can be Calabi-Yau geometry \cite{Driesse:2024feo,Driesse:2024xad} and recently at 5PM-2SF (second order in self-force expansion) order \cite{Driesse:2026qiz}, several of the master integrals in terms of which these Feynman integrals can be expanded contain iterated integrals with K3 periods \cite{Frellesvig:2023bbf,Klemm:2024wtd}\footnote{See \cite{Duhr:2023eld,Duhr:2025lbz,Duhr:2025ppd} for formal developments on CY geometry in the context of Feynman integral.}. This points to a broader fact that increasing precision not only changes the number of integrals to be computed but also demands the invocation of a new class of functions to write the final answer. So one of the main challenges is to identify the space of  functions, and our present study draws motivation from that.

Particle physics encountered similar questions much earlier. Heavy Quark Effective Theory (HQET), for example, separates the large mass of a heavy quark from its softer residual motion. At leading order, the heavy quarks move approximately with constant velocity \cite{Eichten:1989zv, Georgi:1990um,Grozin:2006xm}. This makes their propagators essentially depend linearly on the momenta. Two and three-loop studies of HQET developed efficient methods for calculating physical quantities using this effective description \cite{Broadhurst:1994se,Grozin:2000jv}. When an additional massive flavor is retained, this method combines this linear propagator with the usual massive propagators. The resulting integrals combine velocity-dependent denominators with ordinary massive propagators.  Interested readers are referred to \cite{Grozin:2007ap} for massive three-loop HQET master integrals. However, note that the WQFT integrals considered here are not identical to their HQET counterparts. The fixed velocities enter into the integrand through worldline delta function constraints rather than static propagators. Nevertheless, the comparison may be useful. In both cases, the kinematic velocity factor gives rise to a new class of Feynman integral families, while the additional (internal) masses introduce the scales and branch points that require appropriate analytic handling.

There are also independent physical reasons to study massive exchanges in gravitational scattering. Additional scalar or pseudoscalar fields can mediate finite-range interactions \cite{Kuntz:2019zef, Huang:2018pbu,Zhang:2021mks,Bhattacharyya:2023kbh,Damour:2025oys}. Such an extra massive degree of freedom may modify the  scattering data and open up additional radiative modes. Earlier WQFT calculations investigated the impulse in massive scalar-tensor theory \cite{Bhattacharyya:2024aeq}, and subsequent work has made further progress on the associated integrals and scattering observables at 2PM order \cite{Ghosh:2026gev,Biendarra:2026nig}.\footnote{Related compactified integral representations also appeared in the thesis \cite{Ghosh:2026gev}. See  \cite{Biendarra:2026nig} for a recent numerical analysis of the scattering angle due to the unavailability of the closed-form expression for some of the integrals, and also some interesting implications. In the presence of scalar self-interactions, PM counting depends on the scaling assigned to the additional couplings, rather than on loop order alone.} Generically, massive QFT in position space (in our context, in the impact parameter space) has difficulty because the Fourier transform of a massive propagator to impact-parameter space produces a modified Bessel function. In the context of WQFT, at higher orders, these special functions appear inside further integrations, together with the worldline constraints and multiloop kernels. Therefore, the problem with mass is not that it will just change the propagator, but the introduction of mass, in general, in the Feynman integral, generates structural difficulty as it introduces a new scale in the theory.
Modern methods provide several complementary ways to compute integrals appearing in these problems.  Integration by parts (IBP) allows the many integrals associated with a Feynman diagram to be expressed in terms of a small set of basic integrals, called master integrals \cite{Chetyrkin:1981qh, Kotikov:1990kg}. Subsequently, the Laporta algorithm has systematized this idea by turning it into an efficient computer algebra procedure \cite{Laporta:2000dsw}.  Differential equation (in terms of masses and kinematic invariants provided) methods provide an efficient route to compute this finite set of master integrals \cite{Kotikov:1990kg, Kotikov:1991pm, Remiddi:1997ny, Gehrmann:1999as,Laporta:2000dsw, Henn:2013pwa}. Yet a reduction algorithm does not, by itself, tell us which functions will solve these differential equations. The equal-mass two-loop sunrise diagram provides an interesting example.  Its homogeneous solutions are elliptic periods rather than ordinary logarithms or multiple polylogarithms \cite{Remiddi:1999ew}. The appearance of elliptic geometries (first appeared in Feynman integral literature in \cite{Sabry:1962}, following in many contexts of particle physics \cite{Broadhurst:1990,Bauberger:1994, Caffo:1998du,Laporta:2004rb,MullerStach:2011ru,Adams:2013kgc,Remiddi:2013joa,Bloch:2013tra,Adams:2014vja,Adams:2015gva,Adams:2015ydq,Remiddi:2016gno,Adams:2016xah,Primo:2016ebd,Primo:2017ipr,vonManteuffel:2017hms,Adams:2017ejb,Broedel:2018iwv}) and more general algebraic geometries while computing these master integrals is now recognized in precision particle physics and GR \cite{Pogel:2022ken,Pogel:2022vat,Pogel:2022yat,Duhr:2022dxb,Frellesvig:2023bbf,Klemm:2024wtd,Duhr:2023eld,Forner:2024ojj,Duhr:2024uid,Driesse:2024feo,Driesse:2024xad,Duhr:2025kkq,Pogel:2025bca,Duhr:2025lbz,Duhr:2025ppd,Maggio:2025jel,Driesse:2026qiz}.

Over the past two decades, numerous unexpected connections between scattering amplitudes and geometry have been uncovered, starting from \cite{Witten:2003nn,Arkani-Hamed:2012zlh,Arkani-Hamed:2013jha,Atiyah:2017erd}. A useful way to make the connection with the underlying geometry is to study maximal cuts of the given integral. Maximal cuts yield homogeneous solutions to the system of differential equations used to solve the full uncut master integrals. Following the seminal work by Mizera \cite{Mizera:2017} (based on the connection observed in the context of Picard–Lefschetz theory \cite{Pham:1965}) and further developed in \cite{Mastrolia:2018uzb,Frellesvig:2019kgj,Frellesvig:2019uqt,Mizera:2019vvs,CaronHuot:2021,CaronHuot:2021xqj}, one can use the intersection theory to study the geometric interpretation of this integral reduction (IBP identities). They showed that the relations among different Feynman integrals within a given family can be encoded in terms of the intersection numbers of differential forms.  This viewpoint connects the choice of master integrals to the independent integration contours and the singularities of the integrand, and helps one completely bypass the need to solve a large number of linear algebraic systems. This paper provides a systematic way to investigate both the number of independent cut integrals and the differential equations they satisfy for massive integral families.

Last but not least, this geometric analysis is also motivated by the Landau bootstrap program \cite{Hannesdottir:2024hke, Bargiela:2026lje}. The central idea is to use the singularities and local behaviour of an integral to constrain its expression within an appropriate space of functions. The cut geometry, monodromy, and period pairings studied
here provide natural candidate data for such an approach. We do not perform a bootstrap on the full massive integral family that we study here; rather, this perspective motivates the usefulness of determining their homogeneous analytic structure for future calculations.

In this work, we study a representative two-loop WQFT family with four equal-mass propagators and two classical worldline constraints in the form of two Dirac delta functions. We impose the maximal cut, which places the four quadratic propagators on shell. Using the global Baikov representation \cite{Baikov:1996iu}, after performing two successive integrations, reduces the maximal cut to a single integral over one variable. In four dimensions, the remaining square root expression, containing a fourth-degree polynomial, inside the integrand defines a smooth quartic algebraic curve of genus one. \textcolor{black}{Elliptic geometry at two loops is, of course, familiar from the equal-mass sunrise~\cite{Sabry:1962,Broadhurst:1990}, where one usually studies the two-point function. Here in our analysis, the difference is the four-point kinematics with the classical worldline.} We next identify its branch points, corresponding periods, and associated Picard-Lefschetz monodromy to study some of the subtleties arising at the degeneration of the genus-one surface. In the massless limit, or in the low-velocity (static) limit, one of the two cycles of the torus gets pinched, giving a kinematic genus drop. But note that this simplification applies only to the maximal cut. It does not establish that the full integral can be expressed in terms of polylogarithms in these limits.

The same geometric connection reappears when one tries to solve this using the differential equation method. We further organize the five master integrals (after the cut) required for our case into smaller sets that are easier to solve. We show that their elliptic solutions match those obtained using the Baikov method. We also use the intersection theory to find relations between the cut integrals and derive their differential equations \cite{Mizera:2017,Mastrolia:2018uzb}. At an arbitrary dimension, the maximal cut has five independent contour integrals, which can be expressed as Euler integrals or Lauricella functions. In four dimensions, only two independent elliptic periods remain. We express the three master integrals in the elliptic sector in terms of elliptic polylogarithms up to order \(\epsilon^2\), where the dimensional regularization parameter \(\epsilon=(4-d)/2\). \textcolor{black}{In impact-parameter space, we meet the same family of curves at a different value of its modulus and an algebraic twist. The algebraic Baikov weight is replaced by a modified Bessel function, giving a Bessel-weighted elliptic period. The Bessel curve and topological recursion
are already known \cite{DoNorbury2018}.}

Our contribution in this paper is to connect these structures to the massive WQFT integrals studied here. This work applies some of the geometric techniques discussed above to a massive two-loop WQFT integral family. We identify its elliptic maximal cut, derive the associated homogeneous solutions and relations among cut integrals, and find a Bessel-weighted elliptic structure in impact-parameter space. These results concern the maximal cut and homogeneous equations. Evaluation of the full integrals requires lower sectors, boundary conditions, and inhomogeneous terms. Note that our goal in this paper is not to perform any theory-specific observable calculation, but rather to understand the underlying geometric structure and identify the function space of these integrals. The analysis presented here thus naturally provides a starting point for future calculations of integrals that may appear in the GW observables for theories involving massive fields.

The paper is organized as follows. Sec.~(\ref{sec:twoloop}) discusses the integral family and their elliptic maximal cut.  Sec.~(\ref{sec:cut-de}) derives the differential equation for the cut and its Picard-Fuchs factorization, and Sec.~(\ref{sec:intersec}) redoes the reduction with intersection theory in two cocycle bases and studies the global monodromy. Sec.~(\ref{sec:IBP}) compares the two reductions. Sec.~(\ref{sec:Periodmatr}) constructs the period matrices and the twisted Riemann period relations. Sec.~(\ref{sec:epsilonexpnad}) and (\ref{sec:diffcannon}) give the $\epsilon$-expansion and the canonicalization. Sec.~(\ref{sec9}) examines the Bessel-weighted Fourier cut and its quotient geometry. Sec.~(\ref{sec10}) summarizes and ends with future directions. Appendices~(\ref{app:intersection-entries})-(\ref{app:maximal-cut}) collect the intersection-matrix
entries, the impact-parameter reduction, and the momentum-space derivation of the same elliptic curve. 

\subsection*{A reader's guide:} At the first read, it might be confusing for the reader, as we use different charts to represent the same elliptic curve appearing in different contexts of the article. We state at the outset which object lives where. One elliptic curve runs through the whole paper in different charts. In $u$ it is the quartic \eqref{eq:cut-elliptic-curve} produced directly by the Baikov fibre integrations. Rescaling to $z=u/\sqrt\mu$ puts it in the form $y^{2}=(1-z^{2})(z^{2}+\tau)$ used for the differential equation and the $\epsilon$-expansion. Quotienting by $z\to-z$ gives the cubic $x$-chart used for the period relations. In impact-parameter space, we use a representation (which allows us to do the `$q$' integration more easily) of the same class of curve as $y^{2}=(1-z^{2})(z^{2}-k'^{2})$, whose extra involutions are the subject of Sec.~(\ref{sec9}). Note that in the Fourier integral, the elliptic modulus $\sim$ $k'(\gamma)$ is only a function of the kinematic variable $\gamma$, in contrast with the two-loop elliptic modulus $\sim$ $\tau(s,\gamma)$, which is a function of both $\gamma$ and $-q^2$.

\textcolor{black}{The letter $x$ is used for four unrelated variables, each confined to
one place: the velocity variable $x=\gamma-\sqrt{\gamma^{2}-1}$ of \eqref{eq:new7},
used only in Sec.~(\ref{sec:cut-de}); the local coordinate near a node in
Sec.~(\ref{new:sec2.2.3}); the quotient coordinate $x=z^{2}$ of Sec.~(\ref{Sec:6.4}) and Sec.~(\ref{sec9}); and the Legendre coordinate $x=(1-z^{2})/k^{2}$ of Appendix~(\ref{app:maximal-cut}). From Sec.~(\ref{sec:intersec}) onwards the kinematic variable is always $\tau$, never $x$.}

\section{The two-loop WQFT family and  elliptic curve}
\label{sec:twoloop}

\textcolor{black}{In this paper, we will study the following family of massive two-loop integrals. Until Section~(\ref{sec9}), we will hold the momentum transfer fixed and analyze its maximal cut. Two successive integrations will reduce this cut to a one-dimensional period, whose four-dimensional geometry will turn out to be elliptic.}
\begin{align}
\begin{split}
J_{a_1,\ldots,a_4}
&=
\vcenter{\hbox{
  \includegraphics[width=3.0cm]{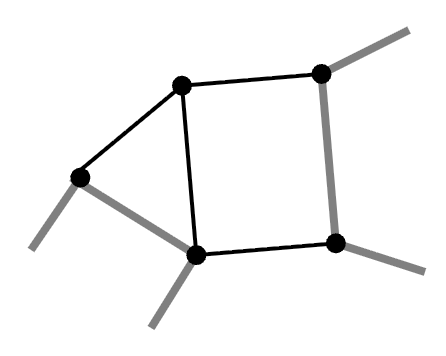}
}}
\;=\;
\int_q e^{iq\cdot b}\,
\hat{\delta}(q\!\cdot\! v_1)\,
\hat{\delta}(q\!\cdot\! v_2)
\int_{\ell_1,\ell_2}
\frac{
\hat{\delta}(\ell_1\!\cdot\! v_1)\,
\hat{\delta}(\ell_2\!\cdot\! v_2)
}{
D_1^{a_1}D_2^{a_2}D_3^{a_3}D_4^{a_4}
}\,.
\end{split}
\label{3.1w}
\end{align}
with inverse propagators
\begin{align}
D_1&=\ell_1^2-m^2\,,&
D_2&=\ell_2^2-m^2\,,&
D_3&=(q-\ell_1)^2-m^2\,,&
D_4&=(\ell_1+\ell_2)^2-m^2\,.
\end{align}
The kinematic variables are chosen such that
\begin{align}
    \begin{split}
        v_1^2=1\,,\,v_2^2=1\,,\,v_1\cdot v_2=\gamma=\frac{1+x^2}{2x}\,.\label{eq:new7}
    \end{split} 
\end{align}
\textcolor{black}{We emphasize at the outset that \eqref{3.1w} is a \emph{representative} family. It is not, as it stands, a term in any particular observable at a specific post-Minkowskian order.  It is chosen because it provides us with the simplest two-loop worldline topology, in which there are two features that we wish to study here i.e all four quadratic propagators carry the mediator mass $m$, so that the integral has a genuine internal scale and the associated branch points, and both the worldline constraints $\hat\delta(\ell_j\!\cdot\!v_j)$ are present, so that the velocities enter linearly.  The geometry we find is therefore expected to be generic to massive two-loop WQFT families rather than specific to this one, and the methods developed in this paper will apply to those members of the class that appear in some observables.}

\textcolor{black}{It is convenient to separate the two-loop integration from the Fourier transform to impact-parameter space,
\begin{equation}
\mathcal F_{a_1\ldots a_4}(q)=\int_{\ell_1,\ell_2}
\frac{\hat\delta(\ell_1\!\cdot\! v_1)\,\hat\delta(\ell_2\!\cdot\! v_2)}
{D_1^{a_1}D_2^{a_2}D_3^{a_3}D_4^{a_4}}\,,
\qquad
J_{a_1\ldots a_4}=\int_q e^{\ii q\cdot b}\,
\hat\delta(q\!\cdot\! v_1)\,\hat\delta(q\!\cdot\! v_2)\,
\mathcal F_{a_1\ldots a_4}(q)\,.
\end{equation}
We reserve $J$ for the full impact-parameter integral \eqref{3.1w} and $\mathcal F$
for the fixed-$q$ two-loop factor; everything from here to
Sec.~(\ref{sec:diffcannon}) concerns $\mathcal F$. The Fourier transform is
reinstated only in Sec.~(\ref{sec9}) and
Appendix~(\ref{app:twoloop})--(\ref{app:maximal-cut}). The distinction also controls
the scaling. With the loop measure $\int_{\ell}\equiv\int d^{d}\ell/\pi^{d/2}$ used
throughout, dimensional analysis gives, in terms of the two dimensionless ratios
$\bar\rho=s/\mu$ and $\gamma$,
\begin{equation}
\mathcal F_{\{a_i\}}(s,\mu,\gamma,\epsilon)
=\mu^{\,d-1-\sum_i a_i}\;
\widetilde{\mathcal F}_{\{a_i\}}(\bar\rho,\gamma,\epsilon)\,,
\label{eq:new8}\end{equation}
whereas $J$ in \eqref{3.1w} carries the further dimensionless variable $m|b|$. That
is why the impact-parameter geometry discussed in the later  Sec.~(\ref{sec9}) is a different twisted problem on the same family of curves, and not the fixed-$q$ geometry studied
here.}

After the IBP reduction, we focus on the sector in which all $a_i=1$, which contains five master integrals.
{\color{black}
We first study the maximal cut of the \(\ell_1,\ell_2\) integral.
In the subsequent Baikov calculation, we use  \(\hat\delta(a)=2\pi\delta(a)\), and
normalize each propagator cut as \((2\pi i)^{-1}\oint\).
Thus
\begin{align}
 \MaxCut(\mathcal{F}_{1})
 =
 \int_{\Gamma_{\textrm{cut}}\subset \mathbb{C}^{2d}}\frac{d^d\ell_1\,d^d\ell_2}{\pi^d}\,
 \frac{\hat\delta(\ell_1\cdot v_1)\hat\delta(\ell_2\cdot v_2)}
 {D_1D_2D_3D_4}\,. \label{eq:maxcut}
\end{align}
Note that the $\Gamma_{\textrm{cut}}$ is an analytically continued contour integral that encircles all poles of $D_1=\cdots=D_4=0$, and its orientation and branches can be explicitly fixed by imposing the Gram constraint that we will discuss shortly. \textcolor{black}{One may think of the maximal cut as localising all four propagators at once. With the normalisation $(2\pi i)^{-1}\oint$ used here, each $1/D_i$ is replaced by its unit residue, i.e.\ effectively by $\delta(D_i)$. The familiar unitary Cutkosky rule $1/(D_i+i0)\to-2\pi i\,\delta^{(+)}(D_i)$ differs from this only by an extra positive energy constraint.}
It is worth noting that, unlike the standard unitarity cut, it is not a product of real, positive energy on-shell delta functions. 
\textcolor{black}{The maximal cut is the homogeneous solution for the
differential equations of the full uncut integral}. As we will see, the study of maximal cuts gives enormous information about the integral family that will help us to concretely investigate the full uncut integrals in the future. 
\subsection{Global Baikov representation and the maximal cut}
The study of maximal cuts can be pursued through several complementary techniques, among which the Baikov representation provides a particularly powerful framework for establishing connections with methods from algebraic geometry. We begin by presenting the maximal cut construction for a general $(L)$-loop integral in Baikov variable, pointing out the worldline constraints. We then specialize the discussion to the four-propagator family mentioned in (\ref{3.1w}). This general formulation also fixes the integration measure and the exponent of the Baikov polynomial before any left-out fiber integration is performed.

The Baikov representation admits two main formulations: the global Baikov representation and the loop-by-loop Baikov representation. For convenience, we use the global Baikov representation to compute the maximal cut.
\subsubsection{General Baikov construction} \label{newsec:1}
Now, we study maximal cuts using the global Baikov representation. \textcolor{black}{We define the maximal cut by contour integration in (analytically continued) complexified
loop-momentum space. Each cut propagator is replaced by a residue around
its pole, and the remaining variables are integrated over a specified
cycle with a chosen branch of the Baikov weight
\cite{Kosower:2011ty,Mastrolia:2018uzb,Frellesvig:2017aai,Bosma:2017ens} with a suitable Gram constraint imposed.
Naturally, the construction applies when the simultaneous on-shell equations do not have a real momentum solution.}

Before turning to the details, let us first recall a few basic aspects of this representation, adapted to the type of Feynman integrals that will appear in this paper. Consider an $L$-loop integral of the form
\begin{align}
\mathcal{F}_{\beta_1,\ldots,\beta_N}
=
\textcolor{black}{\int \prod_{j=1}^{L}\frac{d^d\ell_j}{\pi^{d/2}}}\,
\prod_{j=1}^{L}\hat{\delta}\!\left(\ell_j\cdot v_{a_j}\right)
\prod_{m=1}^{N}\frac{1}{D_m^{\beta_m}} \, .
\end{align}
Here, the $D_m$ denote inverse propagators, while the delta functions impose the classical on-shell constraints associated with the massive external particles. The Baikov representation is obtained by changing variables from the loop momenta to the independent scalar products formed out of the loop momenta and the independent external momenta. Let
\begin{align}
p_j \in \{\ell_1,\ell_2,\ldots,\ell_L,k_1,\ldots,k_E\}
\end{align}
where $k_1,\ldots,k_E$ are independent external momenta. The independent scalar products involving at least one loop momentum are $\ell_i\cdot k_a,\,\,\,
\ell_i\cdot \ell_j$, 
with $i,j=1,\ldots,L$ and $a=1,\ldots,E\,.$ Their total number is $M = LE + \frac{L(L+1)}{2}.$

In the original integral, however, we may have only $N$ propagator denominators together with $L$ delta-function constraints. In general,
$N+L \le M\,.$ Therefore, in order to perform the Baikov change of variables, one completes the set of variables by introducing
$M-N-L$
additional irreducible scalar products, or equivalently, auxiliary denominators. Thus, one obtains a complete set of Baikov variables: 
$z_\alpha\,,\,\, \alpha=1,\ldots,M$
which may be chosen as
\begin{align}
z_\alpha
=
\{D_1,\ldots,D_N,\rho_1,\ldots,\rho_L,x_1,\ldots,x_{M-N-L}\},
\end{align}
where
$\rho_j = \ell_j\cdot v_{a_j}$
are the variables constrained by the delta functions, and the $x_i$ are the remaining irreducible scalar products. After this change of variables, the integral takes the Baikov form \cite{Baikov:1996iu},
\begin{align}
\mathcal{F}_{\beta_1,\ldots,\beta_N}
=
  \textcolor{black}{\mathcal{N}_{\rm B}}
\int_{\Gamma}
\mathcal{B}(z)^{\nu_B}
\,
\frac{
\bigwedge_{\alpha=1}^{M} dz_\alpha
}{
\prod_{m=1}^{N} z_m^{\beta_m}
}
\prod_{j=1}^{L}
\hat{\delta}\!\left(z_{N+j}\right)\,,
\qquad
\nu_B=\frac{d-E-L-1}{2}\,.
\label{3.11g}
\end{align}
Here \(\mathcal B(z)\) is the Baikov polynomial, and the normalization is
\begin{align}
    \begin{split} \label{new1}
\mathbfcal{B}=\frac{\det\,\mathbf{G}_{\ell_1,\cdots,\ell_L,k_1,\cdots k_E}}{\det\,\mathbf{G}_{k_1,\cdots k_E}}\,,\qquad   \textcolor{black}{\mathcal{N}_{\rm B}}=\frac{1}{J_z}\frac{\pi^\frac{L-M}{2}\left(\det\,\boldsymbol{G}_{k_1,k_2\cdots k_E}\right)^{-L/2}}{\prod_{i=1}^L\Gamma\left(\frac{d-L-E+i}{2}\right)}
    \end{split}
\end{align}
where \(  \textcolor{black}{\mathcal{N}_{\rm B}}\) is the overall Jacobian factor and $J_z=\left|\det(\partial z_\alpha/\partial s_\beta)\right|$ is the constant Jacobian from the scalar-product space $\{s_\beta\}$ to the Baikov variable $z_{\alpha}$. The integration domain is obtained by imposing the positivity conditions
\begin{align}
    \Gamma_i:=
    \left\{
    \frac{\det\,\mathbf{G}_{\ell_i,\ldots,\ell_L,k_1,\ldots,k_E}}
    {\det\,\mathbf{G}_{\ell_{i+1},\ldots,\ell_L,k_1,\ldots,k_E}}
    >0
    \right\},
    \qquad
    \Gamma=\Gamma_1\cap\Gamma_2\cap\cdots\cap\Gamma_L \,.\label{2.14hhhh}
\end{align}

In this form, the propagators, classical delta-function constraints, and irreducible scalar products are all treated as coordinates on the same Baikov space. This is particularly useful for studying maximal cuts: on the maximal cut, the propagator variables are localized at $D_m=0$, and the remaining integral is naturally described in terms of the residual Baikov variables and the Baikov polynomial. The maximal cut corresponds to the integration contour:
\begin{align}
    \circlearrowleft_1\wedge \circlearrowleft_2\cdots\wedge\circlearrowleft_N\wedge\mathcal{C},\,\, 
    \qquad\circlearrowleft_a:=\frac{1}{2\pi i}\{|z_a|=\delta\}
\end{align}
where \(\mathcal C=\Gamma\cap\{z_1=\delta\}\cap\cdots\cap\{z_N=\delta\}\). \textcolor{black}{Here, $\delta$ is an infinitesimal radius, not to be confused with the dimensional regulator $\epsilon$ used later on in the paper}. Evaluating \eqref{3.11g} on this contour, with the delta-function variables also localized, gives the maximal cut in Baikov form,
\begin{align}
    \MaxCut\!\left(\mathcal{F}_{\beta_1,\ldots,\beta_N}\right)
    :=
  \textcolor{black}{\mathcal{N}_{\rm B}}\int_\mathcal{C}
    \mathbfcal{B}(\boldsymbol z)^{\nu_B}
    \varphi(\boldsymbol z)\,,\quad \boldsymbol{z}:=(z_{N+1},\cdots,z_{M})
\end{align}
Here \(\mathbfcal B(\boldsymbol z)=\mathbfcal B(z)|_{z_1=\cdots=z_N=0}\), and the residual \((M-N)\)-form is
\begin{align}
   \varphi(\boldsymbol z)
   =
   \mathcal B(\boldsymbol z)^{-\nu_B}
   \oint_{\circlearrowleft_1\wedge\cdots\wedge\circlearrowleft_N}
   \mathcal B(z)^{\nu_B}
 \bigwedge_{\alpha=1}^M \frac{dz_\alpha}{z_\alpha^{\beta_\alpha}}\,,
\end{align}
and  $\varphi(\boldsymbol{z})$ can be thought as the simplest \emph{Grothendieck residue}\footnote{Readers are encouraged to see the excellent textbook \cite{Weinzierl:2022eaz} on Feynman integrals by Stefan Weinzierl for many further in-depth details.} in the multivariate complex variable,
\begin{align}
    \begin{split}
         \varphi(\boldsymbol z)= \mathcal B(\boldsymbol z)^{-\nu_B}
\underset{z_1=\cdots=z_N=0}{\operatorname{Res}}
\left[
\mathcal{B}(z)^{\nu_B}
\bigwedge_{\alpha=1}^{M}
\frac{dz_\alpha}{z_\alpha^{\beta_\alpha}}
\right].
    \end{split}
\end{align}
\subsubsection{Global variables and cut polynomial}
We now return to the family of integrals defined in \eqref{3.1w} from the perspective of the \emph{global Baikov representation}. Since the problem is two-loop with three independent external vectors, the total number of independent scalar products involving the loop momenta is
$\textcolor{black}{M=\frac{L(L+1)}{2}+LE=9}.$
Among these, six are identified with propagator-type variables (including the linear cut conditions coming from the delta functions), while the remaining three serve as auxiliary irreducible scalar products. We choose the Baikov coordinates
\begin{align}
\begin{alignedat}{2}
z_1&=\ell_1^2-m^2\,,
&\qquad
z_2&=\ell_2^2-m^2\,,
\\
z_3&=(\ell_1-q)^2-m^2\,,
&\qquad
z_4&=(\ell_1+\ell_2)^2-m^2\,,
\\
z_5&=\ell_1\cdot v_1\,,
&\qquad
z_6&=\ell_2\cdot v_2
\end{alignedat}
\end{align}
together with $u=\ell_1\cdot v_2\,,
r=\ell_2\cdot v_1\,,
w=\ell_2\cdot q\,.$
The corresponding Baikov polynomial is the ratio of Gram determinants
\begin{align}
\mathcal B(\mathbf z)
=
\frac{\det\{\ell_1,\ell_2,v_1,v_2,q\}}{\det\{v_1,v_2,q\}}\,.
\end{align}
Using the conventions, $\mu=m^2,\, s=-q^2,$ together with,
$v_1^2=v_2^2=1,\,v_1\!\cdot v_2=\gamma,\, q\!\cdot v_1=q\!\cdot v_2=0\,,$ the external Gram determinant is
\begin{align}
\det\{v_1,v_2,q\}
&=
\det
\begin{pmatrix}
1 & \gamma & 0 \\
\gamma & 1 & 0 \\
0 & 0 & -s
\end{pmatrix}
=
s(\gamma^2-1)\,.
\end{align}
The complete polynomial follows from the compact Gram representation
\begin{align}
 \det G_{\ell_1,\ell_2,v_1,v_2,q}
 =\det\begin{pmatrix}
 \mu+z_1&\frac{z_4-z_1-z_2-\mu}{2}&z_5&u&\frac{z_1-z_3-s}{2}\\
 \frac{z_4-z_1-z_2-\mu}{2}&\mu+z_2&r&z_6&w\\
 z_5&r&1&\gamma&0\\
 u&z_6&\gamma&1&0\\
 \frac{z_1-z_3-s}{2}&w&0&0&-s
 \end{pmatrix}.
\end{align}
For scalar products ordered as
\begin{equation}
(\ell_1^2,\ell_2^2,\ell_1\cdot\ell_2,\ell_1\cdot q,
   \ell_1\cdot v_1,\ell_2\cdot v_2,u,r,w)\,, \label{eq:new11}
   \end{equation}
the transformation to \((z_1,\ldots,z_6,u,r,w)\) has \(J_z=4\).
Further, on the maximal cut, we impose
\begin{align}
z_1=z_2=z_3=z_4=z_5=z_6=0\,,
\end{align}
so that the residual geometry is parameterized by the affine coordinates \((u,r,w)\). In this chart, the natural top differential form is simply
\begin{align}
\varphi=du\wedge dr\wedge dw\,.
\end{align}
\textcolor{black}{Now we collect all the conventions that were fixed earlier. The measure is $d^d\ell/\pi^{d/2}$ per loop, the two $\hat\delta$'s each supply $2\pi$ and $J_z=4$. With
$L=2$, $E=3$, $M=9$ this gives
$\pi^{(L-M)/2}(2\pi)^{L}/J_z=\pi^{-7/2}4\pi^{2}/4=\pi^{-3/2}$, so that the \textcolor{black}{integral in (\ref{eq:maxcut})   becomes},
\begin{equation}
\mathrm{MaxCut}(\mathcal{F}_1)=\mathcal{N}_{\rm B}\,[s(\gamma^{2}-1)]^{\epsilon}
\int_{\mathcal{C}}\mathcal{P}_{\rm cut}^{-1-\epsilon}\,du\wedge dr\wedge dw\,,
\qquad
\mathcal{N}_{\rm B}=\frac{1}{\pi^{3/2}\,\Gamma(-\epsilon)\Gamma\!\big(\tfrac12-\epsilon\big)}\,.
\end{equation}}


Here $\mathcal{C}$ is a suitable contour in the complex plane such that the Gram positivity condition in \eqref{2.14hhhh} is maintained, and therefore the normalized Baikov polynomial is $\mathcal{B}_{\textrm{cut}}=\mathcal P_{\textrm{cut}}/(s(\gamma^2-1))$
\begin{align}
\begin{split}
\mathcal P_{\mathrm{cut}}(u,r,w)
= \det G_{\ell_1,\ell_2,v_1,v_2,q}|_{\rm cut}=
\frac{1}{4}\Big[
&r^2 s\left(4\mu+s-4u^2\right)
-4\gamma r s u(\mu+w)
+(\gamma^2-1)\mu s^2
\\
&
+\mu s\left((\gamma^2-1)(3\mu-2w)+4u^2\right)
+4w^2\left((\gamma^2-1)\mu+u^2\right)
\Big].\label{2.25jjjjj}
\end{split}
\end{align}
In general, the maximal-cut contour need not be restricted to a particular region in
which the successive Gram conditions are satisfied. If, however, we choose
a contour associated with the real physical kinematic region, the successive
Gram determinants must satisfy the corresponding positivity conditions.
For the present two-loop problem, these conditions are
\begin{align}
    \Gamma_2
    &=
    \frac{
        \det \mathbf{G}_{\{\ell_2,v_1,v_2,q\}}
    }{
        \det \mathbf{G}_{\{v_1,v_2,q\}}
    }
    >0,
    \qquad
    \Gamma_1
    =
    \frac{
        \det \mathbf{G}_{\{\ell_1,\ell_2,v_1,v_2,q\}}
    }{
        \det \mathbf{G}_{\{\ell_2,v_1,v_2,q\}}
    }
    >0 .
    \label{eq:gram}
\end{align}
More generally, the integration domain $\mathcal C$ may decompose into several admissible subregions that satisfy the Gram positivity condition. Different choices may define independent integration domains, although some of them can be related by homological relations. In the physical kinematic regime, there are five real subregions as shown in Fig.~\eqref{Fig1.1} (All the domains may not generate new independent functions):
\begin{align}
\begin{split}
&\mathcal C_{\mathrm{I}}:\quad
|u|<\frac{\sqrt{3\mu}}{2},
\qquad
r\in\mathbb R,
\qquad
w\in\mathbb R,
\\&
\mathcal C_{\mathrm{II}}:\quad
\frac{\sqrt{3\mu}}{2}<|u|<\sqrt{\mu},
\qquad
r_-(u)<r<r_+(u),
\qquad
\left(
w<w_-(r,u)
\ \text{or}\
w>w_+(r,u)
\right),
\\ &
\mathcal C_{\mathrm{III}}:\quad
 \frac{\sqrt{3\mu}}{2}<|u|<\sqrt{\mu},
\qquad
\left(
r<r_-
\ \text{or}\
r>r_+
\right),
\qquad
w\in\mathbb R,
\\ &
\mathcal C_{\mathrm{IV}}:\quad
 |u|>\sqrt{\mu},
\qquad
r_-<r<r_+,
\qquad
w\in\mathbb R,
\\ &
\mathcal C_{\mathrm{V}}:\quad
 |u|>\sqrt{\mu},
\qquad
\left(
r<r_-
\ \text{or}\
r>r_+
\right),
\qquad
\left(
w<w_-
\ \text{or}\
w>w_+
\right).
\end{split}
\label{eq:cut-regions}
\end{align}
\begin{figure}
    \centering
    \includegraphics[width=0.5\linewidth]{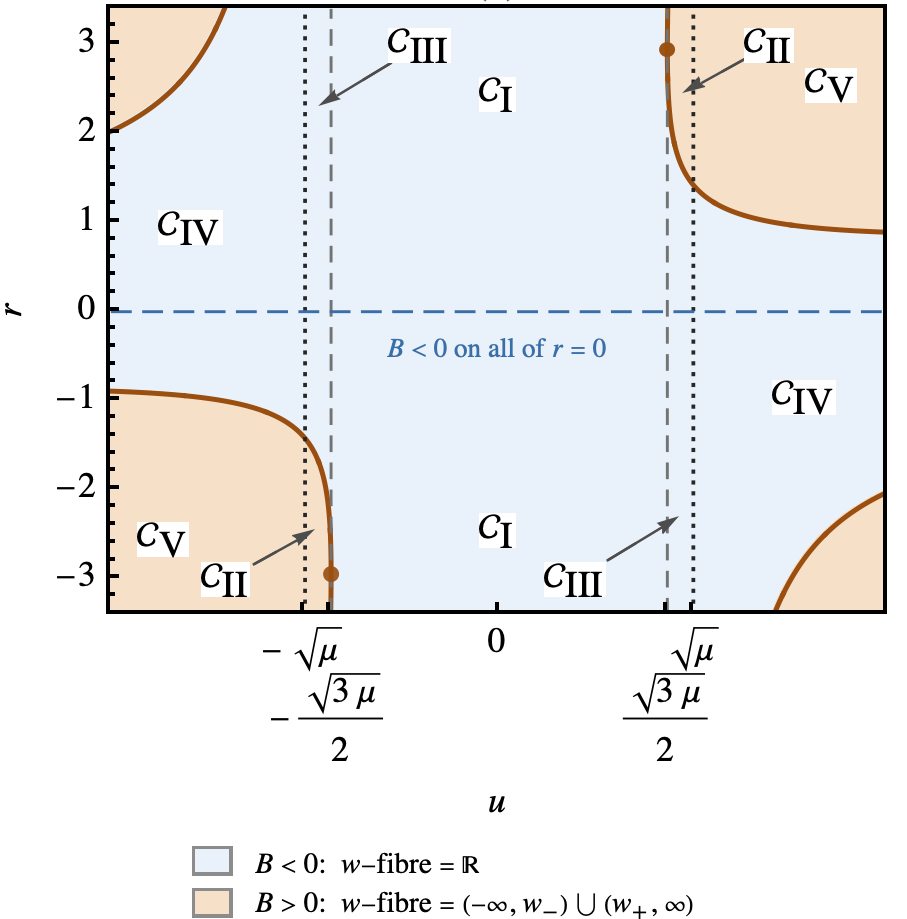}
    \caption{\textcolor{black}{The real chambers $\mathcal C_{\rm I},\dots,\mathcal C_{\rm V}$ of \eqref{eq:cut-regions} over the $(u,r)$ plane, for representative physical kinematics. Blue: $B(r,u)<0$, where $\mathcal P_{\rm cut}>0$ for every real $w$. Orange: $B(r,u)>0$, where $\mathcal P_{\rm cut}>0$ only for $w\notin[w_-,w_+]$. The solid curve is $B=0$; the vertical lines are $|u|=\sqrt{3\mu}/2$ and
$|u|=\sqrt\mu$; the dashed line $r=0$ runs through $\mathcal C_{\rm I}$, $\mathcal
C_{\rm III}$ and $\mathcal C_{\rm IV}$.}}    \label{Fig1.1}
\end{figure}
\textcolor{black}{Fig.~(\ref{Fig1.1}) shows the five chambers over the $(u,r)$ plane.}
On the maximal cut, the first successive Gram ratio can be evaluated
explicitly as
\begin{align}
    \Gamma_2=
    \ell_{2\perp}^{\,2}
    =
    \mu
    +\frac{r^2}{\gamma^2-1}
    +\frac{w^2}{s}.
    \label{eq:Gamma2}
\end{align}
Therefore, in the physical kinematic regime $ \mu>0,\,
    s>0,\,
    \gamma^2>1,$ and for $r,w\in\mathbb R$, one immediately finds
\begin{align}
    \Gamma_2
    =
    \mu
    +\frac{r^2}{\gamma^2-1}
    +\frac{w^2}{s}
    >0\,.
\end{align}
Thus, on the real contour, the first successive Gram condition is
automatically satisfied. Similarly, the second successive Gram ratio is
\begin{align}
    \Gamma_1=
    \frac{
        \mathcal P_{\mathrm{cut}}
    }{
        s(\gamma^2-1)
        \left(
            \mu
            +\dfrac{r^2}{\gamma^2-1}
            +\dfrac{w^2}{s}
        \right)
    }=
    \frac{
        \mathcal P_{\mathrm{cut}}
    }{
        s\mu(\gamma^2-1)
        +s r^2
        +(\gamma^2-1)w^2
    } .
    \label{eq:Gamma1}
\end{align}
Since
\begin{align}
    s\mu(\gamma^2-1)
    +s r^2
    +(\gamma^2-1)w^2
    >0
\end{align}
throughout the real physical region, the second Gram condition is
equivalent to
\begin{align}
    \Gamma_1>0
    \qquad\Longleftrightarrow\qquad
    \mathcal P_{\mathrm{cut}}>0 .
    \label{eq:gram1}
\end{align}
Hence, for a real physical contour, the successive Gram conditions reduce to
\begin{align}
    \Gamma_2>0
    \quad\text{automatically},
    \qquad
    \Gamma_1>0
    \quad\Longleftrightarrow\quad
    \mathcal P_{\mathrm{cut}}>0 .
\end{align}
The allowed integration regions may therefore be viewed as those subregions of the maximal-cut domain that are compatible with
$\mathcal P_{\mathrm{cut}}>0$ and all of the subregions in \eqref{eq:cut-regions} satisfy the Gram condition.  With these constraints in place, we can evaluate the maximal cut
sequentially, integrating one Baikov variable at a time. After imposing all maximal-cut conditions, three Baikov variables $(u,r,w)$ remain, and in what follows, for definiteness, we choose \(C_{\rm II}\) as a real reference chamber. It is useful
to regard the real cut domain as
\begin{align}
    \mathcal D_{\mathrm{cut}}
    =
    \left\{
        (u,r,w)\in \mathbb{R}^3
        \,\middle|\,
        \mathcal P_{\mathrm{cut}}(u,r,w)>0
    \right\}.
\end{align}
The corresponding geometry naturally admits the nested projection
\begin{align}
    \mathcal D_{\mathrm{cut}}
    \xrightarrow{\;w\;}
    \mathcal D_{ur}
    \xrightarrow{\;r\;}
    \mathcal D_u .
    \label{eq:fibration-structure}
\end{align}
At each stage, we first integrate over the fiber of the corresponding
projection and then descend to its base. We refer to each such step as a
\emph{fiber integration}. The first two fibers, associated with the $w$ and $r$ integrations, are
conics. Consequently, these two integrations are elementary and reduce to
beta-function-type integrals; in particular, they introduce no new
transcendental geometry. After performing these two fiber integrations, all nontrivial geometric information is encoded in the remaining one-dimensional
integral over $u$. It is at this final stage that a quartic square root
emerges, making manifest the elliptic curve that governs the maximal cut.

Equivalently, the boundary of the integration domain is determined
by
\begin{align}
    \mathcal P_{\mathrm{cut}}(u,r,w)=0,
\end{align}
and the successive elimination of $w$ and $r$ projects this hypersurface onto
the final $u$-plane. The branch points generated by this projection define
the quartic curve underlying the elliptic maximal-cut geometry.
\subsubsection{Successive fiber integrations}
After imposing the maximal-cut conditions, the three remaining Baikov
variables are integrated in the order $w$, $r$, and $u$. With the
orientation induced by $du\wedge dr\wedge dw$, the cut takes the nested
integration,
\begin{align}
 \MaxCut(f_1)
 &=\mathcal N_{\rm B}[s(\gamma^2-1)]^\epsilon
   \int_{\mathcal C}\mathcal P_{\rm cut}^{-1-\epsilon}
       \,du\wedge dr\wedge dw\nonumber\\
 &=\mathcal N_{\rm B}[s(\gamma^2-1)]^\epsilon
   \int_{\mathcal C_u}du
   \int_{\mathcal C_r(u)}dr
   \int_{\mathcal C_w(r,u)}
       \mathcal P_{\rm cut}(u,r,w)^{-1-\epsilon}\,dw .
 \label{eq:baikov-nested-integral}
\end{align}
The iterative fibre integration goes as follows: $\mathcal C_w(r,u)$ is integrated first at fixed $(u,r)$,
followed by $\mathcal C_r(u)$ at fixed $u$, and finally $\mathcal C_u$. Note that the polynomial \eqref{2.25jjjjj} is quadratic in $w$,
\begin{equation}
 \mathcal P_{\rm cut}(u,r,w)
 =a_w(u)w^2+b_w(r,u)w+c_w(r,u),
\end{equation}
with
\begin{align}
\begin{split}
& a_w(u)=u^2+\mu(\gamma^2-1),\,\,
 b_w(r,u)=-\frac{s}{2}\bigl[\mu(\gamma^2-1)+2\gamma ru\bigr],\\ &
 c_w(r,u)=\frac{s}{4}\Bigl[
 r^2(s+4\mu-4u^2)-4\gamma\mu ru
+\mu\bigl((\gamma^2-1)(3\mu+s)+4u^2\bigr)\Bigr].
\end{split}
\end{align}
For fixed $(u,r)$, its roots are the branch points of the $w$ fiber.
The discriminant factorizes as
\begin{equation}
 \Delta_w(r,u)=b_w(r,u)^2-4a_w(u)c_w(r,u)
             =\frac{s}{4}A(u)B(r,u),
\end{equation}
where
\begin{equation}
\begin{aligned}
 A(u)&=4u^2+(4\mu+s)(\gamma^2-1),\\
 B(r,u)&=4(u^2-\mu)r^2+4\gamma\mu ur
          +\mu\bigl[3\mu(1-\gamma^2)-4u^2\bigr].
\end{aligned}
\end{equation}
This factorization separates the dependence on the final coordinate
$u$ from the residual quadratic dependence on $r$. The $w$ contour is
\begin{equation}
\begin{aligned}
 \mathcal C_w(r,u)&=(-\infty,w_-(r,u))\cup(w_+(r,u),\infty),\\
 w_\pm(r,u)&=\frac{-b_w(r,u)\pm\sqrt{\Delta_w(r,u)}}{2a_w(u)}.
\end{aligned}
\end{equation}
Both components are oriented towards increasing $w$. In the real
reference chamber, $a_w>0$ and $\mathcal P_{\rm cut}>0$ on these exterior
intervals. We fix $\arg\mathcal P_{\rm cut}=0$ on the right tail and
continue through the upper $w$ half-plane. Each root crossing contributes
$e^{-i\pi(1+\epsilon)}$, so the phases on the right tail, inner interval,
and left tail are respectively
\begin{equation}
 1,\qquad e^{-i\pi(1+\epsilon)},\qquad e^{-2\pi i\epsilon}.
\end{equation}
These phases are included in the loaded contour
$\mathcal C_w(r,u)$ in \eqref{eq:baikov-nested-integral}.

After the $w$ integration, the remaining quadratic factor is $B(r,u)$.
For $\sqrt{3\mu}/2<|u|<\sqrt\mu$, we take
$\mathcal C_r(u)=(r_-(u),r_+(u))$, oriented from $r_-$ to $r_+$, with
\begin{equation}
\begin{aligned}
 \Delta_r(u)&=16\mu(4u^2-3\mu)a_w(u),\\
 r_\pm(u)&=\frac{\gamma\mu u}{2(\mu-u^2)}
       \pm\frac{\sqrt{\mu(4u^2-3\mu)a_w(u)}}{2|\mu-u^2|}. \label{eq:rpm}
\end{aligned}
\end{equation}
The range of $u$ also comes from our subregion assumption of $r\in \mathbb{R}$. With this condition the region of $u$ is the following intersection: 
\begin{align}
\begin{split}
\mathcal{C}_u:\{u\in\mathbb R:4u^2-3\mu>0\}
\cap
\{u\in\mathbb R:\mu-u^2>0\}
=
\left(-\sqrt{\mu},-\frac{\sqrt{3\mu}}{2}\right)
\cup
\left(\frac{\sqrt{3\mu}}{2},\sqrt{\mu}\right).
\end{split}
\end{align}
The absolute value preserves the ordering $r_-<r_+$ on both sides of
$u^2=\mu$ whenever the roots are real. In the reference chamber,
\begin{equation}
 B(r,u)=4(\mu-u^2)(r-r_-(u))(r_+(u)-r)>0,
 \qquad r_-<r<r_+.
\end{equation}
The finite endpoints of the first fiber satisfy
$\mathcal P_{\rm cut}(u,r,w_\pm)=0$. At the endpoints of the second
fiber, $B(r_\pm,u)=0$ and hence $\Delta_w=0$, so the two $w$ branch
points coalesce.

\textcolor{black}{\paragraph{Positivity on the five chambers.}
The decomposition \eqref{eq:cut-regions} follows from the two discriminants above.
For $\mu>0$, $s>0$, $\gamma^{2}>1$ and real $(u,r,w)$ one has $a_w(u)>0$ and $A(u)>0$, so
$\mathcal P_{\rm cut}$ is an upward parabola in $w$ whose discriminant $\Delta_w$ has the
sign of $B(r,u)$:
\begin{equation}
B(r,u)<0:\ \ \mathcal P_{\rm cut}>0\ \ \forall\,w\in\mathbb R\,,
\qquad
B(r,u)>0:\ \ \mathcal P_{\rm cut}>0\iff w\notin[w_-,w_+]\,.
\label{eq:Pcut-sign-w}
\end{equation}
$B$ is itself a quadratic in $r$ with leading coefficient $4(u^{2}-\mu)$ and
discriminant $\Delta_r(u)$ of \eqref{eq:rpm}, which has the sign of $4u^{2}-3\mu$.
Hence, band by band in $u$,
\begin{align}
|u|<\tfrac{\sqrt{3\mu}}{2}:&\quad \Delta_r<0,\ \ 4(u^{2}-\mu)<0
\ \Rightarrow\ B<0\ \ \forall\,r
&&\Rightarrow\ \mathcal C_{\mathrm I}\,,\nonumber\\
\tfrac{\sqrt{3\mu}}{2}<|u|<\sqrt\mu:&\quad B>0\ \text{on}\ (r_-,r_+),\ \ B<0\ , \text{outside}
&&\Rightarrow\ \mathcal C_{\mathrm{II}},\ \mathcal C_{\mathrm{III}}\,,\nonumber\\
|u|>\sqrt\mu:&\quad B<0\ \text{on}\ (r_-,r_+),\ \ B>0\ \text{outside}
&&\Rightarrow\ \mathcal C_{\mathrm{IV}},\ \mathcal C_{\mathrm{V}}\,.
\label{eq:chamber-sign-table}
\end{align}
Combined with \eqref{eq:Pcut-sign-w}, this shows that $\mathcal P_{\rm cut}>0$ on each of
$\mathcal C_{\mathrm I},\ldots,\mathcal C_{\mathrm V}$, and that $\mathcal P_{\rm
cut}<0$ exactly on the complementary set $\{B>0,\ w_-<w<w_+\}$. The five chambers are therefore
pairwise disjoint, and their union is $\{\mathcal P_{\rm cut}>0\}\cap\mathbb R^{3}$, up to the
walls $|u|\in\{\sqrt{3\mu}/2,\sqrt\mu\}$, $r=r_\pm$, $w=w_\pm$, which have measure zero.
On $|u|=\sqrt{3\mu}/2$ one has $B=4(u^{2}-\mu)(r-r_0)^{2}\le0$, so $\mathcal P_{\rm cut}\ge0$
there, with equality only on the line $r=r_0=\gamma\mu u/[2(\mu-u^{2})]$, $w=-b_w/(2a_w)$.
On $u^{2}=\mu$, $B$ is linear in $r$.\footnote{As an independent check, we evaluated
$\mathcal P_{\rm cut}$ at 200-digit precision on $1.7\times10^{7}$ points placed inside the
five chambers. The points include points within a relative distance $10^{-30}$ of every
wall, for $10^{-10}\le s/\mu\le10^{10}$ and $10^{-12}\le\gamma-1\le10^{6}$. We also tested
$2.4\times10^{7}$ random real points for the converse. No exception was found.}
The physical-region assumption is essential. For $\gamma^{2}<1$ one has $a_w(u)<0$ when
$u^{2}<\mu(1-\gamma^{2})$, and $\mathcal P_{\rm cut}\to-\infty$ as $|w|\to\infty$ inside the would-be $\mathcal C_{\mathrm I}$.}

The first two integrations in \eqref{eq:baikov-nested-integral} are elementary beta integrals. Their combined result is
\begin{align}
\begin{split}
 \int_{\mathcal C_r(u)}dr
   \int_{\mathcal C_w(r,u)}
       \mathcal P_{\rm cut}(u,r,w)^{-1-\epsilon}\,dw
 &=(1+e^{-2\pi i\epsilon})\,a_w(u)^\epsilon\times
   \mathrm B(-\epsilon,1+2\epsilon)\,
   \mathrm B\!\left(\tfrac12-\epsilon,\tfrac12-\epsilon\right)\\ &\times
   \frac{[r_+(u)-r_-(u)]^{-2\epsilon}}
   {\bigl[\frac{s}{4}A(u)\bigr]^{1/2+\epsilon}
    [4(\mu-u^2)]^{1/2+\epsilon}},
\end{split}
 \label{eq:baikov-loaded-fiber}
\end{align}
where $\mathrm B(x,y)=\Gamma(x)\Gamma(y)/\Gamma(x+y)$ is Euler's
beta function. The real reference integral converges for
$-\tfrac12<\Re\epsilon<0$; the result elsewhere is obtained by
meromorphic continuation. Since
\begin{equation}
 r_+(u)-r_-(u)
 =\frac{\sqrt{\mu(4u^2-3\mu)a_w(u)}}{\mu-u^2},
\end{equation}
the factors of $a_w^\epsilon$ cancel. Substitution into
\eqref{eq:baikov-nested-integral}, followed by the beta duplication
identity, leaves the final integration over $u$:
\begin{align}
 \MaxCut(f_1)
 &=\frac{2^{1+4\epsilon}e^{-i\pi\epsilon}(\gamma^2-1)^\epsilon}
 {\sqrt{\pi s}\,\mu^\epsilon
  \Gamma(1-\epsilon)\Gamma(\frac12-\epsilon)}
 \int_{\mathcal C_u}
 \frac{(\mu-u^2)^{\epsilon-1/2}}
      {A(u)^{\epsilon+1/2}(4u^2-3\mu)^\epsilon}\,du .
 \label{bcut}
\end{align}
This expression exhibits the maximal cut as an iterated period obtained
by two successive fiber integrations. The first integrates the twisted
top form over the $w$-conic, descending from the three-dimensional cut
space to the discriminant cover over $(u,r)$. The second integrates over
the interval between the roots of the residual quadratic in $r$,
leaving a one-parameter family over $u$. Thus the two rational fibers
are integrated out before the final $u$ integration, whose quartic
square root makes the elliptic geometry manifest. The stated real $u$
range belongs to the reference chamber. For other loaded contours,
including the imaginary-axis cycles used later, the $r$ and $w$ fibers
and their branches are continued consistently with $\mathcal C_u$.

\subsection{Elliptic curve and degeneration of the maximal cut}
\subsubsection{Quartic curve and period basis} \label{sec:newzz}
In the previous subsection, we were still left with the $ u$-fiber integration for an arbitrary dimension. Before doing the integration in the general dimension, it will be instructive to look into $d=4$ to investigate the geometric structure of the integral. Setting $\epsilon=0$ removes the unequal dimensional twist from the Baikov measure and leaves a one-dimensional algebraic period integral with the same weight, and interestingly, we observe that the third fibre integration is not quadratic (i.e the curve associated with it is not rational), rather quartic,
\begin{align}
    \MaxCut(\mathcal{F}_1)=
    \textcolor{black}{\frac{4}{\pi\sqrt{s}}}
    \int_{\mathcal C_u}
    \frac{du}{
    \sqrt{
    4(\mu-u^2)\bigl(4u^2+(4\mu+s)(\gamma^2-1)\bigr)}
    }\,.
    \label{eq:maxcut-eps0}
\end{align}
This form of the period integral is particularly useful and contains a richer structure about the integral family on its maximal cut. Introducing   $ \Lambda:=(4\mu+s)(\gamma^2-1),$ the corresponding affine curve is
\begin{align}
    Y^2=4(\mu-u^2)(4u^2+\Lambda)\,.
    \label{eq:cut-elliptic-curve}
\end{align}
Thus, the maximal cut is a period of the holomorphic differential, $  \eta=\frac{du}{Y}$ on the smooth projective compactification of the affine curve
\begin{align}
    \mathcal E^\circ_u:\quad Y^2=4(\mu-u^2)(4u^2+\Lambda)\,.
\end{align}
As we observe, the evaluation of a family of Feynman integrals now reduces to the study of cycles and holomorphic differentials on a family of algebraic curves. The identification solves a major part of the problem of evaluating the uncut integrals, since the maximal cuts provide solutions to the homogeneous differential equations satisfied by the uncut integrals. Moreover, degenerations of the curve identify the kinematic loci where more complicated geometries reduce to simpler ones, a phenomenon sometimes referred to as a kinematic genus drop. We will discuss these aspects shortly. Now, the projection
\begin{align}
    \pi:\overline{\mathcal E}_u\longrightarrow \mathbb {CP}^1_u\,,
    \qquad
    (u,Y)\longmapsto u\,,
\end{align}
realizes $\overline{\mathcal E}_u$ as a double cover of the $u$-sphere, branched at
\begin{align}
    e_1=-\sqrt{\mu}\,,\qquad
    e_2=+\sqrt{\mu}\,,\qquad
    e_3=+\frac{i}{2}\sqrt{\Lambda}\,,\qquad
    e_4=-\frac{i}{2}\sqrt{\Lambda}\,.
    \label{eq:branch-points-cut}
\end{align}
For generic kinematics, these four branch points are distinct. Hence, by the Riemann-Hurwitz formula for a double cover of the sphere branched at four
points,
\[
    2g(\overline{\mathcal E}_u)-2
    =
    2(2g(\mathbb {CP}^1)-2)+4
    =
    0 \implies g(\overline{\mathcal E}_u)=1\,.
\]
The maximal cut is therefore an elliptic period. In the physical region
\[
    \mu>0,\qquad s>0\,,\qquad \gamma^2>1\,,
\]
the branch points $e_1,e_2$ are real, while $e_3,e_4$ form a
complex-conjugate pair. As the torus has two independent periods, we choose the two periods as: $\delta_R:(e_1,e_2)$ and $\delta_I:(e_4,e_3)\,.$ \textcolor{black}{These are drawn in the right panel of Fig.~(\ref{Fig2.2}); the left panel shows the
canonical pair $A$, $B$ of \eqref{new:eq9}.}
Although the two symmetric cycles $\delta_R$ and $\delta_I$ are linearly
independent, they do not form an integral homology basis of the smooth
compact elliptic curve $\mathcal E$, since
\[
    \delta_R\cdot\delta_I=2\,.
\]
They instead generate an index-two sublattice of
$H_1(\mathcal E,\mathbb Z)$, while forming a basis of
$H_1(\mathcal E,\mathbb Q)$. In the torus convention used here, let
$A$ and $B$ denote the meridian and longitude, respectively, oriented
such that $A\cdot B=1$. The symmetric, doubly intersecting cycles then satisfy
\begin{align}
    \delta_R=2A+B\,,
    \qquad
    \delta_I=B\,. \label{new:eq9}
\end{align}
\textcolor{black}{Equivalently, the symplectic cycles can be expressed as half-integer linear combinations of the cycles \(\delta_R\) and \(\delta_I\). Consequently, the doubly intersecting cycles \(\{\delta_R,\delta_I\}\) do not form an integral basis of \(H_1(E,\mathbb Z)\); in particular, a general integral homology class cannot be expressed as an integer linear combination of \(\delta_R\) and \(\delta_I\).}
\begin{figure}
    \centering
    \includegraphics[width=0.8\linewidth]{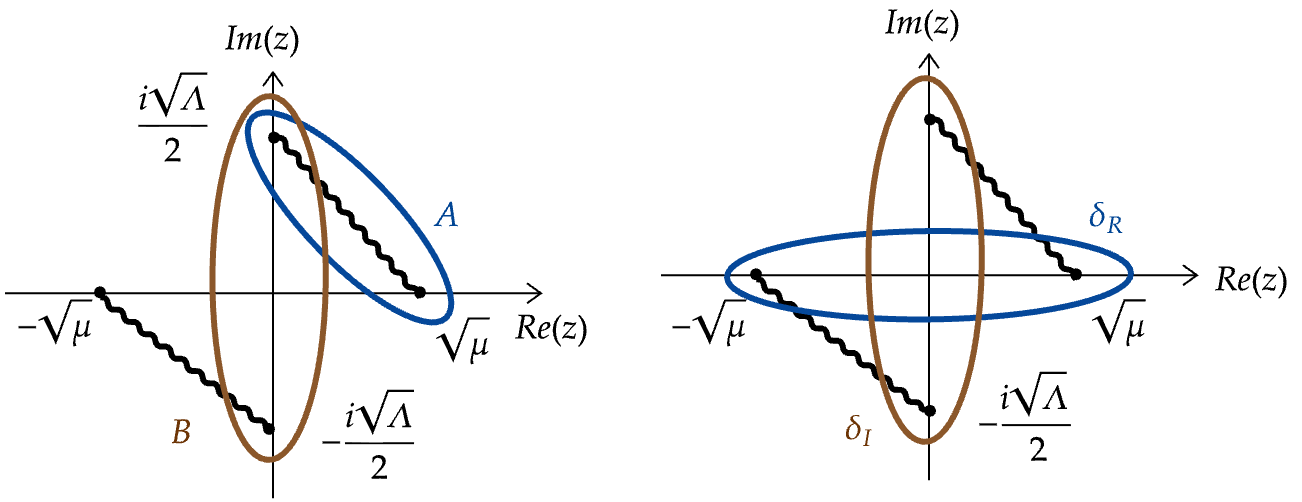}
    \caption{Figure showing possible cut structure and the contour for the period integrals. The Left one shows the usual canonical choice of periods on the torus, and the right one shows doubly intersecting periods. }
    \label{Fig2.2}
\end{figure}
\\
\textbf{\textit{A digression}}: \textcolor{black}{One point about the massless limit is worth making here, because it recurs in Sec.~(\ref{sec:intersec}). The rescaling
$u=\sqrt\mu\,z$ used there is an exact identity, not a limit.
\begin{align}
\oint\frac{du}{\sqrt{(\mu-u^{2})(4u^{2}+\Lambda)}}
=\oint\frac{dz}{\sqrt{(1-z^{2})(4\mu z^{2}+\Lambda)}}
\end{align}
for every $\mu>0$, since $du=\sqrt\mu\,dz$ cancels the $\sqrt\mu$ from
$(\mu-u^{2})^{1/2}$. The two variables, therefore, describe the same period, and the  $\mu\to0$ limit may be taken in either. In $u$, the two real branch points $\pm\sqrt\mu$ coalesce, so the $\delta_R$ cycle shrinks. But the limiting
differential develops a simple pole at the resulting node,
\begin{align}
\frac{du}{\sqrt{(\mu-u^{2})(4u^{2}+\Lambda)}}
\ \xrightarrow[\ \mu\to0\ ]{}\
\frac{du}{i\,u\sqrt{4u^{2}+s(\gamma^{2}-1)}}\,,
\end{align}
whose residue leaves the period finite,
$2\pi/\sqrt{s(\gamma^{2}-1)}$. A vanishing cycle forces a vanishing period only for a differential that stays holomorphic at the node, which is not the case here. This
is why the massless limit of \eqref{eq:maxcut-eps0} stays finite even though $\delta_R$ vanishes.}


\subsubsection{Degeneration of the elliptic curve and physical kinematic limits}
The elliptic curve degenerates when two branch points collide. These
collisions are determined by the discriminant of
\begin{align}
    R(u):=4(\mu-u^2)(4u^2+\Lambda)\,.
\end{align}
A direct computation gives\footnote{The discriminant of a polynomial $P(u)=au^n+\cdots$ with roots $e_1,e_2,\cdots e_n$ is given by
$$\Delta_uP(u)=a^{2n-2}\prod_{i<j}(e_i-e_j)^2.$$
}
\begin{align}
   \Delta_u R(u)
    =a^6\prod_{i<j}(u_i-u_j)^2=
    -2^{18}\,\mu\,\Lambda\,(\Lambda+4\mu)^4\,.
    \label{eq:disc-cut-curve}
\end{align}
Hence, the degeneration divisor is
\begin{align}
    \mathfrak D
    =
    \{\mu=0\}
    \cup
    \{\Lambda=0\}
    \cup
    \{\Lambda+4\mu=0\}\,.
    \label{eq:degeneration-divisor-cut}
\end{align}
Equivalently,
\begin{align}
    \mathfrak D
    =
    \{m^2=0\}
    \cup
    \{(4m^2+s)(\gamma^2-1)=0\}
    \cup
    \{4m^2\gamma^2+s(\gamma^2-1)=0\}\,.
\end{align}
\textcolor{black}{Each component has a simple geometric interpretation \cite{Bloch:2013tra}. On $\{\mu=0\}$, the two real branch points collide, and the $\delta_R$-cycle shrinks down to a point. On the other hand, at 
$\{\Lambda=0\}$ the complex-conjugate pair collides, producing the
degeneration of the complementary cycle ($\delta_I$). \textcolor{black}{On $\{\Lambda+4\mu=0\}$ both collisions happen at once: $e_3\to e_1=-\sqrt\mu$ and
$e_4\to e_2=+\sqrt\mu$ simultaneously, so that
$R(u)\to-16\,(u^2-\mu)^2$ and the fibre becomes reducible, two rational components meeting at the two nodes $u=\pm\sqrt\mu$, rather than an irreducible one-node curve. This locus is also outside the physical region: $4m^2\gamma^2+s(\gamma^2-1)=0$
requires $\gamma^2=s/(4m^2+s)<1$, so it is reached only by analytic continuation to $\gamma^2<1$. In the first two cases, the smooth torus degenerates to an irreducible
nodal curve whose normalization is a genus-zero Riemann surface. In the third case, the normalization is a disjoint pair of genus-zero curves.} Furthermore, the degeneration of the elliptic curve is not just another mathematical artifact, but it has a physical interpretation. For physical kinematics, one has $s>0$ and $m^2\geq0$, so the factor $4m^2+s$ does not vanish in the massive region. Consequently, the physical degeneration limits are the massless limit $m\to0$, and the low-velocity or threshold limit $\gamma\to1$. At these loci, the elliptic cycle collapses and the underlying genus one surface reduces to a genus zero (nodal) surface. This explains precisely the underlying reason why the corresponding limits are expected to be expressible in terms of multiple polylogarithms, whereas generic massive, generic velocity requires elliptic periods and their associated functions. Fig.~(\ref{fig:elliptic-degeneration})
summarizes this degeneration.}

\begin{figure}[t]
\centering
\begin{adjustbox}{max width=\textwidth}
\begin{tikzpicture}[
    >=Stealth,
    line cap=round,
    line join=round,
    guide/.style={line width=0.45pt,gray!65},
    cutA/.style={line width=0.95pt,black},
    cutB/.style={line width=0.95pt,gray!75,dotted},
    motion/.style={->,line width=0.55pt,gray!70},
    curve/.style={line width=0.70pt,black},
    point/.style={circle,fill=black,inner sep=1.25pt}
]
\begin{scope}[xshift=0cm]
    \draw[guide] (-1.18,0)--(1.18,0);
    \draw[guide] (0,-0.92)--(0,0.92);
    \draw[cutA] (-0.86,0)--(0.86,0);
    \draw[cutB] (0,-0.70)--(0,0.70);
    \foreach \p in {(-0.86,0),(0.86,0),(0,0.70),(0,-0.70)}
        \node[point] at \p {};
\end{scope}

\draw[motion] (1.42,0)--(2.02,0);

\begin{scope}[xshift=3.30cm]
    \draw[guide] (-1.18,0)--(1.18,0);
    \draw[guide] (0,-0.92)--(0,0.92);
    \draw[cutB] (0,-0.70)--(0,0.70);
    \draw[cutA,dotted] (-0.12,0)--(0.12,0);
    \node[point] at (-0.12,0) {};
    \node[point] at (0.12,0) {};
    \node[point] at (0,0.70) {};
    \node[point] at (0,-0.70) {};
    \draw[motion] (-0.78,0.34)--(-0.19,0.06);
    \draw[motion] (0.78,0.34)--(0.19,0.06);
\end{scope}

\draw[motion] (4.72,0)--(5.32,0);

\begin{scope}[xshift=6.62cm]
    \draw[curve]
        (0,0.08)
        .. controls (-0.28,0.82) and (-1.05,0.76)..(-1.05,0.08)
        .. controls (-1.05,-0.60) and (-0.28,-0.66)..(0,0.08)
        .. controls (0.28,-0.66) and (1.05,-0.60)..(1.05,0.08)
        .. controls (1.05,0.76) and (0.28,0.82)..(0,0.08);
    \node[point] at (0,0.08) {};
\end{scope}

\draw[motion] (8.04,0)--(8.64,0);

\begin{scope}[xshift=9.72cm]
    \draw[curve,fill=gray!10] (0,0) circle (0.70);
\end{scope}
\end{tikzpicture}
\end{adjustbox}
\caption{Degeneration of the elliptic curve. The generic double
cover of $\mathbb {CP}^1_u$ has branch cuts giving the $A$- and $B$-cycles of
a smooth torus. The solid black segment denotes the $A$-cut between
$e_{1,2}=\mp\sqrt{\mu}$, while the grey dotted segment denotes the $B$-cut
between $e_{3,4}=\pm\tfrac{i}{2}\sqrt{\Lambda}$. The drawing schematically
shows a branch-point collision, the resulting nodal curve, and its
normalization. When $e_1,e_2$ collide the $A$-cycle vanishes, while collision
of $e_3,e_4$ collapses the complementary $B$-cycle. The limiting nodal curve
normalizes to $\mathbb {CP}^1$.}
\label{fig:elliptic-degeneration}
\end{figure}
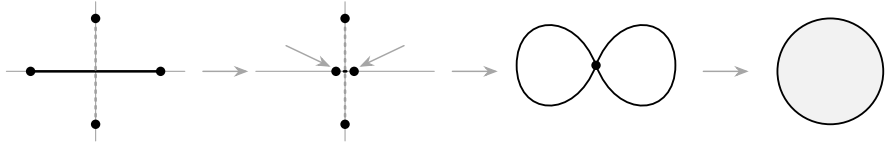

\subsubsection{Picard--Lefschetz monodromy and degeneration} \label{new:sec2.2.3}
Let us be more concrete about the statements made in the previous section. Near a smooth point of the discriminant divisor, the set of kinematic points where a smooth algebraic curve becomes singular, let $\delta$ be the vanishing cycle, the one-cycle on the smooth elliptic curve that
shrinks to the node in the singular fiber. If the kinematic parameters are
analytically continued around this point of the discriminant, the
elliptic curve returns to an isomorphic curve, but the homology basis need not
return to itself, which can be seen from the Picard--Lefschetz formula \cite{PicardSimart1897,Lefschetz1924}:
\begin{align}
    T_\delta(\Gamma)
    =
    \Gamma+(\delta\cdot\Gamma)\,\delta\,,
    \qquad
    \Gamma\in H_1(\overline{\mathcal E}_u,\mathbb Z)\,.
\end{align}
Here $\delta\cdot\Gamma$ is the oriented intersection number of the vanishing
cycle with $\Gamma$. Therefore, a cycle disjoint from $\delta$ is unchanged,
while a cycle intersecting $\delta$ once is shifted by one copy of the
vanishing cycle. For instance, assume $A$-cycle is the vanishing cycle, then $\delta=A\,,$ 
\begin{align}
    T_A(A)=A,\qquad T_A(B)=B+A\,.
\end{align}
The additional non-trivial monodromy is coming from the logarithmic behavior of the period near the degeneration point. The logarithmic behavior can be seen directly from the local geometry of the
pinch. Let $t$ be a local coordinate on kinematic space which vanishes on
the chosen component of the degeneration point. Near the node, the family of curves can be written as,
\begin{align}
    y^2=x^2-t\,.
\end{align}
The vanishing cycle circles the two  branch points
\(x=\pm\sqrt{t}\), whereas the dual cycle passes through the thin neck that opens between them. Locally, the relevant period therefore contains an integral of the form
\begin{align}
    \int^{\Lambda_c}\frac{dx}{\sqrt{x^2-t}}
    =
    \operatorname{arcosh}\!\left(\frac{\Lambda_c}{\sqrt{t}}\right)
    =
    \log\!\left(\frac{2\Lambda_c}{\sqrt{t}}\right)+O(t)
\end{align}
for a fixed cutoff $\Lambda_c$. Thus, the dual period necessarily contains a term proportional to $\log t$, up to normalization and regular terms. This is precisely the origin of the logarithm: when the elliptic curve degenerates, the dual period measures the length of the pinched neck, and that length diverges logarithmically. The same conclusion follows from Picard--Lefschetz monodromy. Analytic continuation around the degeneration point can be written as,
\begin{align}
    t\longmapsto e^{2\pi i}t\,,
    \qquad
    \log(e^{2\pi i}t)=\log t+2\pi i\,.
\end{align}
If the vanishing cycle is \(\delta=A\), let
\begin{align}
    P_A(t)=\oint_A\eta,
    \qquad
    P_B(t)=\oint_B\eta
\end{align}
denote the periods over closed homology cycles. Applying the
Picard--Lefschetz formula explicitly with \(\delta=A\), and using the
symplectic orientation \(A\cdot B=1\), gives
\begin{align}
    T_A(A)
    =
    A+(A\cdot A)A
    =
    A\,,
    \qquad
    T_A(B)
    =
    B+(A\cdot B)A
    =
    B+A\,,
\end{align}
since \(A\cdot A=0\). Thus, the Picard-Lefschetz twist of these cycles is
\begin{align}
    A\longmapsto A\,,\qquad B\longmapsto B+A
\end{align}
and hence the periods transform as
\begin{align}
    P_A(e^{2\pi i}t)=\oint_A\eta=P_A(t)\,,
    \qquad
    P_B(e^{2\pi i}t)=\oint_{B+A}\eta
    =P_B(t)+P_A(t)\,.
\end{align}
Equivalently, in vector form,
\begin{align}
    \begin{pmatrix}
    P_A\\[2pt]
    P_B
    \end{pmatrix}
    \longmapsto
    \begin{pmatrix}
    1&0\\
    1&1
    \end{pmatrix}
    \begin{pmatrix}
    P_A\\[2pt]
    P_B
    \end{pmatrix}.
\end{align}
This monodromy can be reproduced by taking a logarithmic ansatz, and we will see shortly that indeed near the vanishing cycle the dual period behaves logarithmically. 
\begin{align}
    P_B(t)
    =
    \frac{1}{2\pi i}\,P_A(t)\log t+P_B^{\rm reg}(t)\,,
    \qquad
    P_A(t),P_B^{\rm reg}(t)\ \text{regular at }t=0\,,
\end{align}
because one circuit around \(t=0\) shifts \(\log t\) by \(2\pi i\)\,. Hence
\begin{align}
    \operatorname{Disc}_{t=0}P_B=P_A\,,
\end{align}
up to the conventional orientation of the cycle. The period over the
shrinking cycle is regular, while the dual period carries the logarithmic
singularity required by monodromy. The discussion we made above is particularly for the canonical choice of cycle where the modulus of the intersection number is unity. However, note that for our computations, we chose a special symmetric cycle without mixing the real and imaginary 
branch points. For our choice of cycles , $\delta_R$ and $\delta_I$ are twice intersecting (see Fig.~\ref{fig:torus-cycles})), with the convention $\delta_R\cdot \delta_I=2$. Therefore, when $\delta_R$ is the vanishing cycle then Picard--Lefschetz formula gives,
\begin{align}
    T_{\delta_R}(\delta_R)=\delta_R\,,\qquad T_{\delta_R}(\delta_I)=\delta_I+2\delta_R\,,\label{2.65jj}
\end{align}
whereas, when the imaginary cycle $\delta_I$ vanishes, then,
\begin{align}
     T_{\delta_I}(\delta_R)=\delta_R-2\delta_I\,,\qquad T_{\delta_I}(\delta_I)=\delta_I\,,\label{2.66jj}
\end{align}
The negative sign in \eqref{2.66jj} reflects the fact that $\delta_I\cdot \delta_R=-2\,.$
\par
\begin{figure}
    \centering
    \includegraphics[width=0.4\linewidth]{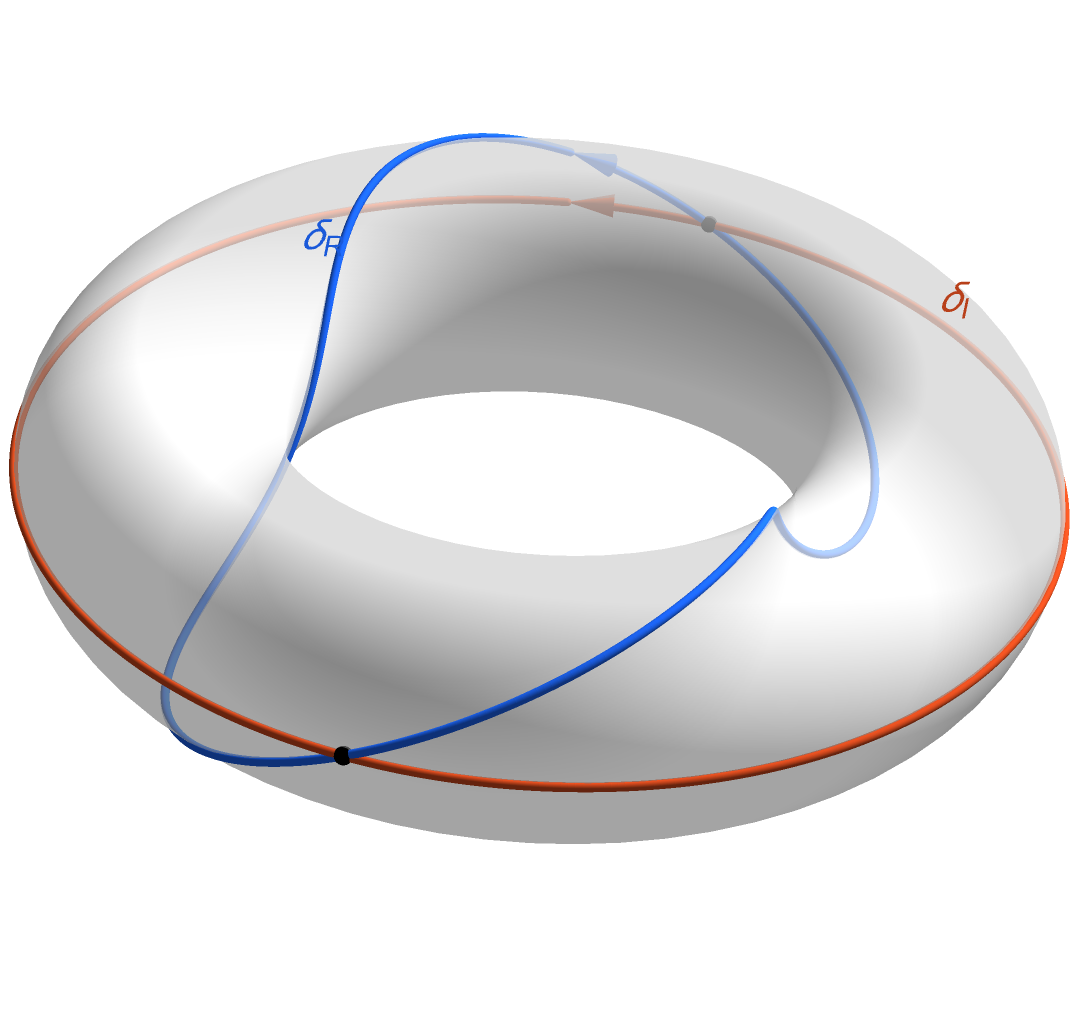}
    \caption{Figure showing a schematic illustration of the doubly intersecting cycles on a genus one surface as discussed in the article.}
    \label{fig:torus-cycles}
\end{figure}
The study of Picard–Lefschetz twist around the vanishing cycle is important to understand some of the beautiful inner structure of the family of integrals, as it not only provides the information about the specific kinematics where the degeneration happens, but it also makes us aware of what exactly is happening at those degeneration points. Therefore, the maximal cut and its monodromy behavior fix the non-analytic behavior that the complete Feynman integral must reproduce, as it is noted that the maximal cuts are solutions of the homogeneous differential equation. In particular, when the kinematics approach a massless, threshold, or degeneration locus, the logarithmic period encodes the non-analytic information, discontinuity, and a possible imaginary part of the amplitude. In the present problem, this explains why special limits such as $m\to 0$ or $\gamma\to 1$ reduce to genus-zero, polylogarithmic behavior, while generic massive finite-velocity kinematics retain genuinely elliptic periods. With this in mind, let us concretely compute the period integrals and see whether the Picard--Lefschetz monodromy really holds what is claimed above.

\textcolor{black}{\begin{align}
\Pi_{\delta_R}=\int_{e_1}^{e_2}\eta=
\frac{K\!\left(-\tfrac{4\mu}{\Lambda}\right)}{\sqrt{\Lambda}}\,,
\qquad
\Pi_{\delta_I}=\int_{e_4}^{e_3}\eta=
\frac{\ii}{2\sqrt{\mu}}K\!\left(-\tfrac{\Lambda}{4\mu}\right),
\qquad \Lambda=(4\mu+s)(\gamma^2-1)\,,\label{eq:baikov-periods}
\end{align}
so that, restoring the prefactor of \eqref{eq:maxcut-eps0},
$\mathrm{MaxCut}(\mathcal{F}_1)\big|_{\delta_{R,I}}=\frac{4}{\pi\sqrt{s}}\,\Pi_{\delta_{R,I}}\,.$ At this point, note that, throughout this paper, the complete and incomplete elliptic integrals are used in the
\emph{parameter} convention,
\begin{align}
F(\phi\,|\,m)=\int_0^{\phi}\frac{d\theta}{\sqrt{1-m\sin^2\theta}}\,,\qquad
E(\phi\,|\,m)=\int_0^{\phi}\sqrt{1-m\sin^2\theta}\;d\theta\,,
\end{align}
with $K(m)=F(\tfrac{\pi}{2}\,|\,m)$ and $E(m)=E(\tfrac{\pi}{2}\,|\,m)$. The arguments above are negative, beginning with $m=-\Lambda/4\mu$ in
\eqref{eq:baikov-periods}, and the defining integrals are perfectly regular there.}

Setting, $ \rho=\frac{4\mu}{\Lambda}\,,$
we get, $\textcolor{black}{\Pi_{\delta_R}}=\Lambda^{-1/2}K(-\rho)$, whereas
$\textcolor{black}{\Pi_{\delta_I}}=i(2\sqrt{\mu})^{-1}K(-1/\rho)$. Near the degeneration
\(\mu=0\), equivalently \(\rho=0\), the asymptotic formula for the complete elliptic function is given by,
\begin{align}
    K(-\rho)=K(0)+O(\rho)\,,
    \qquad
    K(-1/\rho)
    =
    \sqrt{\rho}\left[
        \log\!\left(\frac{4}{\sqrt{\rho}}\right)
        +O(\rho\log\rho)
    \right]\,.
\end{align}
Thus \(\textcolor{black}{\Pi_{\delta_R}}\) is regular, while
\begin{align}
   \textcolor{black}{\Pi_{\delta_I}}
    =
    \frac{i}{\sqrt{\Lambda}}
    \log\!\left(\frac{4}{\sqrt{\rho}}\right)
    +\text{regular}.
\end{align}
Taking $\rho\mapsto e^{2\pi i}\rho$ (with $\rho\to 0$) gives
\begin{align}
    \textcolor{black}{\Pi_{\delta_R}}\longmapsto \textcolor{black}{\Pi_{\delta_R}}\,,
    \qquad
    \textcolor{black}{\Pi_{\delta_I}}\longmapsto \textcolor{black}{\Pi_{\delta_I}}+2\textcolor{black}{\Pi_{\delta_R}}
\end{align}
where we used $K(0)=\pi/2$, with the sign fixed by the orientation of the loop, and it verifies \eqref{2.65jj}. Similarly, near \(\Lambda=0\), writing \(\sigma=\Lambda/(4\mu)\), the roles of the two periods are exchanged:
\begin{align}
    \textcolor{black}{\Pi_{\delta_I}}\ \text{is regular},\qquad
    \textcolor{black}{\Pi_{\delta_R}}
    =
    \frac{1}{2\sqrt{\mu}}
    \log\!\left(\frac{4}{\sqrt{\sigma}}\right)
    +\text{regular}\,,
\end{align}
so that a full circuit \(\sigma\mapsto e^{2\pi i}\sigma\) gives
\begin{align}
    \textcolor{black}{\Pi_{\delta_R}}\longmapsto \textcolor{black}{\Pi_{\delta_R}}-2\textcolor{black}{\Pi_{\delta_I}}\,,
    \qquad
    \textcolor{black}{\Pi_{\delta_I}}\longmapsto \textcolor{black}{\Pi_{\delta_I}}\,.
\end{align}
This gives precisely \eqref{2.66jj}. The remaining degeneration, \(\Lambda+4\mu=0\), is the point where the
arguments of both complete elliptic integrals approach unity; the usual
expansion $K(m)\sim-\frac12\log(1-m)$ gives the corresponding logarithmic monodromy in the analytically continued kinematic parameters.

\section{Differential equation for the maximal cut}
\label{sec:cut-de}
The Baikov representation provides an efficient and intuitive way to identify the geometric structure underlying the integral. However, it is well known that the algebraic periods associated with a family of algebraic curves are related to a special class of linear ordinary differential equations, namely the Picard–Fuchs equations, and to their factorization properties. Nonetheless, the differential-equation technique has proven to be one of the most effective methods for computing a family of Feynman integrals. As mentioned earlier, it provides a complementary approach in which the same cut periods are identified as solutions of the homogeneous IBP system. We choose the following basis of irreducible scalar products (ISPs) in our calculation,
\begin{align}
    \textrm{ISPs}: \{\alpha\}:&(\ell_1\cdot v_2,\ell_2\cdot v_1,\ell_1^2-m^2,\ell_2^2-m^2,(\ell_1-q)^2-m^2,(\ell_2-q)^2, (\ell_1+\ell_2)^2-m^2,\ell_1\cdot v_1,\ell_2\cdot v_2).\label{3.1nn}
\end{align}
\textcolor{black}{We first retain the same choice of differentiation variable $(x)$ defined in (\ref{eq:new7}) as what has been used typically for the massless case. The purpose of this section is to examine the structure of the corresponding differential equation and to identify any additional complications that may arise from the massive kinematics. This analysis will, in turn, provide useful insight into the suitability of the variable $x$ and motivate the choice of a more convenient kinematic variable (if any) when necessary.}
Before going to the explicit computation of the differential equation, we want to emphasize that, unlike the massless case, the massive integrals have one more independent scale as discussed in (\ref{eq:new8}). Following the discussion around (\ref{eq:new8}) we 
can scale out the mass and write the differential equation in terms of  two independent scales $\bar \rho, \gamma$,
\begin{align}
    d\vec{\mathcal{F}}=\mathcal{A_{\gamma}}(\gamma,\bar\rho)\vec{\mathcal{F}}d\gamma+\mathcal{A_{\bar \rho}}(\gamma,\bar\rho)\vec{\mathcal{F}}d\bar \rho\,.
\end{align}
However, in the subsequent sections, we use $x=\gamma-\sqrt{\gamma^2-1}$ as our differentiating variable, and we will study the partial differential equation with respect to $x$ with fixed $\bar \rho$. We suppress the other kinematic-dependent implicit terms and drop them from the notation.
\subsection{Homogeneous cut system} \label{newsec5}
Let \(\vec{\mathcal F}(x,\epsilon)\) denote a basis of master integrals for the
family under consideration. The IBP identities imply a first-order system in
the kinematic variable \(x\),
\begin{equation}
    \partial_x \vec{\mathcal F}(x,\epsilon)
    =\Omega(x,\epsilon)\vec{\mathcal F}(x,\epsilon)\,.
    \label{eq:de-full-system}
\end{equation}
After organizing the master integrals
by sector, the equations in a sector \(s_\mu\) take the form
\begin{equation}
     \partial_x \mathcal{F}^{s_\mu}_i(x)
     =\mathcal{A}_{ij}(x,\epsilon)\mathcal{F}_j^{s_{\mu}}(x)
      +\mathcal{G}_i^{s_{\nu|\mu}}(x,\epsilon)\,.
      \label{eq:de-sector-system}
\end{equation}
\begin{figure}[t]
\centering
\begin{adjustbox}{max width=\textwidth}
\begin{tikzpicture}[
    line cap=round,
    line join=round,
    propagator/.style={black,line width=0.85pt},
    worldline/.style={black!48,line width=1.55pt},
    vertex/.style={circle,fill=black,inner sep=1.25pt},
    raised/.style={circle,fill=black,inner sep=2.05pt},
    basislabel/.style={font=\small}
]
\newcommand{\drawliteredmaster}[2]{%
  \begin{scope}[xshift=#1]
    \coordinate (left#2)  at (-1.05,0.00);
    \coordinate (upper#2) at (-0.43,0.57);
    \coordinate (lower#2) at (-0.36,-0.61);
    \coordinate (rightu#2) at (0.61,0.67);
    \coordinate (rightl#2) at (0.68,-0.51);

    \draw[worldline]
      (-1.42,-0.34)--(left#2)--(lower#2)--(-0.61,-0.96);
    \draw[worldline]
      (0.98,0.94)--(rightu#2)--(rightl#2)--(1.04,-0.73);

    \draw[propagator] (left#2)--(upper#2);
    \draw[propagator] (upper#2)--(lower#2);
    \draw[propagator] (upper#2)--(rightu#2);
    \draw[propagator] (lower#2)--(rightl#2);

    \node[vertex] at (left#2) {};
    \node[vertex] at (upper#2) {};
    \node[vertex] at (lower#2) {};
    \node[vertex] at (rightu#2) {};
    \node[vertex] at (rightl#2) {};

    \ifcase#2\relax
    \or
      \node[raised] at ($(left#2)!0.50!(upper#2)$) {};
    \or
      \node[raised] at ($(upper#2)!0.52!(rightu#2)$) {};
    \or
      \node[raised] at ($(left#2)!0.52!(lower#2)$) {};
    \or
      \node[raised] at ($(rightu#2)!0.52!(rightl#2)$) {};
    \fi

    \node[basislabel] at (-0.16,-1.22) {$B_{#2}$};
  \end{scope}
}

\drawliteredmaster{0.0cm}{0}
\drawliteredmaster{2.8cm}{1}
\drawliteredmaster{5.6cm}{2}
\drawliteredmaster{8.4cm}{3}
\drawliteredmaster{11.2cm}{4}
\end{tikzpicture}
\end{adjustbox}
\caption{The LiteRed master basis on the maximal cut.  From left to
right, \(B_0\) is the undotted mother topology, \(B_1\) and \(B_2\) have a
raised power of \(D_1\) and \(D_2\), and \(B_3\) and \(B_4\) have a raised
reverse-unitarity power of \(\rho_1=\ell_1\!\cdot v_1\) and
\(\rho_2=\ell_2\!\cdot v_2\), respectively.   No line
is pinched, so all five masters have the same graph topology.}
\label{Fig5.5}
\end{figure}
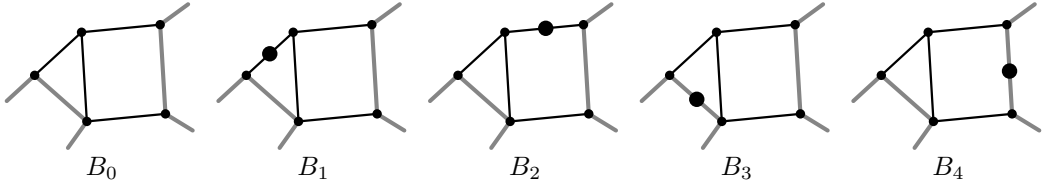

The inhomogeneous term contains only lower-sector integrals. Upon imposing the
maximal cut of \(s_\mu\), these lower subtopologies vanish, and the cut system is
given by the homogeneous block,
\begin{equation}
     \partial_x \bigl(\MaxCut\mathcal{F}^{s_\mu}_i(x)\bigr)
     =\mathcal{A}_{ij}(x,\epsilon)
      \bigl(\MaxCut\mathcal{F}_j^{s_{\mu}}(x)\bigr)\,.
      \label{eq:de-cut-system-general}
\end{equation}
For the equal-mass $\mathbb{Y}$-ladder sector, let $\alpha=(0,0,1,1,1,0,1,1,1)$ in the ordering of ISPs of \eqref{3.1nn}. We choose the ordered cut basis
\begin{equation}
\vec{\mathcal F}^{\rm cut}=
\left(\mathcal F_{\alpha}^{\rm cut},\,
\mathcal F_{\alpha+\mathbf e_3}^{\rm cut},\,
\mathcal F_{\alpha+\mathbf e_4}^{\rm cut},\,
\mathcal F_{\alpha+\mathbf e_8}^{\rm cut},\,
\mathcal F_{\alpha+\mathbf e_9}^{\rm cut}\right)^T\,,
\label{eq:de-cut-basis}
\end{equation}
where $\mathbf e_j$ denotes the unit vector which raises the \(j\)-th propagator power by one and topologies are shown in (\ref{Fig5.5}).
The corresponding cut differential equation is
\begin{equation}
    \partial_x\vec{\mathcal F}^{\rm cut}
    =\mathcal A(x,\epsilon)\vec{\mathcal F}^{\rm cut}\,,
    \qquad
    \mathcal A(x,\epsilon)=\mathcal A^{(0)}(x)+\epsilon\mathcal A^{(1)}(x)+\epsilon^2 \mathcal{A}^{(2)}(x)\,.
    \label{eq:de-cut-system}
\end{equation}
The common denominator factors of the matrix are denoted by
\begingroup
\begin{align}
\begin{split}
&\Delta_1=x(x^2-1)(2m^2+s)\,,\quad
\Delta_2=4m^2(x^2+1)^2+s(x^2-1)^2\,,\quad
\Delta_3=\textcolor{black}{(4m^2+s)(x^2-1)^2+12m^2x^2}\,.
\end{split}
\end{align}
\endgroup
\textcolor{black}{It is useful to recognize these denominators as loci already met in Sec.~(\ref{sec:newzz}). With $\gamma=(1+x^2)/2x$ and $\Lambda=(4\mu+s)(\gamma^2-1)$,
\begin{align}
\Delta_2=4x^2\bigl(\Lambda+4\mu\bigr)\,,\qquad
\Delta_3=4x^2\bigl(\Lambda+3\mu\bigr)\,,
\end{align}
so $\Delta_2$ is the third component $\{\Lambda+4\mu=0\}$ of the degeneration
divisor \eqref{eq:degeneration-divisor-cut}, whereas $\Delta_3=0$ is \emph{not} a
degeneration of the curve, as we will see that the second order Picard-Fuchs equation only have the $\Delta_2$ factor, $\Delta_3$ is absent there. In the variable $\tau=\Lambda/4\mu$ it is the point $\tau=-3/4$ at which the twist punctures $z=\pm i\sqrt\tau$ collide with the two
extra punctures $z=\pm\sqrt3/2$ generated by the Baikov factor $(4z^2-3)$. It is therefore a singularity of the twisted system and not of the underlying geometry and starts coming from $\mathcal{O}(\epsilon)$, and
we return to in the later section Sec.~\eqref{sec:new12}.}\\
The remaining polynomial numerators are denoted by
\begingroup
\small
\begin{equation}
\begin{aligned}
\mathcal P^{(0)}_{23}
&=8m^6(2x^8+13x^6+18x^4+13x^2+2)
  +8m^4s(x^4+1)(x^4+4x^2+1)
  +m^2s^2(x^4-1)^2\,,\\
\mathcal P^{(1)}_{21}
&=8m^4x^2(3x^4+2x^2+3)
  -4m^2s(x^8-x^6+8x^4-x^2+1)
  -s^2(x^2-1)^4\,,\\
  \mathcal P_{21}^{(2)}
&=(x^2-1)^2(2m^2+s)
\left[4m^2(1+x^4)+s(1-6x^2+x^4)\right]\,,\\
\mathcal P^{(1)}_{23}
&=8m^6(2x^8-x^6+6x^4-x^2+2)
  +4m^4s(2x^8-11x^6+10x^4-11x^2+2)
  +m^2s^2(x^2-1)^2(x^4-10x^2+1)\,,\\
\mathcal P^{(1)}_{33}
&=-8m^4(4x^4-6x^2+4)
  -8m^2s(2x^4-5x^2+2)
  -2s^2(x^2-1)^2\,.
\end{aligned}
\label{eq:de-numerators}
\end{equation}
\endgroup
In this basis, the coefficient matrices have the block-diagonal form
\begin{equation}
    \mathcal A^{(r)}(x)=
    \begin{pmatrix}
    B^{(r)}(x)&0\\
    0&C^{(r)}(x)
    \end{pmatrix},
   \qquad r=0,1,2\,.
\label{eq:de-block-form}
\end{equation}
The \(3\times3\) blocks are
\begingroup
\setlength{\arraycolsep}{3pt}
\renewcommand{\arraystretch}{1.35}
\begin{equation}
\resizebox{0.96\textwidth}{!}{$
\displaystyle
B^{(0)}=
\left(
\begin{array}{ccc}
-\dfrac{(x^2+1)(4m^2+s)}{\Delta_1}
&
-\dfrac{2m^2(x^2+1)(4m^2+s)}{\Delta_1}
&
\dfrac{2m^2(x^2+1)(4m^2+s)}{\Delta_1}
\\[1.0em]
\dfrac{2(x^2+1)\left[m^2(x^4+14x^2+1)+4sx^2\right]}{\Delta_1\Delta_2}
&
\dfrac{(x^2+1)(4m^2+s)\left[16m^2x^2-s(x^2-1)^2\right]}{\Delta_1\Delta_2}
&
-\dfrac{2(x^2+1)\mathcal P^{(0)}_{23}}{\Delta_1\Delta_2\Delta_3}
\\[1.0em]
0
&
0
&
-\dfrac{(x^4-1)(4m^2+s)}{x\Delta_3}
\end{array}
\right)\,,
$}
\label{eq:de-B0}
\end{equation}
\vspace{-0.5em}
\begin{equation}
\resizebox{0.96\textwidth}{!}{$
\displaystyle
B^{(1)}=
\left(
\begin{array}{ccc}
\dfrac{2(x^2+1)}{x(x^2-1)}
&
0
&
0
\\[1.0em]
\dfrac{2(x^2+1)\mathcal P^{(1)}_{21}}{\Delta_1\Delta_2\Delta_3}
&
\dfrac{4m^2(x^2+1)^3(4m^2+s)}{\Delta_1\Delta_3}
&
-\dfrac{4(x^2+1)\mathcal P^{(1)}_{23}}{\Delta_1\Delta_2\Delta_3}
\\[1.0em]
\dfrac{2(x^2+1)^3(4m^2+s)}{\Delta_1\Delta_3}
&
\dfrac{2(x^2+1)(4m^2+s)\left[4m^2(x^4+1)+s(x^2-1)^2\right]}{\Delta_1\Delta_3}
&
\dfrac{(x^2+1)\mathcal P^{(1)}_{33}}{\Delta_1\Delta_3}
\end{array}
\right)\,,
$}
\label{eq:de-B1}
\end{equation}
\vspace{-0.5em}
\begin{equation}
    B^{(2)}(x)
=
\begin{pmatrix}
0 & 0 & 0 \\[2mm]
-\dfrac{4(x^2+1)\mathcal P_{21}^{(2)}}
{\Delta_1\Delta_2\Delta_3}
& 0 & 0 \\[4mm]
-\dfrac{2(x^2+1)(x^2-1)(4m^2+s)}
{m^2x\Delta_3}
& 0 & 0
\end{pmatrix}\,.
\end{equation}
\endgroup
The \(2\times2\) blocks are
\begingroup
\setlength{\arraycolsep}{4pt}
\renewcommand{\arraystretch}{1.35}
\begin{equation}
\resizebox{0.74\textwidth}{!}{$
\displaystyle
C^{(0)}=
\left(
\begin{array}{cc}
\dfrac{32m^2x(x^2+1)}{(x^2-1)\Delta_2}
+\dfrac{x^4+6x^2+1}{x-x^5}
&
-\dfrac{4m^2(x^2-1)(x^2+1)^2(4m^2+s)}{\Delta_2\Delta_3}
\\[1.0em]
0
&
-\dfrac{(x^4-1)(4m^2+s)}{x\Delta_3}
\end{array}
\right)\,,
$}
\label{eq:de-C0}
\end{equation}
\vspace{-0.5em}
\begin{equation}
\resizebox{0.74\textwidth}{!}{$
\displaystyle
C^{(1)}=
\left(
\begin{array}{cc}
\dfrac{24m^2x(x^2+1)}{(x^2-1)\Delta_3}
&
-\dfrac{8m^2(x^2-1)(x^2+1)^2(4m^2+s)}{\Delta_2\Delta_3}
\\[1.0em]
-\dfrac{2(x^2-1)(4m^2+s)}{\Delta_3}
&
\dfrac{24m^2x(x^2+1)}{(x^2-1)\Delta_3}
\end{array}
\right)
$}
\label{eq:de-C1}
\end{equation}
\endgroup
and $C^{(2)}$ is a vanishing matrix.
\subsection{Block decomposition and Picard--Fuchs factorization}
As one can immediately notice, the sector is reducible into two sub-sectors. Therefore the integral $\mathcal{F}^{\textrm{cut}}_{\alpha}$ is the part of $3\times 3$ sub-block. Therefore, any integral inside the block satisfies a third-order homogeneous differential equation. After some messy but straightforward algebra, one can show that the integral $\mathcal{F}^{\textrm{cut}}_{\alpha}$ satisfies the following second-order inhomogeneous differential equation at $\epsilon=0$ (however, the source belongs to the same sector),
\begin{align}
    \begin{split}
\mathcal{L}^{(2)}_x\mathcal{F}^{\textrm{cut}}_{\alpha}=-\frac{24m^4(4m^2+s)(1+x^2)^2}{\Delta_2\Delta_3}\mathcal{F}_{\alpha+\mathbf e_4}^{\textrm{cut}}\label{4.13hh}
    \end{split}
\end{align}
where the second-order operator is given by,
\begin{align}
    \begin{split}
        \mathcal{L}_x^{(2)}\equiv \partial_x^2+a_{\textrm{ell}}(x)\partial_{x}+b_{\textrm{ell}}(x)
    \end{split}
\end{align}
with,
\begin{align}
    \begin{split}
        a_{\textrm{ell}}(x)=\partial_x\log\left(\frac{(x^2-1)\Delta_2(x)}{x(1+x^2)}\right),\,\,\,b_{\textrm{ell}}(x)=\frac{(4m^2+s)(1+x^2)^2}{x^2\Delta_2(x)}\,.
    \end{split}
\end{align}
Now, let us concentrate on the third master integral of the first irreducible block: $\mathcal{F}^{\textrm{cut}}_{\alpha+\mathbf{e}_4}$. As one can immediately notice, this integral is completely decoupled and can be solved independently.
\begin{align}
    \partial_x\mathcal{F}^{\textrm{cut}}_{\alpha+\mathbf{e}_4}=-\frac{(x^4-1)(4m^2+s)}{x\Delta_3}\mathcal{F}^{\textrm{cut}}_{\alpha+\mathbf{e}_4}\equiv \partial_x (\log \frac{x}{
\sqrt{\Delta_3(x)}
    })\mathcal{F}^{\textrm{cut}}_{\alpha+\mathbf{e}_4}\implies \mathcal{F}^{\textrm{cut}}_{\alpha+\mathbf{e}_4}=c_3 \frac{x}{
\sqrt{\Delta_3(x)}
    }\,.
\end{align}
Therefore, \eqref{4.13hh} takes the form,
\begin{align}
    \mathcal{L}_x^{(2)}\mathcal{F}_{\alpha}^{\textrm{cut}}=-c_3\frac{24m^4(4m^2+s)x(1+x^2)^2}{\Delta_2\Delta_3^{3/2}}\,.\label{4.17hh}
\end{align}
Therefore, the RHS of \eqref{4.17hh} will be annihilated by the following order one operator,
\begin{align}
    \mathcal{L}_{x}^{(1)}\equiv \partial_x-\vec\partial_x\left(\log \frac{x(1+x^2)^2}{\Delta_3^{3/2}\Delta_2}\right)\,.
\end{align}
Note that $\vec\partial_x$ denotes the fact that the derivative only acts on the first term after it. Therefore, $\mathcal{F}^{\textrm{cut}}_{\alpha}$ satisfies the following third order Picard-Fuchs differential equation,
\begin{align}
    \mathcal{L}^{(3)}_x\mathcal{F}^{\textrm{cut}}_{\alpha}=0,\,\qquad\mathcal{L}_x^{(3)}=\mathcal{L}_x^{(1)}\circ\mathcal{L}_x^{(2)}
\end{align}
and as anticipated, it factorizes into two irreducible differential operators of order one and two, respectively. 
\subsection*{Elliptic period solutions and the matching with the Baikov representation.}
Two of the linearly independent solutions of $\mathcal{L}_x^{(2)}f(x)=0$ can be found by solving the second-order differential equation\footnote{\textcolor{black}{$\mathcal L^{(3)}_x f=0$ has three linearly independent solutions. Two of them solve
$\mathcal L^{(2)}_x f=0$. The third, $f_1$, solves $\mathcal L^{(2)}_x f_1=g$ with
$\mathcal L^{(1)}_x g=0$, i.e.\ $g\propto x(1+x^2)^2/(\Delta_2\Delta_3^{3/2})$, from \eqref{4.17hh}. The master integral can therefore be written as $\mathcal F^{\rm
cut}_\alpha=c_1^{(\alpha)}\Pi_1+c_2^{(\alpha)}\Pi_2+c_3^{(\alpha)}f_1$.}},
\begin{align}
    \Pi_2=i x\frac{K({\tilde\tau}(x))}{\sqrt{{\Delta_2}(x)}},\,\,\quad\Pi_1=\,x\frac{K(1-{\tilde\tau}(x))}{\sqrt{{\Delta_2}(x)}}\,,\quad \tilde\tau(x)=\frac{(1-x^2)^2(4m^2+s)}{{\Delta_2(x)}}\,.\label{4.20j}
\end{align}
We have used the fact that $\Delta_2(x) = 4x^2(\Lambda + 4\mu)$ from Sec.~(\ref{newsec5}). These are the Baikov periods of Sec.~(\ref{new:sec2.2.3}) in disguise. With
$\gamma=(1+x^2)/2x$ one has $\gamma^2-1=(1-x^2)^2/4x^2$, hence
\begin{align}
\Delta_2(x)=4x^2(\Lambda+4\mu)\,,\qquad
\tilde\tau=\frac{\Lambda}{\Lambda+4\mu}=\frac{\tau}{1+\tau}\,,
\qquad \tau=\frac{\Lambda}{4\mu}\,,
\end{align}
Applying the Pfaff transformation $K(-\rho)=(1+\rho)^{-1/2}K\big(\rho/(1+\rho)\big)$,
with $\rho=4\mu/\Lambda$ for $\Pi_1$ and $\rho=\Lambda/(4\mu)$ for $\Pi_2$, then gives,
exactly,
\begin{align}
\Pi_1=\tfrac12\,\Pi_{\delta_R}\,,\qquad \Pi_2=\tfrac12\,\Pi_{\delta_I}\,.
\end{align}
The factor $1/2$ is the overall normalization.

\section{Intersection theory and master integrals on maximal cuts} \label{sec:intersec}
The differential system above may be obtained by performing an IBP reduction using conventional computer algebra software, which primarily relies on the Laporta algorithm. However, the goal of this paper is not only to compute the integrals but also to study the geometric structure of the family of integrals. From this perspective, integration-by-parts identities can be interpreted as equivalence relations between twisted differential forms.  Moreover, the intersection theory of twisted differential forms provides an analytical method for performing the reduction and allows the number of master integrals to be determined purely from geometric considerations.

\subsection{Twisted cohomology as IBP reduction on the maximal cut}
Intersection theory gives a systematic way to identify a minimal basis of master integrals on maximal cuts. In the Baikov representation, once the cut conditions are imposed, the remaining
integral takes the form
\begin{align}
    I=\int_{\mathcal{C}} \mathcal{U}(z)\,\varphi(z)\,,
    \qquad 
    z=(z_1,\ldots,z_m)\,,
    \qquad 
    \varphi(z)=\hat{\varphi}(z)\,d^m z
\end{align}
where $\mathcal{U}(z)$ is the multivalued Baikov weight, typically a non-integer power of the Baikov polynomial, and $\varphi$ is a rational top form. The contour $\mathcal{C}$ is a twisted cycle
specified by the cut and by the branch structure of $\mathcal{U}(z)$. The equivalence follows from the vanishing of total derivatives. For any $(m-1)$-form
$\xi$, assuming no boundary contribution,
\begin{align}
    0=\int_{\mathcal{C}} d(\mathcal{U}\,\xi)
    =
    \int_{\mathcal{C}}\mathcal{U}\left(d+d\log\mathcal{U}\wedge\right)\xi
    =
    \int_{\mathcal{C}}\mathcal{U}\,\nabla_{\omega}\xi\,,\label{4.2jj}
\end{align}
with
\begin{align}
    \nabla_{\omega}:=d+\omega\wedge\,,
    \qquad 
    \omega:=d\log\mathcal{U}\,.\label{4.3aa}
\end{align}
Hence
\begin{align}
    \varphi \sim \varphi+\nabla_{\omega}\xi\,,
    \qquad
    \int_{\mathcal{C}}\mathcal{U}\,\varphi
    =
    \int_{\mathcal{C}}\mathcal{U}\left(\varphi+\nabla_{\omega}\xi\right)\,.
\end{align}
Thus, the relevant objects are twisted cocycles, namely, equivalence classes of differential forms
modulo twisted exact terms,
\begin{align}
    {}_{\omega}\langle \varphi|
    :
    \qquad
    \varphi\sim\varphi+\nabla_{\omega}\xi\,.
\end{align}
As we will be dealing with integrals, we only consider the elements of the twisted cohomology group.

\textcolor{black}{\begin{align}
H^{k}(X,\nabla_{\omega})=Z^{k}(X,\nabla_{\omega})/B^{k}(X,\nabla_{\omega})\,,
\end{align}
with,
\begin{align}
\begin{split}
& Z^{k}(X,\nabla_{\omega})=\{k-\textrm{forms} \,\,\varphi\,\textrm{on} \,X:
\nabla_{\omega}\varphi=0\}\,,\\ &
B^{k}(X,\nabla_{\omega})=\{k-\textrm{forms} \,\,\nabla_{\omega}\xi: \xi \,\,
\textrm{a}\,(k-1)-\textrm{form on }X\}\,.
\end{split}
\end{align}
Note that all twisted cohomology groups are finite-dimensional and only the middle group is non-vanishing, so that it is enough to work with $k=m$. The elements of that group are called twisted cocycles. On the other hand, the integration contours are twisted cycles from
\begin{align}
Z_{m}(X,{L}^\vee)=\{m-\textrm{chains}\,\,\mathcal{C}\otimes
\mathcal{U}|_{\mathcal{C}}:\partial(\mathcal{C}\otimes
\mathcal{U}|_{\mathcal{C}})=0\}
\end{align}
where $\mathcal{C}\otimes \mathcal{U}|_{\mathcal{C}}$ denotes an $m$-chain
$\mathcal{C}$ on $X$ together with a chosen branch of $\mathcal{U}$, and
${L}^{\vee}$ denotes the locally constant sheaf (the \emph{local system})
generated by the multivalued function $\mathcal{U}$. In most situations, we consider closed contours and want to discard boundary terms, so we quotient by
$ B_{m}(X,{L}^{\vee})=\{\partial(\mathcal{C}\otimes\mathcal{U})\ :\
\mathcal{C}\ \textrm{an}\ (m+1)-\textrm{chain}\}$ and take twisted cycles from the
twisted homology group
\begin{align}
H_{m}(X, {L}^{\vee})= Z_{m}(X,{L}^\vee)/B_{m}(X,
{L}^{\vee})\,,\qquad \textcolor{black}{X=\mathbb{CP}^{M-N}\setminus\{\textrm{punctures}\}}\,.
\end{align}}
However, we will be more precise about the definitions in Sec.~(\ref{Sec:6.4}). Also, in this section, we will discuss what really happens if we take open contours with non-vanishing boundary terms. Hence, one, in principle, should be more careful about the choice of twisted cycles.  \\\\
In the language of Feynman integrals, \eqref{4.2jj} is the cohomological version of IBP identities on the maximal cut. The cut integral is then a bilinear pairing between a twisted cocycle and a twisted cycle,
\begin{align}
    \langle \varphi|\mathcal{C}]
    :=
    \int_{\mathcal{C}}\mathcal{U}\,\varphi\,.
\end{align}
If the dimension of the twisted cohomology is $|\chi|$, then $|\chi|$ counts the number of independent master integrals on the cut. Choosing a basis $\{\langle e_i|\}_{i=1}^{|\chi|}$ and a dual basis $\{|h_i\rangle\}_{i=1}^{|\chi|}$, with intersection matrix  $  \mathbf{C}_{ij}:=\langle e_i|h_j\rangle ,$ any cocycle can be projected onto the master basis as
\begin{align}
    \langle \varphi|
    =
    \sum_{i,j=1}^{|\chi|}
    \langle \varphi|h_j\rangle
    \left(\mathbf{C}^{-1}\right)_{ji}
    \langle e_i|\,.
\end{align}
This is precisely the (twisted-) cohomological version of basis integral reduction. The master integrals are therefore represented by the basis cocycles $\langle e_i|$, and the reduction coefficients are computed from
intersection numbers. The bases $\langle e_i|$ and $|h_i\rangle$ can be chosen conveniently. We can choose them in a monomial basis as,
\begin{align}
    \langle e_i|\equiv\varphi_i:=z^{i-1}dz,\,\forall i=1,\cdots,|\chi|\,.
\end{align}
In this basis, $|e_i\rangle=|h_i\rangle$. There are other choices of basis that can be used conveniently. The details can be seen in \cite{Frellesvig:2019kgj}. 
\subsection{Critical points and the basis of master integrals}
However, the most important point is to understand how one determines the dimension of the cocycle basis. For generic kinematics, it is determined by the critical points of the logarithmic Baikov weight. Equivalently, it may be deduced from the punctures of the integration domain, which is generally the full complex plane or its compactification to the Riemann sphere. We will see that applying those criteria, we get five master integrals as anticipated from the previous section.
In that context, let's define a logarithmic one-form
\begin{align}
    \omega:=d\log \mathbfcal{B}(\boldsymbol{z})^{\nu_{B}}\,.
\end{align}
 Remarkably, as noted in \cite{Mastrolia:2018uzb}, the logarithm of $\mathbfcal{B}(\boldsymbol{z})^{\nu_B}$ is a Morse function as its critical points are non-degenerate. Therefore, using the Morse theory, one can deduce that the dimension of the cocycle basis,
\begin{align}
    |\chi|=\#\{z:\,\,\omega(z)=0\}\,.
\end{align}
Also define a set $\mathcal{P}$ as a set of poles of the one form $\omega$,
\begin{align}
    \mathcal{P}: \{z\,|\,z \,\,\textrm{is a pole of } \,\,\omega\}\,.
\end{align}
Also note that if the one form $\omega$ has a nonvanishing residue at $z=\infty$, that also has to be included in the set. With those remarks, let us investigate the cut integral in \eqref{bcut} in arbitrary dimension. We already reduced the integral to a single variable integral. Therefore, we will only deal with one-form and the integral can be recast as the following bilinear pairing with $z=\frac{u}{\sqrt{\mu}}$,
\begin{align}
    \langle \varphi|\mathcal{C}]:=\int_{\mathcal{C}} \mathcal{U}(z|\tau)\varphi_0\,,\qquad\varphi_0=dz 
\end{align}
with, 
\begin{align}
    \mathcal{U}(z|\tau)=\mathcal{N}(\epsilon,\mu,s)\tau^{\epsilon}(1-z^2)^{\epsilon-\frac{1}{2}}(z^2+\tau)^{-\epsilon-\frac{1}{2}}(4z^2-3)^{-\epsilon}\,,\quad \tau=\frac{(4\mu+s)(\gamma^2-1)}{4\mu}
\end{align}
where,
\textcolor{black}{
\begin{align}
    \mathcal{N}(\epsilon,\mu,s)=\frac{2^{4\epsilon}\mu^{-1/2-\epsilon}
       e^{-i\pi\epsilon}\Gamma(1+\epsilon)\sin(\pi\epsilon)}
 {\pi^{3/2}(4\mu+s)^\epsilon\sqrt{s}\,
       \epsilon\,\Gamma(\frac12-\epsilon)}
\end{align}
where it is straightforward to see that at $d=4$, the normalization becomes $\mathcal{N}(0,\mu,s)=1/(\pi\sqrt{\mu s})$ and the subscript $z$ is implied.}
Now define the logarithmic one form as \footnote{Later in the discussion, we will only use the symbol $d\log\mathcal{U}$ in place of $d_z\log\mathcal{U}$.},
\begin{align}
    \begin{split}
        \omega:=d_z\log \mathcal{U}(z|\tau)&=d_z\left[\epsilon \,\log\tau+\left(\epsilon-\frac{1}{2}\right)\log(1-z^2)+\left(-\epsilon-\frac{1}{2}\right)\log(z^2+\tau)-\epsilon\, \log(4z^2-3)\right]\\ &
        =-\left[\frac{2z\left(\epsilon-\frac{1}{2}\right)}{1-z^2}+\frac{2z\left(\epsilon+\frac{1}{2}\right)}{z^2+\tau}+\frac{8\,\epsilon\, z}{4z^2-3}\right]dz\,.\label{4.13}
    \end{split}
\end{align}
From \eqref{4.13}, one can immediately deduce that $\omega=0$ has five non-degenerate (for generic $\epsilon$ and $\tau$) solutions corresponding to,
\begin{align}
    z\,\mathcal{Q}(z^2)=0,\quad \mathcal{Q}(z^2)=8(\epsilon+1)z^4+(4\tau-16\epsilon-10)z^2+(6\epsilon-2\epsilon\,\tau-3\tau+3)\,.
\end{align}
One can deduce the set of poles of the one-form $\omega$:
\begin{align}
    \begin{split}
        \mathcal{P}: \left\{1,-1,\sqrt{-\tau},-\sqrt{-\tau},\frac{\sqrt{3}}{2},-\frac{\sqrt{3}}{2},\infty\right\}\,.
    \end{split}
\end{align}
The number of master integrals is equal to the absolute value of the Euler characteristic $\chi$, of the space $\mathbb{CP}^{1}\backslash\mathcal P$ on which the one form is defined, where the number of poles $|\mathcal{P}|$ is also be related to the number of master integrals as, $|\mathcal{P}|=|\chi|+2$ \footnote{\textcolor{black}{As noted in \cite{Frellesvig:2019kgj}, the relation holds provided none of the residues of $\omega$ is a non-negative integer, which is the case here for generic $\epsilon\,.$ At $\epsilon=0$ one has
$\mathrm{Res}_{z=\infty}\,\omega=2(\epsilon+1)\to2$, a non-negative integer, and
simultaneously the punctures $z=\pm\sqrt3/2$ acquire vanishing residue and cease to
be punctures at all. The count therefore, does not commute with $\epsilon\to0\,.$ and will be discussed Sec.~(\ref{sec:period-d4}) }}.
\subsection{Intersection numbers as basis projection}
Once the dimension of the cocycle basis is fixed, the IBP reduction coefficients of an arbitrary integral can be determined through intersection numbers, which are evaluated as sums of local residues at the singular points of the logarithmic one-form. We therefore briefly introduce the relevant definition before proceeding with the explicit computation.

Given two univariate one-forms, $\varphi_L$ and $\varphi_R$, and a logarithmic one-form $\omega$ defining the twist, their intersection number is defined as \footnote{The general formula for intersection number  of two twisted $m$-form is given by,
\begin{align}
    \langle \varphi_L \mid \varphi_R \rangle_{\omega}=\frac{1}{(2\pi i)^m}\sum_{\alpha,\beta=1}^{|\chi|}\int_{\mathcal{C}_{L,\alpha}} e^{\int \omega}\varphi_L I_{h,\beta\alpha}^{-1} \int_{\mathcal{C_{R,\beta}}} e^{-\int \omega}\varphi_{R}\,, \label{4.18kk}
\end{align}
where $e^{\int \omega}$ is the multivalued Baikov polynomial, the sets of twisted cycles $\{\mathcal{C}_{L,\alpha},\mathcal{C}_{R,\beta}\}$ form their respective homology group, and $I_h$ is the intersection matrix, whose elements are intersection numbers between those cycles. We will later see that the relation in \eqref{4.18kk} is nothing but the Riemann period relation. 
}
\begin{align}
\langle \varphi_L \mid \varphi_R \rangle_{\omega}
:=
\sum_{p\in\mathcal{P}}
\textrm{Res}_{z=p}\left(\psi_p \,\varphi_R\right)\label{4.18jj}
\end{align}
where $\mathcal{P}$ denotes the set of singular points of the twisted differential system. For each $p\in\mathcal{P}$, the function $\psi_p$ is a local solution of the inhomogeneous differential equation
\begin{align}
\nabla_{\omega}\psi_p=\varphi_{L,p}\,,
\qquad
\nabla_{\omega}=d+\omega\wedge\,,
\end{align}
expanded in a neighbourhood of $z=p$. Thus, the global intersection number is obtained by summing the residue contributions computed locally around all singular points.

\subsection{Differential  equation from cohomological projection in non-dlog basis}
In order to construct the differential equation for the master integrals, we first define the monomial cocycle basis (master integrals):
\begin{align}
    \begin{split}
      I_i(\tau)= \int_{\mathcal{C}} \mathcal{U}(z|\tau) \varphi_i\,.
    \end{split}
\end{align}
As one can notice, the basis cocycle $\varphi_i$ does not depend on $\tau$ explicitly \footnote{Also note that now we have kinematic variable $(\mu,s,\tau)$ unlike the previous analysis we used $(\mu,s,x)$, as we go on we can see the $\tau$ parameter is a more suitable choice for calculation. In the previous section we study the differential equation w.r.t the kinematic parameter $x$. In this section we will study the more convenient $\tau$ parameter as the choice of parameter in the differential system.}. Therefore, the differential equation with respect to the kinematic variable $\tau$ takes the following form\footnote{An important point is that the $\delta_I$-cycle surrounds the branch points
$z=\pm i\sqrt{\tau}$, which move when the kinematic variable $\tau$ changes.
Therefore, when differentiating the corresponding period with respect to
$\tau$, one must include both the change of the integrand and the change of
the integration contour. Writing
\begin{align}
    \mathcal{U}\varphi_i=f_i\,dz\,,
    \qquad
    v=\left.\partial_\tau z\right|_t=\frac{z}{2\tau}\,,\nonumber
\end{align}
the derivative is
\begin{align}
    \partial_\tau I_i
    =
    \oint_{\mathcal{\delta}_I(\tau)}
    \left[
        \partial_\tau\!\left(\mathcal{U}\varphi_i\right)
        +
        \mathcal{L}_v\!\left(\mathcal{U}\varphi_i\right)
    \right]\,.\nonumber
\end{align}
The first term describes the explicit $\tau$-dependence of the integrand,
while the second term accounts for the motion of the contour. For the
one-form  $ \alpha:=\mathcal{U}\varphi_i=f_i\,dz,$
Cartan's formula gives
\begin{align}
    \mathcal{L}_v\alpha=d_z(vf_i)\,.\nonumber
\end{align}
For an ordinary positively oriented loop $\delta_I$, based at $z_0$, the
integral of this total derivative does not generally vanish because $vf_i$
is multivalued. Instead,
\begin{align}
    \oint_{\delta_I}d_z(vf_i)
    =
    \left(M_I-1\right)(vf_i)(z_0)\,,
    \qquad
    M_I
    =
    e^{2\pi i\,2\left(-\epsilon-\frac{1}{2}\right)}
    =
    e^{-4\pi i\epsilon}\,.\nonumber
\end{align}
For generic $\epsilon$, one has $M_I\neq1$, and hence this contribution is nonzero. Therefore, the $B$-cycle must be defined together with the
appropriate branch information so that the total twisted derivatives integrate
to zero:
\begin{align}
    d_z(vf_i)
    &=
    \mathcal{U}\,
\nabla_\omega\!\left(\iota_v\varphi_i\right)\,,   
    \int_{\delta_I}
    \mathcal{U}\,
    \nabla_\omega\!\left(\iota_v\varphi_i\right)=0\,.\nonumber
\end{align}
This is the prescription used in
\cite{Mastrolia:2018uzb,Frellesvig:2019kgj} and stated in \eqref{4.2jj}. At $\epsilon=0$, the monodromy becomes trivial, $M_I=1,$ and $vf_i=vz^{i-1}/y$ becomes single-valued on the elliptic curve  $ y^2=(1-z^2)(z^2+\tau).$ The integral of the total derivative then vanishes. In the calculations below, however, we derive the differential equation using the physical cycle $[-1,1]$, whose endpoints do not depend on $\tau$. Therefore, the
complication associated with the moving $\delta_I$-cycle does not arise.}
\begin{align}
   \partial_{\tau}I_i=\int_{\mathcal{C}} \mathcal{U}(z|\tau)\, \partial_{\tau}\log\mathcal{U}(z|\tau)\varphi_i=\int_{\mathcal{C}}\mathcal{U}(z|\tau)\,\Phi_i\,,\quad \Phi_i=\partial_{\tau}\log\mathcal{U}(z|\tau)\varphi_i\,.\label{4.20jj}
\end{align}
Now the primary task is to reduce the modified cocycle $\Phi_i$ to the original basis cocycle i.e, we want to find a matrix $\Omega$ such that (with explicit bra-ket notation),
\begin{align}
    \langle\Phi_i|=\sum_{j=1}^{|\chi|=5}\Omega_{ij}\langle\varphi_j|\,.
\end{align}
Now taking the inner product with the dual cocycles, we get,
\begin{align}
    \begin{split}
        \langle\Phi_i|\varphi_k\rangle:=\mathbf{F}_{ik}=\sum_{j=1}^{|\chi|=5}\Omega_{i,j}\langle\varphi_j|\varphi_k\rangle=\sum_{j=1}^{|\chi|=5}\Omega_{ij} \mathbf{C}_{jk}\implies  \Omega=\mathbf{F}\,\mathbf{C}^{-1}\,.
    \end{split}
\end{align}
So computing the differential equation now boils down to computing the following two sets of intersection numbers: $    \{\langle\Phi_i|\varphi_k\rangle_{\omega}\}$ and $ \{\langle\varphi_i|\varphi_k\rangle_{\omega}\}$. Now, we have more simplification,
\begin{align}
 \Phi_i=   \partial_{\tau} \log\mathcal{U}(z|\tau)\varphi_i =\left(\frac{\epsilon}{\tau}-\frac{\left(\epsilon+\frac{1}{2}\right)}{z^2+\tau}\right)z^{i-1} dz\,.\label{4.29jj}
\end{align}
Therefore, we only need to compute the following intersection number (with $a,b\ge 0$),
\begin{align}
    \begin{split}
    I(a,b|\lambda_1,\lambda_2):=   \left\langle \frac{z^a}{(z^2+\tau)^{\lambda_1}}\Bigg | \frac{z^b}{(z^2+\tau)^{\lambda_2}}\right\rangle_{\omega}=\langle \widetilde{\varphi}_L^{a,\lambda_1}|\widetilde{\varphi}_{R}^{b,\lambda_2}\rangle\,.
    \end{split}
\end{align}
To compute the intersection numbers, we first need to know the $0$-form $\psi_p$ corresponding to the one form $\widetilde{\varphi}$ such that,
\begin{align}
    \nabla_{\omega} \psi^{a,\lambda}_{(p)}=\widetilde{\varphi}^{a,\lambda}_{(p)}\,.
\end{align}
While one can, in principle, solve the differential equation directly to find the function $\psi$. But, it would be easier to consider a Laurent expansion of $\psi^{a,\lambda}$  around the poles of $\omega$, $z=p$, denoted as $\psi^{a,\lambda}_{(p)}$  and isolate the correct pole to compute the residue. Now, defining $\zeta=z-p$ and taking the following ansatz,
\begin{align}
    \psi_p=\sum_{j=\textrm{ord}_p(\widetilde{\varphi}_L^{a,\lambda_1})+1}^{-\textrm{ord}_p(\widetilde{\varphi}_R^{b,\lambda_2})-1}\psi_{(p)}^{(j)}\,\zeta^j+\cdots\,.
\end{align}
Also, it is important to note that a given pole $p$ will contribute if $j_\textrm{min}\le j_\textrm{max}$ and the Laurent expansion exists if $\textrm{Res}_{z=p}(\omega)$ is not a non-positive integer \cite{Frellesvig:2019kgj}. Now, for $p\in \mathcal{P}$ , and $\lambda_1=0$, $\textrm{ord}_p(\widetilde{\varphi}_{L,R})=0$. Therefore, $j_{\textrm{min}}>j_{\max}$, implying every finite pole of $p\in \mathcal{P}$ does not contribute to the Laurent expansion. However, because of the presence of factors like $(z^2+\tau)^{\lambda_1}$, computations got more intricate. Now for $\lambda_{1}=1$,  $\textrm{ord}_{p=\pm\sqrt{-\tau}}(\widetilde{\varphi}_{L,R})=-1$ implying $j_{\min}=0>j_{\max}=-1$. \textcolor{black}{Therefore, the denominator factor does not alter the fact that the finite poles do not contribute. For generic $\lambda_1$ and $\lambda_2$ one has $j_{\min}=1-\lambda_1$ and $j_{\max}=\lambda_2-1$, so the contribution condition
$j_{\min}\le j_{\max}$ reads $\lambda_1+\lambda_2\ge2\,.$ A single Laurent coefficient survives when it is exactly equal. In our computation, we need only
$\lambda_1=1,\lambda_2=0$, for which $\lambda_1+\lambda_2=1\,.$ The finite poles therefore give no contribution, and the only contributing pole is the one at infinity.} \\\\
\textbf{Local expansion of $\widetilde{\varphi}_{L}^{a,\lambda_1}$ around $z=\infty$:}\\
Doing the change of variable : $\varpi=\frac{1}{z}$ the one form takes the following form,
\begin{align}
    \widetilde{\varphi}_{L}^{a,\lambda_1}=- \varpi^{-a-2+2\lambda_1}(1+\tau\varpi^2)^{-\lambda_1}d\varpi\,.
\end{align}
Also it is worth noting that the one form $\omega=d\log\mathcal{U}$  has the following behaviour around $z=\infty$:
\begin{align}
    \begin{split}
        \omega=\left[\frac{2(\epsilon+1)}{\varpi}+\sum_{k\ge 1}\beta_k \varpi^{2k-1}\right]d\varpi,\qquad \beta_k=(1-2\epsilon)-(2\epsilon+1)(-1)^{k-1}\tau^k+2\epsilon \left(\frac{3}{4}\right)^k\,.
    \end{split}
\end{align}
\textbf{Solving $\nabla_{\omega}\psi_{\infty}^{a,\lambda_1}=\widetilde{\varphi}^{a,\lambda_1}$:} We need to solve for $\psi_{\infty}$ from the following.
\begin{align}
\nabla_{\omega}\psi_{\infty}^{a,\lambda_1}=\widetilde{\varphi}^{a,\lambda_1},\qquad \nabla_{\omega}=d+\omega\wedge\,.\label{4.28jj}
\end{align}
The idea is to substitute the ansatz for $\psi_\infty$ in the LHS and match the power of $\varpi$ on both sides to identify the coefficients. Let,
\begin{align}
    \psi_\infty^{a,\lambda_1}=\sum_{j=j_{\min}}^{j_{\max}} \psi_\infty^{(j)} \varpi^{j},\,\quad j_{\min}=-a-1+2\lambda_1,\,j_{\max}=-\mathrm{ord}_{\infty}\!\left(\widetilde\varphi_R^{b,\lambda_2}
\right)-1=b+1\,,\ \lambda_2=0\,.\label{4.29hh}
\end{align}
The LHS of \eqref{4.28jj} becomes,
\begin{align}
    \nabla_\omega \psi_{\infty}^{a,\lambda_1}=\left(\frac{d\psi_\infty^{a,\lambda_1}}{d\varpi}+\left[\frac{2(\epsilon+1)}{\varpi}+\sum_{k\ge 1}\beta_k \varpi^{2k-1}\right]\psi_{\infty}^{a,\lambda_1}\right)d\varpi\,.\label{4.30gg}
\end{align}
Now, using the generalized binomial theorem, we have the following expansion of the one form $\widetilde{\varphi}_L^{a,\lambda_1}$.
\begin{align}
    \begin{split}
        \widetilde{\varphi}_L^{a,\lambda_1}=-\varpi^{-a-2+2\lambda_1}\sum_{j\ge0} (-1)^j\frac{(\lambda_1)_j}{j!}(\tau \varpi^2)^j \,d\varpi\,.
    \end{split}
\end{align}
Therefore, the new reduced ansatz is given by,
\begin{align}    \psi_{\infty}^{a,\lambda_1}=\sum_{j=0}^{\left\lfloor\frac{j_{\max}-j_{\min}}{2}\right\rfloor}c_j^{(a,\lambda_1)}\,\varpi^{2j+j_{\min}}\,,\quad c_j^{(a,\lambda_1)}=\psi_\infty^{(2j+j_{\textrm {min})}}\,.\label{4.33jj}
\end{align}
Now, substituting \eqref{4.33jj} into \eqref{4.30gg} we get,
\begin{align}
\begin{split}
 &   \sum_{j=0}c_j^{(a,\lambda_1)}(2j+j_{\textrm{min}})\varpi^{2j+j_{\min}-1}+2(\epsilon+1)\sum_{j=0}c_j^{(a,\lambda_1)}\,\varpi^{2j+j_{\min}-1}+\sum_{j=0}\sum_{k\ge 1}\beta_k c_j^{(a,\lambda_1)}\,\varpi^{2j+j_{\min}+2k-1}\\ &
 =-\sum_{j\ge0} (-1)^j\frac{(\lambda_1)_j}{j!}\tau^j( \varpi)^{2j+j_{\min}-1}\,.
 \end{split}
\end{align}
Now comparing the coefficient of $\varpi^{2j+j_{\min}-1}$ we get the following recursion relation,
\begin{align}
    \begin{split}
    \boxed{
c_j^{(a,\lambda_1)}=\frac{(-1)^{j+1}\frac{(\lambda_1)_j}{j!}\tau^j-\sum_{k=1}^j\beta_kc_{j-k}^{(a,\lambda_1)}}{2j+j_{\textrm{min}}+2\epsilon+2}\,.}\label{4.42hh}
       \end{split}
\end{align}
We next compute the two intersection matrices $F$ and $C$.  Their nonzero entries obey $C_{ij}=F_{ij}=0$ for $i+j$ odd.  We denote these entries by $c_{ij}$ and $f_{ij}$, with their explicit rational forms collected in Appendix~(\ref{app:intersection-entries}).
Then the matrices in the monomial basis are 
\begin{equation}
C=
\begin{pmatrix}
c_{11}&0&c_{13}&0&c_{15}\\
0&c_{22}&0&c_{24}&0\\
c_{31}&0&c_{33}&0&c_{35}\\
0&c_{42}&0&c_{44}&0\\
c_{51}&0&c_{53}&0&c_{55}
\end{pmatrix},
\qquad
F=
\begin{pmatrix}
f_{11}&0&f_{13}&0&f_{15}\\
0&f_{22}&0&f_{24}&0\\
f_{31}&0&f_{33}&0&f_{35}\\
0&f_{42}&0&f_{44}&0\\
f_{51}&0&f_{53}&0&f_{55}
\end{pmatrix}.
\label{eq:intersection-matrices}
\end{equation}
Therefore, the $\Omega$ matrix can be written as,
\begin{align}
    \begin{split}
        \Omega=FC^{-1}:=A_0(\tau)+\epsilon A_1(\tau)\label{4.43mm}
    \end{split}
\end{align}
where, 
\begin{align}
    \begin{split}
        A_0=\frac{1}{g(\tau)}\left(
\begin{array}{ccccc}
 -2 \tau -\frac{7}{2} & 0 & 2-\frac{11}{2 \tau } & 0 & \frac{6}{\tau } \\
 0 & 2-2 \tau  & 0 & -4 & 0 \\
 -\frac{3}{2} & 0 & \frac{11}{2}-2 \tau  & 0 & -6 \\
 0 & \frac{1}{2} (-11 \tau -3) & 0 & 4 \tau  & 0 \\
 \frac{3 \tau }{2} & 0 & -9 \tau -\frac{3}{2} & 0 & 6 \tau  \\
\end{array}
\right),\quad A_1=\frac{1}{g(\tau)}\left(
\begin{array}{ccccc}
 0 & 0 & \frac{8}{\tau }+4 & 0 & -\frac{4}{\tau } \\
 0 & \frac{3}{\tau }-1 & 0 & 4 & 0 \\
 0 & 0 & \frac{3}{\tau }-1 & 0 & 4 \\
 0 & \tau -3 & 0 & \frac{3}{\tau }+7 & 0 \\
 0 & 0 & \tau -3 & 0 & \frac{3}{\tau }+7 \\
\end{array}
\right)
    \end{split}
\end{align}
where, $g(\tau)=(\tau+1) (4 \tau+3)$. \textcolor{black}{As we observe, choosing $\tau$ as a more suitable differentiation variable leads to a differential equation that is linear in $\epsilon$ when derived
through cohomological reduction, in contrast to the form obtained from the
\texttt{LiteRed}-based reduction. Moreover, the Baikov representation naturally
suggests a more convenient parametrization of the kinematics. In particular,
the variables $(\tau,\mu,s)$ provide an independent kinematic coordinate system
in which the differential equations take a significantly simpler form.} Before going to the explicit computation of the master integrals on the maximal cuts, we look at the differential equation in a different cocycle basis, namely the log-basis.
\subsection{Gauss-Manin differential system in the orthonormal dlog basis}
Before going to the discussion on the period matrix, let's discuss another choice of cocycle basis that has been extensively used in the literature: the so-called orthonormal dlog basis. One of the advantages of this choice is that the $C$-matrix becomes orthonormal here. The algorithm of the choice goes as follows: Out of the set of poles of the logarithmic one form $\omega$: $\mathcal{P}=\{z_1,\cdots,z_{\nu+1},z_{\nu+2}\}$, $\nu$ be the number of masters, pick two special poles, say $z_{\nu+1}, z_{\nu+2}$. Then one can construct the $\nu$ one form as,
\begin{align}
    \langle e_i|:=d\log\frac{z-z_i}{z-z_{\nu+1}}\,,\qquad |h_i\rangle:=\textrm{Res}_{z=z_i}(\omega)d\log\frac{z-z_i}{z-z_{\nu+2}}\,.
\end{align}
For convenience, we choose the order of poles as
\begin{align}
    (z_1,\cdots,z_7)=\left(-1,-\frac{\sqrt{3}}{2},\frac{\sqrt{3}}{2},i\sqrt{\tau},-i\sqrt{\tau},\infty,1\right)\,.\label{5.2kk}
\end{align}
Therefore, as evident from \eqref{5.2kk}, we choose $z_6=\infty$ and $z_7=1$ as the special poles. The residues, $r_a:=\textrm{Res}_{z=z_a}\omega$, are as follows,
\begin{align}
    (r_1,\cdots,r_7)=\left(\epsilon-\frac{1}{2},-\epsilon,-\epsilon,-\epsilon-\frac{1}{2},-\epsilon-\frac{1}{2},2(\epsilon+1),\epsilon-\frac{1}{2}\right)\,.
\end{align}
The cocycle and dual cocycle basis for our case is,
\begin{align}
\begin{split}
    \langle e_i|=d\log\frac{z-z_i}{z-z_6}\equiv \frac{dz}{z-z_i},\qquad |h_i\rangle=r_i\left(\frac{1}{z-z_i}-\frac{1}{z-1}\right)dz,\quad i=1,\cdots,5\,.
\end{split}    
\end{align}
Now if \eqref{4.18jj} are intersection numbers between two logarithmic one-forms, then the formula reduces to a simpler form,
\begin{align}
    \langle \varphi_L|\varphi_R\rangle=\sum_{p\in\mathcal{P}}\frac{\res_{p}\, \varphi_L\,\res_{p}\varphi_R}{\res_p\, \omega}\,.
\end{align}
Therefore, for given logarithmic co-cycles and dual-cocycles, we have the following intersection matrix,
\begin{equation}
 C_{ij}
 :=
 \langle e_i|h_j\rangle_\omega
 =
 \sum_{p\in\mathcal P}
 \frac{
 \operatorname{Res}_{p}e_i\,
 \operatorname{Res}_{p}h_j
 }{
 \operatorname{Res}_{p}\omega
 }
 =
 \frac{r_j}{r_i}\,\delta_{ij}
 =
 \delta_{ij}\,.
 \label{4.50jj}
\end{equation}
The proof goes as follows:
\begin{align}
    \res_{z_p}e_i=\delta_{pi}-\delta_{p6},\quad \res_{z_p}h_j=r_j(\delta_{pj}-\delta_{p7}),\quad p=1,\cdots 7
\end{align}
which implies the result of \eqref{4.50jj}, satisfying the fundamental constraint on the $C-$ matrix in the dlog basis. Now, define the master integral in the dlog basis as follows,
\begin{align}
    \begin{split}
     J_i(\tau):=\int_{\mathcal{C}} \mathcal{U}(z|\tau)e_i  ,\qquad \vec J=(J_1,\cdots, J_5)^T\,.
    \end{split}
\end{align}
Taking derivative w.r.t the kinematic variable $\tau$ gives,
\begin{align}
    \begin{split}
        \partial_{\tau}J_i=\int_{\mathcal{C}} \mathcal{U}\hat{\Phi}_i,\qquad \hat{\Phi}_i:=\partial_{\tau}e_i+(\partial_{\tau}\log \mathcal{U})e_i\,,\qquad \partial_\tau e_i=\frac{v_i dz}{(z-z_i)^2}\label{4.53hh}
    \end{split}
\end{align}
where,
\begin{align}
    v_i:=\partial_{\tau}z_i =\left(0,0,0,\frac{z_4}{2\tau},\frac{z_5}{2\tau}\right)\,.
\end{align}
Moreover, for convenience, \eqref{4.13} and \eqref{4.29jj} can be rewritten as,
\begin{align}
\omega=\sum_{a\in{1,\cdots,5,7}}r_a\frac{dz}{z-z_a}\,,\qquad  \partial_\tau \log\mathcal{U}=\frac{\epsilon}{\tau}-\sum_{a\in\{1,\cdots,5,7\}}\frac{r_av_a}{z-z_a}\,.
\end{align}
Now the double pole in \eqref{4.53hh} can be further rewritten by employing the following cohomological  equivalence,
\begin{align}
    \begin{split}
        0\simeq_\omega \nabla_{\omega} \frac{1}{z-z_i}= (r_i-1)\frac{dz}{(z-z_i)^2}+\sum_{a\ne i} r_a\frac{dz}{(z-z_a)(z-z_i)}\,.\label{4.56kk}
    \end{split}
\end{align}
Now using \eqref{4.56kk} the one form $\hat{\Phi}_i$ can be written as,
\begin{align}
    \begin{split}
        \hat{\Phi}_i\simeq_{\omega} \left(\frac{\epsilon}{\tau}+\sum_{a\ne i}\mathbf{u}_{ia}\right)e_i-\sum_{a\ne i}\mathbf{u}_{ia}\vartheta_a\,,\qquad \vartheta_a=\frac{dz}{z-z_a}\,,\qquad \mathbf{u}_{ia}=r_a\frac{v_i-v_a}{z_i-z_a}\,.
    \end{split}
\end{align}
Note that the newly defined one form $\vartheta_a$ has all elements of $e_i$ except $\vartheta_7$ as the extra piece, which can be further removed by the following,
\begin{align}
    \begin{split}
        0\simeq_{\omega} \omega\equiv \nabla_\omega \mathbf{1}=\sum_{j=1}^5r_j e_j+r_7 \vartheta_7\,,\qquad \vartheta_7\simeq_{\omega}-\sum_{j=1}^5 \frac{r_j}{r_7} e_j\,.\label{4.58mm}
    \end{split}
\end{align}
One can convince \eqref{4.58mm} by noting that,
\begin{align}
    \int_{\mathcal{C}} \mathcal{U}\omega=\int_{\mathcal{C}} d(\mathcal{U})=0\,.
\end{align}
Therefore, the Gauss-Manin matrix elements are given by,
\begin{equation}
\boxed{
\begin{aligned}
\Omega^{\textrm{dlog}}_{ij}
=
F_{ij}
=
\langle \hat{\Phi}_i|h_j\rangle
=
\begin{cases}
\displaystyle
-\mathbf{u}_{ij}
+\frac{r_j}{r_7}\mathbf{u}_{i7},
& i\neq j\,,
\\[9pt]
\displaystyle
\frac{\epsilon}{\tau}
+\sum_{a\neq i}\mathbf{u}_{ia}
+\frac{r_i}{r_7}\mathbf{u}_{i7}\,,
& i=j\,.
\end{cases}
\end{aligned}
}
\label{4.60mm}
\end{equation}
As an aside, one can, in principle, find the transformation between two bases. The idea behind the computation of the transformation matrix is to search for the residual function $q_i(z)$ and the coefficients $R_{ik}$ satisfying the following
\footnote{\textcolor{black}{Another important point is that the loop around the physical cut $[-1,1]$ encircles
four punctures, $z=\pm1$ and $z=\pm\sqrt3/2$, since $\sqrt3/2<1$. Its monodromy is
therefore
$\exp\!\big(2\pi i\sum_a r_a\big)
=\exp\!\big(2\pi i\,[\,2(\epsilon-\tfrac12)+2(-\epsilon)\,]\big)=e^{-2\pi i}=1\,.$ The $\epsilon$-dependence of the endpoint pair cancels against that of the two
interior punctures, so $M_A=1$ identically in $\epsilon$, in contrast to the
$B$-cycle, for which $M_I=e^{-4\pi i\epsilon}\ne1$. This is the same cancellation that will make the loaded combination
$\mathcal C_1+\mathcal C_2+\mathcal C_3$ of Sec.~(\ref{sec:Periodmatr}) a closed
twisted cycle which will be discussed later. Moreover, the residual functions are single-valued along the $A$-cycle, so no analogous ambiguity arises in discarding the boundary term.}},
\begin{align}
    \begin{split}
e_i=\nabla_{\omega}q_i(z)+\sum_{k}R_{ik}\varphi_k\,.
    \end{split}
\end{align}
In \texttt{Mathematica}, the search can be done, and we get the following residual functions $q_i(z)$,
\begin{align}
\begin{split}
    q_{i}(z)=\Bigg\{&-\frac{(z-1) \left(4 z^2-3\right) \left(\tau +z^2\right)}{(\tau +1) (2 \epsilon -1)},-\frac{8 \left(2 \sqrt{3} z-3\right) \left(z^2-1\right) \left(\tau +z^2\right)}{3 (4 \tau +3) \epsilon }\,,\\ &\frac{8 \left(2 \sqrt{3} z+3\right) \left(z^2-1\right) \left(\tau +z^2\right)}{3 (4 \tau +3) \epsilon }\,,\frac{i \left(4 z^4-7 z^2+3\right) \left(z+i \sqrt{\tau }\right)}{\sqrt{\tau } (\tau +1) (4 \tau +3) (2 \epsilon +1)}\,,-\frac{(z^2-1) \left(4 z^2-3\right) \left(\sqrt{\tau }+i z\right)}{\sqrt{\tau } \left(4 \tau ^2+7 \tau +3\right) (2 \epsilon +1)}\Bigg\}\,,
    \end{split}
\end{align}
and the coefficients $R_{ik}$ in the matrix form is given by,
\begingroup
\setlength{\arraycolsep}{3.0pt}
\renewcommand{\arraystretch}{1.75}
\begin{equation}
\resizebox{0.99\textwidth}{!}{$\displaystyle
R=
\left(
\begin{array}{ccccc}
\dfrac{-1-4\tau+2\epsilon(1+\tau)}{\Delta_-}
&
-\dfrac{4(-1+2\epsilon+\tau)}{\Delta_-}
&
\dfrac{-7+8\epsilon+8\tau}{\Delta_-}
&
\dfrac{8(\epsilon-1)}{\Delta_-}
&
\dfrac{4(3-2\epsilon)}{\Delta_-}
\\
\dfrac{2[\epsilon(3+4\tau)-8\tau]}{\sqrt3\,\Delta_0}
&
\dfrac{8-20\epsilon-8\tau}{\Delta_0}
&
\dfrac{8(-4+5\epsilon+4\tau)}{\sqrt3\,\Delta_0}
&
\dfrac{16(\epsilon-1)}{\Delta_0}
&
\dfrac{16(3-2\epsilon)}{\sqrt3\,\Delta_0}
\\
\dfrac{2[8\tau-\epsilon(3+4\tau)]}{\sqrt3\,\Delta_0}
&
\dfrac{8-20\epsilon-8\tau}{\Delta_0}
&
-\dfrac{8(-4+5\epsilon+4\tau)}{\sqrt3\,\Delta_0}
&
\dfrac{16(\epsilon-1)}{\Delta_0}
&
\dfrac{16(2\epsilon-3)}{\sqrt3\,\Delta_0}
\\
\dfrac{iP_R}{\sqrt\tau\,\Delta_+}
&
\dfrac{4[-1+\tau+2\epsilon(2+\tau)]}{\Delta_+}
&
-\dfrac{iQ_R}{\sqrt\tau\,\Delta_+}
&
-\dfrac{8(\epsilon-1)}{\Delta_+}
&
\dfrac{4i(2\epsilon-3)}{\sqrt\tau\,\Delta_+}
\\
-\dfrac{iP_R}{\sqrt\tau\,\Delta_+}
&
\dfrac{4[-1+\tau+2\epsilon(2+\tau)]}{\Delta_+}
&
\dfrac{iQ_R}{\sqrt\tau\,\Delta_+}
&
-\dfrac{8(\epsilon-1)}{\Delta_+}
&
-\dfrac{4i(2\epsilon-3)}{\sqrt\tau\,\Delta_+}
\end{array}
\right)
$}
\label{4.63kk}
\end{equation}
\endgroup
with, 
\begin{equation}
\begin{aligned}
 &\Delta_-=(2\epsilon-1)(1+\tau)\,,\quad
 \Delta_0=\epsilon(3+4\tau)\,,\quad
 \Delta_+=(2\epsilon+1)(1+\tau)(3+4\tau)\,,\\&
 P_R=\tau(7+4\tau)
       +2\epsilon(3+7\tau+4\tau^2)\,,\quad
 Q_R=-11+4\tau+8\epsilon(2+\tau)\,.
\end{aligned}
\label{4.64kkk}
\end{equation}
One can check that the $R$-matrix has a non-zero determinant and therefore can be inverted, and it is straightforward to show that the $\Omega$-matrix computed in these two bases is related as,
\begin{align}
    \Omega_{\textrm{dlog}}=(\partial_\tau R+R\,\Omega_{\textrm{mon}})R^{-1}\,.\label{4.65jj}
\end{align}
We also performed an independent cross-check by directly substituting the monomial basis matrix computed in \eqref{4.43mm} and the dlog basis matrix in \eqref{4.60mm} into \eqref{4.65jj} and found perfect agreement. The corresponding checks are provided with the \texttt{arXiv} submission with \texttt{\textbf{02-orthogonal-dlog-gauss-manin-checks.nb}}
\subsection{Global monodromy of differential equation and twisted Riemann bilinear relation} \label{sec:global-monodromy}
As we have seen, the solutions of multiloop Feynman integrals generically exhibit analytic functions having complicated branching structures. Those structures are associated with the monodromy group of the integrals. As previously, we have noted that the monodromy plays a key role while interpreting the solutions of these integrals. However, the monodromy studied there is in the space of twisted co-cycles. However, it is more interesting to investigate the monodromy of integrals in kinematic space. As we know, the differential equation encodes all the information regarding the individual integrals; it is intuitive to study the monodromy of the differential equation itself. \textcolor{black}{In this subsection, we will study the monodromy relations for the differential equation for the master basis which is in the even $(z \rightarrow -z )$, i.e., the $3 \times 3$ elliptic block of (\ref{eq:intersection-matrices})}. First, we write the matrix in the Fuchsian form 
\begin{align}
    \begin{split}
        A^{(135)}(\tau,\epsilon)=\frac{R_0}{\tau}+\frac{R_{-3/4}}{\tau+\frac{3}{4}}+\frac{R_{-1}}{\tau+1}
    \end{split}
\end{align}
with, 
\begin{align}
    \begin{split}
        R_0=\left(
\begin{array}{ccc}
 0 & \frac{8 \epsilon }{3}-\frac{11}{6} & 2-\frac{4 \epsilon }{3} \\
 0 & \epsilon  & 0 \\
 0 & 0 & \epsilon  \\
\end{array}
\right)\,,\quad R_{-3/4}=\left(
\begin{array}{ccc}
 -2 & -\frac{4}{3} (5 \epsilon -7) & \frac{16 \epsilon }{3}-8 \\
 -\frac{3}{2} & 7-5 \epsilon  & 4 \epsilon -6 \\
 -\frac{9}{8} & -\frac{3}{4} (5 \epsilon -7) & 3 \epsilon -\frac{9}{2} \\
\end{array}
\right)\,,\quad R_{-1}=\left(
\begin{array}{ccc}
 \frac{3}{2} & 4 \epsilon -\frac{15}{2} & 6-4 \epsilon  \\
 \frac{3}{2} & 4 \epsilon -\frac{15}{2} & 6-4 \epsilon  \\
 \frac{3}{2} & 4 \epsilon -\frac{15}{2} & 6-4 \epsilon  \\
\end{array}
\right)\,.
    \end{split}
\end{align}
As we see, the matrix has simple poles in the kinematic space $\tau$, at $\tau=0,-3/4,-1,\infty$ with the residue matrices written as $R_0,\,R_{-3/4},\,R_{-1},R_\infty=-\sum_{i\in \textrm{finite poles}} R_i$ respectively. \textcolor{black}{Of these four points, only three are degenerations of the elliptic curve:
$\tau=0$ is $\{\Lambda=0\}$, $\tau=-1$ (which does not lie in the physical kinematic region) is $\{\Lambda+4\mu=0\}$, and $\tau=\infty$ is
the massless limit $\{\mu=0\}$ of \eqref{eq:degeneration-divisor-cut}. The remaining
point $\tau=-3/4$ belongs to the twist rather than to the curve: it is where the pair
$z=\pm i\sqrt{\tau}$ meets the pair $z=\pm\sqrt3/2$. It is not an apparent singularity, since the local exponents, which are the eigenvalues of $R_{-3/4}\,,$  are $0,0$ and $\tfrac12-2\epsilon\,.$ So a half-integer
exponent survives even at $\epsilon=0\,.$ The corresponding square root is carried by an algebraic third solution $\propto\sqrt{3+4\tau}$ which we will discuss later in Sec.~(\ref{sec:diffcannon}), not by an elliptic period.  We also note for later use
that $R_{-1}$ is nilpotent, $R_{-1}^2=0$, so the local monodromy at $\tau=-1$ is
unipotent, while the eigenvalues of $R_\infty$ are $\tfrac12,\tfrac12,-\tfrac32$
independently of $\epsilon$.}

Therefore the monodromy generator \cite{Lee:2025rik}, $\mathcal{M}_a$ corresponding to a loop around the pole $a$ similar to $\exp(2\pi i R_a)$ with $\zeta=e^{i\pi d}$,
\begin{align}
    \mathcal{M}^{\textrm{loc}}_0\sim e^{2\pi i R_0}\sim  \textrm{diag}(1,\zeta^{-1},\zeta^{-1}),\,\,\mathcal{M}^{\textrm{loc}}_{-\frac{3}{4}}\sim e^{2\pi i R_{-3/4}}\sim \textcolor{black}{\textrm{diag}(1,1,-\zeta^2)}\,.
    \end{align}
We note that spectrum of $R_{\infty}=\left(\frac{1}{2},\frac{1}{2},-\frac{3}{2}\right)$ and we see the $R_{\infty}$ has resonances as $\frac{1}{2}-\left(-\frac{3}{2}\right)=2\in \mathbb{Z}$. Therefore we can not write $\mathcal{M}^{\textrm{loc}}_{\infty}\sim e^{2\pi i R_{\infty}}$.   To compute the local monodromy around the special singular point $\tau=\infty$, we take use of the Jordan data, suitable projectors, and balance transformation, which can be computed using the publicly available Mathematica package \texttt{Libra}\cite{Lee:2020zfb}. The goal is the following: the residue matrix $R_\infty$ has spectrum $\left(\frac{1}{2},\frac{1}{2},-\frac{3}{2}\right)$. We want to find a (or a series of)  rational transformation such that the spectrum does not have the resonances. We find two consecutive balance transformations $T_1$ and $T_2$,
\begin{align}
    \vec {\mathcal{J}}=T_1(\tau)T_2(\tau)\vec Y\,.
\end{align}
Consequently, the new matrix takes the form,
\begin{align}
    \bar A=T^{-1} AT-T^{-1} \partial_\tau T,\qquad T=T_1\cdot T_2\cdot T_3\cdot T_4\,.
\end{align}
where the explicit forms of the balance transformation matrices are given by,
\begin{align}
T_1(\tau,\epsilon)=
\begin{pmatrix}
1 & 0 & 0\\
0 & 1 & 0\\[2pt]
\dfrac{3\tau}{4(3-2\epsilon)}
&
\dfrac{(8\epsilon-15)\tau}{4(3-2\epsilon)}
&
1+\tau
\end{pmatrix}, \quad T_2(\tau,\epsilon)=
\begin{pmatrix}
1 & 0 & 0\\
0 & 1+\tau & 0\\[2pt]
0
&
\dfrac{(7-8\epsilon)\tau}{4(3-2\epsilon)}
&
1
\end{pmatrix}.
\end{align}
\begin{equation}
T_3(\tau,\epsilon)=
\begin{pmatrix}
1 & 0 & 0\\[4pt]
-\dfrac{\tau}{(1-2\epsilon)(1+\tau)}
&
1+\dfrac{(8-10\epsilon)\tau}
        {3(1-2\epsilon)(1+\tau)}
&
-\dfrac{4(3-2\epsilon)\tau}
         {3(1-2\epsilon)(1+\tau)}
\\[10pt]
-\dfrac{(7-8\epsilon)\tau}
         {4(3-2\epsilon)(1-2\epsilon)(1+\tau)}
&
\dfrac{(7-8\epsilon)(8-10\epsilon)\tau}
       {12(3-2\epsilon)(1-2\epsilon)(1+\tau)}
&
1-\dfrac{(7-8\epsilon)\tau}
        {3(1-2\epsilon)(1+\tau)}
\end{pmatrix}.
\end{equation}

\begin{equation}
T_4(\tau,\epsilon)=
\begin{pmatrix}
1 & 0 & 0\\[4pt]
-\dfrac{\tau}{3(1-2\epsilon)(1+\tau)}
&
1-\dfrac{(56\epsilon^2-23\epsilon-16)\tau}
        {9\epsilon(1-2\epsilon)(1+\tau)}
&
-\dfrac{4(3-2\epsilon)(2+5\epsilon)\tau}
         {9\epsilon(1-2\epsilon)(1+\tau)}
\\[10pt]
-\dfrac{(8-10\epsilon)\tau}
         {12(3-2\epsilon)(1-2\epsilon)(1+\tau)}
&
-\dfrac{(8-10\epsilon)(56\epsilon^2-23\epsilon-16)\tau}
         {36\epsilon(3-2\epsilon)(1-2\epsilon)(1+\tau)}
&
1-\dfrac{(8-10\epsilon)(2+5\epsilon)\tau}
        {9\epsilon(1-2\epsilon)(1+\tau)}
\end{pmatrix}.
\end{equation}
The new matrix system takes the form,
\begin{align}
    \partial_\tau \vec Y=\bar{A}\vec Y,\qquad  &\bar{A}=\frac{\bar R_0}{\tau}
 +\frac{\bar R_{-3/4}}{\tau+3/4}
 +\frac{\bar R_{-1}}{\tau+1},
\end{align}
The new residue matrices are given by,
\begin{align}
\begin{split}
\bar R_0
&=
\begin{pmatrix}
0 & \dfrac{16\epsilon-11}{6}
  & \dfrac{2(3-2\epsilon)}{3}\\[4pt]
0 & \epsilon & 0\\
0 & 0 & \epsilon
\end{pmatrix},
\\[1.5em]
\bar R_{-3/4}
&=
\begin{pmatrix}
0 & 0 & 0\\[6pt]
\dfrac{1+\epsilon}{1-2\epsilon}
&
\dfrac{2(41\epsilon^3+6\epsilon^2-22\epsilon-5)}
      {3\epsilon(1-2\epsilon)}
&
\dfrac{(3-2\epsilon)(28\epsilon^2+26\epsilon+5)}
      {3\epsilon(1-2\epsilon)}
\\[12pt]
\dfrac{(4-5\epsilon)(1+\epsilon)}
      {2(3-2\epsilon)(1-2\epsilon)}
&
\dfrac{(4-5\epsilon)(41\epsilon^3+6\epsilon^2-22\epsilon-5)}
      {3\epsilon(3-2\epsilon)(1-2\epsilon)}
&
\dfrac{(4-5\epsilon)(28\epsilon^2+26\epsilon+5)}
      {6\epsilon(1-2\epsilon)}
\end{pmatrix},
\\[1.5em]
\bar R_{-1}
&=
\begin{pmatrix}
\dfrac16
&
\dfrac{56\epsilon^2-23\epsilon-16}{18\epsilon}
&
\dfrac{2(3-2\epsilon)(2+5\epsilon)}{9\epsilon}
\\[12pt]
-\dfrac{3+10\epsilon}{6(1-2\epsilon)}
&
-\dfrac{(3+10\epsilon)(56\epsilon^2-23\epsilon-16)}
       {18\epsilon(1-2\epsilon)}
&
-\dfrac{2(3-2\epsilon)(2+5\epsilon)(3+10\epsilon)}
       {9\epsilon(1-2\epsilon)}
\\[12pt]
\dfrac{14\epsilon^2-7\epsilon-3}
      {3(3-2\epsilon)(1-2\epsilon)}
&
\dfrac{(14\epsilon^2-7\epsilon-3)
       (56\epsilon^2-23\epsilon-16)}
      {9\epsilon(3-2\epsilon)(1-2\epsilon)}
&
\dfrac{4(2+5\epsilon)(14\epsilon^2-7\epsilon-3)}
      {9\epsilon(1-2\epsilon)}
\end{pmatrix}.
\end{split}
\end{align}
The first two consecutive balance transformations shift the residue
eigenvalue at infinity by one unit at each step,
\begin{align}
-\frac{3}{2}
\;\longrightarrow\;
-\frac{1}{2}
\;\longrightarrow\;
\frac{1}{2}\,.
\end{align}
Consequently, the transformed residue at infinity has the normalized
spectrum
\begin{align}
\operatorname{Spec}\!\left(\overline{R}_{\infty}\right)
=
\left\{
\frac{1}{2},
\frac{1}{2},
\frac{1}{2}
\right\}
\end{align}
and explicitly $\bar R_{\infty}$ is given by,
\begin{align}
    \begin{split}
        \bar{R}_\infty=\frac{1}{2}I+N_{\infty}\,,\quad N_\infty
=
-\begin{pmatrix}
\dfrac23
&
\dfrac{4(13\epsilon^2-7\epsilon-2)}{9\epsilon}
&
\dfrac{4(3-2\epsilon)(1+4\epsilon)}{9\epsilon}
\\[12pt]
\dfrac{3-4\epsilon}{6(1-2\epsilon)}
&
\dfrac{(3-4\epsilon)(13\epsilon^2-7\epsilon-2)}
      {9\epsilon(1-2\epsilon)}
&
\dfrac{(3-2\epsilon)(3-4\epsilon)(1+4\epsilon)}
      {9\epsilon(1-2\epsilon)}
\\[12pt]
\dfrac{13\epsilon^2-17\epsilon+6}
      {6(3-2\epsilon)(1-2\epsilon)}
&
\dfrac{(13\epsilon^2-17\epsilon+6)
       (13\epsilon^2-7\epsilon-2)}
      {9\epsilon(3-2\epsilon)(1-2\epsilon)}
&
\dfrac{(1+4\epsilon)(13\epsilon^2-17\epsilon+6)}
      {9\epsilon(1-2\epsilon)}
\end{pmatrix}.
    \end{split}
\end{align}
One can check $N_{\infty}^2=0$. Therefore the local monodromies around other punctures are similar to,
\begin{align}
     \mathcal M_0^{'\mathrm{loc}}\sim e^{2\pi i\bar R_0},\qquad  \mathcal M_{-3/4}^{'\mathrm{loc}}\sim e^{2\pi i\bar R_{-3/4}}\qquad
 \mathcal M_{-1}^{'\mathrm{loc}}\sim I+2\pi i\bar R_{-1},\qquad
 \mathcal M_\infty^{'\mathrm{loc}}\sim -(I+2\pi iN_\infty).
 \label{4.76jj}
\end{align}
With $\zeta=e^{i\pi d}$, the spectra of the transformed residues are respectively $\{1,\zeta^{-1},\zeta^{-1}\}$, $\{1,1,-\zeta^2\}$, $\{1,1,1\}$ and $\{-1,-1,-1\}$.
Now we want to construct the global monodromy around some fixed base point $\tau=0$. We use the connection matrix construction of \cite{Lee:2025rik} to find the global monodromy matrices. We first numerically construct the monodromy matrices and then use the \textbf{\texttt{PSLQ}} algorithm to reconstruct the analytic functions. For that we choose the dimension: $d\equiv -\frac{i}{\pi}$ such that $\zeta=e^{i\pi d}=e$. This is important as $e$ is transcendental. We numerically compute the matrices at $\zeta=e$ and reconstruct their rational dependence on a symbolic $\zeta$. One may use \textbf{\texttt{Libra}} to construct a local power series solution around the singular points. However, one series cannot reach all the singular points because of the finite convergence disk. Therefore, we introduce intermediate regular points as,
\begin{align}
    \left\{-1,\,\textcolor{blue}{-\frac{7}{8}},\,-\frac{3}{4},\,\textcolor{blue}{-\frac{1}{2}},0\right\}.\label{4.76h}
\end{align}
Now, the convergence region of a series centered at $a$ is limited by its nearest singular point i.e,
\begin{align}
    \begin{split}
        D_a=\{\tau\in \mathbb{C}: |\tau-a|<r_a\}
    \end{split}
\end{align}
where $r_a$ (the convergence radius) is the distance from center $a$ to the nearest singularity point. Therefore, the convergence radii are computed straightforwardly,
$$
    \left\{\frac{1}{4},\frac{1}{8},\frac{1}{4},\frac{1}{4},\frac{3}{4}\right\}\,.$$
We take the matching points that are placed inside two consecutive convergence disks:
\begin{align}
  -\frac{11}{12},\,\textcolor{blue}{-\frac{5}{6}},\,-\frac{5}{8},\,-\frac{3}{8}\,.
\end{align}
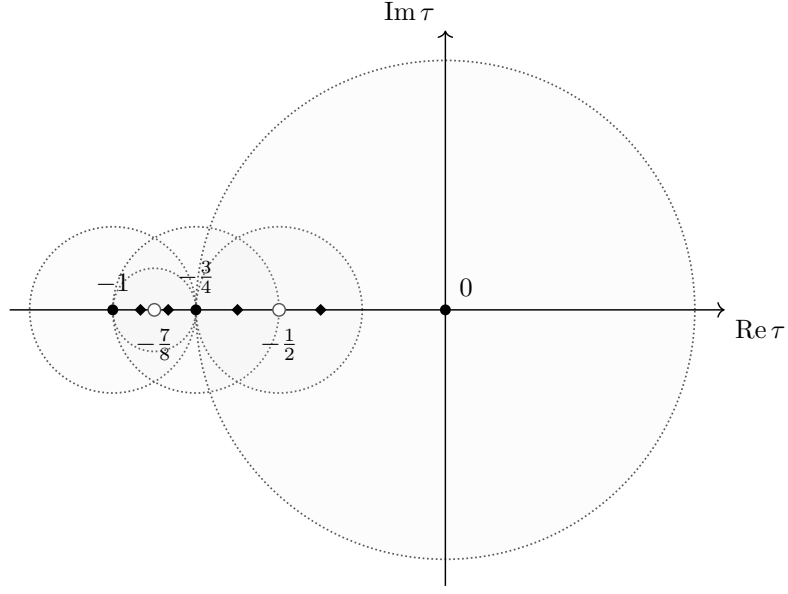
\begin{figure}
\hspace{3 cm}
\begin{tikzpicture}[
  scale=4.4,
  every node/.style={font=\small},
  disk/.style={draw=black!65, densely dotted, line width=0.65pt,
               fill=gray!20, fill opacity=0.10},
  singular/.style={circle, fill=black, inner sep=1.45pt},
  regular/.style={circle, draw=black!65, fill=white,
                  line width=0.55pt, inner sep=1.65pt}
]
  \draw[disk] (-1,0) circle[radius=1/4];
  \draw[disk] (-7/8,0) circle[radius=1/8];
  \draw[disk] (-3/4,0) circle[radius=1/4];
  \draw[disk] (-1/2,0) circle[radius=1/4];
  \draw[disk] (0,0) circle[radius=3/4];
  \draw[->, line width=0.55pt] (-5/4-0.06,0) -- (3/4+0.09,0)
    node[below right] {$\operatorname{Re}\tau$};
  \draw[->, line width=0.55pt] (0,-3/4-0.08) -- (0,3/4+0.09)
    node[above left] {$\operatorname{Im}\tau$};
  \node[singular] at (-1,0) {};
  \node[singular] at (-3/4,0) {};
  \node[singular] at (0,0) {};
  \node[regular] at (-7/8,0) {};
  \node[regular] at (-1/2,0) {};
  \node[above=2.5pt] at (-1,0) {$-1$};
  \node[below=3.2pt] at (-7/8,0) {$-\frac78$};
  \node[above=2.5pt] at (-3/4,0) {$-\frac34$};
  \node[below=3.2pt] at (-1/2,0) {$-\frac12$};
  \node[above right=1.5pt] at (0,0) {$0$};
  \foreach \x in {-11/12,-5/6,-5/8,-3/8}{
    \fill[black] (\x,0.018) -- (\x+0.018,0)
      -- (\x,-0.018) -- (\x-0.018,0) -- cycle;
  }
\end{tikzpicture}
\caption{Figure illustrates the series of
overlapping convergence disks used to analytically continue the
fundamental matrix along the real $\tau$ axis. The filled circles
denote the finite singular points $\tau=-1,-3/4,0$, around which generalized
Frobenius solutions are constructed, whereas the open circles denote
the auxiliary regular centers $\tau=-7/8,-1/2$. The radius of each
disk is fixed by the nearest singular point of the differential
system. The black diamonds are matching points lying inside
two consecutive disks. Both local series, therefore, converge at each
matching point, allowing the corresponding constant connection matrix
to be determined. Successive multiplication of these connection
matrices transports the local fundamental solution from one singular
region to another.}
\label{fig4}
\end{figure}
Near each point `$a$' belonging to \eqref{4.76h}, \textbf{\texttt{Libra}} constructs the fundamental matrix,
\begin{align}
    F_a(\tau)=H_a(\tau)(\tau-a)^{R_a}
\end{align}
where, $H_a(\tau)$ is analytic near the point $a$. Suppose, $F_L(\tau)$ and $F_{R}(\tau)$ are local series solution centered at neighboring points. Then at their common intersection region,
\begin{align}
    F_{L}=F_{R} U(R,L)\implies U(R,L)=F_{R}^{-1} F_{L}
\end{align}
where $U(R,L)$ is called the connection matrix. From the information of $F_{L,R}$, one can construct different connection matrices numerically. For example, the basis at $-\frac{3}{4}$ and the basis at $0$ can be connected by the following successive operation of the connection matrix,
\begin{align}
    U\left(0,-\frac{1}{2}\right) U\left(-\frac{1}{2},-\frac{3}{4}\right)=F_0^{-1}F_{-3/4}\implies U\left(-\frac{3}{4},0\right)=F_{-3/4}^{-1}F_0\,.
\end{align}
A pictorial representation of the convergence disks is schematically shown in Fig.~\eqref{fig4}.
Now, as we discussed earlier, all the local monodromies are calculated as $\mathcal{M}^{\textrm{loc}}_a= e^{2\pi i \bar R_a}$. Therefore, with the connection matrix, the global monodromies can be computed as shown in Fig.~\eqref{Fig7.7},
\begin{align}
    \begin{split}
        \rho_a=U(a,0)^{-1} \mathcal{M}_a^{'\textrm{loc}} U(a,0)\,.
    \end{split}
\end{align}
\begin{figure}
    \centering
    \includegraphics[width=0.4\linewidth]{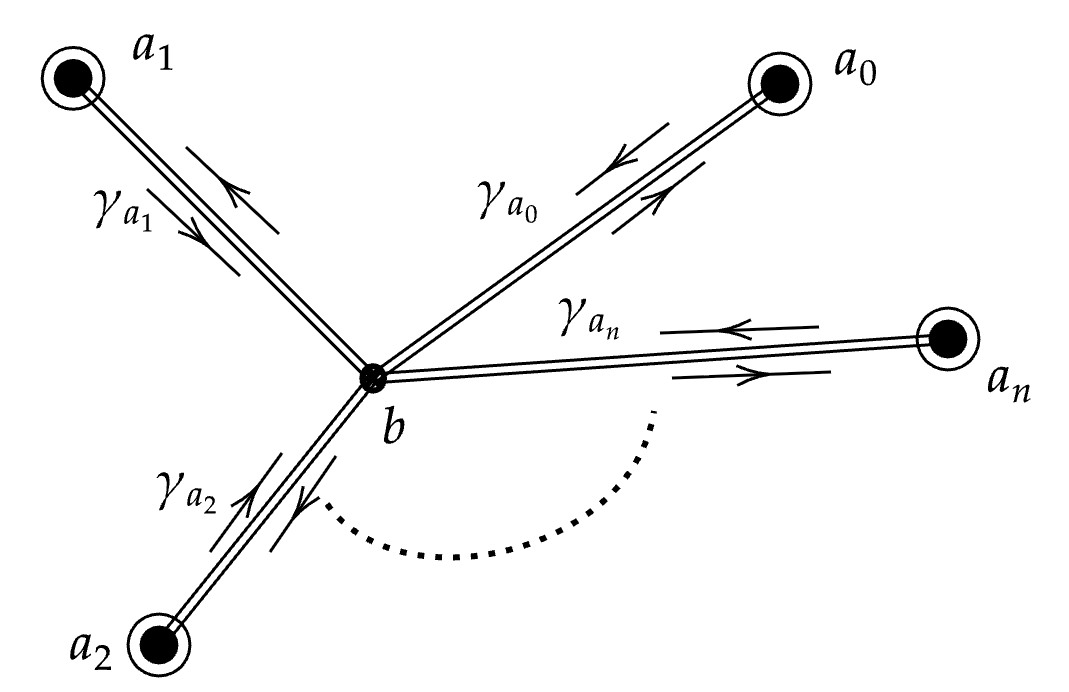}
    \caption{Pictorial representation of the computation of global monodromy group generators. With respect to a base point $b$, the contours goes along the ray $\gamma_{a_i}$ encircles $a_i$ and comes back to the base point.}
    \label{Fig7.7}
\end{figure}
With the caveat at $\infty$, we reconstruct $\rho_\infty$ using connection from $\infty\to 2\to \frac{1}{2}\to 0$. Sometimes it is useful to go to the orbit basis of Lee and Pomeransky~\cite{Lee:2025rik}, built
from the global monodromies: first find the eigenvector $v_1$ satisfying
\begin{align}
    \begin{split}
        \mathcal{M}_0^{'\textrm{loc}}v_1=v_1\,\implies \rho_0v_1=v_1.
    \end{split}
\end{align}
Then constructs the following matrix,
\begin{align}
    C_L=(v_1\quad \rho_{-\frac{3}{4}}v_1\qquad \rho_{-1}v_1)\,.
\end{align}
Consequently, the monodromy matrices in the basis are given by,
\begin{align}
    \hat{\rho}_a=C_L^{-1} \rho_a C_L\,.
\end{align}
At this point, we know the monodromy matrices numerically. Now, assume an entry of the matrix has the form of,
\begin{align}
    f(\zeta)=-\frac{a_0+a_1\zeta+a_2\zeta^2}{b_0+b_1\zeta+b_2\zeta^2}\,.
\end{align}
Now at $\zeta=e$ we have,
\begin{align}
    a_0+a_1e+a_2e^2+f(e)(b_0+b_1e+b_2e^2)=0\,.
\end{align}
Now using \textbf{\texttt{PSLQ}} algorithm, implemented in Mathematica as \texttt{FindIntegerNullVector[basis]}, we can reconstruct the matrices. The reconstructed matrices are given as,
\begin{equation}
\boxed{
\begin{aligned}
\hat \rho_0&=
\left(
\begin{array}{ccc}
 1 & \frac{(\zeta -1) (2 \zeta +1)}{\zeta } & -\frac{\zeta +3}{\zeta } \\
 0 & \frac{1}{\zeta } & 0 \\
 0 & 0 & \frac{1}{\zeta }
\end{array}
\right),
\\[4pt]
\hat\rho_{-\frac{3}{4}}&=
\left(
\begin{array}{ccc}
 0 & \zeta ^2 & \frac{1}{\zeta } \\
 1 & 1-\zeta ^2 & -\frac{1}{\zeta } \\
 0 & 0 & 1
\end{array}
\right),
\qquad
\hat\rho_{-1}=
\left(
\begin{array}{ccc}
 0 & \zeta ^2-\zeta -1 & -1 \\
 0 & 1 & 0 \\
 1 & -\zeta ^2+\zeta +1 & 2
\end{array}
\right),
\\[4pt]
\hat\rho_{\infty}&=
\left(
\begin{array}{ccc}
 \frac{\zeta ^2+\zeta -1}{\zeta ^2}
 &
 -\frac{(\zeta +1) (2 \zeta -1)(\zeta ^2-\zeta -1)}{\zeta ^2}
 &
 \frac{(\zeta +1) (\zeta +2) (2 \zeta -1)}{\zeta ^2}
 \\[4pt]
 \frac{1}{\zeta ^2}
 &
 -\frac{(\zeta -1) (2 \zeta +1)}{\zeta ^2}
 &
 \frac{\zeta +2}{\zeta ^2}
 \\[4pt]
 -\frac{1}{\zeta}
 &
 \frac{\zeta ^2-\zeta -1}{\zeta}
 &
 -\frac{2(\zeta +1)}{\zeta}
\end{array}
\right)\,.
\end{aligned}
}
\end{equation}
One can show that the monodromy generators satisfy the so-called \textit{sphere-relation} of monodromy group\footnote{One can, in principle, compute $\hat\rho_{\infty}$ by directly invoking the sphere relation after computing other finite pole $\hat\rho$'s. However, we provide an independent computation to calculate $\hat \rho_\infty$ and show it certainly satisfies the sphere relation, which provides a non-trivial cross-check of our calculation.  },
\begin{align}
    \hat \rho_0\hat\rho_{-\frac{3}{4}}\hat\rho_{-1}\hat\rho_{\infty}=\mathbb{I}\,.
\end{align}
Remarkably, all the elements of the monodromy group are represented by matrices in
$GL(3,\mathbb{Z}[\zeta,\zeta^{-1}])$: their entries are Laurent polynomials in
$\zeta^{k}$, with integer coefficients, and their determinants are $\det\hat\rho_a=\pm  \zeta^{k}$, $k\in \mathbb{Z}\,.$
Furthermore, one may find an invertible matrix $B(\zeta)$ such that,
\begin{align}
  \hat\rho_a(\zeta^{-1})^T  B(\zeta)\hat\rho_a(\zeta)=B(\zeta)\,,\label{4.90j}
\end{align}
which is known as the twisted Riemann bilinear relation. Solving \eqref{4.90j} we get the following,
\begin{align}
\begin{split}
   & B(\zeta)=\left(
\begin{array}{ccc}
 -\left((\zeta -1) \zeta ^2\right) & -2 \zeta ^4+\zeta ^3+\zeta ^2 & \zeta ^2 (\zeta +3) \\
 -((\zeta -1) \zeta  (\zeta +2)) & -(\zeta -1) \zeta ^2 & \zeta  (\zeta +1) (\zeta +2)-2 \\
 -\zeta ^2 (3 \zeta +1) & \zeta ^2 (\zeta  (2 (\zeta -1) \zeta -3)-1) & -(\zeta -1) \zeta ^2 \\
\end{array}
\right)\,,\\ &\det(B)=4 (\zeta -1) \zeta ^5 (\zeta +1)^4\,.
\end{split}
\end{align}
Note that the matrix $B(\zeta)$ is independent of the choice of contour. The purpose of studying the global monodromy of the differential equation is not just a beautiful mathematical artifact. The differential equation is the building block for the set of physical observables (directly depends on the Feynman integrals) over the kinematic space, and the monodromy relations provide the global analytic structure of those observables. It specifies how independent solutions are glued together in the full kinematic space. The global structure is important to know, as the local structure provides some basis-dependent analytic structure. Moreover, the existence of basis-independent invariant quantities (bilinear identities) reveals the additional algebraic structure that is invisible to the local singularity analysis of the differential equation.  The exact sphere relation and invariant-form checks are provided in \texttt{04-exact-monodromy-bilinear-checks.nb};   local Frobenius-series and monodromy reconstruction using \texttt{Libra} is given in \texttt{05-libra-monodromy-independent-check.nb} with the \texttt{arXiv} submission.
\section{IBP reduction: Computer algebra tools vs Intersection theory, a crosscheck } \label{sec:IBP}
Before proceeding further, it is important to perform IBP reduction using standard computer algebra tools like \texttt{LiteRed} and the intersection theory method. To avoid any kind of index ordering ambiguity, we use the order that we used when defining the global Baikov variable. 
\begin{align}
    \mathbf{W}=(D_1,D_2,D_3,D_4,\rho_1,\rho_2,u,r,w)\,,\textcolor{black}{\rho_1=\ell_1\cdot v_1}\,,\rho_2=\ell_2\cdot v_2\,,
\textcolor{black}{u=\ell_1\!\cdot\!v_2\,, r=\ell_2\!\cdot\!v_1\,,w=\ell_2\!\cdot\! q}
\end{align}
\textcolor{black}{Note that this is the Baikov ordering of (\ref{eq:new11}), not the ISP ordering of (\ref{3.1nn})}. 

Next, we define a set of integrals as,
\begin{align}
    \begin{split}
        \mathcal{K}_n:=\textrm{MaxCut}\,\mathcal{F}_{(1,1,1,1,1,1,-n,0,0)},\,\,n\ge 0\,.
    \end{split}
\end{align}
Since $\mathcal K_n$ has the numerator $u^n=(\ell_1\cdot v_2)^n$ and $u=\sqrt{\mu}\,z$,
the integral we need to evaluate is the following,
\begin{align}
    \mathcal{K}_n=\mu^{n/2} J_n,\qquad J_n:=\int_{\mathcal{C}}\mathcal{U}(z|\tau) z^n dz\,.
\end{align}
We want to reduce a non-master cocycle $\hat\varphi_5=z^5 dz$ in different methods. Below, we provide the logical flow between the checks, which also acts as a computational cross-check of our previous and further calculations and results.
\paragraph{Logical flow of the three-way reduction check.}
As the \texttt{LiteRed} basis and the intersection cocycle basis are not the same, we have to be careful about the notion of `comparison'. The comparison proceeds as follows:
\begin{enumerate}
    \item \textit{Common integral and basis.}
    We first fix an integral that must be expressed in terms of master integrals. For simplicity, we choose
    \begin{align}
        \mathcal K_5
        =
        \operatorname{MaxCut}
        \mathcal F_{(1,1,1,1,1,1,-5,0,0)}
    \end{align}
    in the ordered master basis
    \begin{align}
        \mathbf K
        =
        (\mathcal K_0,\mathcal K_1,\mathcal K_2,
        \mathcal K_3,\mathcal K_4)^T\,.
    \end{align}

    \item \textit{One-dimensional representation.}
  As we previously saw,  after performing the two fibre integrations and setting
    \(u=\sqrt{\mu}\,z\), all cut integrals take the form
    \[
        \mathcal K_n
        =
     \mu^{n/2}J_n\,,
        \qquad
        J_n=\int_{\mathcal C}\mathcal U(z|\tau)z^n\,dz\,,
    \]

    \item \textit{ Intersection-theory reduction via derived recursion relation.} In this step, we reduce the target integral into the master integral using the intersection theory and also by explicitly invoking the recursion relation that we derived in \eqref{4.42hh}. Now, in the monomial basis  
      $  (1,z,z^2,z^3,z^4)\,dz\,,$  we construct the basis intersection matrix and the target row,
    \begin{align}
        C_{a+1,b+1}
        =
        \langle z^a dz\mid z^b dz\rangle_\omega\,,
        \qquad
        F^{(5)}_{b+1}
        =
        \langle z^5dz\mid z^b dz\rangle_\omega\,.
    \end{align}
    The reduction coefficients are then obtained from
    \begin{align}
        \boldsymbol c_{\rm rec}=F^{(5)}C^{-1}\,.
    \end{align}

    \item \textcolor{black}{\textit{Direct twisted-cohomology check.} Independently of any residue
data, we can choose the IBP generating function
\begin{equation}
\mathscr D(z):=(1-z^2)(z^2+\tau)(4z^2-3)\,,
\end{equation}
for which one gets
\begin{equation}
\nabla_\omega\mathscr D(z)=
\Big(\big[(d+7)\tau-3d+9\big]z+2\big[4d-6\tau-7\big]z^{3}-4d\,z^{5}\Big)\,dz\,.
\end{equation}
Since $\int_{\mathcal C}\mathcal U\,\nabla_\omega\mathscr D=0$, the reduction coefficients can be read off directly, without computing a single
intersection number:
\begin{equation}
\boldsymbol c_{\rm exact}=
\Big(0,\ \frac{(d+7)\tau-3d+9}{4d},\ 0,\ \frac{4d-6\tau-7}{2d},\ 0\Big)\,.
\end{equation}
This is certainly independent of step 3: it uses only the vanishing of a
total twisted derivative, whereas step 3 uses the residue recursion
\eqref{4.42hh} and the intersection matrix $C$.}  

    \item \textit{Independent LiteRed reduction.} Now it is time to find the reduction using the computer algebra package \texttt{LiteRed}.  We reduce the same integral with LiteRed before performing the
    one-dimensional integration. After imposing the maximal cut, all
    lower sectors vanish. LiteRed's top-sector basis
    \(\mathbf B\) is related to the monomial basis by
    \[
        \mathbf K=T\mathbf B,
        \qquad \det T\neq0
    \]
    for generic dimension and kinematics. The LiteRed result is
    therefore transformed to the common basis using \(T^{-1}\), giving
    the coefficient row \(\boldsymbol c_{\rm LR}\).

    \item \textit{Symbolic comparison.}
    After accounting for the powers of \(\mu\) generated by
    \(u=\sqrt{\mu}\,z\), all three methods give
    \[
        \boldsymbol c_{\rm rec}
        =
        \boldsymbol c_{\rm exact}
        =
        \boldsymbol c_{\rm LR}
        =
        \left(
        0,\,
        \frac{\mu^2\big[(d+7)\tau-3d+9\big]}{4d},\,
        0,\,
        \frac{\mu\big[4d-6\tau-7\big]}{2d},\,
        0
        \right).
    \]
    Hence,
    \[
        \mathcal K_5
        =
        \frac{\mu^2\big[(d+7)\tau-3d+9\big]}{4d}\,
        \mathcal K_1
        +
        \frac{\mu\big[4d-6\tau-7\big]}{2d}\,
        \mathcal K_3\,.
    \]
\end{enumerate}
The three calculations are logically independent: the intersection theory uses residue data from our derived recursion relation and also finds an independent IBP generator to construct the same reduction; on the other hand, LiteRed derives loop-momentum IBP identities and derives the same reduction after a suitable basis transformation.
In summary, their agreement is symbolically tested using \texttt{Mathematica}, only after imposing the same maximal-cut projection,
normalization, and basis convention. The cross verifications of the reduction coefficients stated above are supplied in the ancillary notebook with the \texttt{arXiv} submission {\texttt{{01-litered-intersection-ibp-crosscheck.nb}}}.
\section{Period matrix in non-dlog basis in arbitrary dimension} \label{sec:Periodmatr}
The period matrix is defined as the bilinear pairing of the minimal basis of chosen cycles and cocycles. Each column of the period matrix is the solution of the same differential equation. Therefore, a non-singular period matrix acts as the fundamental matrix of an ordinary matrix differential equation, which we will discuss in the next section in more detail.
The first-order system of differential equations in \eqref{4.43mm} implies that
each master integral $I_i$ satisfies a scalar homogeneous differential equation
of order $|\chi|$. A basis of solutions to this scalar equation can be obtained
from the periods
\begin{equation}
    \mathbf{P}_{ij}
    =
    \langle e_i|\mathcal{C}_j]\,,
    \qquad
    i,j=1,\ldots,|\chi|\,,
\end{equation}
where $\{e_i\}$ is a basis of twisted cocycles and $\{\mathcal{C}_j\}$ is a basis
of twisted cycles. More explicitly,
\begin{equation}
    \mathcal{C}_j
    \in
    H_1\!\left(\mathbb{CP}^1\setminus\mathcal{P},{L}\right),
\end{equation}
with ${L}$ denoting the local system associated with the multivalued
Baikov weight $\mathcal{U}$. We will refine these definitions as needed at the appropriate points in the discussion.
\subsection{Choice of twisted cycles}
In general dimension, namely at finite $\epsilon$, each branch point of the Baikov weight may be regarded as a puncture or singular point of the integration domain. Consequently, a valid basis of integration cycles must keep track not only of the branch points, but also of the associated branch cuts and the integration contours defined with respect to them. Below, we provide a prescription for choosing a set of integration paths that spans the relevant twisted homology group. The choice of a basis of twisted cycles is not unique. In the following, we list the criteria for the choice of twisted cycles:
\begin{itemize}
    \item the endpoints of each twisted cycle are punctures of the twist
    one-form $\omega$;
    \item the interior of each cycle avoids all other punctures;
    \item the chosen cycles form a basis of the twisted homology group.
\end{itemize}
For finite $\epsilon$, the finite punctures are
\begin{equation}
    p_m\in
    \left\{
    -1,\,
    -h,\,
    h,\,
    1,\,
    i r,\,
    - i r
    \right\}\,,
    \qquad
    h=\frac{\sqrt{3}}{2}\,,
    \qquad
    r=\sqrt{\tau}\,.
\end{equation}
The four real punctures obey the ordering
$    -1<-h<h<1\,.$
A convenient basis of twisted cycles is therefore
\begin{equation}
\begin{aligned}
    \mathcal{C}_1 &= (-1,-h)\,,&
    \mathcal{C}_2 &= (-h,h)\,,&
    \mathcal{C}_3 &= (h,1)\,,&
    \mathcal{C}_4 &= (-1,i r)\,,&
    \mathcal{C}_5 &= (-1,-i r)\,.
\end{aligned}
\end{equation}
\textcolor{black}{We now make two remarks on this basis.  The second criterion above is what forces the four real punctures to be split into the three cycles $\mathcal C_1,\mathcal C_2,\mathcal C_3$ rather than the single interval
$[-1,1]\,.$ At finite $\epsilon$ the points $z=\pm h$ carry non-trivial monodromy, and an arc through them is not a well-defined chain.  The interval $[-1,1]$ that we will use 
from Sec.~(\ref{sec:epsilonexpnad}) onwards is therefore not a new contour but the loaded combination $\mathcal C_1+\mathcal C_2+\mathcal C_3$, and the $i0$-prescriptions given there are the statement of how that combination is loaded across the two interior punctures.  At $d=4$, the factor $(4z^{2}-3)^{-\epsilon}$ becomes unity, $\pm h$ cease to be punctures, and the combination becomes a closed cycle on the elliptic curve, which is why only two cycles survive in Sec.~(\ref{sec:period-d4}).}

We denote the endpoints of $\mathcal{C}_j$ by
 $   \mathcal{C}_j=(L_j,R_j)\,,\,\,
    \Delta_j=R_j-L_j .$ and the entries of the period matrix are given by
\begin{equation}
    \mathbf{P}_{ij}
    =
    \int_{\mathcal{C}_j}
    \mathcal{U}(z|\tau)\,z^{i-1}\,dz\,.
\end{equation}
To bring each contour integral to the interval $[0,1]$, we use the
parametrization
\begin{equation}
    z=L_j+\Delta_j t\,,
\qquad
t\in[0,1]\,.
\end{equation}
Then
\begin{equation}
    \mathbf{P}_{ij}
=
\Delta_j
\int_0^1
\mathcal{U}(L_j+\Delta_j t|\tau)
\left(L_j+\Delta_j t\right)^{i-1}
dt\,.
\end{equation}
The exponents of the Baikov weight at the finite punctures are
\begin{equation}
    \alpha_{\pm 1}
    =
    \epsilon-\frac{1}{2}\,,
    \qquad
    \alpha_{\pm h}
    =
    -\epsilon\,,
    \qquad
    \alpha_{\pm i r}
    =
    -\epsilon-\frac{1}{2}\,.
\end{equation}
\subsection{Euler representation and Lauricella functions}
In the previous subsection, we saw that, through a suitable change of variables, the branch points can be mapped to the unit interval, thereby transforming the integral into an Euler-type form. Now, for a fixed cycle $\mathcal{C}_j=(L_j,R_j)$, we separate the two endpoint
singularities from the remaining finite punctures:
\begin{equation}
    \mathcal{U}(z|\tau)
    =
    \widehat{N}\,
    (z-L_j)^{\alpha_{L_j}}
    (z-R_j)^{\alpha_{R_j}}
    \prod_{m\neq L_j,R_j}
    (z-p_m)^{\alpha_m},\qquad  \widehat{N}
    = \textcolor{blue}{4^{-\epsilon}\,}
    \tau^\epsilon
    \mathcal{N}(\epsilon,\mu,s)
    e^{i\pi(\epsilon-1/2)}\,.
\end{equation}
For each non-endpoint puncture (for a particular twisted cycle) $p_m\neq L_j,R_j$, define
\begin{equation}
    X_{j,m}
    =
    -\frac{\Delta_j}{L_j-p_m}\,.
\end{equation}
Now, a direct substitution of the contour parametrization gives,
\begin{equation}
    z-L_j=\Delta_j t\,,
\qquad
z-R_j=-\Delta_j(1-t)\,,
\qquad
z-p_m=(L_j-p_m)\left(1-X_{j,m}t\right)
\end{equation}
 and then we obtain
\begin{equation}
 \mathcal{U}(z|\tau)\,dz
=
\kappa_j\,
t^{\alpha_{L_j}}
(1-t)^{\alpha_{R_j}}
\prod_{m\neq L_j,R_j}
\left(1-X_{j,m}t\right)^{\alpha_m}
dt\,,
\end{equation}
where the cycle-dependent prefactor is
\begin{equation}
   \kappa_j
    =
    \widehat{N}\,
    e^{i\pi\alpha_{R_j}}
    \Delta_j^{1+\alpha_{L_j}+\alpha_{R_j}}
    \prod_{m\neq L_j,R_j}
    (L_j-p_m)^{\alpha_m}\,.
\end{equation}
Therefore, the period matrix in an arbitrary dimension is
\begin{equation}
  \mathbf{P}_{ij}
=
\kappa_j
\int_0^1
t^{\alpha_{L_j}}
(1-t)^{\alpha_{R_j}}
\prod_{m\neq L_j,R_j}
\left(1-X_{j,m}t\right)^{\alpha_m}
\left(L_j+\Delta_j t\right)^{i-1}
dt\,.
\end{equation}
Furthermore, using the binomial expansion,
\begin{align}
    \begin{split}
      (L_j+\Delta_j t)^{i-1}=\sum_{n=0}^{i-1} L_j^{i-1-n} \Delta_j^n t^n\,,
    \end{split}
\end{align}
the period matrix element can be written as,
\begin{align}
    \begin{split}
      \mathbf{P}_{ij}=\kappa_j\sum_{n=0}^{i-1} L_j^{i-1-n} \Delta_j^n \int_0^1
t^{\alpha_{L_j}+n} (1-t)^{\alpha_{R_j}}\prod_{m\ne L_j,R_j}
(1-X_{j,m}t)^{\alpha_m}dt\,.
    \end{split}
\end{align}
The resulting integral can further be written in terms of the Lauricella hypergeometric function.
\begin{align}
    \begin{split}  
    \boxed{\mathbf{P}_{ij}=\kappa_j\sum_{n=0}^{i-1} L_j^{i-1-n} \Delta_j^n B(\alpha_{L_j}+n+1,\alpha_{R_j}+1)
F_D^{(n_{\textrm{var}})}
\left(
\alpha_{L_j}+n+1;
\{-\alpha_m\};
\alpha_{L_j}+\alpha_{R_j}+n+2;
\{X_{j,m}\}
\right)}
    \end{split}
\end{align}
where $n_{\textrm{var}}$ is the number of variables inside the product i.e,
 $ n_{\textrm{var}}=\# \{m: m\ne L_j,R_j\}\,.$
As we see in the general dimension, the twisted period matrix elements are generalized Hypergeometric functions. 
\subsection{Period matrix in $d=4$} \label{sec:period-d4}
In the preceding discussion, we focused primarily on the computation of period matrices in arbitrary dimension, with $d\notin\mathbb{Z}$. Naturally the question arises that what will happen if we take integer valued dimension or more precisely $d=4$? Does a naive substitution of $\epsilon=0$ at the computed period matrices work? The answer is the limit cannot be taken naively by simply setting $\epsilon\to0$, because the topology of the underlying Riemann surface changes discontinuously at the exact integer dimension. In particular, punctures with trivial monodromy no longer define independent twisted cycles.

One must therefore distinguish between the period matrix obtained by first working at finite $\epsilon$ and subsequently taking the limit $\epsilon\to0$, and the period matrix constructed directly at $\epsilon=0$. The latter has rank two, in agreement with the two independent homology cycles of an elliptic curve. To make this distinction explicit, let us first recall the form of the Baikov weight at $\epsilon=0$. 

\begin{align}
    \mathcal{U}(z|\tau)\Big|_{\epsilon=0}=\frac{
    1}{\textcolor{black}{\pi\sqrt{\mu s}}}\Big(-(z^2-1)(z^2+\tau)\Big)^{-1/2}\,.
\end{align}
Note that in $d=4$, the Baikov weight has trivial monodromy around $z=\infty$ (scales as $1/z^2$) as opposed to general dimension where it scales as $\frac{1}{z^{2+2\epsilon}}$. Therefore, naively counting the solutions of $\omega=0$ for exact dimension may give a wrong result, as it may mistakenly count $\infty$ as a puncture of the one-form $\omega$. To make resemblance with our previous discussion, let us redefine the integral $\mathcal{U}(z|\tau)\varphi_0$ as  $\widetilde{U}(z|\tau)\tilde \varphi_0$ with,
\begin{align}
    \begin{split}
        \widetilde{U}(z|\tau)=\frac{1}{\textcolor{black}{\pi\sqrt{\mu s}}}\left(-\frac{z^2-1}{z^2+\tau}\right)^{-1/2},\,\,\,\quad \tilde \varphi_0=\frac{\varphi_0}{z^2+\tau}
    \end{split}
\end{align}
which leaves the combination $\mathcal{U}(z|\tau)\varphi_0=\widetilde{\mathcal{U}}(z|\tau)\tilde \varphi_0$ invariant. Consequently, let's define the redefined logarithmic one form as,
\begin{align}
    \begin{split}
        \widetilde{\omega}=d\log \widetilde{U}(z|\tau)=\textcolor{black}{d\log\widetilde{\mathcal{U}}(z|\tau)
=\left(\frac{z}{z^{2}+\tau}-\frac{z}{z^{2}-1}\right)dz}\,.
    \end{split}
\end{align}
Therefore $\tilde\omega=0$ has two solutions at $z=0,\infty$ (unlike the generalized case where $z=\infty$ is a solution of $\omega=0$ ). Similarly, $\tilde \omega$ has four non degenerate poles: $\{\pm 1,\pm i\sqrt{\tau}\}$, therefore for $d=4$ we have two independent cycles and cocyles and we also expect that there must be two independent (co-)cycles as for $\epsilon=0$, the underlying curve becomes pure elliptic curve, having two independent elliptic period. Correspondingly, the chosen cycles are,
\begin{align}
    |\mathcal{C}_1]=(-1,1)\,,\,\,\,\,|\mathcal{C}_2]=(-i\sqrt{\tau},i\sqrt{\tau})\,.
\end{align}
Therefore, the $2\times 2$ period matrix in $d=4$ is given by,
\begin{align}
P^{(0)}(\tau)
=
\begin{pmatrix}
\displaystyle \int_{\mathcal{C}_1}\frac{dz}{y}
&
\displaystyle \int_{\mathcal{C}_2}\frac{dz}{y}
\\[1.2em]
\displaystyle \int_{\mathcal{C}_1}\frac{z^2\,dz}{y}
&
\displaystyle \int_{\mathcal{C}_2}\frac{z^2\,dz}{y}
\end{pmatrix}\,,\qquad y^2=(1-z^2)(z^2+\tau)\,.
\end{align}
The elements of the period matrix can be written in terms of complete elliptic functions. \textcolor{black}{Below we provide the explicit closed form expression for the period matrix elements where the common normalization factor $1/(\pi\sqrt{\mu s})$ has been stripped off. }
\begin{align}
    \begin{split}
        &{P}^{0}_{11}=\frac{2 K\left(-\frac{1}{\tau }\right)}{\sqrt{\tau }}\,,{P}^{0}_{12}=2i{K(-\tau)}\,,
        {P}^{0}_{21}=
        2 \sqrt{\tau } \left(E\left(-\frac{1}{\tau }\right)-K\left(-\frac{1}{\tau }\right)\right)\,, {{P}}^{0}_{22}=2i(K(-\tau)-E(-\tau))\,.    \label{6.23yy} 
    \end{split}
\end{align}
\textcolor{black}{These four entries will reappear twice: as the segment periods \eqref{eq:wqft-elliptic-segment-periods} of the $x$-chart in Sec.~(\ref{Sec:6.4}),
and as the first two columns of the fundamental matrix
\eqref{eq:exact-elliptic-fundamental} of Sec.~(\ref{sec:diffcannon}). The agreement
of the three computations, and in particular of the Wronskian $\det P=-2\pi i$, is a
check on the cycle orientations and on the branch conventions used in each chart.}
\subsubsection{Period transformation and canonicalization of differential equation}

\textcolor{black}{
The period matrix plays an important role in the canonicalization of the differential system. For instance, the differential equation for the original integrals in the top sector usually satisfies the following differential equation,
\begin{align}
    \partial_\tau \vec I_{\rm top}=\Omega(\tau,\epsilon)\vec I_{\rm top}+\tilde\Omega(\tau, \epsilon) \vec I _{\rm lower}\,.
\end{align}
Previously, we studied maximal cuts as they project the set of master integrals onto its top sector. As we noticed from our previous analysis, the $\Omega$ matrix takes the following form,
\begin{align}
    \begin{split}
        \Omega(\tau,\epsilon)=A_{0}(\tau)+\epsilon\,A_1(\tau)\,.
    \end{split}
\end{align}
and the matrix is not purely canonicalized. However, doing the period transformation makes the differential system canonical. We know that the period matrix satisfies the same differential equation as the master integral basis. 
\begin{align}
    \begin{split}
        \partial_{\tau} \mathbf{P}=\Omega(\tau,\epsilon)\mathbf{P}=(A_0+\epsilon A_1)\mathbf{P}\,.\label{5.28hh}
    \end{split}
\end{align}
Now take $\epsilon\to 0$ in the both side of \eqref{5.28hh} (note that here we take $\epsilon=0$ after computing the period matrix for finite $\epsilon$, which is not same as $\mathbf{P}^{(0)}$). We denote the period matrix at $\mathbf{P}|_{\epsilon=0}$ as $\mathbf{P}_{0}$. Now doing the following basis change from $\vec I_{\rm top}\to \vec J=\mathbf{P}_{0}^{-1}\vec I_{\rm top}$, we get the following $\epsilon$ factorized form of the differential equation for the transformed basis:
\begin{align}
    \begin{split}
        \partial_{\tau}\vec J=\epsilon \mathbf{P}_0^{-1}A_1\mathbf{P}_0 \vec J\,.\label{6.27jj}
    \end{split}
\end{align}}
\subsubsection{Derivative basis and Wronskian}
Another useful choice of de Rham basis is obtained from the holomorphic differential and its derivative with respect to the kinematic variable. For the elliptic curve
\begin{align}
    y^2=(1-z^2)(z^2+\tau)\,,
\end{align}
we define
\begin{align}
   \hat \omega=\frac{dz}{y}\,,
    \qquad
    I_i(\tau)=\int_{\mathcal{C}_i}\hat \omega\,,
    \qquad i=1,2
\end{align}
where $\mathcal{C}_1$ and $\mathcal{C}_2$ are the two independent homology cycles. \textcolor{black}{Modulo exact differentials, $\partial_\tau\hat\omega$ provides the second independent de Rham class}. The corresponding period matrix in the derivative basis is therefore
\begin{align}
P^{(0)}(\tau)
=
\begin{pmatrix}
\displaystyle \int_{\mathcal{C}_1}\frac{dz}{y}
&
\displaystyle \int_{\mathcal{C}_2}\frac{dz}{y}
\\[1.2em]
\displaystyle \int_{\mathcal{C}_1}
\partial_\tau\!\left(\frac{dz}{y}\right)
&
\displaystyle \int_{\mathcal{C}_2}
\partial_\tau\!\left(\frac{dz}{y}\right)
\end{pmatrix}
=
\begin{pmatrix}
I_1(\tau) & I_2(\tau)
\\
\partial_\tau I_1(\tau) & \partial_\tau I_2(\tau)
\end{pmatrix}\,.
\end{align}
Here, the cycles are understood to be transported according to the Gauss--Manin connection. In this basis, the period matrix is a Wronskian matrix, and its determinant is
\begin{align}
    \det P^{(0)}(\tau)
    =
    I_1(\tau)\,\partial_\tau I_2(\tau)
    -
    I_2(\tau)\,\partial_\tau I_1(\tau)\,.
\end{align}
Thus, $\det P^{(0)}(\tau)$ is the Wronskian of the scalar Picard--Fuchs equation satisfied by the periods $I_i(\tau)$ and measures their linear independence. The derivative basis and the monomial de Rham basis are related modulo exact differentials. In the present case, one finds
\begin{align}
    \partial_\tau\!\left(\frac{dz}{y}\right)
    =
    -\frac{1}{2(1+\tau)}\frac{dz}{y}
    -\frac{1}{2\tau(1+\tau)}\frac{z^2\,dz}{y}
    +
    d\left(
    \frac{z(z^2-1)}
    {2\tau(1+\tau)y}
    \right)\,.
\end{align}
Since the exact differential does not contribute when integrated over a closed cycle, the periods in the derivative basis are algebraically related to those in the monomial basis
\begin{align}
    \left\{
    \frac{dz}{y},
    \frac{z^2\,dz}{y}
    \right\}\,.
\end{align}
The Wronskian plays an important role in constructing the initial basis used in the canonicalization of the uncut master integrals \cite{Dlapa:2022wdu}.

\subsection{Period matrix and twisted Riemann period relation}\label{Sec:6.4}

In this subsection, we end our analysis of the period matrix by investigating the relation between the intersection theory of twisted cocycles and the twisted Riemann period relation by invoking the connection between the intersection theories of twisted cycles and twisted cocycles \cite{Duhr:2024rxe}. \textcolor{black}{The relation derived in this subsection should not be confused with the
monodromy invariant $B(\zeta)$ of Sec.~(\ref{sec:global-monodromy}). There, the
bilinear form is invariant under the monodromy of the differential equation in
kinematic space. On the other hand, here it pairs the period matrices of the twisted local system and its dual at fixed kinematics. The two are logically independent, and we use
different letters for them throughout.}

We now follow the convention of \cite{Cho_Matsumoto_1995}. Let $\mathcal{P}=\{x_0,\ldots,x_n\}$ be $n+1$ distinct marked points on $\mathbb{CP}^1$, with $x_0=\infty$. We work with generic nonresonant exponents. In the affine $x$-chart, let
\begin{align}
    \omega=\sum_{j=1}^n r_j\frac{dx}{x-x_j}\,,\qquad r_0=\operatorname{Res}_{x=\infty}\omega=-\sum_{j=1}^n r_j
\end{align}
be a connection one-form. The first twisted cohomology group,
\begin{align}
    H^1(U,L)\simeq \mathbf{H}^{1}(\mathbb{CP}^1,(\Omega,\log D,\nabla_\omega)), U=\mathbb{CP}^1\backslash\{x_0,\cdots x_n\}
\end{align}
with the connection defined in \eqref{4.3aa} is known to be isomorphic to,
\begin{align}
    \begin{split}
        \Gamma(\mathbb{CP}^1,\Omega^1(\log D))/\boldsymbol{C}\cdot \omega,\qquad D:=x_0+\cdots+x_n
    \end{split}
\end{align}
where,
\begin{align}
    L:=\textrm{Ker}(\nabla_\omega|_{U}:\mathcal{O}_U\rightarrow \Omega_{\mathbb{CP}^1}^1(\log D)|_U)
\end{align}
is a local system on $U$ defined by the connection. The dual of the cohomology group $H^1(U,L)$ is given by the cohomology group with compact support $H_c^1(U,L^\vee)$, where $L^\vee$ is the local system defined by the connection $\nabla_{-\omega}:=d-\omega\wedge$ dual to $\nabla_\omega$. In \cite{Cho_Matsumoto_1995}, it is shown that the dual cohomology group is isomorphic to $\Gamma(\mathbb{CP}^1,\Omega^1(\log D))/\boldsymbol{C}\cdot (-\omega)\,.$ Since there exists a dual pairing between two cohomology groups, it is natural that there should exist a bilinear form on the spaces $\Gamma(\mathbb{CP}^1,\Omega^1(\log D))/\boldsymbol{C}\cdot \omega$ and $\Gamma(\mathbb{CP}^1,\Omega^1(\log D))/\boldsymbol{C}\cdot (-\omega)$. By using the following  logarithmic form,
\begin{align}
    \begin{split}
        \sigma_j:= \frac{dx}{x-x_j}-\frac{dx}{x-x_{j+1}}, 1\leq j\leq n-1\label{6.40aa}
    \end{split}
\end{align}
we defined the bases for the spaces defined earlier as,
\begin{align}
    \begin{split}
        \sigma_j^{+} \in \Gamma(\mathbb{CP}^1,\Omega^1(\log D))/\boldsymbol{C}\cdot \omega\,,\quad \sigma_j^{-}\in \Gamma(\mathbb{CP}^1,\Omega^1(\log D))/\boldsymbol{C}\cdot (-\omega)\,,
    \end{split}
\end{align}
where $\sigma_j^+$ and $\sigma_j^-$ are the images of $\sigma_j$ by projections from the space of global meromorphic one-forms on the Riemann sphere $\mathbb{CP}^1$. Now the intersection numbers for the chosen base cocycles are \footnote{We have slightly modified the notation for the intersection numbers. In the literature on period relations, an additional factor of \(2\pi i\) is included; apart from this convention, the two definitions are equivalent.
 },
\begin{align}
    \langle\sigma_i^+,\sigma_j^-\rangle:=\sum_{p\in\mathcal P}\frac{\operatorname{Res}_p\sigma_i\,\operatorname{Res}_p\sigma_j}{\operatorname{Res}_p\omega}=\left[\delta_{ij}\left(\frac{1}{r_j}+\frac{1}{r_{j+1}}\right)-\frac{\delta_{i+1,j}}{r_{j}}-\frac{\delta_{i,j+1}}{r_{j+1}}\right]\,.
\end{align}
Now, the Riemann period relation states that there exists a relation between the intersection numbers of twisted homologies, twisted cohomologies, and the pairing of twisted homologies and twisted cohomologies. Let us define,
\begin{align}
    \gamma_j^+\in H_1(U,L^\vee),\quad \delta_j^{-}\in H_1(U,L),\quad j=1,\cdots,n-1\,.
\end{align}
be the bases of twisted cycles and 
\begin{align}
    \begin{split}
        \xi_i^{+}\in\Gamma(\mathbb{CP}^1,\Omega^1(\log D))/\boldsymbol{C}\cdot \omega,\quad \eta_i^{-}\in \Gamma(\mathbb{CP}^1,\Omega^1(\log D))/\boldsymbol{C}\cdot (-\omega)\,,
    \end{split}
\end{align}
With that, one can define the homology/cohomology intersection matrices as,
\begin{align}
    I_{ij}^{h}=\langle \gamma_i^+,\delta_j^-\rangle,\quad I_{ij}^{ch}=\langle \xi_i^+,\eta_j^-\rangle,\qquad  1\leq i,j\leq n-1\,.
\end{align}
Now let's assume that that  all the poles $\{x_j\}$ are real and are arranged as,
\begin{align}
    x_1<\cdots<x_n\,,\qquad x_0=\infty\,,
\end{align}
and $u_0(x)$ be a branch of the multivalued function (which is nothing but the weighted Baikov polynomial for our case) $\mathcal{U}=\prod (x-x_j)^{r_j}$ defined on the lower half $x-$ plane. Now according to \cite{Kita1994IntersectionTF,Cho_Matsumoto_1995}, we choose special cycles by,
\begin{align}
\begin{split}
   & \gamma_j^+=I_j \otimes u_0+\frac{1}{c_j-1} S_j\otimes u_0-\frac{1}{c_{j+1}-1}S_{j+1}\otimes u_0\,,\\ &
   \gamma_j^-=I_j \otimes u_0^{-1}-\frac{c_j}{c_j-1} S_j\otimes u_0^{-1}+\frac{c_{j+1}}{c_{j+1}-1}S_{j+1}\otimes u_0^{-1}\,,\quad c_j=e^{2\pi i r_j}\,.
    \end{split}
\end{align}
Here $I_j$ is oriented from $p_j=x_j+\rho$ to $q_{j+1}=x_{j+1}-\rho$, with $\rho$ smaller than half the neighboring puncture separations. Both endpoint circles are counterclockwise: the left one starts at $p_j$, and the right one at $q_{j+1}$. Their initial germs are inherited from the interval. In the notation of the duality pairing, we set $\delta_j^-=\gamma_j^-$. Let's take a moment to explain why the chosen cycle and dual cycle are special. The claim is that the cycles are chosen in such a way that any boundary term vanishes.
\begin{align}
    \int_{\gamma_{j}^+} u_0\,\nabla_\omega\xi=0\,,\quad \int_{\gamma_j^-} u_0^{-1}\,\nabla_{-\omega}\xi=0\,.
\end{align}
\textbf{Proof:} 
\begin{align}
    \begin{split}
        \int_{\gamma_j^+}u_0 \nabla_{\omega} \xi=\int_{\gamma_j^+}d(u_0\xi)&=\int_{I_j}d(u_0\xi)+\frac{1}{c_{j}-1} \int_{S_{j}}d(u_0\xi)-\frac{1}{c_{j+1}-1}\int_{S_{j+1}}d(u_0\xi)\\&
        =(u_0\xi)(q_{j+1})-(u_0\xi)(p_j)+\frac{1}{c_j-1}(c_{j}-1)(u_0\xi)(p_j)\\ &\hspace{1 cm}-\frac{1}{c_{j+1}-1}(c_{j+1}-1)(u_0\xi)(q_{j+1})=0\,.
    \end{split}
\end{align}
Here we used the fact that $u_0(p_j)\xrightarrow[]{S_j} c_j u_0(p_j)$. Similarly, for the dual branch,
\begin{align}
\begin{split}
    \int_{\gamma_{j}^-} u_0^{-1}\nabla_{-\omega}\xi=\int_{\gamma_j^{-}} d(u_0^{-1}\xi)&=(u_0^{-1}\xi)(q_{j+1})-(u_0^{-1}\xi)(p_j)-\frac{c_j}{c_j-1}(c_j^{-1}-1)(u_0^{-1}\xi)(p_j)\\ &
    +\frac{c_{j+1}}{c_{j+1}-1}(c_{j+1}^{-1}-1)(u_0^{-1}\xi)(q_{j+1})=0\,.
    \end{split}
\end{align}
Here we also used the fact that the dual function has local monodromy: $u_0^{-1}\xrightarrow[]{S_j}c_j^{-1}u_0^{-1}$. Thus, for a single-valued meromorphic function $\xi$ with no pole on the integration paths, the total derivative integrates to zero on the closed twisted cycle. The endpoint-circle construction is itself the regularization of the open chamber. Here, we do not have to assume that the unregularized endpoint values vanish identically. Now let's compute the intersection numbers for the twisted cycles \cite{Kita1994IntersectionTF}. The computation of intersection numbers for cycles is well known in the mathematical literature, e.g., \cite{Kita1994IntersectionTF}. Here, for the reader's advantage, we present a pictorial representation to extract the intersection number of twisted cycles. The intersection number depends on the orientation of intersecting paths, the weights of the paths where they are intersecting, and the branching of the co-cycle. Moreover, sometimes two branches overlap, so one can not assign a unique intersection number. However, we can slightly deform any of the contours without cutting any branch point. 
\begin{figure}[htb!]
    \centering
    \includegraphics[width=0.5\linewidth]{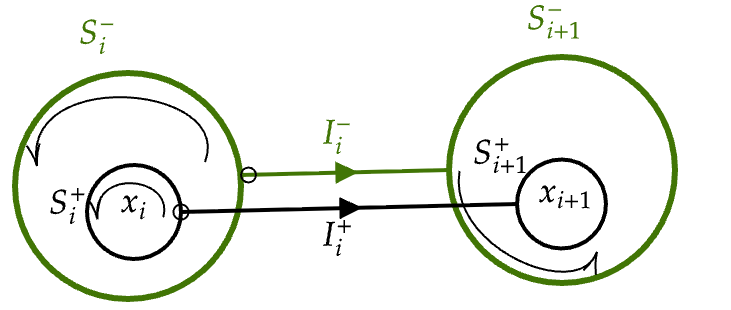}
    \caption{Cycle of integration for computing the intersection number: $\langle \gamma_i^+,\gamma_i^-\rangle$}
    \label{fig5}
\end{figure}
As we previously saw, the cycles can be written as,
\begin{align}
    \gamma_{j}^{+}=\sum_a A_a^{+} \,\mathscr{P}_a^{+}\otimes u_0,\quad \gamma_{j}^{-}=\sum_a A_a^{-} \,\mathscr{P}_a^{-}\otimes u_0^{-1}\,,
\end{align}
where $A_a^{\pm}$, $\mathscr{P}^{\pm}$ are the weights and paths of the cycle, respectively, for the positive and negative branches. The intersection number is defined as,
\begin{align}
    \langle \gamma_i^+,
    \gamma_j^-\rangle:=\sum_{p\in \gamma_i^+\cap\tilde{\gamma}_j^-} \epsilon(p) A^+_p A^-_p  u_{p}^{+}\,u_{p}^{-}
\end{align}
where, $\epsilon(p)$ denotes the orientation sign at the intersection point $p$, 
$A_p^{\pm}$ are the weights carried by the corresponding segments at $p$, and 
$u_p^+$ and $u_p^-$ are the local multivalued Baikov polynomials transported along the two intersecting segments, and the angular brackets count the relative local monodromies that are carried by the two multivalued functions with respect to a reference point. It is determined by their relative winding, not by the number of geometric crossings or by a circuit completed after the crossing. With this convention, we have,
\begin{align}
\begin{split}
    \langle \gamma_i^+,\gamma_i^-\rangle&=(+1)\langle I_i^+, S_i^-\rangle_{x_i}\langle \!\langle u_0,c_i^{-1}u_0^{-1}\rangle \!\rangle+(-1)\langle I_i^+, S_{i+1}^-\rangle_{x_{i+1}}\langle \!\langle u_0,u_0^{-1} \rangle \!\rangle\\ &=-\frac{c_i}{d_i}\times \frac{1}{c_i}+(-1)\frac{c_{i+1}}{d_{i+1}}(1)=-\frac{c_{i}c_{i+1}-1}{d_id_{i+1}},\quad d_i=c_i-1\,.
    \end{split}
\end{align}
Fig.~(\ref{fig5}) shows a transverse deformation avoiding the punctures, with the the suitable branch weights. The angles in the following calculation are continuous lifts of the argument along these loaded paths, not principal arguments reset modulo $2\pi$. At the left intersection (which is just the right side of the branch point $x_i$), the positive branch comes at the intersection point with an angle of  $\theta_+=0$ and the negative branch wraps around the branch point $x_i$ once, carrying $\theta_-=2\pi$. Therefore the branch factor gives $e^{ir_i0}(e^{2\pi i r_i})^{-1}=\frac{1}{c_i}$. On the other hand, the right intersection happens on the left side of the branch point $x_{i+1}$ where the positive branch carries $\theta_+=-\pi$ (note that $u_0$ is a branch of $\mathcal{U}$ lying below the real line) and the negative branch also carries $\theta_-=-\pi$ having unity branch factor.
Similarly, one can deduce that $\langle \gamma_i^+,\gamma_{i+1}^-\rangle =\frac{1}{d_{i+1}}$ and $\langle \gamma_{i+1}^+,\gamma_{i}^-\rangle =\frac{c_{i+1}}{d_{i+1}}$ which can be seen from Fig.~\eqref{fig6} and Fig~\eqref{fig7}. Also, it is easy to see that there is no topological intersection between the positive and the negative path when $|i-j|>1$ i.e $\langle \gamma_i^+,\gamma_j^-\rangle =0,\ |i-j|>1$. Note that the intersection number compute above is independent of the choice of deformation \footnote{For example, if we have a path as below, the intersection number $\langle \gamma_i^+,\gamma_i^-\rangle=1-\frac{c_i}{d_i}-\frac{c_{i+1}}{d_{i+1}}=-\frac{c_ic_{i+1}-1}{d_id_{i+1}}$.
\centering
\includegraphics[width=0.3\textwidth]{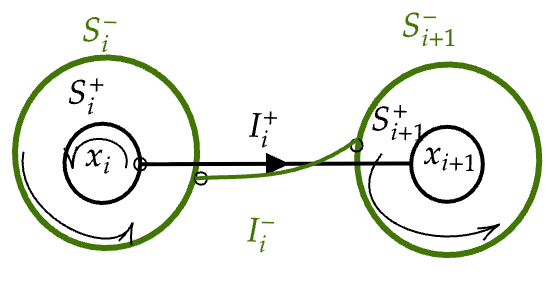}
}. 
\begin{figure}[htb!]
    \centering
    \includegraphics[width=0.5\linewidth]{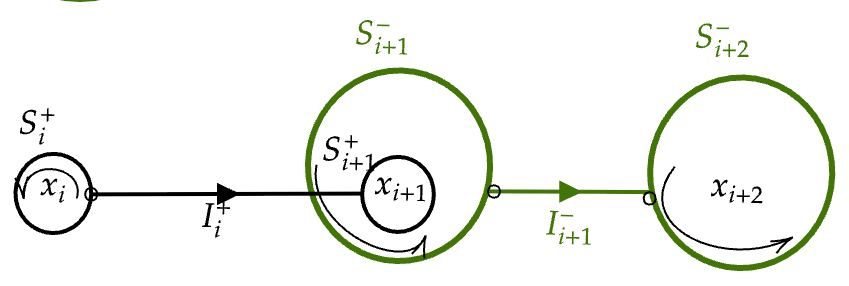}
    \caption{Choice of deformed cycle to compute the intersection number $\langle \gamma_i^+,\gamma_{i+1}^{-}\rangle.$ }
    \label{fig6}
\end{figure}
\begin{figure}[htb!]
    \centering    \includegraphics[width=0.5\linewidth]{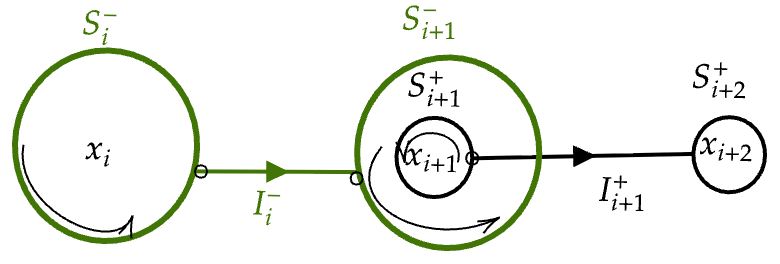}
    \caption{Choice of deformed cycle to compute the intersection number $\langle \gamma_{i+1}^+,\gamma_{i}^{-}\rangle.$}
    \label{fig7}
\end{figure}
\textcolor{black}{We now specialize to our integral.  The relevant variable is $x=z^{2}$, and this is not a convenience. It is the quotient by the reflection $z\to-z$ whose $\pm$ eigenspaces produced the $3\oplus2$ block decomposition of \eqref{eq:intersection-matrices}.  On the quotient, the six finite punctures $\{\pm1,\pm h,\pm i\sqrt\tau\}$ become the four points $\{-\tau,0,\tfrac34,1\}$, the Jacobian $dz=dx/2\sqrt x$ supplies the extra exponent $r_2=-\tfrac12$ at $x=0$, and three independent twisted cycles and cocycles remain.  These are the three master integrals of the elliptic block; the two master integrals of the odd block are invisible in $x$ and are not needed for the period relations below.  The curve $\mathcal E_x$ that appears at $\epsilon=0$ in \eqref{eq:wqft-compact-x-curve}
is the same elliptic quotient that the extra involution will produce later in Sec.~(\ref{sec9}).  With this, the multivalued Baikov polynomial up to an overall normalization is,}
\begin{align}
    \begin{split}
        \hat{\mathcal{U}}(x,\tau)=(x+\tau)^{-\epsilon-\frac{1}{2}} x^{-1/2}(4x-3)^{-\epsilon}(1-x)^{\epsilon-\frac{1}{2}}\,.
    \end{split}
\end{align}
The $x^{-1/2}$ factor comes from the Jacobian of the transformation.
An overall $x$-independent normalization is omitted. It cancels from the bilinear period relation when the positive twist and its inverse are normalized reciprocally, even if it depends on the kinematic variables. Therefore, we have a branch point:
\begin{align}
  x_1=-\tau<x_2=0<x_3=\frac34<x_4=1\,,\qquad\tau>0
\end{align}
with, $x_0=\infty$ which is also marked point on $\mathbb{CP}^1$. To remind, in $x$ variable, the connection form is given by,
\begin{align}
    \omega=d\log u_+=\sum_{j=1}^4 r_j\frac{dx}{x-x_j}
\end{align}
where, $r_1=-\epsilon-\frac{1}{2},\,r_2=-\frac{1}{2},\,r_3=-\epsilon, \,r_4=\epsilon-\frac{1}{2}$ and expanding $\omega$ around $\infty$ we find $r_{0}=\epsilon+3/2$. Therefore, $\sum_{j=0}^nr_j=0$ is satisfied trivially. With the choice of \eqref{6.40aa}, the cocycles are given by,
\begin{align}
    \sigma_1=\frac{dx}{x+\tau}-\frac{dx}{x},\,\,\sigma_2=\frac{dx}{x}-\frac{dx}{x-\frac{3}{4}},\,\,\sigma _3=\frac{dx}{x-\frac{3}{4}}-\frac{dx}{x-1}\,.
\end{align}
The intersection numbers of twisted co-cycles are given by,
\begin{align}
\begin{split}
  &  \langle \sigma_1,\sigma_1\rangle=-2+\frac{1}{-\epsilon-\frac{1}{2}}=-4\left(\frac{\epsilon+1}{2\epsilon+1}\right)\,,\quad
  \langle \sigma_1,\sigma _2 \rangle= 2,\quad \langle\sigma_2,\sigma_2\rangle= -\left(\frac{2\epsilon+1}{\epsilon}\right)\,,\\ &
  \langle \sigma_2,\sigma _3 \rangle =\frac{1}{\epsilon},
  \langle \sigma_3,\,\sigma_3\rangle =\frac{1}{\epsilon(2\epsilon-1)}\,, \langle \sigma_2,\sigma_1\rangle=2,\,\langle \sigma_3,\sigma_2\rangle=\frac{1}{\epsilon}
    \end{split}
\end{align}
and all other entries vanish. We make the superscripts $+$ and $-$ on the co-cycles implicit in these pairings and can be understood. We define the residue-normalized intersection matrix for twisted co-cycles: $I_{ch}^{ij}=\langle\sigma_i,\sigma_j\rangle$. Similarly, one can compute the intersection number of twisted cycles with the method discussed earlier, given by,
\begin{align}
    \begin{split}
        I_h=\begin{pmatrix}
            -\frac{\zeta-1}{2(\zeta+1)} & -\frac{1}{2}& 0\\
            \frac{1}{2}& -\frac{\zeta+1}{2(\zeta-1)}& \frac{1}{\zeta-1}\\
            0& \frac{\zeta}{\zeta-1}& -\frac{2\zeta}{\zeta^2-1}
        \end{pmatrix}\,.
    \end{split}
\end{align}
Here $\zeta=e^{-2\pi i\epsilon}$, so $(c_1,c_2,c_3,c_4)=(-\zeta,-1,\zeta,-\zeta^{-1})$. We next compute the period matrices for the positive and inverse twists. Define the period matrices as,
\begin{align}
    P^+_{ij}=\int_{\gamma_j^+} u_0\, \sigma_i\,,\quad P_{ij}^-=\int_{\gamma_j^-}u_0^{-1} \sigma_i\,.
\end{align}
Now we have (with the conventional chosen branch),
\begin{align}
 u_+(x):=   u_0=(x+\tau)^{-\epsilon-\frac{1}{2}} x^{-1/2}(4x-3)^{-\epsilon}(1-x)^{\epsilon-\frac{1}{2}}\,,\quad u_{-}(x):=u_+(x)^{-1}\,.
\end{align}
The powers for positive (and negative) cocycles can be written as,
\begin{align}
\mathbf{r}^{+}=\left(-\epsilon-\frac{1}{2},-\frac{1}{2},-\epsilon,\epsilon-\frac{1}{2}\right):=(r_1^+,r_2^+,r_3^+,r_4^+),\quad \mathbf{r}^-=- \mathbf{r}^{+}\,.
\end{align}
The chosen oriented chambers are,
\begin{align}
    \begin{split}
    \ell_1=(-\tau,0)\,,\,\,\ell_2=\left(0,\frac{3}{4}\right)\,,\,\,\ell_3=\left(\frac{3}{4},1\right)\,.
    \end{split}
\end{align}
\textbf{Branch convention of the twisted cycle.} We choose the branch on the lower half $x$-plane. Consequently, a negative factor of $(x-x_a)$, evaluated on the lower boundary (i.e approaching a point on the real $x$-axis from the lower half of the complex $x$-plane) contributes,
\begin{align}
    (-\rho-i0)^a=e^{-i\pi a}\rho^a\,.
\end{align}
Now in all three chambers, $1-x>0$, therefore this factor will not carry any extra (phase appearing because of the chosen lower half branch) non-trivial monodromy. Equivalently, on this branch $u_{\mathbf r}=4^{r_3}e^{i\pi r_4}\prod_{a=1}^4(x-x_a)^{r_a}$. This $x$-independent factor accounts for writing $(1-x)^{r_4}$ instead of $(x-1)^{r_4}$; it is inverted for the dual twist. In the chamber $\ell_1$, the factors (for the positive cocycles ) $x$ and $x-\frac{3}{4}$ carries the phase: $e^{-i\pi(r_2+r_3)}$, on $\ell_2$ the factor $x-\frac{3}{4}$ carries the phase $e^{-i\pi r_3}$ and in $\ell_3$ the none of the factors will carry any non trivial phase. The phases of the multivalued function in different chambers may be listed as follows,
\begin{equation}
\begin{array}{c|cccc|c}
\text{chamber}
& x+\tau
& x
& x-\dfrac{3}{4}
& 1-x
& \text{phase of }\mathcal U
\\ \hline
\ell_{1}=(-\tau,0)
& + & - & - & +
& e^{-i\pi(r_{2}+r_{3})}
\\[1mm]
\ell_{2}=\left(0,\dfrac{3}{4}\right)
& + & + & - & +
& e^{-i\pi r_{3}}
\\[1mm]
\ell_{3}=\left(\dfrac{3}{4},1\right)
& + & + & + & +
& 1
\end{array}
\end{equation}
Moreover, with the chosen chambers, we can define the regularised contour of integration as we discussed earlier,
\begin{align}
    \gamma_j^{(\mathbf{r})}=I_j\otimes u_{\mathbf{r}}+ \frac{S_j\otimes u_{\mathbf{r}}}{c_{j}(\mathbf{r})-1}-\frac{S_{j+1}\otimes u_{\mathbf{r}}}{c_{j+1}(\mathbf{r})-1},\quad c_j(\mathbf{r})=e^{2\pi i r_j}\,.\label{6.65mm}
\end{align}
Accordingly, for $\mathbf{r}=\boldsymbol r^+$, $\gamma_j^{(\mathbf{r})}$ is $\gamma_j^+$ and follows for the negative branch. \\\\
\textbf{Meromorphic continuation and endpoint regularisation.} Let's consider the chamber $\ell_2$. Now around left endpoint i.e $x=0$, the integrand (for e.g $P_{22}^{+}\sim\int u_+\sigma_2$),
\begin{align}
  h_0 x^{-3/2}dx+O(x^{-1/2})dx\,,\qquad h_0\ne0\,.\label{6.66kk} 
\end{align}
Now, a naive integral over the $\ell_2$ chamber will cause a divergence at the $x=0$ point. However, a careful analysis of the regulated cycle in \eqref{6.65mm} will take care of the divergence. Assume around $x=0$, there is a closed circle $S_0$ with radius $\rho$, and the expansion in \eqref{6.66kk} valid upto a fixed point $\delta$ such that $0<\rho<\delta$.  Now the integral around the circle gives,
\begin{align}
  \frac{1}{c_2-1}\int_{S_0}h_0x^{r_2-1}dx=\frac{h_0\rho^{r_2}}{r_2}\,.\label{6.67jj}
\end{align}
Now for the straight contour, the integral (around the vicinity of $x=0$) becomes,
\begin{align}
    \int_\rho^\delta h_0x^{r_2-1}dx+O(1)=\frac{h_0}{r_2}(\delta^{r_2}-\rho^{r_2})+O(1)\,.\label{6.68jj}
\end{align}
As we see by adding \eqref{6.67jj} and \eqref{6.68jj}, the chosen contour $\gamma_j^+$, cancels the endpoint divergence at $\rho\to 0$. Therefore, while doing the period integral in Euler representation (as we see shortly) the chamber integral can be thought of as a meromorphic continuation of the ordinary integral,
\begin{align}
    \int_0^\delta h_0 x^{r_2-1}dx=\frac{h_0\delta^{r_2}}{r_2}\,,\quad \textrm{Re}(r_2)>0\,,
\end{align}
which is exactly what we get from the circular regularization around the endpoint. Therefore, one can deduce that with the contour prescription, the standard Euler integral can be computed by meromorphic continuation to the region $\textrm{Re}(r_2)\le0$. The same analogy holds for each chamber and each endpoint. With this in mind, we can directly compute the period matrices using the Euler parametrization that we discussed in the previous section. More generally, a local expansion $t^{r_k-1}\sum_{n\geq0}h_nt^n\,dt$ has primitive $\sum_{n\geq0}h_nt^{r_k+n}/(r_k+n)$. The endpoint circle cancels each cutoff-dependent term in this primitive, with the corresponding endpoint orientation. This proves the cancellation at either end of all three chambers. One may first establish the Euler formulas for independent exponents with positive real parts, and then continue meromorphically in those exponents to $\mathbf r^\pm$; this is not an assertion that the divergent real interval integral converges at the physical exponents. At resonance, the resulting meromorphic family must be treated by its Laurent expansion. We define the period matrix in this sense as,
\begin{align}
    \begin{split}
        P_{ij}^{\pm}=\mathcal{MC}\int_{\ell_j} u_{\pm}(x)\sigma_i(x)=\mathcal K_j(\mathbf r^\pm)Q_{ij}^\pm\,,
    \end{split}
\end{align}
where, 
\begin{align}
\begin{split}
&\mathcal{K}_1(\mathbf{r})
=
e^{-i\pi(r_2+r_3)}
\tau^{r_1+r_2}
L^{r_3}T^{r_4}\,,\quad
\mathcal{K}_2(\mathbf{r})
=
e^{-i\pi r_3}
\tau^{r_1}
\left(\frac{3}{4}\right)^{r_2}
3^{r_3}\,,\quad 
\mathcal{K}_3(\mathbf{r})
=
\left(\frac{L}{4}\right)^{r_1}
\left(\frac{3}{4}\right)^{r_2}
4^{-r_4}\,.
\end{split}
\end{align}
Here $L=3+4\tau$ and $T=1+\tau$. These factors follow from the affine substitutions $x=-\tau+\tau t$, $x=3t/4$, and $x=(3+t)/4$, respectively. For example, in the first chamber,
\begin{align}
 u_{\mathbf r}=\mathcal K_1(\mathbf r)t^{r_1}(1-t)^{r_2}
 (1-X_1t)^{r_3}(1-Y_1t)^{r_4}\,,\qquad
 \sigma_1=\frac{dt}{t(1-t)}\,.
\end{align}
The continued Euler identity
\begin{align}
 \mathcal{MC}\int_0^1t^{a-1}(1-t)^{b-1}(1-Xt)^{-c}(1-Yt)^{-d}\,dt
 =B(a,b)F_1(a;c,d;a+b;X,Y)
\end{align}
therefore gives $Q_{11}(\mathbf r)=B(r_1,r_2)F_1(r_1;-r_3,-r_4;r_1+r_2;X_1,Y_1)$. Pulling back the other two cocycles and applying the same identity yields the entries below.
Note that every exponent changes sign under $\mathbf{r}^+ \mapsto\mathbf{r}^-=-\mathbf{r}^+$. Therefore, we have $\mathcal{K}_j(\mathbf{r}^-)=\mathcal{K}_j(\mathbf{r}^+)^{-1}$. We note down all nine components of $Q_{ij}^\pm$. For the positive branch, we have
{\small
\allowdisplaybreaks[2]
\begin{align}
\begin{split}
&Q_{11}^{+} =
B\!\left(-\epsilon-\frac12,-\frac12\right)
\mathcal{A}_{1}\!\left(
-\epsilon-\frac12;\epsilon,\frac12-\epsilon;-\epsilon-1
\right)\,,\,\,
Q_{21}^{+}
=
-\frac{3}{L}
B\!\left(\frac12-\epsilon,-\frac12\right)
\mathcal{A}_{1}\!\left(
\frac12-\epsilon;1+\epsilon,\frac12-\epsilon;-\epsilon
\right)\,,\\&
Q_{12}^{+}
=
-B\!\left(-\frac12,1-\epsilon\right)
\mathcal{A}_{2}\!\left(
-\frac12;\frac32+\epsilon,\frac12-\epsilon;\frac12-\epsilon
\right)\,,\,\,
Q_{22}^{+}=
B\!\left(-\frac12,-\epsilon\right)
\mathcal{A}_{2}\!\left(
-\frac12;\frac12+\epsilon,\frac12-\epsilon;-\frac12-\epsilon
\right)\,,\\&
Q_{32}^{+}
=
-\frac14
B\!\left(\frac12,-\epsilon\right)
\mathcal{A}_{2}\!\left(
\frac12;\frac12+\epsilon,\frac32-\epsilon;\frac12-\epsilon
\right)\,,\,\, 
Q_{13}^{+}
=
-\frac{4\tau}{3L}
B\!\left(1-\epsilon,\epsilon+\frac12\right)
\mathcal{A}_{3}\!\left(
1-\epsilon;\frac32+\epsilon,\frac32;\frac32
\right),\\&
Q_{23}^{+}
=
-B\!\left(-\epsilon,\epsilon+\frac12\right)
\mathcal{A}_{3}\!\left(
-\epsilon;\frac12+\epsilon,\frac32;\frac12
\right)\,,\quad 
Q_{33}^{+}
=
B\!\left(-\epsilon,\epsilon-\frac12\right)
\mathcal{A}_{3}\!\left(
-\epsilon;\frac12+\epsilon,\frac12;-\frac12
\right)\,,\\&
Q_{31}^{+}
=
-\frac{\tau}{LT}
B\!\left(\frac12-\epsilon,\frac12\right)
\mathcal{A}_{1}\!\left(
\frac12-\epsilon;1+\epsilon,\frac32-\epsilon;1-\epsilon
\right)\,.
\end{split}
\end{align}
}
Similarly, for the negative branch, we have,
{\small
\allowdisplaybreaks[2]
\begin{align}
\begin{split}
&Q_{11}^{-}
=
B\!\left(\epsilon+\frac12,\frac12\right)
\mathcal{A}_{1}\!\left(
\epsilon+\frac12;-\epsilon,\epsilon-\frac12;\epsilon+1
\right)\,,\,\,
Q_{21}^{-}
=
-\frac{3}{L}
B\!\left(\epsilon+\frac32,\frac12\right)
\mathcal{A}_{1}\!\left(
\epsilon+\frac32;1-\epsilon,\epsilon-\frac12;\epsilon+2
\right)\,,\\&
Q_{12}^{-}
=
-B\!\left(\frac12,\epsilon+1\right)
\mathcal{A}_{2}\!\left(
\frac12;\frac12-\epsilon,\epsilon-\frac12;\epsilon+\frac32
\right)\,,\,\,
Q_{22}^{-}
=
B\!\left(\frac12,\epsilon\right)
\mathcal{A}_{2}\!\left(
\frac12;-\frac12-\epsilon,\epsilon-\frac12;\epsilon+\frac12
\right)\,,\\&
Q_{32}^{-}
=
-\frac14
B\!\left(\frac32,\epsilon\right)
\mathcal{A}_{2}\!\left(
\frac32;-\frac12-\epsilon,\epsilon+\frac12;\epsilon+\frac32
\right)\,,\,\,
Q_{13}^{-}
=
-\frac{4\tau}{3L}
B\!\left(1+\epsilon,\frac32-\epsilon\right)
\mathcal{A}_{3}\!\left(
1+\epsilon;\frac12-\epsilon,\frac12;\frac52
\right)\,,\\&
Q_{23}^{-}
=
-B\!\left(\epsilon,\frac32-\epsilon\right)
\mathcal{A}_{3}\!\left(
\epsilon;-\frac12-\epsilon,\frac12;\frac32
\right)\,,\,\,
Q_{33}^{-}
=
B\!\left(\epsilon,\frac12-\epsilon\right)
\mathcal{A}_{3}\!\left(
\epsilon;-\frac12-\epsilon,-\frac12;\frac12
\right)\,,\\&
Q_{31}^{-}=
-\frac{\tau}{LT}
B\!\left(\epsilon+\frac32,\frac32\right)
\mathcal{A}_{1}\!\left(
\epsilon+\frac32;1-\epsilon,\epsilon+\frac12;\epsilon+3
\right)
\end{split}
\end{align}
}
with, $L=3+4\tau$ and $T=1+\tau$ and $\mathcal A_j(a;b,b';c)=F_1(a;b,b';c;X_j,Y_j)$, where,
\begin{align}
    \begin{array}{c|cc}
j & X_j & Y_j \\
\hline
1 & \dfrac{4\tau}{L} & \dfrac{\tau}{T} \\[6pt]
2 & -\dfrac{3}{4\tau} & \dfrac{3}{4} \\[6pt]
3 & -\dfrac{1}{L} & -\dfrac{1}{3}
\end{array}
\end{align}

\paragraph{Twisted Riemann period relation.}
We now combine the two intersection matrices with the explicit positive and negative periods, which gives the period relation explicitly. Rows of $P^\pm$ label cocycles, while columns label the three ordered
chambers/cycles. The period relation is given by
\begin{align}
(P^-)^T I_{ch}^{-1}P^+=2\pi i I_h^T\,,
 \qquad
 P^+I_h^{-T}(P^-)^T=2\pi i I_{ch}\,.
 \label{eq:wqft-trpr}
\end{align}
For completeness, the transpose and normalization can be fixed directly
by Poincaré duality. Write the compactly supported inverse-twist dual of
$\gamma_j^+$ as \cite{Cho_Matsumoto_1995}
\begin{align}
 \theta_c(\gamma_j^+)=\sum_k\Theta_{kj}\sigma_k^-.
\end{align}
Pairing this equation with the positive cocycles and with the negative
cycles, respectively, gives
\begin{align}
 P^+=(2\pi i I_{ch})\Theta,
 \qquad I_h=\Theta^T P^-.
\end{align}
Eliminating $\Theta$ and transposing proves the first identity in
\eqref{eq:wqft-trpr}; inversion gives the second. In particular, replacing
$I_h^T$ by $I_h$ while keeping the same period ordering would change the
off-diagonal relations. For the particular family of integrals, the inverse of the cocycle intersection matrix is given by,
\begin{align}
 I_{ch}^{-1}
 =\frac{1}{2\epsilon+3}
 \begin{pmatrix}
 -(2\epsilon+1)&-\dfrac{2\epsilon+1}{2}
     &\dfrac{(2\epsilon-1)(2\epsilon+1)}{2}\\[5pt]
 -\dfrac{2\epsilon+1}{2}&-(\epsilon+1)
     &(\epsilon+1)(2\epsilon-1)\\[5pt]
 \dfrac{(2\epsilon-1)(2\epsilon+1)}{2}
     &(\epsilon+1)(2\epsilon-1)&(2\epsilon-1)(2\epsilon+1)
 \end{pmatrix}.
 \label{eq:wqft-ich-inverse}
\end{align}
Using $\zeta=e^{-2\pi i\epsilon}$, the complete quadratic relation becomes
\begin{align}
 \boxed{
 (P^-)^T I_{ch}^{-1}P^+
 =\pi\begin{pmatrix}
 -\tan(\pi\epsilon)&i&0\\
 -i&\cot(\pi\epsilon)&-\cot(\pi\epsilon)+i\\
 0&-\cot(\pi\epsilon)-i&2\csc(2\pi\epsilon)
 \end{pmatrix}.}
 \label{eq:wqft-trpr-explicit}
\end{align}
Every entry on the left is a specified sum of products of the
beta and Appell functions. Although each period depends on $\tau$,
all nine combinations in \eqref{eq:wqft-trpr-explicit} are independent
of $\tau$. It is important to note that the inverse twist reverses \emph{all four}
exponents, not merely $\epsilon$. Indeed, for the branch used here,
\begin{align}
 u_-(x;\epsilon)
 =x(1-x)(x+\tau)\,u_+(x;-\epsilon)\,.
 \label{eq:wqft-dual-twist-shift}
\end{align}
Consequently, $P^-(\epsilon)$ cannot be replaced by $P^+(-\epsilon)$ without also transforming the cocycles and matching the loaded cycles. Also, an overall $x$-independent normalization, which we omit in the calculation, does not change the overall result:
$u_+\mapsto\mathcal N(\tau,\epsilon)u_+$ and
$u_-\mapsto\mathcal N(\tau,\epsilon)^{-1}u_-$ multiply $P^+$ and $P^-$ by reciprocal factors. Both identities in \eqref{eq:wqft-trpr} are therefore unchanged. The symbolic intersection-matrix identities and independent high-precision checks of the beta--Appell period relations are provided in \texttt{06-twisted-riemann-period-relations.nb}.

\paragraph{Quadratic identities between Feynman integral and its corresponding `dual'.}
Let $\mathbf Q_j^\pm=(Q_{1j}^\pm,Q_{2j}^\pm,Q_{3j}^\pm)^T$ denote the
columns already given above, and set
$$D=\operatorname{diag}(\mathcal K_1(\mathbf r^+),
\mathcal K_2(\mathbf r^+),\mathcal K_3(\mathbf r^+))\,.$$
Since $P^+=Q^+D$ and $P^-=Q^-D^{-1}$, the period relation takes the form,
\begin{align}
 (Q^-)^T I_{ch}^{-1}Q^+=2\pi i D I_h^T D^{-1}\,.
 \label{eq:wqft-q-relation}
\end{align}
The diagonal identities are particularly simple:
\begin{align}
 &(\mathbf Q_1^-)^T I_{ch}^{-1}\mathbf Q_1^+
   =-\pi\tan(\pi\epsilon)\,,\quad
   (\mathbf Q_2^-)^T I_{ch}^{-1}\mathbf Q_2^+
   =\pi\cot(\pi\epsilon)\,,\quad
 (\mathbf Q_3^-)^T I_{ch}^{-1}\mathbf Q_3^+=2\pi\csc(2\pi\epsilon)\,. \label{eq:wqft-q-diagonal-one}
\end{align}
For example, substituting \eqref{eq:wqft-ich-inverse} into the first identity gives the explicit quadratic relation
\begin{align}
 &-(2\epsilon+1)Q_{11}^-Q_{11}^+
 -\frac{2\epsilon+1}{2}
       (Q_{11}^-Q_{21}^++Q_{21}^-Q_{11}^+)
 -(\epsilon+1)Q_{21}^-Q_{21}^+\notag\\
 &\quad+\frac{(2\epsilon-1)(2\epsilon+1)}{2}
       (Q_{11}^-Q_{31}^++Q_{31}^-Q_{11}^+)
 +(\epsilon+1)(2\epsilon-1)
       (Q_{21}^-Q_{31}^++Q_{31}^-Q_{21}^+)\notag\\
 &\quad+(2\epsilon-1)(2\epsilon+1)Q_{31}^-Q_{31}^+
 =-\pi(2\epsilon+3)\tan(\pi\epsilon)\,.
 \label{eq:wqft-expanded-quadratic}
\end{align}
Notably, the right-hand side is independent of the kinematic parameter $\tau\,.$ The particular nonadjacent chambers give two vanishing identities,
\begin{align}
 (\mathbf Q_1^-)^T I_{ch}^{-1}\mathbf Q_3^+=0\,,
 \qquad
 (\mathbf Q_3^-)^T I_{ch}^{-1}\mathbf Q_1^+=0\,.
 \label{eq:wqft-cross-orthogonality}
\end{align}
Thus, the Appell functions are evaluated at the distinct argument pairs $(X_1,Y_1)$ and $(X_3,Y_3)$ that obey exact bilinear cancellation relations. For adjacent chambers, the phases in the ratios of $\mathcal K_j$ cancel the phases in $I_h^T$. The four resulting identities are
\begin{equation}
\boxed{
\begin{aligned}
(\mathbf Q_1^-)^T I_{ch}^{-1}\mathbf Q_2^+
&=
-\frac{\pi\sqrt3}{2\sqrt{\tau T}}
\left(\frac{3T}{L}\right)^\epsilon\,,
\qquad
(\mathbf Q_2^-)^T I_{ch}^{-1}\mathbf Q_1^+
=
-\frac{2\pi\sqrt{\tau T}}{\sqrt3}
\left(\frac{L}{3T}\right)^\epsilon\,,
\\[6pt]
(\mathbf Q_2^-)^T I_{ch}^{-1}\mathbf Q_3^+
&=
-\frac{\pi}{4}\sqrt{\frac{L}{\tau}}
\left(\frac{L}{3\tau}\right)^\epsilon
\csc(\pi\epsilon)\,,
\qquad
(\mathbf Q_3^-)^T I_{ch}^{-1}\mathbf Q_2^+
=
-4\pi\sqrt{\frac{\tau}{L}}
\left(\frac{3\tau}{L}\right)^\epsilon
\csc(\pi\epsilon)\,.
\end{aligned}
}
\label{eq:wqft-adjacent-appell}
\end{equation}
Here $L=3+4\tau$, $T=1+\tau$. Equations \eqref{eq:wqft-q-diagonal-one}--\eqref{eq:wqft-adjacent-appell}
give all nine entries of \eqref{eq:wqft-q-relation}. As a corollary, we may derive the determinant identities.
The determinants of the two intersection matrices are
\begin{align}
 \det I_{ch}=\frac{4(2\epsilon+3)}{\epsilon(4\epsilon^2-1)}\,,
 \qquad
 \det I_h=-\frac{\zeta}{2(\zeta^2-1)}
          =-\frac{i}{4\sin(2\pi\epsilon)}\,.
\end{align}
Taking determinants in \eqref{eq:wqft-trpr} therefore yields
\begin{align}
 \det P^+\det P^-
 =\det Q^+\det Q^-
 =-\frac{8\pi^3(2\epsilon+3)}
 {\epsilon(4\epsilon^2-1)\sin(2\pi\epsilon)}\,.
 \label{6.94kk}
\end{align}
This is a kinematics-independent product identity for two explicitly
known $3\times3$ beta--Appell determinants. In particular, at generic $\epsilon$, neither period matrix can lose rank at a nonsingular value of $\tau$ in the chosen chamber. 
\textcolor{black}{The $\epsilon\to0$ behaviour of \eqref{6.94kk} is itself important:
$\det P^{+}\det P^{-}\sim 12\pi^{2}/\epsilon^{2}$, so the two period matrices cannot both stay finite and invertible in that limit. This is the quantitative
form of the statement made in Sec.~(\ref{sec:period-d4}), that the rank of the
period matrix drops from three to two at exactly $d=4\,.$ The marked point
$x=3/4$ ceases to be a branch point, and two of the three chambers merge.}

\paragraph{Corollary-I. \textit{The beta reflection identity}.}
The first chamber,$(-\tau,0)$, shrinks when $\tau\to0^+$. After extracting its explicit prefactor, $X_1,Y_1\to0$, so every Appell function in its column becomes unity. The beta recurrence relation ($B(a+1,b)=\frac{a}{a+b}B(a,b)$) gives
\begin{align}
 &\mathbf Q_1^+\longrightarrow
 B\!\left(-\epsilon-\tfrac12,-\tfrac12\right)
 \begin{pmatrix}1\\[-2pt]-\dfrac{\epsilon+1/2}{\epsilon+1}\\[2pt]0\end{pmatrix}\,,\quad
 \mathbf Q_1^-\longrightarrow
 B\!\left(\epsilon+\tfrac12,\tfrac12\right)
 \begin{pmatrix}1\\[-2pt]-\dfrac{\epsilon+1/2}{\epsilon+1}\\[2pt]0\end{pmatrix}\,.
\end{align}
The quadratic form on this vector equals
$-(2\epsilon+1)/(4(\epsilon+1))\,.$
Consequently, the first equation of \eqref{eq:wqft-q-diagonal-one} reduces to
\begin{align}
\boxed{
 B\!\left(-\epsilon-\tfrac12,-\tfrac12\right)
 B\!\left(\epsilon+\tfrac12,\tfrac12\right)
 =\frac{4\pi(\epsilon+1)}{2\epsilon+1}\tan(\pi\epsilon)\,.}
 \label{eq:wqft-beta-degeneration}
\end{align}
This is the beta reflection identity, here recovered as a
degeneration of the kinematics-dependent Appell relation, \textit{which also serves as a crosscheck against our period calculation.}

\paragraph{Corollary-II. \textit{The Elliptic Legendre relation}.}
A direct substitution of $d=4$ i.e $\epsilon=0$ will make an immediate disaster, as the geometric structure changes with exact four dimension as $x=3/4$ is no longer a branch point of the Baikov polynomial. At $\epsilon=0$ the twist becomes
\begin{equation}
 u_0(x)=\frac{1}{\sqrt{x(x+\tau)(1-x)}}\,,
 \qquad \mathcal E_x:\quad Y^2=x(x+\tau)(1-x)\,,
 \qquad \tau>0\,.
 \label{eq:wqft-compact-x-curve}
\end{equation}
The smooth compactification of $\mathcal E_x$ is a genus-one (quotient-) curve (we will comment on this in great detail in Sec.~\eqref{sec9} ),
branched at $-\tau,0,1,\infty$. The point $x=3/4$ is unramified and no longer carries any non-trivial monodromy. Therefore, we can only choose two logarithmic cocycles. We choose, $\sigma_2+\sigma_3=dx/x-dx/(x-1)$
has no pole at $x=3/4$, whereas the two summands separately do.
Thus, the $\epsilon\to 0$ limit requires a choice of regular classes; it can not be
obtained by a naive substitution of  $\epsilon=0$ into every entry of the generic
$3\times3$ period matrix. We use the twisted de -Rham basis \footnote{One may use the standard logarithmic basis: $\rho_1=\sigma_1,\,\,\rho_2=\sigma_2+\sigma_3$ to deduce the same conclusion. However, to make the calculation less heavy, we choose the standard de Rham basis. It is straightforward to show that these two bases are related by a rational transformation that depends only on the kinematic parameter $\tau$. }
\begin{equation}
 \alpha=\frac{dx}{Y}\,,\qquad \beta=\frac{x\,dx}{Y}\,.
 \label{eq:wqft-compact-x-forms}
\end{equation}
Here $\alpha$ is holomorphic and $\beta$ is a differential of the
second kind, with a double pole at the single point at infinity.

The lower-half-plane convention fixes the forward-sheet values to be
\begin{equation}
 Y(x)=
 \begin{cases}
 \sqrt{x(x+\tau)(1-x)}\,,&0<x<1\,,\\[3pt]
 -i\sqrt{(-x)(x+\tau)(1-x)},&-\tau<x<0\,,
 \end{cases}
 \label{eq:wqft-compact-x-branches}
\end{equation}
where the square roots on the right are positive. In particular,
$1/Y$ is positive imaginary on $(-\tau,0)$.
Let $A$ be the closed lift of $[0,1]$ and $B$ the closed lift of
$[-\tau,0]$: each follows its interval from left to right on the
specified sheet and returns on the opposite sheet. The $A$ interval
joins the second and third chambers across the now regular point
$3/4$, while the $B$ interval is the first chamber. These definitions
fix the compact column ordering to be $(A,B)$.

The two closed lifts meet at the ramification point above $x=0$.
In the local coordinate $s=\sqrt{x}$, their oriented tangents point
along the positive real and positive imaginary directions, respectively.
Consequently,
\begin{equation}
 A\cdot B=+1\,,\qquad
 I_h^{\mathrm{ell}}=\begin{pmatrix}0&1\\-1&0\end{pmatrix}\,.
 \label{eq:wqft-compact-x-homology}
\end{equation}
This is a canonical choice of basis of $H_1(\mathcal E_x,\mathbb Z)\,.$ We define the period matrix in the (twisted-) de Rham basis by
\begin{equation}
 P=
 \begin{pmatrix}
 \displaystyle\int_0^1\alpha&\displaystyle\int_{-\tau}^0\alpha\\[5pt]
 \displaystyle\int_0^1\beta&\displaystyle\int_{-\tau}^0\beta
 \end{pmatrix}.
\end{equation}
The substitutions $x=\sin^2\theta$ on $[0,1]$ and
$x=-\tau\sin^2\theta$ on $[-\tau,0]$ give
\begin{align}
\begin{split}
 \int_0^1\alpha
   &=2\int_0^{\pi/2}\frac{d\theta}{\sqrt{\tau+\sin^2\theta}}
     =\frac{2}{\sqrt\tau}K(-1/\tau)\,,\\
 \int_0^1\beta
   &=2\int_0^{\pi/2}\frac{\sin^2\theta\,d\theta}
                              {\sqrt{\tau+\sin^2\theta}}
     =2\sqrt\tau\,\Big[E(-1/\tau)-K(-1/\tau)\Big]\,,\\
 \int_{-\tau}^0\alpha
   &=2i\int_0^{\pi/2}\frac{d\theta}{\sqrt{1+\tau\sin^2\theta}}
     =2iK(-\tau)\,,\\
 \int_{-\tau}^0\beta
   &=-2i\tau\int_0^{\pi/2}\frac{\sin^2\theta\,d\theta}
                                   {\sqrt{1+\tau\sin^2\theta}}
     =2i\Big[K(-\tau)-E(-\tau)
     \Big]\,.
     \end{split}
\end{align}
\textcolor{black}{Here $K$ and $E$ are the complete elliptic integrals in the \emph{parameter}
convention fixed in Sec.~(\ref{new:sec2.2.3}). Hence,}
\begin{equation}
 P=
 \begin{pmatrix}
 \dfrac{2}{\sqrt\tau}K(-1/\tau)&2iK(-\tau)\\[5pt]
 2\sqrt\tau\,[E(-1/\tau)-K(-1/\tau)]
     &2i[K(-\tau)-E(-\tau)]
 \end{pmatrix}\,.
 \label{eq:wqft-elliptic-segment-periods}
\end{equation}
To compute the intersection number of the twisted cocycles, use the definition of \eqref{4.18jj},
$t=x^{-1/2}$ at infinity. Choosing the sign of $t$ so that
$Y=it^{-3}(1+O(t^2))\,,$ we find
\begin{equation}
 \alpha=2i\,(1+O(t^2))\,dt,\qquad
 \beta=\left(\frac{2i}{t^2}+O(1)\right)dt\,.
\end{equation}
A local primitive of $\alpha$ is $\psi_\alpha=2it+O(t^3)\,,$ and hence
\begin{equation}
 \langle\alpha,\beta\rangle_{ch}
 =\operatorname{Res}_{t=0}(\psi_\alpha\beta)
 =\operatorname{Res}_{t=0}\left(-\frac4t\,dt+O(t)dt\right)=-4\,.
\end{equation}
There are no other poles contributing to this intersection number. Therefore, we get,
\begin{equation}
 I_{ch}^{\mathrm{ell}}
 =-4\begin{pmatrix}0&1\\-1&0\end{pmatrix}\,.
 \label{eq:wqft-compact-x-cohomology}
\end{equation}
The ordinary Riemann bilinear relation now reads (with $\Pi=2P$)\footnote{At $\epsilon=0$ the two positive and negative local systems are isomorphic, i.e $u_-$ and $u_+$ are related by a single-valued rational transformation. Therefore, one can choose a gauge transformation to the logarithmic basis to obtain the same period matrix for the negative branch. To express the 
period relation using a single de Rham period matrix, we pass to the
gauge-adapted dual basis
\[
\widetilde{\rho}_a^{-}=g^{-1}\rho_a^{+}\,,
\qquad
g(x)=x(x+\tau)(1-x).
\]
Then
\[
u_{-}\widetilde{\rho}_a^{-}
=u_{+}\rho_a^{+}\,,
\]
so the plus and gauge-transformed minus periods are represented by the
same closed-cycle matrix $\Pi$. If the same logarithmic basis
are retained in both twisted cohomology spaces, the two period matrices
must instead be kept distinct. For general epsilon, the positive and negative branches are not connected by the rational transformation, and we can not find a rational gauge transformation that retains the integrand invariant. Hence, the distinction is needed in the general dimension.},
\begin{equation}
 \Pi\,(I_h^{\mathrm{ell}})^{-T}\Pi^T
 =2\pi i I_{ch}^{\mathrm{ell}}\,.
\end{equation}
Using $\Pi J\Pi^T=(\det\Pi)J$ for
$J=\bigl(\begin{smallmatrix}0&1\\-1&0\end{smallmatrix}\bigr)$ gives
\begin{equation}
 \det\Pi=-8\pi i\,,\qquad
 \det P=-2\pi i\,.
 \label{eq:wqft-compact-x-determinant}
\end{equation}
Equivalently,
\begin{align}
\boxed{\frac{K(-1/\tau)E(-\tau)}{\sqrt\tau}
  +\sqrt\tau\,K(-\tau)E(-1/\tau)
  -\left(\sqrt\tau+\frac1{\sqrt\tau}\right)
       K(-1/\tau)K(-\tau)=\frac\pi2\,.}
 \label{eq:wqft-elliptic-legendre-check}
\end{align}
This is the classical Legendre relation in the two reciprocal negative
parameters natural to these $x$-intervals. It provides a consistency check after selecting the regular two-dimensional
cohomology sector.  The corresponding checks are included in the Mathematica notebooks \texttt{03-elliptic-fundamental-matrix-checks.nb} and \texttt{06-twisted-riemann-period-relations.nb} with the \texttt{arXiv} submission.

\section{$\epsilon$-expansion of master integrals on the maximal cut} \label{sec:epsilonexpnad}
In this section, we will explicitly compute the master integrals in their $\epsilon$-expansion. Note that the maximal cut admits an exact closed-form representation at generic $\epsilon$ in terms of the generalized Hypergeometric function, reconstructing the full Feynman integral requires its Laurent
expansion around the physical dimension, since the maximal cut provides the
homogeneous input to a differential equation coupled to lower sectors. In the
elliptic case, expanding the $\epsilon$-dependent Appell $F_1$ function
generates logarithmic insertions over the elliptic curve, which naturally
organize into Elliptic Multiple Polylogs (eMPLs) \cite{Broedel:2017kkb,Broedel:2017siw} and, after passing to the modular parameter, into iterated
integrals of modular forms. Thus, the Appell function efficiently
encodes the all-orders result, while the $\epsilon$-expanded form makes the
elliptic and modular structure needed for solving the full inhomogeneous
system explicit. Therefore, to determine the full function space, it is important to study the epsilon expansion rather than a formal and  general Appell representation. It is worth emphasizing that the usefulness of the eMPL representation does
not arise from replacing the integral by an elementary function, since eMPLs are themselves defined as iterated integrals. Their advantage is that they provide a functional basis adapted to the elliptic geometry and closed
under the operations required in the differential-equation approach. In
particular, differentiation removes the first integration kernel, while
integration against a kernel of the elliptic alphabet simply increases the
length of the iterated integral. This makes the recursive inclusion of the
lower-sector inhomogeneities considerably more systematic than working
directly with parameter derivatives of Appell $F_1$ functions generated by
the $\epsilon$-expansion. Moreover, the eMPL representation makes the
transcendental weight, the relevant marked points, and the underlying elliptic
alphabet manifest. After uniformization in terms of the modular parameter,
the same structure can be organized in terms of iterated integrals of modular
forms, providing a natural language for solving the complete differential
system order by order in $\epsilon$. Before going to the detailed computations, we would like to make a detour to some aspects of elliptic polylogs defined on the genus-one surface. A quartic elliptic curve defined on a genus one surface has the following form,
\begin{align}
    y^2=P_{4}(z)=-(z-a_1)(z-a_2)(z-a_3)(z-a_4)
\end{align}
We define the branch point vector as $\vec a=(a_1,a_2,a_3,a_4)$, with $a_1=-1,a_2=1,a_3=-ir, a_4=ir$. The fundamental periods on this elliptic curve can be written (as defined in \cite{Broedel:2017kkb}),
\begin{align}
    \omega_1=2c_4\int_{a_2}^{a_3} \frac{dz}{y}=\textcolor{black}{-c_4(P^{0}_{11}+P^{0}_{12})},\qquad \omega_2=2c_4\int_{a_1}^{a_2} \frac{dz}{y}=\textcolor{black}{2c_4 P^{0}_{11}},\quad c_4=\frac{1}{2}\sqrt{a_{13}a_{24}}=\textcolor{black}{\frac{i+r}{2}}
\end{align}
Any other periods can be written as a linear combination of the fundamental periods with integer coefficients. 
\begin{align}
    2c_4\int_{a_i}^{a_j} \frac{dz}{y}=m_{ij}\omega_1+n_{ij}\omega_2, \qquad m_{ij},\,n_{ij}\in \mathbb{Z}.
\end{align}
Moreover, the quasi-periods can be defined as,
\begin{align}
    \eta_1=-\frac{1}{2}\int_{a_2}^{a_3}dz \,\,\widetilde{\Phi}_4(z),\,\,\,\eta_2=-\frac{1}{2}\textcolor{black}{\int_{a_1}^{a_2}}dz \,\,\widetilde{\Phi}_4(z)
\end{align}
where we define,
\begin{align}
    \widetilde{\Phi}_4(z):=\frac{1}{c_4 y
    }\left(z^2-\frac{s_1}{2}z+\frac{s_2}{6}\right)
\end{align}
where, $s_n(\vec a)$ is the elementary symmetric function of degree $n$. \textcolor{black}{The normalizations are fixed by the Legendre-type bilinear relation between periods
and quasi-periods, which for the conventions above reads
$
\omega_1\eta_2-\omega_2\eta_1=i\pi\,.$}
\subsection{Polylogs associated with a quartic elliptic curve}
For generic kinematics, the compactified quartic curve is associated with genus-one Riemann surface and its periods are associated with a complex torus. The modular parameter $\tau_{\mathcal{E}}$  is defined as,
\begin{align}
 \tau_{\mathcal E}:=\frac{\omega_2}{\omega_1}\,,
 \qquad \operatorname{Im}\tau_{\mathcal E}>0\,,
 \qquad
 v(z):=\frac{1}{\omega_1}\int_{a_1}^{z}\frac{c_4\,dt}{y(t)}
 \quad \bmod\bigl(\mathbb Z+\tau_{\mathcal E}\mathbb Z\bigr)\,.
\end{align}
The Abel map $v(z)$ is defined along a chosen path on the curve. One can, in principle, change the path that may shift $v$ by a period.
As the main ingredient to construct the elliptic kernel of the iterated integrals is the \emph{Jacobi theta} function may be introduced as \cite{Broedel:2017kkb},
\begin{align}
\mathfrak{\vartheta}_1(v\mid\tau_{\mathcal E})
 &=-i\sum_{m\in\mathbb Z}(-1)^m
   e^{\pi i\tau_{\mathcal E}(m+1/2)^2+2\pi i(m+1/2)v}\,.
\end{align}
and correspondingly the Eisenstein-Kronecker (EK) function defines the kernels
$g^{(n)}$ as,
\begin{align}
 F(v,\alpha\mid\tau_{\mathcal E})
 &=\frac{\vartheta_1'(0\mid\tau_{\mathcal E})\,
          \vartheta_1(v+\alpha\mid\tau_{\mathcal E})}
         {\vartheta_1(v\mid\tau_{\mathcal E})\,
          \vartheta_1(\alpha\mid\tau_{\mathcal E})}
  =\frac{1}{\alpha}\sum_{n=0}^{\infty}
       g^{(n)}(v\mid\tau_{\mathcal E})\,\alpha^n=\frac{1}{\alpha}\exp\left(-\sum_{j\ge1}\frac{(-\alpha)^j}{j}(E_j(z)-G_j)\right)
 \label{7.8ii}
\end{align}
where $E_j$ and $G_j$ are Eisenstein functions and series defined as,
\begin{align}
    E_j(z):=\sum_{(m,n)\ne0}\frac{1}{(z+m\omega_1+n\omega_2)^j}\,,\quad G_j:=E_{j}(0)\,.
\end{align}
In particular,
$g^{(0)}=1$ and
$g^{(1)}(v\mid\tau_{\mathcal E})
=\partial_v\log\vartheta_1(v\mid\tau_{\mathcal E})$.
Since $g^{(1)}(v)\sim1/v-\frac{E_2(\tau_{\mathcal{E}})}{3}v+\mathcal{O}(v^2)$ near $v=0$, this is the torus analogue of the
ordinary multiple-polylogarithm kernel $d\log(z-c)$.  The individual kernels
need not be doubly periodic: for example,
$g^{(1)}(v+\tau_{\mathcal E})=g^{(1)}(v)-2\pi i$.
At fixed $\tau_{\mathcal{E}}$, iterating the kernels provides the elliptic multiple polylogs on the covering surface. Now, with $v_0:=v(0)$ and $m_i\ge0$, the elliptic polylogs are generated by the following iterative integral,
\begin{align}    \widetilde\Gamma\!\left(
  \begin{matrix}m_1&\cdots&m_k\\ b_1&\cdots&b_k\end{matrix};v
 \right)
 &=\int_{0}^{v}d\xi\,
   g^{(m_1)}(\xi-b_1\mid\tau_{\mathcal E})
\widetilde\Gamma\!\left(
    \begin{matrix}m_2&\cdots&m_k\\ b_2&\cdots&b_k\end{matrix};\xi
   \right)\,,\qquad \widetilde{\Gamma}(;v)=1\,.
 \label{7.9oo}
\end{align}
For computational purposes, we use the coordinate $z$ instead of $v$. Consequently we use the kernels $\psi_n'$s (discussed shortly) instead of $g^{(m)}$. In general, $g^{(m)}(v|\tau_{\mathcal{E}})dv$ and $\psi_{n}dz$ describe the same elliptic iterated integrals in different coordinate systems. For the simplest example,
\begin{align}
    dv=\frac{c_4}{\omega_1}\frac{dz}{y(z)}\implies \psi_0(0,z)dz=\omega_1 g^{(0)}(v)dv\,, \qquad g^{(0)}=1\,.
\end{align}
In the algebraic coordinate $z$, the elliptic polylogs associated with the quartic polynomial are defined by the iterative integral as,
\begin{align}
    \mathrm{E}_4\left(\begin{matrix}
n_1 & \cdots & n_k \\
c_1 & \cdots & c_k
\end{matrix}; z, \vec{a}\right)=\int_0^zdt\,\psi_{n_1}(c_1,t,\vec a) \mathrm{E}_4\left(\begin{matrix}
n_2 & \cdots & n_k \\
c_2 & \cdots & c_k
\end{matrix}; t, \vec{a}\right)\,,
\end{align}
with $n_i\in \mathbb{Z}$ and $c_i\in \hat{\mathbb{C}}$ and the recursion starts at $E_4(;z,\vec a)=1$. The integration kernels $\psi_n$'s take the following form,
\begin{align}
\begin{split}
&    \psi_0(0,z)=\frac{c_4}{y},\,\,\psi_1(c,z)=\frac{1}{z-c},\,\,\psi_{{-1}}(c,z)=\frac{y_c}{y(z-c)},\,\,\psi_1(\infty,z)=\frac{c_4}{y}Z_4^{(1)}(z)\,,\\&
\psi_{-1}(\infty,z)=\frac{z}{y},\,\,\psi_{-n}(\infty,z)=\frac{z}{y}Z_4^{(n-1)}(z)-\frac{\delta_{n,2}}{c_4},\,\,\psi_n(c,z)=\frac{Z_4^{(n-1)}(z)}{z-c}-\delta_{n,2}\Phi_4(z)\,,\\ &
\psi_n(\infty,z)=\frac{c_4}{y}Z_{4}^{(n)}(z),\,\, \psi_{-n}(c,z)
=\frac{y_c}{y(z-c)}Z_4^{(n-1)}(z)\,,\end{split}
\end{align}
where, $Z^{(0)}_4:=1$,
\begin{align}
    \Phi_4(z):=\widetilde{\Phi}_4(z)+4c_4\frac{\eta_1}{\omega_1}\frac{1}{y},\qquad Z_4\equiv Z_4^{(1)}(z):=\int_{a_1}^z dz'\,\,\Phi_4(z')\,.
\end{align}
All of the above definitions are a direct analytic continuation of the definition given in \cite{Broedel:2017kkb}. For example, it is easy to check that $Z_4$ has a simple pole at infinity and regular elsewhere as noted in \cite{Broedel:2017kkb}.\\\\
\textit{Check:} For our choice of branch points, one can immediately note that $s_1(\vec a)=0$ and $s_2(\vec a)=\tau-1$. Therefore,
\begin{align}
    \widetilde{\Phi}_4(z)=\frac{1}{c_4y}\left(z^2+\frac{\tau-1}{6}\right)\xrightarrow[]{z\to \infty} \frac{i}{c_4}\left(-1
+\frac{ (\tau -1)}{3 z^2}+\textcolor{black}{\mathcal{O}(z^{-4})}\right)\,.
\end{align}
Similarly,
\begin{align}
    \Phi_4(z)\xrightarrow[]{z\to \infty} 
\textcolor{black}{-\frac{i}{c_4}+\frac{i(\tau-1)}{3c_4z^2}+4c_4\frac{\eta_1}{\omega_1}\left(-\frac{i}{z^2}+\mathcal{O}(1/z^4)\right)}=-\frac{-i}{c_4}+\mathcal{O}(1/z^2)\,.
\end{align}
\textcolor{black}{Therefore,
\begin{align}
    Z_4(z)=\int^z \left[\frac{-i}{c_4}+\mathcal{O}(1/z'^2)\right]dz'\xrightarrow[]{z\to \infty}\frac{-iz}{c_4}+C_{4,0}+\mathcal{O}(z^{-1})\,.\label{6.11hh}
\end{align}}
where $C_{4,0}$ is a finite constant term coming from the Laurent expansion of $Z_4(z)$ around $z=\infty$.
Therefore, from \eqref{6.11hh} it is evident that $Z_4^{(1)}(z)$ has a simple pole at $z\to \infty$. Similarly, it is straightforward to show that $Z_4^{(1)}$ is regular otherwise. Now we are ready to compute the master integral with a given choice of period in $\epsilon$-expansion. The integrals are defined as,
\begin{align}
    I_i(\tau,\epsilon)=\int_{\mathcal{C}} \mathcal{U}(z|\tau)z^{i-1}dz
\end{align}
where,
\begin{align}
    \mathcal{U}(z|\tau)=\mathcal{N}\tau^\epsilon (1-z^2)^{\epsilon-\frac{1}{2}}(z^2+\tau)^{-\epsilon-\frac{1}{2}} (4z^2-3)^{-\epsilon}\,.
\end{align}
Now, the integrals can be rewritten as,
\begin{align}
    I_i(\tau,\epsilon)
 &=\frac{1}{\pi\sqrt{\mu s}}\int_{\mathcal C}\frac{z^{i-1}\,dz}{y}
 \left[1+\epsilon\mathcal L(z)
 +\epsilon^2\left(\frac{\mathcal L(z)^2}{2}-\frac{\pi^2}{3}\right)
 +O(\epsilon^3)\right]\nonumber\\
 &=I_i^{(0)}(\tau)+\epsilon I_i^{(1)}(\tau)
   +\epsilon^2 I_i^{(2)}(\tau)+O(\epsilon^3)
\end{align}
where,
{\color{black}
\begin{align}
   \mathcal L(z)
 &=\log\left(\frac{4\tau}{\mu(4\mu+s)}\right)
   -2\gamma_{\!E}-i\pi\nonumber\\
 &\quad+\log(1-z^2)-\log(z^2+\tau)-\log(4z^2-3),
 \qquad \gamma_{\!E}\ \text{is Euler's constant}\,.
\end{align}
}
While expanding in $\epsilon$, the underlying elliptic curve has two independent periods.  
{\color{black}
\begin{align}
    \begin{split}
        I_i^{(0)}(\tau)&=\frac{1}{\pi\sqrt{s\mu}}
 \int_{\mathcal C}\frac{z^{i-1}\,dz}{y}\,,\\
 I_i^{(1)}(\tau)&=\frac{1}{\pi\sqrt{s\mu}}
 \int_{\mathcal C}\frac{z^{i-1}\mathcal L(z)}{y}\,dz\,,\\
 I_i^{(2)}(\tau)&=\frac{1}{\pi\sqrt{s\mu}}
 \int_{\mathcal C}\frac{z^{i-1}}{y}
 \left(\frac{\mathcal L(z)^2}{2}-\frac{\pi^2}{3}\right)dz\,.
    \end{split}
\end{align}
}
For, $\tau>0$, we choose the cycle $|\mathcal{C}]=[-1,1]$ such that $y$ is real along the contour. Moreover, it is important to note that, after expanding in $\epsilon$, the richer geometric structure is no longer present: the five independent cycles reduce to the two fundamental cycles of the torus. Also note that symmetry along  $|\mathcal{C}]$ ensures $I_i(\tau)=0$ for $i=2,4$. Before going to the computation, let's discuss an important subtlety about the $\epsilon$-expansion and the integral convergence.
\subsubsection*{Can we interchange between the \(\epsilon\)-expansion and integration?}
Termwise $\epsilon$-expansion of an improper Euler integral can not be done trivially.  Let's take a very simple example to understand the idea.
\begin{align}
    J(\epsilon)
    &=
    \int_0^1 x^{\epsilon-1}(1-x)^{-\epsilon}\,dx
     =
    \frac{\pi}{\sin(\pi\epsilon)}
     =
    \frac{1}{\epsilon}
    +\frac{\pi^2}{6}\epsilon
    +\mathcal{O}(\epsilon^3),
    \qquad
    0<\Re\epsilon<1\,,
\end{align}
the integrand has the formal expansion
\begin{align}
    x^{\epsilon-1}(1-x)^{-\epsilon}
    =
    \frac{1}{x}
    \left[
        1+\epsilon\bigl(\log x-\log(1-x)\bigr)
        +\mathcal{O}(\epsilon^2)
    \right],
\end{align}
whose coefficient integrals already diverge at leading order.  Equivalently,
the substitution $t=x/(1-x)$ gives
\begin{align}
    J(\epsilon)
    =
    \int_0^\infty\frac{t^{\epsilon-1}}{1+t}\,dt\,,
    \qquad
    \operatorname{Newt}(1+t)=\operatorname{conv}\{0,1\}\,.
\end{align}
The convergence condition
\(\Re\epsilon\in\operatorname{relint}\operatorname{Newt}(1+t)=(0,1)\)
places \(\epsilon=0\) on the boundary of the Newton polytope, in agreement
with the \(1/\epsilon\) pole shown in Fig.~\eqref{fig:newton-polytope-J}.

\begin{figure}[t]
\centering
\begin{tikzpicture}[
    >=Stealth,
    x=5.0cm,
    y=1.0cm,
    every node/.style={font=\footnotesize},
    polytope/.style={blue!65!black,line width=2.2pt},
    axis/.style={black!55,line width=0.55pt,-{Stealth[length=1.8mm]}}
]
    \node at (0.5,1.20)
        {\(\operatorname{Newt}(1+t)=\operatorname{conv}\{0,1\}\)};
    \draw[polytope] (0,0.55)--(1,0.55);
    \fill[blue!65!black] (0,0.55) circle[radius=1.45pt];
    \fill[blue!65!black] (1,0.55) circle[radius=1.45pt];
    \node[above=3pt] at (0,0.55) {\(0\)};
    \node[above=3pt] at (1,0.55) {\(1\)};

    \draw[axis] (-0.16,-0.35)--(1.18,-0.35)
        node[right] {\(\Re\epsilon\)};
    \draw[polytope,line width=1.4pt] (0,-0.35)--(1,-0.35);
    \draw[blue!65!black,fill=white,line width=0.8pt]
        (0,-0.35) circle[radius=1.75pt];
    \draw[blue!65!black,fill=white,line width=0.8pt]
        (1,-0.35) circle[radius=1.75pt];
    \node[below=3pt] at (0,-0.35) {\(0\)};
    \node[below=3pt] at (1,-0.35) {\(1\)};
    \node[below=14pt] at (0.5,-0.35)
        {\(0<\Re\epsilon<1\)};

    \fill[red!75!black] (0.42,-0.35) circle[radius=1.45pt];
    \node[above=3pt] at (0.42,-0.35) {\(\Re\epsilon\)};
    \draw[-{Stealth[length=1.8mm]},red!75!black,line width=0.7pt]
        (0.38,0.02) to[bend right=18]
        node[above,font=\scriptsize] {\(\epsilon\to0^+\)}
        (0.02,-0.29);
\end{tikzpicture}
\caption{The Newton polytope of \(1+t\) (top) and its relative interior,
which is the convergence domain of \(J(\epsilon)\) (bottom).  The limit
\(\epsilon\to0\) reaches the boundary, where \(J\) develops a pole.}
\label{fig:newton-polytope-J}
\end{figure}
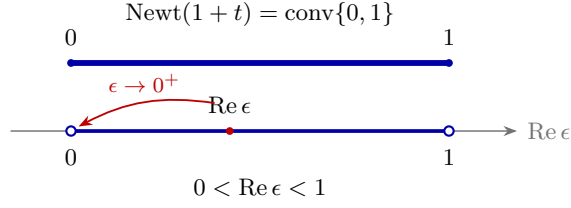

We now consider
\begin{align}
\begin{split}
    I_i(\tau,\epsilon)
    &=
    \int_{-1}^{1}F_i(z,\tau,\epsilon)\,dz,
    \qquad
    F_i(z,\tau,\epsilon)
    =
    \mathcal{U}(z|\tau)z^{i-1}\,,
    \\
    \mathcal{U}(z|\tau)
    &=
    \mathcal{N}(\epsilon)\,
    \tau^\epsilon
    (1-z^2)^{\epsilon-\frac12}
    (z^2+\tau)^{-\epsilon-\frac12}
    (4z^2-3)^{-\epsilon}\,.
    \end{split}
\end{align}
Assume that $\tau>0$ is fixed, $i\in\mathbb{Z}_{\geq1}$,
\(\mathcal{N}\) is holomorphic for $|\epsilon|<1/2$, and the boundary-value
branches on the contour are fixed independently of \(\epsilon\).  (The
normalization used here satisfies this assumption.)  The only singular
behaviors are
\begin{align}
    F_i\,dz
    \sim
    x^{\epsilon-\frac12}\,dx
    \quad (z\to\pm1)\,,
    \qquad
    F_i\,dz
    \sim
    x^{-\epsilon}\,dx
    \quad
    \left(z\to\pm\frac{\sqrt3}{2}\right)\,,
\end{align}
so the integral is absolutely convergent for
\begin{align}
    -\frac12<\Re\epsilon<1\,.
\end{align}
As we see $\epsilon=0$ lies inside the convergence region.  For any $0<\rho<1/2$, however, one has the
uniform bound
\begin{align}
    \bigl|F_i(z,\tau,\epsilon)\bigr|
    \leq
    C_{\tau,\rho}\,
    (1-z^2)^{-\frac12-\rho}
    \lvert4z^2-3\rvert^{-\rho}\,,
    \qquad
    |\epsilon|\leq\rho\,,
\end{align}
and the right-hand side belongs to \(L^1[-1,1]\).  Holomorphic dominated
convergence therefore gives
\begin{align}
    I_i(\tau,\epsilon)
    &=
    \sum_{n=0}^{\infty}\frac{\epsilon^n}{n!}
    \int_{-1}^{1}
   \partial_\epsilon^n F_i(z,\tau,\epsilon)|_{\epsilon=0}\,dz,
    \qquad |\epsilon|<\rho\,.
\end{align}
Thus, the $\epsilon$-expansion may be integrated term by term for fixed
\(\tau>0\). However, one has to be careful while taking limit $\tau\to 0$, because
$C_{\tau,\rho}$ diverges in that limit.  The obstruction is already visible
in the $i=1$ local model
\begin{align}
    A(\tau,\epsilon)
    =
    \tau^\epsilon
    \int_0^\delta
    (z^2+\tau)^{-\epsilon-\frac12}\,dz\,.
\end{align}
At \(\epsilon=0\),
\begin{align}
    A(\tau,0)
    =
    \operatorname{arsinh}\left(\frac{\delta}{\sqrt{\tau}}\right)
    =
    -\frac12\log\tau+\log(2\delta)+\mathcal{O}(1)\,,
\end{align}
whereas, for \(\Re\epsilon>0\),
\begin{align}
    \lim_{\tau\to0}A(\tau,\epsilon)
    =
    \int_0^\infty(1+u^2)^{-\epsilon-\frac12}\,du
    =
    \frac{\sqrt{\pi}\,\Gamma(\epsilon)}
         {2\,\Gamma\left(\epsilon+\frac12\right)}
    =
    \frac{1}{2\epsilon}+\log2+\mathcal{O}(\epsilon)\,.
\end{align}
The two procedures, therefore, probe different singular asymptotic regimes;
The fixed-\(\tau\) Taylor expansion does not furnish a joint expansion near
\((\tau,\epsilon)=(0,0)\).  For \(i>1\), the corresponding one-sided local
term becomes
\begin{align}
    A_i(\tau,\epsilon)
    =
    \tau^{\frac{i-1}{2}}
    \int_0^{\delta/\sqrt{\tau}}
    u^{i-1}(1+u^2)^{-\epsilon-\frac12}\,du\,,
\end{align}
for which
\(A_i(\tau,0)\to\delta^{i-1}/(i-1)\), while
\(A_i(\tau,\epsilon)\to0\) at fixed \(\Re\epsilon>0\).  On the full symmetric
contour this contribution is multiplied by \(1+(-1)^{i-1}\), so the even
moments cancel.  The precise degeneration is therefore \(i\)-dependent and
should be analyzed separately rather than expressed as a general inequality
between iterated limits.  If \(\mathcal{N}\) is meromorphic rather than
holomorphic at \(\epsilon=0\), its Laurent prefactor must likewise be kept
outside the dominated-convergence argument and expanded separately.
\subsubsection{Solutions of the masters up to $\epsilon^2$}
\textit{$\epsilon^0$-order solutions:}
At leading order, the solutions can be trivially integrated in \texttt{Mathematica} as,
\begin{align}
 I_1^{(0)}(\tau)
 &=\frac{2}{\pi \sqrt{s\mu\tau}}K\left(-\frac{1}{\tau}\right),\nonumber\\
 I_3^{(0)}(\tau)
 &=\frac{2\sqrt{\tau}}{\pi \sqrt{s\mu}}
 \left[E\left(-\frac{1}{\tau}\right)-K\left(-\frac{1}{\tau}\right)\right],\nonumber\\
 I_5^{(0)}(\tau)
 &=-\frac{2\sqrt{\tau}}{3\pi\sqrt{s\mu}}
 \left[(1-2\tau)K\left(-\frac{1}{\tau}\right)
 +2(\tau-1)E\left(-\frac{1}{\tau}\right)\right].
 \label{eq:sec6-I-order0-complete}
\end{align}
\textit{$\epsilon^1$-order solutions:} In next-to-leading order, the solutions can not be trivially integrated. We will invoke the formalism of the elliptic polylogarithm discussed above.
\begin{align}
    I_1^{(1)}(\tau)=\frac{2}{\pi \sqrt{s\mu}}\int_0^1 \frac{\mathcal{L}(z)dz}{y}\,.\label{6.18nn}
\end{align}
In the above integral, we encounter the following general integral of the form,
\begin{align}
    \begin{split}
    \int_0^1 \frac{dz}{y}\log\left(\frac{z-c}{-c}\right)=\frac{1}{c_4}\int_0^1dz \,\psi_0(0,z)\int_0^z dt\,\psi_1(c,t)  E_4(;t,\vec a)&=\frac{1}{c_4}\int_0^1dz \,\psi_0(0,z)\mathrm{E}_4\left(\begin{matrix}
1\\
c 
\end{matrix}; z, \vec{a}\right)\\ &
=\frac{1}{c_4}\mathrm{E}_4\left(\begin{matrix}
0\,\,1\\
0\,\,c 
\end{matrix}; 1, \vec{a}\right)\,.
    \end{split}
\end{align}
Therefore the integral in \eqref{6.18nn} can be written as,
\begin{align}
\begin{split}
    I_1^{(1)}&=\frac{2}{\pi c_4\sqrt{s\mu}}\left[\int_0^1 dz\,\psi_0(0,z)\sum_{c\in \mathcal{P'}}\lambda_c\int_0^z dt\, \psi_1(c,t)E_4(;t,\vec a)+\mathcal L_0\int_0^1dz \,\psi_0(0,z)\right],\\&=\frac{2}{\pi c_4\sqrt{s\mu}} \left[\sum_{c\in \mathcal{P'}}\lambda_c\,\mathrm{E}_4\left(\begin{matrix}
0\,\,1\\
0\,\,c 
\end{matrix}; 1, \vec{a}\right)+\mathcal{L}_0\mathrm{E}_4\left(\begin{matrix}
0\\
0
\end{matrix}; 1, \vec{a}\right)\right]
    \end{split} \label{new-eqq1}
\end{align}
where $\mathcal{P'}=\{-1,1,-ir,ir,-\frac{\sqrt{3}}{2},\frac{\sqrt{3}}{2}\}$,
$\lambda_{\pm1}=1$, $\lambda_{\pm ir}=\lambda_{\pm\sqrt{3}/2}=-1$, and
\begin{equation}
\mathcal L_0=\mathcal L(0)
 =\log\!\left(\frac{4}{3\mu(4\mu+s)}\right)-2\gamma_{\!E}+\ell_{\rm br}\,,
 \qquad
 \ell_{\rm br}=\begin{cases}-2\pi i,&(4z^2-3+i0)^{-\epsilon}\,,\\[2pt]
 0,&(4z^2-3-i0)^{-\epsilon}\,,\end{cases}
 \label{eq:L0-correct}
\end{equation}
 Equivalently, $\mathcal L_0$ is fixed modulo $2\pi i$. Now, let us move on to the computation of the other two integrals. The differential $z^2 dz/y$ can be written as,
\begin{align}
     \frac{z^2dz}{y}=c_4\,dZ_{4}(z)-\alpha_0\psi_0(0,z)\,dz,\qquad \alpha_0=\frac{\tau-1}{6c_4}+4c_4\frac{\eta_1}{\omega_1}\,.
\end{align}
{\color{black}
\begin{align}
    \begin{split}
        I_3^{(1)}(\tau)&=\frac{2}{\pi \sqrt{s\mu}}\int_0^1 \left(c_4 dZ_{4}(z)-\alpha_0\psi_0(0,z)dz\right)\left(\mathcal{L}_0+\sum_{c\in \mathcal{P'}}\lambda_c\int_0^z dt\psi_1(c,t)E_4(;t,\vec a)\right)\\ &
        =\frac{2}{\pi \sqrt{s\mu}}\mathcal{L}_0\left(c_4\Delta Z_4-\alpha_0\mathrm{E}_4\left(\begin{matrix}
0\\
0
\end{matrix}; 1, \vec{a}\right)\right)+\frac{2 c_4}{\pi \sqrt{s\mu}}\sum_{c\in \mathcal{P'}}\lambda_c\int_0^1 dZ_4(z)\log\left(\frac{z-c}{-c}\right)\\ &
\hspace{3 cm}-\frac{2\alpha_0}{\pi \sqrt{s\mu}}\sum_{c\in \mathcal{P'}}\lambda_c\int_0^1 dz \,\psi_0(0,z)\mathrm{E}_4\left(\begin{matrix}
1\\
c
\end{matrix}; z, \vec{a}\right)\label{6.23hh},\qquad \Delta Z_4=Z_4(1)-Z_4(0)
    \end{split}
\end{align}
}
After integration by parts of the second term in \eqref{6.23hh} reduces to,
\begin{align}
    \begin{split}
           \int_0^1 dZ_4(z) \log\left(\frac{z-c}{-c}\right)= Z_4(1)\log\left(\frac{1-c}{-c}\right)-\int_0^1 dz \frac{Z_4(z)}{z-c}\,.\label{6.24kk}
    \end{split}
\end{align}
From \eqref{6.24kk}, it seems that the integral has an endpoint divergence at $c=1$. However, one can show that the divergence will cancel in the second term on the RHS of \eqref{6.24kk}. More generally, we have,
{\color{black}
\begin{align}
 \int_0^1dZ_4(z)\log\!\left(\frac{z-c}{-c}\right)
 &=-\int_0^1\frac{Z_4(z)-Z_4(1)}{z-c}\,dz
 \label{6.25hh}\\
 &=Z_4(1)\mathrm E_4\!\left(\begin{matrix}1\\c\end{matrix};1,\vec a\right)
 -\mathrm E_4\!\left(\begin{matrix}2\\c\end{matrix};1,\vec a\right)
 -\Delta Z_4\,.\nonumber
\end{align}
At $c=1$, this denotes the limit with a common upper limit \(b\to1^{-}\) in both terms, or the corresponding finite part.
The integral is finite because \(Z_4(z)-Z_4(1)=O(\sqrt{1-z})\).
For \(c=h\), the contour retains the  $i0$-prescription.
}
$\Delta Z_4$ can be explicitly computed as,
\begin{align}
\begin{split}
    \Delta Z_4&=\int_{-1}^1 dz\,\Phi_4(z)-\int_{-1}^0 dz\,\Phi_4(z)=\int_0^1 dz\,\Phi_4(z
    )\\ &=\frac{\sqrt{\tau }}{c_4} \left[E\left(-\frac{1}{\tau }\right)-K\left(-\frac{1}{\tau }\right)\right]+\left(\frac{\tau-1}{6c_4}+4c_4\frac{\eta_1}{\omega_1}\right)\frac{K\left(-\frac{1}{\tau }\right)}{\sqrt{\tau }}\,.
    \end{split}
\end{align}
Therefore, the integral in \eqref{6.23hh} can be written as,
{\color{black}
\begin{align}
 I_3^{(1)}
 =\frac{2}{\pi \sqrt{s\mu}}\Bigg\{&
 \mathcal L_0\left[c_4\Delta Z_4
 -\alpha_0\mathrm E_4\!\left(\begin{matrix}0\\0\end{matrix};1,\vec a\right)\right]
 \nonumber\\
 &+c_4\sum_{c\in\mathcal P'}\lambda_c
 \left[
 Z_4(1)\mathrm E_4\!\left(\begin{matrix}1\\c\end{matrix};1,\vec a\right)
 -\mathrm E_4\!\left(\begin{matrix}2\\c\end{matrix};1,\vec a\right)
 -\Delta Z_4\right]\nonumber\\
 &-\alpha_0\sum_{c\in\mathcal P'}\lambda_c
 \mathrm E_4\!\left(\begin{matrix}0&1\\0&c\end{matrix};1,\vec a\right)
 \Bigg\}.
\end{align}
}
We are now left with the last integral,
\begin{align}
I_5^{(1)}=\frac{2}{\pi \sqrt{s\mu}}\int_0^1 \frac{z^4dz}{y}\mathcal{L}(z)\,.
\end{align}
Note that the differential $z^4dz/y$ is not an independent differential; it depends on the other two independent differential $z^2dz/y$ and $dz/y$ as,
\begin{align}
    \begin{split}
        \frac{z^4 dz}{y}=\frac{2(1-\tau)}{3}\frac{z^2dz}{y}+\frac{\tau}{3}\frac{dz}{y}-\frac{1}{3}d(zy)\,.
    \end{split}
\end{align}
Therefore,
{\color{black}
\begin{align}
    \begin{split}
        I_5^{(1)}=\frac{2(1-\tau)}{3} I_3^{(1)} +\frac{\tau}{3}I_1^{(1)}-\frac{2}{3\pi\sqrt{s\mu}}\int_0^1 d(zy)\mathcal{L}(z)\,.\label{6.31kk}
    \end{split}
\end{align}
}
The last term of the integration in \eqref{6.31kk} can further be simplified by invoking the integration by parts,
\begin{align}
 -  \int d(zy) \mathcal{L}(z)=\int_0^1 zy\,d\mathcal{L}(z),\quad 
    d\mathcal{L}(z)=-\frac{2z\,dz}{1-z^2}-\frac{2z\,dz}{z^2+\tau}-\frac{8z\,dz}{4z^2-3}\,.
\end{align}
\textcolor{black}{The boundary term vanishes at both ends: $zy\,\mathcal L\sim\sqrt{1-z^{2}}\,
\log(1-z^{2})\to0$ as $z\to1^{-}$, and $zy=0$ at $z=0$. The logarithmic singularity of $\mathcal L$ at the interior point $z=\sqrt3/2$ is integrable and does not obstruct the integration by parts, provided the $+i0$ prescription is kept fixed throughout. The same remark applies to the $\epsilon^{2}$ reduction shown later, where the integrand carries $\mathcal L\,d\mathcal L\,.$}
Therefore, the integral becomes,
\begin{align}
    \begin{split}
       & -2\int_0^1 dz  \,z^2y\left(\frac{1}{1-z^2}+\frac{1}{z^2+\tau}+\frac{4}{4z^2-3}\right)\\ &=-2\int_0^1 \frac{dz}{y}\,z^2\left(z^2+\tau\right)-2\int_0^1\frac{dz}{y}z^2(1-z^2)-8\int_0^1 \frac{dz}{y(4z^2-3)}z^2(1-z^2)(z^2+\tau)\\ &
       =-2\int_0^1 \frac{dz}{y}\,z^2\left(z^2+\tau\right)-2\int_0^1\frac{dz}{y}z^2(1-z^2)-8\int_0^1\frac{dz}{y}\left(\frac{\tau }{16}-\frac{z^4}{4}+\left(\frac{1}{16}-\frac{\tau }{4}\right) z^2+\frac{3 (4 \tau +3)}{64 \left(4 z^2-3\right)}+\frac{3}{64}\right)\,.\label{6.33hh}
    \end{split}
\end{align}
In \eqref{6.33hh}, all the integrals are primitive except,
{\color{black}
\begin{align}
\begin{split}
    \int_0^1\frac{dz\,(4\tau+3)}{16 y(4z^2-3+i0)}&=\frac{y_c}{4\sqrt{3}}\int_0^1 \frac{{dz\,y_c}}{y}\left(\frac{1}{z-c}-\frac{1}{z+c}\right),\qquad y_c=y_{-c}=\frac14\sqrt{4\tau+3}\\ &=\frac{y_c}{4\sqrt{3}}\int_0^1 dz (\psi_{-1}(c,z)-\psi_{-1}(-c,z))E_4(;z,\vec a)\,,\quad c=\frac{\sqrt{3}}{2}\\ &
    =\frac{y_c}{4\sqrt{3}}\left[\mathrm{E}_4\left(\begin{matrix}
-1\\
\frac{\sqrt{3}}{2}
\end{matrix}; 1, \vec{a}\right)-\mathrm{E}_4\left(\begin{matrix}
-1\\
-\frac{\sqrt{3}}{2}
\end{matrix}; 1, \vec{a}\right)\right]\,.\label{6.34hhhh}
    \end{split}
\end{align}
}
With $D=4\tau+3$ and
$h=\sqrt3/2$, we have the following identity,
\begin{equation}
 zy\,d\mathcal L
 =\frac{dz}{y}\left[
 2z^4-\frac52z^2-\frac{D}{8}
 -\frac{3D}{8(4z^2-3+i0)}\right].
 \label{eq:sec6-zydL-reduction}
\end{equation}
Writing
\begin{equation}
 \Delta\mathrm E_{-1}:=
 \mathrm E_4\!\left(\begin{matrix}-1\\ h\end{matrix};1,\vec a\right)
 -\mathrm E_4\!\left(\begin{matrix}-1\\ -h\end{matrix};1,\vec a\right),
 \label{eq:sec6-delta-Eminusone}
\end{equation}
with the prescribed values, one has
$\Delta\mathrm E_{-1}=\sqrt{3D}\int_0^1
dz/[y(4z^2-3+i0)]$.  Therefore, the result is
{\color{black}
\begin{align}
 I_5^{(1)}={}&\frac{2(1-\tau)}{3}I_3^{(1)}
 +\frac{\tau}{3}I_1^{(1)}\nonumber\\
 &+\frac{2}{3\pi\sqrt{s\mu}}\left[
 \frac{(64\tau^2+64\tau-18)K(-1/\tau)
 -8\tau(8\tau+7)E(-1/\tau)}{48\sqrt\tau}
 -\frac{\sqrt{3(4\tau+3)}}8\,\Delta\mathrm E_{-1}
 \right].
 \label{eq:sec6-I5-order1-corrected}
\end{align}
}
\textit{An important note:} In the integral \eqref{6.34hhhh} is that the factor $(4z^2-3)^{-1}$ has two poles on the real integration contour,
\[
z=\pm c\,,\qquad c=\frac{\sqrt{3}}{2}\,.
\]
Hence integrals containing this factor are not ordinary real integrals; they must be defined with an $i0$-prescription.
We have,
\begin{align}
\mathrm E_4\!\left(
\begin{matrix}
-1\\ a
\end{matrix}
;1,\vec a
\right)
:=
\int_0^1 dz\,\frac{y_a}{y(z)}\,\frac{1}{z-a}\,,
\end{align}
{\color{black}
where the contour $i0$-prescription must be specified when a pole lies on the integration path
}. For a Feynman prescription (which may appear in the context of eikonal phase computation),
\begin{align}
\frac{1}{4z^2-3+i0}
=
\frac{1}{8c}
\left[
\frac{1}{z-c+i0}
-
\frac{1}{z+c-i0}
\right].
\end{align}
Using the evenness of $y(z)$ on the real cycle, the integral over $(-1,1)$ becomes
\begin{align}
\int_{-1}^{1}\frac{dz\,y_c}{y(z)\,[4z^2-3+i0]}
&=
\frac{1}{2\sqrt3}
\left\{
\operatorname{PV}\!\left[
\mathrm E_4\!\left(
\begin{matrix}
-1\\ c
\end{matrix}
;1,\vec a
\right)
-
\mathrm E_4\!\left(
\begin{matrix}
-1\\ -c
\end{matrix}
;1,\vec a
\right)
\right]
-i\pi
\right\}.
\end{align}
Here, the two-pole contributions add to the imaginary part. For a retarded prescription (which will be useful while computing the observable, such as scattering angle and waveform) one instead shifts the integration variable as
$z\longrightarrow z+i0,$ so that
\begin{align}
\frac{1}{4(z+i0)^2-3}
=
\frac{1}{8c}
\left[
\frac{1}{z-c+i0}
-
\frac{1}{z+c+i0}
\right].
\end{align}
In this case, the two poles contribute imaginary parts with opposite signs. Since $y(c)=y(-c)$, the imaginary parts cancel on the full symmetric contour $(-1,1)$. Hence
\begin{align}
\int_{-1}^{1}\frac{dz\,y_c}{y(z)\,[4(z+i0)^2-3]}
&=
\frac{1}{2\sqrt3}\,
\operatorname{PV}\!\left[
\mathrm E_4\!\left(
\begin{matrix}
-1\\ c
\end{matrix}
;1,\vec a
\right)
-
\mathrm E_4\!\left(
\begin{matrix}
-1\\ -c
\end{matrix}
;1,\vec a
\right)
\right].
\end{align}
Thus, the $i0$-prescription encodes the physical boundary condition. {\color{black}
The two prescriptions, therefore, have the same principal value but
different residue contributions. Here they specify different boundary
conditions of the generalized cut, or more precisely, of the \textit{homogeneous} differential equation. However, for a more precise identification with a Feynman or
retarded prescription for the original uncut integral requires a
separate continuation of that integral's causal contour.}\\\\
\textit{$\epsilon^2$-order solutions:} The subleading corrections can be computed as before, and invoking the Shuffle algebra of iterative integral with elliptic kernel,
\begin{align}
    \begin{split}
        E_4(\vec c,z)E_4(\vec d,z)=\sum_{\vec w\in \vec c \shuffle \vec d } {E_4}(\vec w,z),\qquad \vec c=\begin{pmatrix}
           & n_1\cdots n_k\\ &
            c_1\cdots c_k
        \end{pmatrix}\,.
    \end{split}
\end{align}
\textcolor{black}{No regularisation is needed at $z=0$: none of the letters in $\mathcal P'$ (defined below (\ref{new-eqq1})) vanishes, and $\psi_0(0,z)=c_4/y$ is regular there. At $z=1$ the letter $c=1$ gives following form.$G(1,\ldots;z)\sim\log^{k}(1-z)$, which is integrable against $\psi_0(0,z)\,dz$ and against $dZ_4$. The products of $G$'s are multiplied pointwise for $z<1$, so the shuffle identity holds there without regularisation. The general integral has the following form.}
{\color{black}
\begin{align}
    \begin{split}
        I_i^{(2)}=&\frac{2}{\pi c_4\sqrt{s\mu}}\int_0^1dz \,{z^{i-1}}\psi_0(0,z)\\ &\left[\left(\frac{\mathcal{L}_0^2}{2}-\frac{\pi^2}{3}\right)+\mathcal{L}_0\sum_{c\in \mathcal{P'}}\lambda_c \mathrm{E}_4\left(\begin{matrix}
1\\
c
\end{matrix}; z, \vec{a}\right)+\frac12\sum_{c_1,c_2\in \mathcal{P'}}\lambda_{c_1}\lambda_{c_2}\mathrm{E}_4\left(\begin{matrix}
1\\
c_1
\end{matrix}; z, \vec{a}\right)\mathrm{E}_4\left(\begin{matrix}
1\\
c_2
\end{matrix}; z, \vec{a}\right)\right]
    \end{split}
\end{align}
}
\textcolor{black}{The relative weights here are fixed by
$$\tfrac12\mathcal L(z)^{2}
=\tfrac12\mathcal L_0^{2}+\mathcal L_0\sum_{c\in\mathcal P'}\lambda_c G(c;z)
+\tfrac12\sum_{c_1,c_2\in\mathcal P'}\lambda_{c_1}\lambda_{c_2}G(c_1;z)G(c_2;z)\,,$$ together with
$\mathcal L(z)-\mathcal L_0=\sum_{c\in\mathcal P'}\lambda_c\log\!\big((z-c)/(-c)\big)$
for the alphabet and weights listed above.}
Now using the shuffle algebra for the last term, the integral can be written as,
{\color{black}
\begin{align}
\begin{split}
    I_i^{(2)}=\frac{2}{\pi c_4\sqrt{s\mu}}\int_0^1 dz \,z^{i-1}\psi_0(0,z)\left[\cdots+\frac12\sum_{c_1,c_2\in \mathcal{P'}}\lambda_{c_1}\lambda_{c_2}\left(\mathrm{E}_4\left(\begin{matrix}
1\,\,\,\,\,1\\
c_1\,c_2
\end{matrix}; z, \vec{a}\right)+\mathrm{E}_4\left(\begin{matrix}
1\,\,\,\,1\\
c_2\,c_1
\end{matrix}; z, \vec{a}\right)\right)\right]\,.
\end{split}
\end{align}
}
The first integral takes the following form,
{\color{black}
\begin{align}
    \begin{split}
        I_1^{(2)}=\frac{2}{\pi c_4\sqrt{s\mu}}\Bigg[\left(\frac{\mathcal{L}_0^2}{2}-\frac{\pi^2}{3}\right)\mathrm{E}_4\left(\begin{matrix}
0\\
0
\end{matrix}; 1, \vec{a}\right)&+\mathcal{L}_0\sum_{c\in \mathcal{P'}}\lambda_c\mathrm{E}_4\left(\begin{matrix}
0\,\,\,\,1\\
0\,\,\,c
\end{matrix}; 1, \vec{a}\right)\\ &+\frac12\sum_{c_1,c_2\in \mathcal{P'}}\lambda_{c_1}\lambda_{c_2}\left(\mathrm{E}_4\left(\begin{matrix}
0\,\,\,1\,\,\,\,1\\
0\,\,c_1\,c_2
\end{matrix}; 1, \vec{a}\right)+c_1 \leftrightarrow c_2\right)\Bigg]\,.
    \end{split}
\end{align}
}
We now complete the other two second-order masters.  It is convenient to
write
\begin{equation}
 G(c_1,\ldots,c_k;z):=
 \mathrm E_4\!\left(\begin{matrix}1&\cdots&1\\c_1&\cdots&c_k\end{matrix};
 z,\vec a\right)\,,
 \qquad  G(\varnothing;z):=1\,,
\end{equation}
and to introduce
\begin{equation}
 \mathcal T(c_1,\ldots,c_k):=
 \int_0^1dZ_4(z)\,G(c_1,\ldots,c_k;z)\,,
 \qquad \mathcal T(\varnothing)=\Delta Z_4\,.
 \label{eq:sec6-Tword-definition}
\end{equation}
Integration by parts, together with
$Z_4(z)/(z-c)=\psi_2(c,z)+\Phi_4(z)$, evaluates this integral recursively:
\begin{align}
 \mathcal T(c_1,\ldots,c_k)={}&
 Z_4(1)\textcolor{black}{G}(c_1,\ldots,c_k;1)
 -\mathrm E_4\!\left(
 \begin{matrix}2&1&\cdots&1\\c_1&c_2&\cdots&c_k\end{matrix};
 1,\vec a\right)
 -\mathcal T(c_2,\ldots,c_k)\,,\qquad k\geq1\,.
 \label{eq:sec6-Tword-recursion}
\end{align}
For example,
{\color{black}
\begin{align}
 \mathcal T(c)&=Z_4(1)G(c;1)
 -\mathrm E_4\!\left(\begin{matrix}2\\c\end{matrix};1,\vec a\right)
 -\Delta Z_4\,,\nonumber\\
 \mathcal T(c_1,c_2)&=Z_4(1)G(c_1,c_2;1)
 -\mathrm E_4\!\left(\begin{matrix}2&1\\c_1&c_2\end{matrix};1,\vec a\right)
 -\mathcal T(c_2)\,.
 \label{eq:sec6-Tword-examples}
\end{align}
}
\textcolor{black}{For $c_1=1$ the first two terms on the right of \eqref{eq:sec6-Tword-recursion} diverge separately at $z=1$, both as $Z_4(1)\log(1-z)$. The recursion is then understood with a common upper limit in all terms, exactly as below \eqref{6.25hh}. The divergences cancel because $Z_4(z)-Z_4(1)=O(\sqrt{1-z})$. Letters $c_j=1$ with $j\ge2$ need no such care.}  Hence, no new type of elliptic integral is needed. Next we define $A_{\mathcal L}=\mathcal L_0^2/2-\pi^2/3$, 
\begin{align}
\begin{split}
 \mathcal U_2={}A_{\mathcal L}
 \mathrm E_4\!\left(\begin{matrix}0\\0\end{matrix};1,\vec a\right)
 &+\mathcal L_0\sum_{c\in\mathcal P'}\lambda_c
 \mathrm E_4\!\left(\begin{matrix}0&1\\0&c\end{matrix};1,\vec a\right)
\\ &+\frac12\sum_{c_1,c_2\in\mathcal P'}\lambda_{c_1}\lambda_{c_2}
 \left[
 \mathrm E_4\!\left(\begin{matrix}0&1&1\\0&c_1&c_2\end{matrix};1,\vec a\right)
 +\mathrm E_4\!\left(\begin{matrix}0&1&1\\0&c_2&c_1\end{matrix};1,\vec a\right)
 \right]\,,
 \label{eq:sec6-U2}
 \end{split}
\end{align}
and,
\begin{align}
 \mathcal V_2={}&A_{\mathcal L}\Delta Z_4
 +\mathcal L_0\sum_{c\in\mathcal P'}\lambda_c\mathcal T(c)
+\frac12\sum_{c_1,c_2\in\mathcal P'}\lambda_{c_1}\lambda_{c_2}
 \left[\mathcal T(c_1,c_2)+\mathcal T(c_2,c_1)\right]\,.
 \label{eq:sec6-V2}
\end{align}
Thus, the full second-order first and third masters are {\color{black}
\begin{equation}
 {
 I_1^{(2)}=\frac{2}{\pi c_4\sqrt{s\mu}}\mathcal U_2\,,
 \qquad
 I_3^{(2)}=\frac{2}{\pi \sqrt{s\mu}}
 \left(c_4\mathcal V_2-\alpha_0\mathcal U_2\right).}
 \label{eq:sec6-I1-I3-order2-complete}
\end{equation}
}
For the fifth master, using $d(\mathcal L^2/2-\pi^2/3)=\mathcal L\,d\mathcal L$
and  \eqref{eq:sec6-zydL-reduction}, we define the single pole word
\begin{align}
 \mathcal P_{\mathcal L}:={}&
 \int_0^1\frac{\mathcal L(z)\,dz}{y(4z^2-3+i0)}\nonumber\\
 ={}&\frac{1}{4\sqrt3\,y_h}\Bigg\{
 \mathcal L_0\Delta\mathrm E_{-1}
 +\sum_{c\in\mathcal P'}\lambda_c\left[
 \mathrm E_4\!\left(\begin{matrix}-1&1\\h&c\end{matrix};1,\vec a\right)
 -\mathrm E_4\!\left(\begin{matrix}-1&1\\-h&c\end{matrix};1,\vec a\right)
 \right]\Bigg\}\,,
 \quad y_h=\frac{\sqrt{4\tau+3}}4\,.
 \label{eq:sec6-PL-E4}
\end{align}
For the $+i0$ choice, the two outer $-1$ kernels in
\eqref{eq:sec6-PL-E4} inherit the correlated shifts
$(z-h+i0)^{-1}$ and $(z+h-i0)^{-1}$ from the Feynman partial fraction
displayed above.
Equivalently, keep a common $\eta>0$ throughout the reduction and all
integrations by parts, and take $\eta\to0^+$ only at the end. As previously discussed, the $i0$-prescription will be important here, as it dictates how the boundary conditions are invoked in the system. For now, we keep it formal without again discussing the different boundary conditions. Therefore, the full second-order result is
{\color{black}
\begin{equation}
 \begin{aligned}
 I_5^{(2)}={}&\frac{2(1-\tau)}3I_3^{(2)}+\frac\tau3I_1^{(2)}
 +\frac23I_5^{(1)}-\frac56I_3^{(1)}\\
 &-\frac{4\tau+3}{24}I_1^{(1)}
 -\frac{(4\tau+3)}{4\pi\sqrt{s\mu}}\mathcal P_{\mathcal L}\,.
 \end{aligned}
 \label{eq:sec6-I5-order2-complete}
\end{equation}
}
Therefore, \eqref{eq:sec6-I1-I3-order2-complete} and \eqref{eq:sec6-I5-order2-complete} gives the full direct-integral expansion through $\epsilon^2\,.$ Before moving forward to the next discussion, we intend to summarize the primary findings of this section.
\paragraph{Elliptic function space.} Let us summarize the main outcome of the computations above.
The $\epsilon$-expansion of the three non-trivial elliptic maximal-cut
master integrals $I_1$, $I_3$, and $I_5$ is organized in terms of elliptic
multiple polylogarithms $E_4$ associated with
\begin{equation}
 y^2=(1-z^2)(z^2+\tau)
 =-(z+1)(z-1)(z+ir)(z-ir)\,,
 \qquad
 r=\sqrt{\tau}\,,
 \qquad
 \vec a=(-1,1,-ir,ir)\,.
\end{equation}
The logarithmic insertion can be written as
\begin{equation}
 \mathcal L(z)=\mathcal L_0+\sum_{c\in\mathcal P'}\lambda_c\,G(c;z)\,,
 \qquad
 G(c;z)=\log\!\left(\frac{z-c}{-c}\right)\,,
 \qquad \mathcal L_0=\mathcal L(0)\,,
\end{equation}
where
\begin{equation}
 \mathcal P'=\{-1,1,-ir,ir,-h,h\}\,,
 \qquad
 h=\frac{\sqrt3}{2}\,,
 \qquad
 \lambda_{\pm1}=1\,,
 \qquad
 \lambda_{\pm ir}=\lambda_{\pm h}=-1\,.
\end{equation}
Thus, each additional logarithmic insertion introduces only the
kernels $\psi_1(c,z)=1/(z-c)$ with $c\in\mathcal P'$. Together with these logarithmic kernels, the reduction of the
polynomial numerators require
\begin{equation}
 \psi_0(0,z)=\frac{c_4}{y}\,,
 \qquad
 \psi_2(c,z)=\frac{Z_4(z)}{z-c}-\Phi_4(z)\,,
 \qquad
 \psi_{-1}(c,z)=\frac{y_c}{y(z-c)}\,,
 \qquad
 dZ_4(z)=\Phi_4(z)\,dz\,,
\end{equation}
with the negative-index kernels occurring at $c=\pm h$.
Ordinary multiple polylogarithms form the all-$1$ subfamily,
\begin{equation}
 G(c_1,\ldots,c_k;z)
 =
 E_4\!\left(
 \begin{matrix}
  1&\cdots&1\\
  c_1&\cdots&c_k
 \end{matrix};z,\vec a\right),
 \qquad
 G(\varnothing;z)=1\,.
\end{equation}
They also appear as boundary terms in the reduction of the
second-kind differential. For example,
\begin{equation}
 \int_0^1 dZ_4(z)\,G(c;z)
 =
 Z_4(1)G(c;1)
 -
 E_4\!\left(
 \begin{matrix}2\\c\end{matrix};1,\vec a\right)
 -
 \Delta Z_4\,,
 \qquad
 \Delta Z_4=Z_4(1)-Z_4(0)\,.
\end{equation}

Let $\mathcal R_{\rm ell}$ denote the coefficient ring generated
by the algebraic kinematic factors, the period, and the quasi-period
prefactors entering the reduction, $\mathcal L_0$, and the constants
arising from the scalar prefactor of the cut integral.
In particular, the required combinations of $c_4$,
$\eta_1/\omega_1$, $Z_4(1)$, and $\Delta Z_4$ are included in
$\mathcal R_{\rm ell}$, while ordinary MPL words are displayed
explicitly.
It is convenient to give order by order spanning classes by extrapolating the pattern upto $\mathcal{O}(\epsilon^2)$ and the corresponding argument will be provided later in this section.
\begin{equation}
 \boxed{
 \begin{aligned}
 \mathscr V_N
 :=\operatorname{Span}_{\mathcal R_{\rm ell}}\Bigg\{
 &G(\mathbf c;1),
 &&0\leq k\leq N,
 \\[2pt]
 &E_4\!\left(
 \begin{matrix}
  0&1^k\\
  0&\mathbf c
 \end{matrix};1,\vec a\right),
 &&0\leq k\leq N,
 \\[2pt]
 &E_4\!\left(
 \begin{matrix}
  2&1^k\\
  c_0&\mathbf c
 \end{matrix};1,\vec a\right),
 &&0\leq k\leq N-1,
 \\[2pt]
 &E_4\!\left(
 \begin{matrix}
  -1&1^k\\
  \sigma h&\mathbf c
 \end{matrix};1,\vec a\right),
 &&0\leq k\leq N-1
 \Bigg\}.
 \end{aligned}
 }
\end{equation}
Here $1^k$ denotes a string of $k$ indices equal to $1$,
$\mathbf c=(c_1,\ldots,c_k)$,
$c_0,c_j\in\mathcal P'$, and $\sigma=\pm1$.
The last two families are absent for $N=0$, and
\begin{equation}
 I_1^{(N)},\ I_3^{(N)},\ I_5^{(N)}\in\mathscr V_N\,.
\end{equation}
This spanning class is not intended to be minimal. The length bounds follow from the numerator reduction.
The differential $dz/y$ produces words headed by $\psi_0$, whereas the reduction of $z^2dz/y$ through $dZ_4$ generates
words headed by $\psi_2$ and MPL boundary terms.
For the remaining numerator,
\begin{equation}
 \frac{z^4\,dz}{y}
 =
 \frac{2(1-\tau)}{3}\frac{z^2\,dz}{y}
 +\frac{\tau}{3}\frac{dz}{y}
 -\frac13\,d(zy)\,.
\end{equation}
Multiplication by $\mathcal L(z)^N$ and integration by parts produce
$N \mathcal L(z)^{N-1}d\mathcal L(z)$, whose reduction introduces the pole kernels
at $\pm h$ with at most $N-1$ logarithmic letters. In particular, we provide the first two non-trivial spanning classes emerging from the explicit calculations.
\begin{align}
 \mathscr V_1
 =\mathscr V_0+
 \operatorname{Span}_{\mathcal R_{\rm ell}}
 \Bigg\{
 &G(c;1),
 \quad
 E_4\!\left(
 \begin{matrix}0&1\\0&c\end{matrix};1,\vec a\right)\,,
 \quad
 E_4\!\left(
 \begin{matrix}2\\c\end{matrix};1,\vec a\right)\,,
 \quad
 E_4\!\left(
 \begin{matrix}-1\\\sigma h\end{matrix};1,\vec a\right)
 \Bigg\}\,,
 \\[3pt]
 \mathscr V_2
 =\mathscr V_1+
 \operatorname{Span}_{\mathcal R_{\rm ell}}
 \Bigg\{
 &G(c_1,c_2;1),
 \quad
 E_4\!\left(
 \begin{matrix}0&1&1\\0&c_1&c_2\end{matrix};1,\vec a\right)\,,
 \nonumber\\[-1pt]
 &
 E_4\!\left(
 \begin{matrix}2&1\\c_1&c_2\end{matrix};1,\vec a\right)\,,
 \quad
 E_4\!\left(
 \begin{matrix}-1&1\\\sigma h&c\end{matrix};1,\vec a\right)
 \Bigg\},
\end{align}
where $c,c_1,c_2\in\mathcal P'$. At leading order, the three polynomial-numerator integrals reduce
to combinations of $K(-1/\tau)$ and $E(-1/\tau)$.
Higher orders increase the maximal required logarithmic word
length without introducing a different elliptic curve or
additional finite marked points. The scalar prefactor contributes
only coefficients and lower-length terms, so the expansion is
naturally filtered by maximal word length rather than being
uniform in length.

All iterated integrals are evaluated with the contour prescription
inherited from the maximal cut. Endpoint-singular terms at $z=1$ are combined using a common endpoint or shuffle regularization,
while the marked point $h\in(0,1)$ retains the original $i0$
prescription throughout the reduction.
\begingroup
\color{black}
\section{Differential equation and  epsilon factorization} \label{sec:diffcannon}
In the previous section, we obtained an $\epsilon$-expansion of the
master integrals. It is therefore instructive to develop a systematic
method for canonicalizing the corresponding system of differential
equations. We emphasize that such a construction also provides valuable
information for the canonicalization of the full, uncut integral family.

If the lower sectors do not contain any elliptic blocks, their
canonicalization is expected to be straightforward by rational transformation. On the other hand,
if elliptic structures appear in off-diagonal blocks, the
nontrivial part of the analysis can be restricted to those blocks.
The primary diagonal elliptic diagonal blocks are already brought into canonical
form once we identify a transformation that canonicalizes the
differential equations of the maximal-cut system. Consequently, the
maximal cut isolates the elliptic part of the problem and provides the essential input required to construct a canonical basis
for the full system. The essential ingredient in the diagonal elliptic sector is to find a three-dimensional fundamental matrix (the period matrix) that satisfies the same differential equation. Note that previously, when we computed the period matrix before in $\epsilon=0$, we only had a two-dimensional matrix. But in finite $\epsilon$, to canonicalize the genuine elliptic sector we need a three-dimensional fundamental matrix even at $\epsilon=0$ as mentioned in \eqref{6.27jj}.\par
Throughout this section, we fix $\mu$ and $s$ when differentiating with respect to $\tau$. In particular, the common normalization
$\mathcal N(\epsilon,\mu,s)$ is independent of $\tau$ and affects boundary
data, but not the differential equation.
We use
column vectors
\begin{equation}
 \vec{\mathcal J}^{(135)}=(I_1,I_3,I_5)^T\,,
 \qquad
 \vec{\mathcal J}^{(24)}=(I_2,I_4)^T\,,
 \qquad g(\tau)=(1+\tau)(3+4\tau)\,.
\end{equation}
Both blocks are affine in $\epsilon$,
\begin{equation}
 \partial_\tau\vec{\mathcal J}^{(a)}
 =\bigl(A_0^{(a)}+\epsilon A_1^{(a)}\bigr)
       \vec{\mathcal J}^{(a)},\qquad a=135,24\,.
 \label{eq:exact-affine-system}
\end{equation}
An explicit, $\epsilon$-independent matrix satisfying
$\mathcal P_0^{(a)\prime}=A_0^{(a)}\mathcal P_0^{(a)}$ therefore yields
\begin{equation}
 \vec{\mathcal J}^{(a)}=\mathcal P_0^{(a)}\vec g^{(a)},
 \qquad
 \partial_\tau\vec g^{(a)}=\epsilon\,B^{(a)}\vec g^{(a)},
 \qquad
 B^{(a)}=(\mathcal P_0^{(a)})^{-1}A_1^{(a)}\mathcal P_0^{(a)}.
 \label{6.4mm}
\end{equation}
The goal of this section is to precisely construct $\mathcal{P}^{(a)}_0$, and it is also expected that the transformation matrix that canonicalizes the matrix involves non-rational functions, unlike the polylog case.
\subsection{The elliptic block and the third solution} \label{sec:new12}
In this subsection, we will explicitly compute the third solution for the fundamental matrix. We will focus on the even block ($A^{(135)}$) as it contains the genuine elliptic solution. The matrix for this sector is given by,
\begin{align}
     A_0^{(135)}&=\frac1{g(\tau)}
 \begin{pmatrix}
 -2\tau-\frac72&2-\frac{11}{2\tau}&\frac6\tau\\[3pt]
 -\frac32&\frac{11}2-2\tau&-6\\[3pt]
 \frac{3\tau}2&-9\tau-\frac32&6\tau
 \end{pmatrix}\,,\nonumber\\[3pt]
 A_1^{(135)}&=\frac1{g(\tau)}
 \begin{pmatrix}
 0&\frac8\tau+4&-\frac4\tau\\[3pt]
 0&\frac3\tau-1&4\\[3pt]
 0&\tau-3&\frac3\tau+7
 \end{pmatrix}\,.
 \label{eq:exact-even-connection}
\end{align}
These matrices can also be directly obtained by the exact twisted identity with,
\begin{align}
    \partial_\tau\log\mathcal U
 =\frac\epsilon\tau-\frac{\epsilon+\tfrac12}{z^2+\tau}\,,
 \qquad
 q(z,\tau)=\frac{z(1-z^2)(4z^2-3)}{2g(\tau)}
\end{align}
and,
\begin{align}
    \left[
 (\partial_\tau\log\mathcal U)
 \begin{pmatrix}1\\z^2\\z^4\end{pmatrix}
 -(A_0^{(135)}+\epsilon A_1^{(135)})
 \begin{pmatrix}1\\z^2\\z^4\end{pmatrix}
 \right]dz
 =\nabla_\omega\left[
 q(z,\tau)\begin{pmatrix}\tau^{-1}\\-1\\\tau\end{pmatrix}
 \right]\,.
 \label{eq:exact-connection-certificate}
\end{align}
Now, come to our main discussion. At $\epsilon=0$, the elliptic covering  curve is: $y^2=(1-z^2)(z^2+\tau)$ and the differential identity 
\begin{align}
    \frac{z^4\,dz}{y}
 =\frac{\tau}{3}\frac{dz}{y}
 +\frac{2(1-\tau)}3\frac{z^2\,dz}{y}
 -\frac13\,d(zy),
 \label{eq:exact-moment-identity}
\end{align}
shows why the complete elliptic periods in $\epsilon=0$ alone do not provide the full $3\times 3$ invertible matrix that can canonicalize the differential equation.  On a compact cycle, the last term may vanish. Therefore, we define a transverse quantity $W$ as,
\begin{align}
     W=I_5-\frac\tau3 I_1-\frac{2(1-\tau)}3 I_3,
 \qquad
 W'\big|_{\epsilon=0}=\frac{2}{3+4\tau}W
 \label{eq:exact-transverse-mode}
\end{align}
and also the differential system does not know about the cocycle we choose to integrate, therefore, in general, we have three independent solutions. To construct the third solution, we integrate the three differentials from $0 $ to $h$ on the Riemann sheet with positive $y$. Now the boundary term in \eqref{eq:exact-moment-identity} is no longer zero. \textcolor{black}{Note that \eqref{eq:exact-transverse-mode} already fixes the third solution up
to normalization: $W'=2W/(3+4\tau)$ integrates to $W\propto\sqrt{3+4\tau}$, and the endpoint term computed below has exactly this $\tau$-dependence. The
open-path construction supplies the constant, not the functional form.}
\begin{align}
     w(\tau)=-\frac{h\,y(h)}3
 =-\frac{\sqrt3\sqrt{3+4\tau}}{24}.
 \label{eq:exact-third-endpoint}
\end{align}

We made the particular choice of domain of integration because, from \eqref{eq:exact-connection-certificate}, it is clear that the boundary term vanishes in the domain of integration, as $q(z)$ explicitly contains a factor of $z(4z^2-3)$. Therefore, we do not need to solve an inhomogeneous equation to find the third solution. Explicitly, the solutions can be written as,
\begin{align}
    p_R&=\frac{2}{\sqrt\tau}K(-1/\tau)\,,&
 q_R&=2\sqrt\tau\,[E(-1/\tau)-K(-1/\tau)]\,,\nonumber\\
 p_I&=2iK(-\tau)\,,&
 q_I&=2i[K(-\tau)-E(-\tau)],\nonumber\\
p_h&=\frac1{\sqrt\tau}F(\pi/3\mid-1/\tau)\,,&
 q_h&=\sqrt\tau\,[E(\pi/3\mid-1/\tau)-F(\pi/3\mid-1/\tau)]\,.
 \label{eq:exact-elliptic-columns}
\end{align}
Here $F(\phi\mid m)$ and $E(\phi\mid m)$ are the incomplete elliptic
integrals of the first and second kind; $K(m)=F(\pi/2\mid m)$ and
$E(m)=E(\pi/2\mid m)$. Explicitly, the fundamental matrix is given by,
\begin{align}
    \boxed{\;
 \mathcal P_0^{(135)}=T \textcolor{blue}{\,S } \,,\qquad
 T=\begin{pmatrix}
 1&0&0\\0&1&0\\\frac\tau3&\frac{2(1-\tau)}3&1
 \end{pmatrix}\,,\qquad
 S=\begin{pmatrix}
 p_R&p_I&p_h\\q_R&q_I&q_h\\0&0&w
 \end{pmatrix}\,.
 \;}
 \label{eq:exact-elliptic-fundamental}
\end{align}
One can verify by using direct differentiation that $\mathcal P_0^{(135)\prime}=A_0^{(135)}\mathcal P_0^{(135)}$. One can also check that with the classical Legendre relation,
\begin{align}
    \Delta:=p_Rq_I-p_Iq_R=-2\pi i\,,
 \qquad
 \det\mathcal P_0^{(135)}=\Delta w
 =\frac{i\pi\sqrt3\sqrt{3+4\tau}}{12}\ne0
 \quad(\tau>0)\,.
 \label{eq:exact-elliptic-determinant}
\end{align}
Therefore, the transformation matrix is invertible. The direct differential-equation substitution, determinant, and complete(and incomplete) period calculation checks are provided with the \texttt{arXiv} submission in \texttt{03-elliptic-fundamental-matrix-checks.nb}. \textcolor{black}{Also note that, the quantity $\Delta$ is the Wronskian already obtained in Sec.~(\ref{Sec:6.4}):
$\Delta=\det P=-2\pi i$ of \eqref{eq:wqft-compact-x-determinant}, so
\eqref{eq:exact-elliptic-determinant} adds only the transverse normalization $w$.}
\subsection{The transformed basis and kernel of path ordered exponents}
We focus on the non-trivial $3\times 3$ differential system with a non-elliptic transformation matrix. The ``$B$'' matrix can be written as,
\begin{align}
    B^{(135)}=S^{-1}QS,
\end{align}
where
\begin{align}
& S^{-1}=
 \begin{pmatrix}
 \dfrac{q_I}{\Delta}&-\dfrac{p_I}{\Delta}
   &\dfrac{p_Iq_h-q_Ip_h}{\Delta w}\\[6pt]
 -\dfrac{q_R}{\Delta}&\dfrac{p_R}{\Delta}
   &\dfrac{q_Rp_h-p_Rq_h}{\Delta w}\\[6pt]
 0&0&\dfrac1w
 \end{pmatrix}\,,\nonumber\\&
 Q=\frac1{9\tau(1+\tau)(3+4\tau)}
 \begin{pmatrix}
 -12\tau&12(4+5\tau)&-36\\
 12\tau^2&3(9+5\tau-8\tau^2)&36\tau\\
 \tau(1+\tau)(9+8\tau)&-\tau(1+\tau)(11+16\tau)
     &3(1+\tau)(9+8\tau)
 \end{pmatrix}\,.
 \label{eq:exact-kernel-factors}
\end{align}
In particular the $Q$ matrix is defined as $Q=T^{-1}A_1^{(135)}T$. On the other hand, the differential equation and the canonicalization of the $2\times 2$ odd sector is elementary i.e the canonicalization procedure can be done using elementary functions. Its matrices are,
\begin{align}
    A_0^{(24)}=\frac1{g(\tau)}
 \begin{pmatrix}
 2-2\tau&-4\\[2pt]-(11\tau+3)/2&4\tau
 \end{pmatrix},\qquad
 A_1^{(24)}=\frac1{g(\tau)}
 \begin{pmatrix}
 3/\tau-1&4\\[2pt]\tau-3&3/\tau+7
 \end{pmatrix}.
 \label{eq:exact-odd-connection}
\end{align}
Now, let us introduce two elementary functions,
\begin{align}
     \chi(\tau)=\arcsin\frac1{2\sqrt{1+\tau}}\,,
 \qquad v(\tau)=\frac{\sqrt{3+4\tau}}8
\end{align}
and with that, one can find the transformation matrix that canonicalizes the odd block:
\begin{align}
    \mathcal P_0^{(24)}=
 \begin{pmatrix}
 1&\chi\\[2pt]
 \dfrac{1-\tau}2&\dfrac{1-\tau}2\chi+v
 \end{pmatrix},\qquad
 \mathcal P_0^{(24)\prime}=A_0^{(24)}\mathcal P_0^{(24)}\,,
 \qquad \det\mathcal P_0^{(24)}=v\,.
 \label{eq:exact-odd-fundamental}
\end{align}
For reference, list down the $B$-matrix for this odd sector.
\begin{align}
    \begin{split}
        B^{(24)}_{11}&=\frac1\tau-\frac6{3+4\tau}
                  +\frac{8\tau\chi}{(3+4\tau)^{3/2}}\,,\\
 B^{(24)}_{12}&=\frac1{2(1+\tau)\sqrt{3+4\tau}}
                  -\frac{4\chi}{3+4\tau}
                  +\frac{8\tau\chi^2}{(3+4\tau)^{3/2}}\,,\\
 B^{(24)}_{21}&=-\frac{8\tau}{(3+4\tau)^{3/2}}\,,\\
 B^{(24)}_{22}&=\frac1\tau-\frac2{3+4\tau}
                  -\frac{8\tau\chi}{(3+4\tau)^{3/2}}\,.
 \label{eq:exact-odd-kernel}
    \end{split}
\end{align}
An immediate question that may arise: how does one choose the fundamental matrix that canonicalizes the block $(24)$? Well, this is not magic, and the logic is simple, which goes as follows. The master integrals in this block solve to a simple rational function at $\epsilon=0$,
\begin{align}
   p(\tau)=\begin{pmatrix}
        1\\
        \frac{1-\tau}{2}
    \end{pmatrix}.
\end{align}
Now to find the second column write  it as $q=\chi p+(0,v)^{T}$ and substituting into $q'=A_0^{(24)}q$ we get,
\begin{align}
    \frac{v'}{v}=\Tr{A_0^{(24)}},\qquad \chi'=-\frac{4v}{(1+\tau)(3+4\tau)},
\end{align}
and correspondingly, the solutions are,
\begin{align}
    v=C_v \sqrt{3+4\tau},\qquad \chi =8C_v\arcsin\frac{1}{2\sqrt{1+\tau}}+C_{\chi}
\end{align}
For any non zero choice of $C_v, C_{\chi}$ (we choose $C_v=1/8,\,C_{\chi}=0$) the constructed fundamental matrix canonicalize the system and we get \eqref{eq:exact-odd-fundamental}.
Note that for the odd block, the transformation matrix elements are made of elementary functions (not strictly rational, but one can always make a variable transformation properly to make the matrix look rational in kinematic variables). Combining the two results from the two blocks, we can write down the canonicalized differential equation in a compact form,
\begin{equation}
 \begin{pmatrix}\vec{\mathcal J}^{(135)}\\\vec{\mathcal J}^{(24)}\end{pmatrix}
 =\operatorname{diag}(\mathcal P_0^{(135)},\mathcal P_0^{(24)})
 \begin{pmatrix}\vec g^{(135)}\\\vec g^{(24)}\end{pmatrix},
 \qquad
 \partial_\tau
 \begin{pmatrix}\vec g^{(135)}\\\vec g^{(24)}\end{pmatrix}
 =\epsilon\operatorname{diag}(B^{(135)},B^{(24)})
 \begin{pmatrix}\vec g^{(135)}\\\vec g^{(24)}\end{pmatrix}.
 \label{eq:exact-five-master-form}
\end{equation}
With this in hand, we can write down the all-order solution formally as,
\begin{align}
     \vec{\mathcal J}^{(a)}(\tau,\epsilon)
 =\mathcal P_0^{(a)}(\tau)
 \mathscr P\exp\!\left[\epsilon\int_{\tau_*}^{\tau}
 B^{(a)}(t)\,dt\right]
 (\mathcal P_0^{(a)}(\tau_*))^{-1}
 \vec{\mathcal J}^{(a)}(\tau_*,\epsilon)\,.
 \label{8.19hh}
\end{align}
where, $\tau_{*}$ is a kinematic point where we can do the integrals that does not contain any non-rational information of the underlying curve i.e we choose $\tau_*$ in such a way that the boundary integrals can be evaluated in terms of elementary functions for example we can choose the degeneration points (say $\tau_*=0$), where we find that due to kinematic degeneration the elliptic curves drops down to a rational curve, as our boundary data.
\subsection{The boundary data}
To fix the boundary data, we need to take more care as $\tau=0$ gives rise to a singularity in several instances. Therefore, to use $\tau=0$ as boundary data, the singularity factor must be carefully handled. Particularly for the elliptic block,
\begin{align}
    A^{(135)}=\frac{R(\epsilon)}\tau+\sum_{k\ge0}M_k(\epsilon)\tau^k,
 \qquad
 R(\epsilon)=\begin{pmatrix}
 0&\dfrac{16\epsilon-11}{6}&2-\dfrac{4\epsilon}{3}\\[4pt]
 0&\epsilon&0\\0&0&\epsilon
 \end{pmatrix},
 \label{eq:exact-residue}
\end{align}
and the analytic part begins with,
\begin{align}
    M_0=\begin{pmatrix}
 -\frac76&\frac{89}{18}-\frac{44\epsilon}{9}
             &-\frac{14}3+\frac{28\epsilon}{9}\\[3pt]
 -\frac12&\frac{11}6-\frac{8\epsilon}{3}
             &-2+\frac{4\epsilon}{3}\\[3pt]
 0&-\frac12-\epsilon&0
 \end{pmatrix}.
 \label{eq:sec7-regular-part}
\end{align}
Around the vicinity of the singular point $\tau=0$, one may define the fundamental matrix as,
\begin{align}  \Phi^{(135)}=H(\tau,\epsilon)\tau^{R(\epsilon)},\qquad
 H=\mathbf1+\sum_{n\ge1}H_n(\epsilon)\tau^n,
 \qquad \Phi^{(135)\prime}=A^{(135)}\Phi^{(135)}.\label{8.22kk}
\end{align}
Furthermore, the coefficient matrices $H_n$ follow the following recursion relation.
\begin{align}
    \bigl(n\,\mathrm{Id}-\operatorname{ad}_{R}\bigr)H_n
 =\sum_{k=0}^{n-1}M_kH_{n-1-k},\qquad
 H_0=\mathbf1,\qquad \operatorname{ad}_R(X)=RX-XR.
 \label{eq:sec7-wasow-recursion}
\end{align}
For $|\epsilon|<1$ this operator is invertible for all $n\ge1$,
including $\epsilon=0$, and $H$ is analytic for $|\tau|<3/4$.
Setting $a_\epsilon=(16\epsilon-11)/6$ and
$b_\epsilon=2-4\epsilon/3$, one obtains
\begin{equation}
 \tau^{R(\epsilon)}=
 \begin{pmatrix}
 1&\dfrac{a_\epsilon}{\epsilon}(\tau^\epsilon-1)
   &\dfrac{b_\epsilon}{\epsilon}(\tau^\epsilon-1)\\[5pt]
 0&\tau^\epsilon&0\\0&0&\tau^\epsilon
 \end{pmatrix}.
 \label{7.16hh}
\end{equation}
The limit $\epsilon\to0$ gives
$\tau^{R(0)}=\mathbf1+R(0)\log\tau$. This Frobenius matrix and
$\mathcal P_0^{(135)}$ at $\epsilon=0$ differ only by an invertible
constant factor, they need not have identical normalizations. Now the original integral has the form,
\begin{align}
     I_{2n+1}(\tau,\epsilon)=2\mathcal N\tau^\epsilon\int_0^1
 \frac{z^{2n}(1-z^2)^{\epsilon-1/2}}
 {(z^2+\tau)^{\epsilon+1/2}(4z^2-3+i0)^\epsilon}\,dz\,,
 \qquad n=0,1,2\,.
 \label{7.17k}
\end{align}
As $\tau\to 0^+$, two separate regions of integration have to be investigated carefully.  The factor $z^2+\tau$ changes behavior around $z\sim \sqrt \tau$. We therefore have two distinct regions, and we will use the method of regions to compute the boundary integrals\cite{Beneke:1997zp},
\begin{align}
    \textrm{soft:} \,\,z\sim \sqrt{\tau} ,\qquad \textrm{hard:} \,\,z\sim \mathcal{O}(1)
\end{align}
In the soft region set $z=\sqrt{\tau}\bar z$ and the integral takes the form,
\begin{align}
    I_{2n+1}^{\textrm{ soft}}=2\mathcal{N}\tau^n(-3+i0)^{-\epsilon} \int_0^\infty d\bar z\frac{\bar z^{2n}}{(1+\bar z^2)^{\epsilon+\frac{1}{2}}}+\cdots=\mathcal{N}\tau^n(-3+i0)^{-\epsilon} B\left(n+\frac{1}{2},\epsilon-n\right)+\cdots
\end{align}
Therefore, at $n=0$, the boundary integral takes the explicit form as,
\begin{align}
    \begin{split}
        I_{1}^{\textrm{soft}}\equiv \mathcal{S}(\epsilon)=\mathcal N(-3+i0)^{-\epsilon}
   \frac{\sqrt\pi\,\Gamma(\epsilon)}{\Gamma(\epsilon+\tfrac12)}\,.
    \end{split}
\end{align}
For $n=1,2$, the soft term starts from $\tau $ and $\tau^2$, respectively. i.e
\begin{align}
    I_3^{\textrm{soft}}\sim \mathcal{O}(\tau),\qquad   I_5^{\textrm{soft}}\sim \mathcal{O}(\tau^2)\,.
\end{align}
Therefore, they do not contribute to the leading approximation. However, we are still left with the hard region integrals. As we discussed, in the hard region $z\sim \mathcal{O}(1)$ and the leading contribution takes the form,
\begin{align}
\begin{split}
    I_{2n+1}^{\textrm{hard}}&= 2\mathcal{N} \tau^\epsilon \int_0^1 z^{2n-2\epsilon-1}(1-z^2)^{\epsilon-\frac{1}{2}}(4z^2-3+i0)^{-\epsilon}dz+\cdots\\ &
    =\tau^{\epsilon}\mathcal{B}_n(\epsilon)+\cdots
    \end{split}
\end{align}
where,
\begin{align}
    \mathcal{B}_n(\epsilon)=\mathcal N(-3+i0)^{-\epsilon}
 B\!\left(n-\epsilon,\epsilon+\tfrac12\right)
 {}_2F_1\!\left(\epsilon,n-\epsilon;n+\tfrac12;
                         \tfrac43+i0\right),\qquad n=0,1,2\,.
\end{align}
Now combining both regions of integration, the total boundary integral takes the form,
\begin{align}
    \begin{split}
       & I_1(0^{+},\epsilon)=\mathcal{S}(\epsilon)+\tau^\epsilon \mathcal{B}_0(\epsilon)+\cdots\,,\\&
        I_3(0^{+},\epsilon)=\tau^\epsilon \mathcal{B}_1(\epsilon)+\mathcal{O}(\tau)\,,\\ &
I_5(0^+,\epsilon)=\tau^\epsilon\mathcal{B}_2(\epsilon)+\mathcal{O}(\tau^2)\,.\label{8.32jj}
    \end{split}
\end{align}
\textcolor{black}{The three constants \eqref{8.32jj} are not independent. Writing
$a_\epsilon=(16\epsilon-11)/6$ and $b_\epsilon=2-4\epsilon/3$ for the two
off-diagonal entries of the residue matrix \eqref{eq:exact-residue}, the hard-region
integrals satisfy
\begin{align}
\mathcal B_0(\epsilon)
=\frac{a_\epsilon}{\epsilon}\,\mathcal B_1(\epsilon)
+\frac{b_\epsilon}{\epsilon}\,\mathcal B_2(\epsilon)
\label{eq:hard-region-identity}
\end{align}
as a ${}_2F_1$ contiguity relation at fixed argument $\tfrac43+i0$. This is what
makes the $\tau\to0^{+}$ limit below finite.} Inserting \eqref{7.16hh} one finds
\begin{align}
\tau^{-R(\epsilon)}\vec{\mathcal J}^{(135)}(\tau,\epsilon)\Big|_{\tau\to0^{+}}
=\begin{pmatrix}
\mathcal S+\frac{a_\epsilon}{\epsilon}\mathcal B_1+\frac{b_\epsilon}{\epsilon}\mathcal
B_2\\
\mathcal B_1\\ \mathcal B_2\end{pmatrix}
=\begin{pmatrix}\mathcal S+\mathcal B_0\\ \mathcal B_1\\ \mathcal B_2\end{pmatrix}\,.
\end{align}
\textcolor{black}{By \eqref{eq:hard-region-identity} the explicit $1/\epsilon$ factors in the first entry recombine  into $\mathcal B_0$. Its simple pole $-1/\epsilon$ cancels the pole $+1/\epsilon$ of $\mathcal S$ (see the footnote below), while $\mathcal B_1,\mathcal B_2$ are themselves regular at
$\epsilon=0$. The identity \eqref{eq:hard-region-identity} is the integral over $0<t<1$ of the total derivative
\begin{align}
\frac{d}{dt}\Big[t^{-\epsilon}(1-t)^{\epsilon+\frac12}(4t-3+i0)^{1-\epsilon}\Big]
=t^{-\epsilon-1}(1-t)^{\epsilon-\frac12}(4t-3+i0)^{-\epsilon}
\Big[3\epsilon-\big(8\epsilon-\tfrac{11}{2}\big)t+(4\epsilon-6)t^{2}\Big]\,,
\end{align}
whose boundary terms vanish for $-\tfrac12<\Re\epsilon<0$. Here $t=z^{2}$ is the
variable in which
$\mathcal B_n\propto\int_0^1t^{n-\epsilon-1}(1-t)^{\epsilon-\frac12}(4t-3+i0)^{-\epsilon}dt\,.$}


\textcolor{black}{Since $\tau=0$ is a singular point, $\tau_*$ in \eqref{8.19hh} must be taken at a small positive value. It must lie inside the disk $|\tau|<\tfrac34$ where the Frobenius series \eqref{eq:sec7-wasow-recursion} converges. There we construct the vector $\vec {\mathcal{J}}^{(135)}(\tau_*,\epsilon)$ as},
\begin{align}
    \vec {\mathcal{J}}^{(135)}(\tau_*,\epsilon)=H({\tau_*},\epsilon)\tau_*^{R(\epsilon)}\vec c(\epsilon), \quad \vec c(\epsilon)=\lim_{\tau\to 0^+} \tau^{-R(\epsilon)}H^{-1}(\tau, \epsilon)\vec {\mathcal{J}}(\tau,\epsilon)=\begin{pmatrix}\mathcal S+\mathcal B_0\\\mathcal B_1\\\mathcal B_2\end{pmatrix}\,.
\end{align}
Therefore, the full integral can be written as,
\begin{equation}
\boxed{
\begin{aligned}
\vec{\mathcal J}^{(135)}(\tau,\epsilon)
={}&
\mathcal P_{0}^{(135)}(\tau)\,
\mathscr P\exp\left[
\epsilon\int_{\tau_*}^{\tau}
B^{(135)}(t)\,dt
\right]
\left(\mathcal P_{0}^{(135)}(\tau_*)\right)^{-1}
\\
&\times
H(\tau_*,\epsilon)\,
\tau_*^{R(\epsilon)}
\begin{pmatrix}
\mathcal S(\epsilon)+\mathcal B_0(\epsilon)\\
\mathcal B_1(\epsilon)\\
\mathcal B_2(\epsilon)
\end{pmatrix}.
\end{aligned}
}
\label{eq:epsilon-factorized-frobenius-solution}
\end{equation}
Remarkably, one may see that at $\epsilon=0$ the poles of $\mathcal{S}(\epsilon)$ and $\mathcal{B}_0$ cancels\footnote{Writing
$F(\epsilon)=\mathcal N(\epsilon,\mu,s)(-3+i0)^{-\epsilon}$,
with $F(0)=\mathcal N_0$, one finds
$\mathcal S=F[\,\epsilon^{-1}+2\log2+\mathcal O(\epsilon)\,]$ and
$\mathcal B_0=F[\,-\epsilon^{-1}+2\log2+\mathcal O(\epsilon)\,]$,
where
${}_2F_1(\epsilon,-\epsilon;\frac12;\frac43+i0)
=1+\mathcal O(\epsilon^2)$.
Thus, the poles cancel and
$\lim_{\epsilon\to0}(\mathcal S+\mathcal B_0)
=4\mathcal N_0\log2=\mathcal N_0\log16$.}, which is important as the original integral does not contain any pole as $1/\epsilon$. Therefore, a naive $\tau\to 0$ limit would make a disaster, giving rise to divergences in epsilon. Now, we choose $\tau_*$ very close to $\tau\sim 0$ and numerically compute the boundary term $\sim\mathcal{P}_0^{-1}\mathbf{1}\tau^{R}\begin{pmatrix}
..\\..\\..
\end{pmatrix} 
$ numerically in $\tau_*$ and then compute the path ordered exponential order by order in $\epsilon$. We want to again emphasize that taking $\tau_*\to 0$ limit directly as a boundary condition will give unwanted $1/\epsilon$ divergence, which is in general absent in this particular family of integrals. The affine dependence on \(\epsilon\), both fundamental matrices, the transformed kernels, the finite-\(\epsilon\) cut equation, and the soft and hard boundary limits are checked together in \texttt{07-exact-epsilon-factorization-checks.nb}.
\endgroup
\subsection{Numerical validation of the two $\epsilon$-expansion} \label{sec:numval}
In this subsection, we provide a three-fold numerical check relating Sec.~(\ref{sec:diffcannon}) and Sec. ~(\ref{sec:epsilonexpnad}). For the numerical
analysis we fix $\mu=s=\tau=1$, which corresponds to $\gamma=\sqrt{9/5}$. We first compare the truncated series $I_i^{(0)}+\epsilon I_i^{(1)}+\epsilon^{2}I_i^{(2)}$ against a direct numerical evaluation of \eqref{7.17k} at two values of $\epsilon$. The results are listed in Table~(\ref{Tab1}).
\begin{table}[ht]
\centering
\renewcommand{\arraystretch}{1.15}
\begin{tabular}{c c c c}
\toprule
& $i=1$ & $i=3$ & $i=5$ \\
\midrule
$I_i^{(0)}$
& $0.8346268$
& $0.3813799$
& $0.2782090$ \\

$I_i^{(1)}$
& $-2.536458 - 4.486630\,i$
& $-1.414760 - 1.705471\,i$
& $-1.166333 - 1.113662\,i$ \\

$I_i^{(2)}$
& $-10.469484 + 12.408307\,i$
& $-1.782155 + 5.539008\,i$
& $-0.009726 + 4.135944\,i$ \\
\midrule

$\epsilon = 0.05,\ \text{truncated}$
& $0.681630 - 0.193311\,i$
& $0.306187 - 0.071426\,i$
& $0.219868 - 0.045343\,i$ \\

$\epsilon = 0.05,\ \text{direct}$
& $0.685509 - 0.191331\,i$
& $0.307120 - 0.071423\,i$
& $0.220228 - 0.045745\,i$ \\

$\text{relative deviation}$
& $6.1\times 10^{-3}$
& $3.0\times 10^{-3}$
& $2.4\times 10^{-3}$ \\
\midrule

$\epsilon = 0.02,\ \text{truncated}$
& $0.779710 - 0.084769\,i$
& $0.352372 - 0.031894\,i$
& $0.254878 - 0.020619\,i$ \\

$\epsilon = 0.02,\ \text{direct}$
& $0.779950 - 0.084628\,i$
& $0.352429 - 0.031892\,i$
& $0.254900 - 0.020645\,i$ \\

$\text{relative deviation}$
& $3.6\times 10^{-4}$
& $1.6\times 10^{-4}$
& $1.3\times 10^{-4}$ \\
\bottomrule
\end{tabular}
\caption{\textcolor{black}{The $\epsilon$-expansion of the three elliptic maximal-cut masters at
$\tau=1$ (with $\mu=s=1$, i.e.\ $\gamma=\sqrt{9/5}$), compared with direct numerical
integration of \eqref{7.17k}. All entries carry the common normalization
$\mathcal N(\epsilon,\mu,s)$ and the $+i0$ prescription of \eqref{eq:L0-correct}.
The residual falls by a factor $17$--$19$ between the two values of $\epsilon$,
against $(0.05/0.02)^{3}=15.6$, i.e.\ it is the expected $O(\epsilon^{3})$ truncation
error and not a failure of the term-by-term integration established around
\eqref{eq:sec6-I-order0-complete}.}}
\label{Tab1}
\end{table}
Moreover, as we mentioned, it is worth exhibiting a comparison with the results that can be found by the epsilon expansion after the canonicalization of the differential equation, and we directly evaluate the path-ordered exponent of \eqref{eq:epsilon-factorized-frobenius-solution} numerically with suitable boundary data. Taking, $\tau_*=0.05$, we subsequently compute the boundary constants \eqref{8.32jj}, the Frobenius factor $H(\tau_*,\epsilon)\tau_*^{R(\epsilon)}$, the fundamental matrix of \eqref{eq:exact-elliptic-fundamental} and the path ordered exponential of the canonicalised matrix $B^{(135)}$ and compare the results with \eqref{7.17k} (see Table~(\ref{Tab2})). 
\begin{table}[t!]
\centering
\renewcommand{\arraystretch}{1.15}
\begin{tabular}{c c c c c}
\toprule
$\epsilon$ & $\tau$ & $I_1$ & $I_3$ & $I_5$ \\
\midrule

$0.02$ & $0.8$
& $0.8435135 - 0.0918983\,i$
& $0.3749050 - 0.0340438\,i$
& $0.2699030 - 0.0219140\,i$ \\

$0.02$ & $1.5$
& $0.6698573 - 0.0722062\,i$
& $0.3104168 - 0.0279303\,i$
& $0.2261729 - 0.0182439\,i$ \\

$0.05$ & $0.8$
& $0.7400956 - 0.2074565\,i$
& $0.3258525 - 0.0760609\,i$
& $0.2325194 - 0.0484271\,i$ \\

$0.05$ & $1.5$
& $0.5903104 - 0.1636192\,i$
& $0.2715826 - 0.0627740\,i$
& $0.1962631 - 0.0405868\,i$ \\
\bottomrule
\end{tabular}
\caption{\textcolor{black}{The $\epsilon$-factorized solution
\eqref{eq:epsilon-factorized-frobenius-solution}, transported from $\tau_*=0.05$ with the boundary constants \eqref{8.32jj}, evaluated at two values of $\tau$ and two values of $\epsilon$ ($\mu=s=1$). Direct numerical integration of \eqref{7.17k}
reproduces every entry to the digits shown; the largest relative deviation over the
four rows is $3\times10^{-10}$. The check is simultaneous on the fundamental matrix \eqref{eq:exact-elliptic-fundamental}, the transformed kernel \eqref{eq:exact-kernel-factors}, the Frobenius factor \eqref{8.22kk}-\eqref{7.16hh}, and the soft and hard boundary data.}}
\label{Tab2}
\end{table}
Moreover, as we promised earlier, it is also important to match the numerical results obtained here with the explicit $\epsilon-$expansion in terms of the elliptic polylogs. We use the numerical symbolic calculator {\texttt{\textbf{GiNaC}}} \cite{Bauer:2000cp} to perform the iterated elliptic integrations. We use upto 1024 quadrature nodes to perform the numerics. The resulting coefficients are compared in Table~(\ref{Tab3}). For numerical evaluation one has to be careful about the (apparent) singularities of the integrands, at $z=1$ the integrand contains a square root singularity $y(z)\sim \sqrt{1-z}$. Therefore, we set \(z=1-u^2\). After reversing the
integration limits, the elliptic measure becomes
\begin{align}
\frac{dz}{\sqrt{(1-z^2)(z^2+\tau)}}
\;\longmapsto\;
\frac{2\,du}{\sqrt{2-u^2}\sqrt{(1-u^2)^2+\tau}},
\end{align}
removing the square-root endpoint singularity at \(z=1\); any remaining
logarithmic singularity is integrable and faster converging. The branch point
\(h=\sqrt{3}/2\) maps to \(u=\sqrt{1-h}\), where we split the integration
interval while retaining the prescribed $+i0$. Near $z=1$, the integrand at $\epsilon^k$ behaves roughly as,
\begin{align}
    \frac{\log^k(1-z)}{\sqrt{1-z}}dz
\end{align}
The integral is convergent for $\Re{k}>-1$
and gives $(-1)^k 2^{k+1} \Gamma (k+1)$. However numerically the integral is converging slowly. So, that we use the $u$-variable that maps the integrand as,
\begin{align}
    \frac{\log^k(1-z)}{\sqrt{1-z}}dz\mapsto2(2\log u)^k du,
\end{align}
which converges faster. For one-letter factors we use the exact identity
\(\mathrm G(c;z)=\log(1-z/c)\), avoiding a slowly convergent endpoint
series evaluation. As an additional numerical check, we evaluated representative
length-three elliptic words
$E_4(0,1,1;0,c,d;1)=\int_0^1 dz\,\psi_0(0,z)G(c,d;z)$,
using {\texttt{\textbf{GiNaC}}} for the depth-two \(G\) functions and endpoint-adapted
quadrature for the outer elliptic integral. At \(\tau=1\), the words
with \((c,d)=(-1,+i)\) and \((+i,-1)\) satisfy their shuffle relation
to \(6\times10^{-16}\). We also checked the weighted sum of all
\(36\) ordered depth-two words at \(z=0.95\), across the real-letter
branch point, against \(\mathcal L(z)^2/2-\pi^2/3\) to \(3.7\times10^{-15}\). These tests validate the word ordering and branch implementation used in the \(\epsilon^2\) expansion. 
\begin{table}[t!]
\centering
\renewcommand{\arraystretch}{1.15}
\begin{tabular}{l l r}
\toprule
\textbf{Coefficient}
& \multicolumn{1}{c}{\bfseries\begin{tabular}[b]{@{}c@{}}Reconstructed value\\ using \texttt{GiNaC}\end{tabular}}
& \multicolumn{1}{c}{\bfseries\begin{tabular}[b]{@{}c@{}}Abs.\ difference from\\ direct quadrature\end{tabular}} \\
\midrule

$I_1^{(0)}$
& $0.834626841674074$
& $3.1\times 10^{-15}$ \\

$I_3^{(0)}$
& $0.381379881750907$
& $2.7\times 10^{-15}$ \\

$I_5^{(0)}$
& $0.278208947224691$
& $2.4\times 10^{-15}$ \\[0.3em]

$I_1^{(1)}$
& $-2.53645776491280 - 4.48663017553707\,i$
& $2.7\times 10^{-12}$ \\

$I_3^{(1)}$
& $-1.41476030544982 - 1.70547120997813\,i$
& $2.8\times 10^{-12}$ \\

$I_5^{(1)}$
& $-1.16633317835460 - 1.11366208393436\,i$
& $2.8\times 10^{-12}$ \\[0.3em]

$I_1^{(2)}$
& $-10.4694837710689 + 12.4083073129956\,i$
& $1.2\times 10^{-10}$ \\

$I_3^{(2)}$
& $-1.78215425279251 + 5.53900817126084\,i$
& $1.0\times 10^{-10}$ \\

$I_5^{(2)}$
& $-0.00972513521293283 + 4.13594377952148\,i$
& $8.8\times 10^{-11}$ \\
\bottomrule
\end{tabular}
\caption{\textcolor{black}{The nine expansion coefficients of Table~(\ref{Tab1}) recomputed from the explicit elliptic-polylogarithm expressions of Sec.~(\ref{sec:epsilonexpnad}), with the iterated integrals evaluated in \texttt{GiNaC} on $1024$ quadrature nodes. The last column is the absolute difference from the full-precision direct quadrature of
\eqref{7.17k} that underlies Table~(\ref{Tab1}). Table~(\ref{Tab1}) shows these
values rounded to seven digits. The two routes to the same numbers are independent, so the agreement tests the word ordering, the $i0$ prescription at $z=\pm\sqrt3/2$, and the shuffle regularization at $z=1$.}}
\label{Tab3}
\end{table}
Moreover, it is worth explaining how we implement the branch structure of the function $(4z^2-3+i0)^{-\epsilon}$ in the numerical calculation, as the branch points lie inside the chosen contour. These subtleties start coming from the $\mathcal{O}(\epsilon)$. There are three branches where the function takes the form,
\begin{align}
    (4z^2-3+i0)^{-\epsilon}=\begin{cases}
        (4z^2-3)^{-\epsilon},\qquad -1<z<-h\,\,\textrm{or,}\,\,h<z<1,\\
         e^{-i\pi\epsilon}(3-4z^2)^{-\epsilon},\qquad -h<z<h
    \end{cases}
\end{align}
We implement the $+i0$ by introducing a small positive quantity $\eta>0$ and shifting the two letters $h$ and $-h$ as,
\begin{align}
    c_{+}=h-i\eta,\qquad c_-=-h+i\eta,
\end{align}
Therefore,
\begin{align}
    4(z-c_+)(z-c_-)=4z^2-3+4\eta^2+8ih\eta\to 4z^2-3+i0
\end{align}
Now, in general we have the following integrand (at first order of $\epsilon$) \(F(z)\) which is even in $z$, the logarithmic contribution
on the full integral domain is evaluated as
\begin{equation}
\int_{-1}^{1}F(z)\log(4z^2-3+i0)\,dz
=2\left[
 \int_0^h F(z)\bigl(\log(3-4z^2)+i\pi\bigr)\,dz
 +\int_h^1 F(z)\log(4z^2-3)\,dz
\right].
\label{eq:branch-split-numerics}
\end{equation}
In numerical reconstruction, the positive
half-contour is split at \(z=h\). With \(z=1-\mathfrak{u}^2\), the two quadrature
intervals are \(0<\mathfrak{u}<\sqrt{1-h}\), corresponding to \(h<z<1\), and
\(\sqrt{1-h}<\mathfrak{u}<1\), corresponding to \(0<z<h\). The branch phase is
implemented through the one-letter identity
\(G(c;z)=\log(1-z/c)\), using the correlated shifted letters
\(c_+=h-i\eta\) and \(c_-=-h+i\mathfrak{\eta}\), with \(\eta\to0^+\).
The reproducible code for the calculation in \texttt{\textbf{C++}}  can be found in the ancillary files with the \texttt{\textbf{arXiv}} submission as \texttt{\textbf{ginac-empl-validation.cpp}}.
\section{Bessel-twisted differential and genus drop}\label{sec9}
In all previous discussions, we mainly focused on the two-loop integral; however, in classical gravitational scattering, the observables can be computed by Fourier transforming in impact-parameter space. Then the natural question comes up: what happens if we consider the maximal cut of the full Fourier integral? It turns out that, in the momentum space computation, the cut differential takes the form\footnote{We evaluate the Fourier-transformed maximal cut directly in momentum space. This is convenient because the Fourier transform integrates over \(s=-q^2\), while the Baikov modulus $\tau(s)=\frac{(4m^2+s)(\gamma^2-1)}{4m^2}$ depends on \(s\). The direct calculation avoids integrating over a family of curves with varying modulus. \textcolor{black}{The resulting differential lives on the same family of curves as the Baikov cut, at the
single value $\mathfrak t=-k'^{2}$ discussed below. It is the cut of the impact-parameter integral, on which the transverse loop momenta localise at imaginary values (Appendix~(\ref{app:maximal-cut})). It should not be confused with the Fourier transform over real $\bm q_\perp$ of the fixed-$q$ cut
\eqref{eq:maxcut-eps0}, which falls off as a power of $b$ instead of growing like
$e^{2mb/k'}$.}},
\begin{align}
    \widetilde{\Omega}_{\textrm{cut}}(z,b)\sim I_0^2\left(\frac{m\,b}{z}\right) \frac{dz}{z\sqrt{(1-z^2)(z^2-k'^2)}}\,,\qquad k'=\frac{\sqrt{\gamma^2-1}}{\gamma}\,.\label{9.1kk}
\end{align}
\textcolor{black}{From \eqref{9.1kk}, it is evident that the underlying algebraic geometry is
genuinely elliptic. Nevertheless, if one attempts to evaluate the integral by
expanding the Bessel function in the small-mass parameter $m$, the resulting
terms do not appear to generate elliptic periods; instead, they reduce to
elementary functions. This raises a natural question: what is the mathematical mechanism responsible for this apparent simplification? In this section, we address this question precisely. We show that the disappearance of the elliptic
periods is a consequence of additional discrete symmetries of the underlying
curve and of the differential being integrated.}  
\textcolor{black}{It is worth noting how \eqref{9.1kk} sits relative to the fixed-$q$ analysis. We write
the family of curves met throughout this paper as
\begin{align}
\mathcal E_{z,\mathfrak t}:\qquad y^{2}=(1-z^{2})(z^{2}+\mathfrak t)\,,
\end{align}
with $\mathfrak t$ a parameter of the family rather than a kinematic variable.  From
Sec.~(\ref{sec:intersec}) to Sec.~(\ref{sec:diffcannon}) the cut sits at
$\mathfrak t=\tau(s)>0$, which varies with $s$ as recorded in the footnote above.
After the $q$ integration, this is no longer the case. The curve in \eqref{9.1kk} is
$\mathcal E_{z,\mathfrak t}$ at the single value
\begin{align}
\mathfrak t=-k'^{2}=-\frac{\gamma^{2}-1}{\gamma^{2}}\ \in(-1,0)
\qquad\text{for}\qquad \gamma\in(1,\infty)\,,
\end{align}
fixed by $\gamma$ alone.  Both endpoints of the interval are degenerations of the curve, and they are the two components of \eqref{eq:degeneration-divisor-cut} other than $\{\mu=0\}$.  At $\mathfrak t\to0^{-}$, i.e.\ $\gamma\to1$, the real pair $\pm k'$ collides at the
origin; this is the threshold degeneration $\{\Lambda=0\}$, where for fixed $q$ the colliding pair is instead the imaginary one $\pm i\sqrt{\tau}$.  At $\mathfrak t\to-1^{+}$, i.e.\ $\gamma\to\infty$, one has $y^{2}\to-(1-z^{2})^{2}$ and
the fibre becomes reducible, which is the locus $\{\Lambda+4\mu=0\}$.  The second case deserves a bit more emphasis. In the fixed-$q$ variables that locus requires
$\gamma^{2}<1$ and is reached only by analytic continuation, whereas here it is the
physical ultrarelativistic limit. }

\subsection{Additional involution symmetry, quotient curves and genus drop} \label{newsec4}
It turns out that the parity split of the cut differential is an instance of the extra involution mechanism introduced in \cite{Marzucca:2023gto}, which corresponds to a genus drop. Note that the genus drop referred to here is not the kinematic genus drop that comes from the branch point collision. Before going to the discussion for our case, let us take a moment to review some basic features of the \textit{internal genus drop}.
\subsubsection{Hyperelliptic curves with an extra involution symmetry}
Consider a smooth hyperelliptic curve,
\begin{align}
    \mathcal{H:\qquad} y^2=P_{2g+2}(z)\,.
\end{align}
Besides the hyperelliptic involution,
  $  e_0:\quad (z,y)\mapsto (z,-y)$, suppose a Mobius transformation,
\begin{align}
    z=\gamma(\hat z)=\frac{A\hat z+B}{C\hat z+D},\qquad \hat y=(C\hat z+D)^{g+1}y
\end{align}
puts the equation in the even form,
\begin{align}
    \hat y^2=\widehat{P}(\hat z),\qquad \widehat{P}(\hat z)=(C\hat z+D)^{2g+2}P_{2g+2}(\gamma(\hat z))\,.
    \end{align}
 Then the roots of $\widehat P$ occurs in pairs: $\{\hat \alpha_i,-\hat \alpha_i\}$ and the curve has two addition involution map:
 \begin{align}
     e_1: (\hat z,\hat y) \mapsto (-\hat z,\hat y),\qquad e_2:=e_1 e_0: (\hat z,\hat y)\mapsto (-\hat z,-\hat y) \,.
 \end{align}
 Additionally, the involution maps: $\{\mathbb{I},e_0,e_1,e_2\}$ form a Klein Four-Group ($V_4$). A hyperelliptic curve   is said to possess an extra involution $e_1\in \textrm{Aut}(\mathcal{H})$ if there exists a Mobius transformation $\gamma \in \textrm{PGL}_2(\mathbb{C})$, such that,
 \begin{align}
     \widehat{P}(\hat z)=c (\hat z^2-\hat\alpha_1^2)\cdots (\hat z^2-\hat\alpha_{\textrm{g}+1}^2)
 \end{align}
we have, with $x=\hat z^2$, their quotient curves are,
 \begin{align}
     \begin{split}
         &\mathcal{H}_1:\quad v_1^2=Q(x)\,,\qquad \pi_1(\hat z,\hat y)=(x,v_1)=(\hat z^2, \hat y)\,,\\ &
         \mathcal{H}_2:\quad v_2^2=x\, Q(x)\,,\qquad \pi_2(\hat z,\hat y)=(x,v_2)=(\hat z^2, \hat z\hat y)\,,\label{9.6a}
     \end{split}
 \end{align}
 which gives the genus structure of the quotient curves,
 \begin{align}
     \begin{split}
         g(\mathcal{H}_1)=\left\lfloor \frac{g}{2}\right\rfloor,\quad g(\mathcal{H}_2)=\left\lceil\frac{g}{2}\right\rceil\,.
     \end{split}
 \end{align}
 Thus, we see that the additional involution symmetry separates the original genus-g geometry into two quotient geometries. Moreover, the existence of this coordinate system can directly be tested from the branch point structure of the underlying curve. The strategy goes as follows: partition the $2g+2$ roots into pairs $(p_1^i,p_2^i)$, $i=,1,\cdots, g+1$. A Mobius transformation which sends every pair to the opposite point exists iff there exists a vector $\vec \nu=(\nu_1,\nu_2,\nu_3)$ such that,
 \begin{align}
     2p_1^i p_2^i\,\nu_1+(p_1^i+p_2^i)\nu_2+2\nu_3=0\,,\qquad \nu_2^2-4\nu_1 \nu_3\ne 0\label{9.8h}
 \end{align}
 for every $i$. Thus, the search for extra involution eventually boils down to linear null space computation. The splitting between two quotient curves can be investigated at the level of periods. A holomorphic basis on the curve $\mathcal{H}$ is,
 \begin{align}
     \bar\omega_j=\frac{\hat z^j d\hat z}{\hat y}\,,\,\,\,j=0,\cdots,g-1\,.
 \end{align}
 Now from \eqref{9.6a} gives,
 \begin{align}
 \begin{split}
   &  \bar \omega_{2r+1}=\frac{1}{2}x^r \frac{dx}{v_1}\,,\qquad r=0,1,\cdots,\left\lfloor\frac{g}{2}\right\rfloor-1\,,\\&\bar\omega_{2r}=\frac{1}{2} x^r \frac{dx}{v_2}\,,\qquad r=0,1,\cdots,\left\lceil\frac{g}{2} \right\rceil-1\,.
     \end{split}
 \end{align}
For our case, in momentum space, the underlying curve is given by,
\begin{align}
    \mathcal{H}_z:\quad y^2=(1-z^2)(z^2-k'^2)\,.
\end{align}
The curve equation is already in even form; no initial Mobius transformation is needed, or more precisely, the Mobius transformation is trivial.
\begin{align}
    z=\gamma(\hat z)=\hat z,\qquad \gamma=\begin{pmatrix}
        1\quad0\\
        0\quad 1
    \end{pmatrix}.
\end{align}
Now solving \eqref{9.8h} we get $\vec \nu=(0,1,0)$ such that $\nu_2^2-4\nu_1\nu_3\ne 0$. Therefore, with $x=z^2$, the two quotient curves are,
\begin{align}
\begin{split}
  &  \mathcal{R}_x:\quad v_1^2=(1-x)(x-k'^2)\,,\qquad \pi_{\mathcal{R}}:\,(z,y)\mapsto(z^2,y)=(x,v_1)\,,\\ &
  \mathcal{E}_x: \quad v_2^2=x(1-x)(x-k'^2)\,,\qquad \pi_\mathcal{E}:\,(z,y)\mapsto(z^2,z\,y)=(x,v_2)\,.
    \end{split}
\end{align}
The first curve is a genus-zero rational curve, and the second one is a genus-one elliptic curve.
\begin{center}
\boxed{\begin{tikzpicture}[baseline=(current bounding box.center),>=Latex]
  \node (H) at (0,0.8) {$\mathcal H_z:\ y^2=Q(z^2)$};
  \node (R) at (-3.2,-0.8) {$\mathcal R_x:\ v_1^2=(1-x)(x-k'^2)$};
  \node (E) at (3.2,-0.8) {$\mathcal E_x:\ v_2^2=x(1-x)(x-k'^2)$};
  \draw[->] (H) -- node[above left] {$\pi_{\mathcal R}$} node[below right] {$/\langle e_1\rangle$} (R);
  \draw[->] (H) -- node[above right] {$\pi_{\mathcal E}$} node[below left] {$/\langle e_2\rangle$} (E);
\end{tikzpicture}}
\end{center}
We see that, as anticipated, the extra involutions $e_1,\,e_2$ map the original genus one curve into two quotient curves having one genus one ($\mathcal{E}_x$), and the other is genus-zero ($\mathcal{R}_x$). Note that for a genus-one curve, there is only one holomorphic basis of a differential one form. However, we can study meromorphic differential equations on the curve. A differential, which is invariant under $e_1$  and $e_2$, factors into a genus-zero and a genus-one curve, respectively. To concretize the argument, let's consider the following meromorphic differential on the curve $\mathcal{H}_z$:
\begin{align}
    \eta_m=\frac{dz}{z^my}\,.
\end{align}
The pullback map $e_i^*$ corresponding to the involution map $e_i$ acts on the meromorphic differential as,
\begin{align}
    e_1^{*}\eta_{m}=(-1)^{m+1}\eta_m\,,\qquad e_2^*\eta_m= (-1)^m \eta_m\,.\label{9.16jj}
\end{align}
Let's divide the label `$m$' into even and odd sector as: $\eta_{2n}$ and $\eta_{2n+1}$. With $x=z^2$, $v_1=y$, $v_2=z\,y$, the pullback identities in \eqref{9.16jj} gives,
\begin{align}
    e_2^*\,\eta_{2n}=\eta_{2n}\,,\qquad e_1^*\,\eta_{2n+1}=\eta_{2n+1}\,.\label{9.17jjj}
\end{align}
Therefore, the even sector is invariant under $e_2$ and picks up the elliptic quotient. On the other hand, the odd sector is invariant under $e_1$ and picks up the genus-zero quotient.
\subsection{The Fourier cut integral and complete genus drop} \label{newsec3}
Now it is a good time to revisit our original Fourier cut integral,
\begin{align}
    \textrm{FT}_2=\int I_0^2\left(\frac{m\,b}{z}\right)\frac{dz}{z\sqrt{(1-z^2)(z^2-k'^2)}}\,.\label{9.18bb}
\end{align}
\begin{figure}
    \centering
    \includegraphics[width=0.7\linewidth]{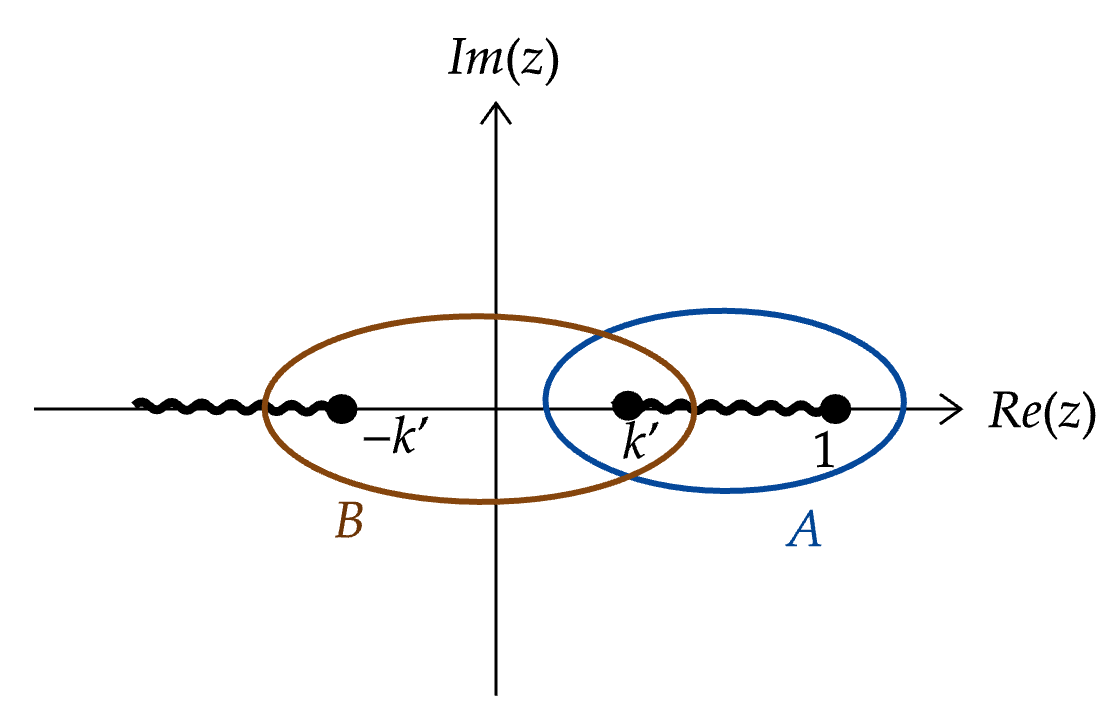}
    \caption{\textcolor{black}{Branch cuts of the curve $y^2=(1-z^2)(z^2-k'^2)$ of the Fourier cut, on $[-1,-k']$ and $[k',1]$, and the canonical cycles used in line with Sec.~(\ref{newsec3}): the $B$-cycle encircles the branch points $-k'$ and $k'$, the $A$-cycle encircles $k'$ and $1$. The $B$-cycle is the  contour $\mathcal C_B$ of Appendix~(\ref{app:maximal-cut}), and $|A\cdot B|=1\,.$}}
    \label{fig:bessel-cut}
\end{figure}
The branch points are $-1,-k',k',1$. The chosen symplectic cycles are \footnote{\textcolor{black}{These are not the cycles used in the Baikov
maximal-cut analysis of Sec.~(\ref{sec:newzz}), whose intersection number is 2\,.}}: 
\begin{align}
    {A}:(k',1) ,\qquad {B}:(-k',k')\,
\end{align}
with $|{A}\cdot {B}|=1$ as shown in Fig.~\eqref{fig:bessel-cut}. Let us consider the ${A}-$cycle carefully,
\begin{align}
    \begin{split}
         \textrm{FT}_2&=\int_{k'}^1 I_0^2\left(\frac{m\,b}{z}\right)\frac{dz}{z\sqrt{(1-z^2)(z^2-k'^2)}}\,,\\ &
         =\sum_{n=0}^{\infty} \frac{(2n)!(mb)^{2n}}{4^n (n!)^4}\int_{k'}^1\frac{dz}{z^{2n+1}y}\,.
    \end{split}
\end{align}
As evident from \eqref{9.17jjj}, the Fourier differential is invariant under the pullback map $e_1^*$ therefore picks up the rational quotient curve $\mathcal{R}_x$ which can be seen via direct integration,
\begin{align}
 \textcolor{black}{\textrm{FT}_2=\sum_{n=0}^{\infty}\frac{(2n)!(mb)^{2n}}{4^n(n!)^4}\int_{k'}^{1}\eta_{2n+1}
=\frac{\pi}{2k'}\sum_{n=0}^{\infty}\frac{(2n)!}{4^n(n!)^4}
\left(\frac{m^2b^2}{k'}\right)^{\! n} P_{n}\!\left(\frac{1}{2k'}+\frac{k'}{2}\right)}\,.\label{9.21kkkk}
\end{align}
As anticipated from the study of the quotient curve, a direct calculation \eqref{9.21kkkk} also shows that no elliptic function comes up in the Fourier integral, which is remarkable. It shows only integral results and does not always show the underlying geometry of the Feynman integral.\textcolor{black}{On the other hand, the even sector gives, on the same open interval,}
\begin{align}
   W_n:= \int_{k'}^1 \eta_{2n}= \frac{\pi}{2}\,{}_2F_1\left(n+\frac{1}{2},\frac{1}{2},1,1-k'^2\right)
\end{align}
where, explicitly one can check \textcolor{black}{$W_0=K(k^2),\,\,W_1=E(k^2)/k'^2$ with $k^2=1-k'^2$, in the convention of Sec.~(\ref{new:sec2.2.3})}. \textcolor{black}{The corresponding closed cycle is twice this,
$\oint_{\mathcal C_A}\eta_{2n}=2W_n$, as used in
Appendix~(\ref{app:maximal-cut})\,. The recurrence follows from the contiguity relations of ${}_2F_1$ and reproduces $W_2,\dots,W_5$ from $W_0,W_1$\,.}
\begin{align}
\textcolor{black}{W_n=\frac{2(n-1)\big(1+k'^{2}\big)\,W_{n-1}-(2n-3)\,W_{n-2}}{(2n-1)\,k'^{2}}\,.}
\end{align}
Also, as expected, the even sector picks up the \textcolor{black}{the elliptic quotient} $\mathcal{E}_x$ and the resulting integrals are all elliptic integrals. \textcolor{black}{A differential with both parities would therefore mix periods of $\mathcal R_x$ and
$\mathcal E_x$. The Fourier cut \eqref{9.18bb} is not of that kind: its exact Bessel series \eqref{9.21kkkk} contains only the odd sector, so its genus drop is complete.
The large-$\beta$ expansion below brings in both parities of the algebraic measure,
but not an alternating function space.}
This can be seen in a different physical limit of \eqref{9.18bb}. 
\textcolor{black}{For $\beta=mb\gg1$ the modified Bessel function has the asymptotic form
\begin{equation}
I_0^{2}\!\left(\frac{\beta}{z}\right)=\frac{z}{2\pi\beta}\,e^{2\beta/z}
\sum_{r\ge0}c_r\left(\frac{z}{\beta}\right)^{\!r},\qquad
c_0=1,\ c_1=\tfrac14,\ \ldots
\end{equation}
so that
\begin{equation}
\mathrm{FT}_2\;\sim\;\frac{1}{2\pi\beta}\sum_{r\ge0}c_r\,\beta^{-r}
\int_{k'}^{1}e^{2\beta/z}\,z^{\,r}\,\frac{dz}{y}\, .\label{eq:new10}
\end{equation}
The algebraic factor $z^{r}dz/y$ is $\eta_{-r}$, so its parity alternates with $r$: even $r$ is $e_2$-even and pulls back from $\mathcal E_x$, odd $r$ is $e_1$-even and pulls back from $\mathcal R_x$. The exponential weight, however, is not parity-even - $e^{2\beta/z}\mapsto e^{-2\beta/z}$ under $z\to-z\,.$ So the individual terms of the asymptotic series are not periodic on either quotient.
What alternates with $r$ is the quotient from which the algebraic measure is pulled back, not the function space of the answer. Every term here is dominated by the endpoint $z=k'$ and carries the same factor $e^{2\beta/k'}\,.$  The exact Bessel series \eqref{9.21kkkk} is therefore the only regime in which the internal genus drop is complete.}

\textcolor{black}{Before we end, it is worth noting that \eqref{eq:new10} is an asymptotic expansion of a homogeneous
solution, not of an observable for which one needs the full integral. Every term is dominated by the endpoint $z=k'$, so $\mathrm{FT}_2$ grows like $e^{2mb/k'}$ at large $b$ rather than showing the exponential suppression a finite-range interaction produces in the impulse.  That suppression is a property of the full integral, in which this solution enters
weighted by boundary data.}


\section{Summary and conclusion} \label{sec10}
In this work, we investigate the geometric (and analytic) structure of a representative two-loop family of the type generated by worldline QFT with a massive mediator. One of the preliminary results of our analysis is that massive loops in classical scattering may produce elliptic geometry at two loops. This is in contrast with the known massless integrals arising in scattering problems in General Relativity, where irrational geometry starts appearing from three loops and onwards. \textcolor{black}{The novelty is not the ellipticity itself, which also appears in other contexts \cite{Sabry:1962,Broadhurst:1990}, but that here the modulus and its degenerations are fixed by the post-Minkowskian kinematics.} The family of integrals considered here provides us with the leverage to study the geometric structure, differential equation, and monodromy properties in great detail with a relatively simpler setting. Let us summarize the central results obtained in the article.\\\\
\textit{Summary:}
\begin{itemize}
    \item Starting from the global Baikov representation, we evaluated the maximal cut by successively eliminating the rational fibers of the cut geometry in $d=4$ dimensions. The remaining one-dimensional integral is a period of the elliptic curve (with suitable transformation):
    \begin{align}
       \mathcal E_z:\quad y^2=(1-z^2)(z^2+\tau)\,.
    \end{align}
\textcolor{black}{We also derive the cut of the impact-parameter integral directly in momentum space (see also Appendix~(\ref{app:maximal-cut})). Its algebraic part lives on the same family of curves as the Baikov cut, at $\mathfrak t=-k'^{2}$.} This agreement also verifies the fact that the emergence of the elliptic curve is an intrinsic property of the Feynman integral and not an artifact of the particular parametrization. Furthermore the discriminant of the quartic curve is given by $\Delta=-16\tau(\tau+1)^4$, which tells how the elliptic curve degenerates into nodal curve (whose normalization is rational) in special physical kinematic limit, $m^2=0$ (which may seem to be absent in this particular parametrization of the curve as the $z$ variable is normalized by $m$) and $\gamma^2=1$, where the loci of the branch point collides. The
associated Picard–Lefschetz transformations explain why the period of the vanishing cycle
remains regular while its dual period develops a logarithm. Consequently, the same analysis
fixes the local branching and discontinuity structure that must be reproduced by the full uncut
Feynman integral. This is the first of the two genus drop mechanisms encountered in this
work: it is a kinematic degeneration of the curve itself.
\item We have solved the related Integral-by-parts (IBP) relation and found that there are five master integrals in this particular family. Furthermore, the reflection symmetry of the Baikov variable reduces the five integral sectors into $3\oplus2$ sectors, where one of the sectors is associated with the elliptic curve and the other one is just rational, having standard MPLs as function space. We derived their differential equation using two independent ways: by directly invoking the IBP identities using \texttt{LiteRed} and by invoking the intersection theory of twisted differential form (with a different choice of bases: monomial and dlog). We also explicitly constructed the rational transformation between two different cohomological bases, thereby verifying that their Gauss-Manin connection describes the same system. We further make suitable crosschecks between the reduction from these two pathways. We also showed the explicit Picard-Fuchs factorization of the differential equation and got the expected elliptic structure out of it.
\item Furthermore, we have also computed the Period matrices in different complementary ways. We first compute the Period matrices in a monomial basis in a general dimension. The relevant periods are twisted Euler integrals and can be expressed in terms of the Lauricella function. The period matrices play an important role in the canonicalization procedure of the differential equation. Furthermore, we derive the twisted Riemann period relation, which provides a non-trivial quadratic relation between the Feynman integral and its dual counterpart. In this regard, we chose the standard dlog-basis while computing the period matrices in the positive and negative branches. We also analyze the global monodromy structure of the differential equation around kinematic singular points. Using that, we explicitly derive Riemann's bilinear relation. The global structure, in principle, provided us with more information about the system than studying the local monodromy around the singular point.
\item We further explicitly compute the master integrals for the maximal cut in the $\epsilon$-expansion up to $\mathcal{O}(\epsilon^2)$. For
fixed $\tau>0$, the convergence domain contains $\epsilon=0$ in its interior, and a uniform integrable
bound justifies interchanging the $\epsilon$-expansion with integration. The resulting logarithmic
insertions are naturally organized as elliptic multiple polylogarithms on the same quartic curve. Their alphabet is fixed by the four branch points together with the marked points generated
by the Baikov factor $4z^2-3$. Thus, higher orders in $\epsilon$ increase the length of the elliptic iterated
integrals without introducing a new elliptic curve or a new set of marked points. The expansion
is not uniform at a degeneration such as $\tau\to 0$; the dimensional and kinematic limits must
therefore be treated as an asymptotic problem in that region. This analysis enables us to concretely define the spanning classes of the function space in arbitrary order of the dimensional regularisation parameter $\epsilon$. 
\item Finally, we consider the Fourier transformation of the maximal cut integral in impact parameter space. The corresponding one form is not purely algebraic but a Bessel-twisted differential. We see that although the covering curve is elliptic, it possesses additional involution symmetry  $(z,y)\mapsto (-z,\pm y)$ along with the standard hyperelliptic involution symmetry. The involution produces two quotient curves: one rational and the other elliptic. The parity of the meromorphic differential determines the quotient through which its periods factor. In particular, in the small mass limit, the Taylor expansion of the Bessel function selects only the rational quotient curves and thereby produces a complete internal genus drop. \textcolor{black}{By contrast, at a large impact parameter, the Bessel weight is not parity-even, so the two quotients alternate only as the source of the algebraic measure order by order in $1/mb\,.$ The individual terms are not periods on either quotient, and the genus drop is not complete there}. The internal genus drops are caused by automorphisms and therefore must not be identified with the genus drop caused by branch point collision in some particular kinematic limit, causing the kinematic genus drop.
\end{itemize}
With this investigation, we believe the study initiates a program of computing the Feynman integral in classical gravitational scattering with a massive mediator. Let us end the article with possible extensions.
\\\\
\textit{Outlook and future direction:}
\begin{itemize}
    \item The primary next step is the computation of the complete uncut two-loop family by solving the differential equation, which has non-zero inhomogeneity coming from the lower sector. The maximal cut obtained here
provides the fundamental matrix of the homogeneous system.  After adding the lower sectors, the uncut masters may schematically be written
as
\begin{equation}
    \vec J(\tau,\epsilon)
    =P(\tau,\epsilon)
    \left[
        \vec C(\epsilon)
        +\int_{\tau_0}^{\tau}
        P^{-1}(t,\epsilon)\,
        \vec S_{\rm lower}(t,\epsilon)\,dt
    \right],
\end{equation}
where \(P\) is constructed from the periods, \(\vec S_{\rm lower}\) contains
the inhomogeneous lower-sector contributions, and \(\vec C\) is fixed by
boundary data.  The period matrices, Wronskians, and basis transformations
derived in this work provide the main ingredients for carrying out the canonicalization procedure.  A central technical goal
is to find a basis in which all non-algebraic homogeneous dependence is
absorbed into \(P\), while the remaining differential system is proportional
to \(\epsilon\) and can be integrated recursively.  The finite eMPL alphabet
identified on the maximal cut is the building block for the function space generated by this recursion.  Completing this program will also require a systematic treatment of sub-topologies, ultraviolet or infrared poles (if any) of the uncut family, and boundary constants in a region where the physical Feynman or retarded prescription is unambiguous.  Degeneration points, at which the elliptic curve that reduces to a rational one should provide especially useful boundary conditions, although the non-commutativity of the limits
\(\epsilon\to0\) and \(\tau\to0\) must then be taken care of with a suitable strategy of region analysis.
\item The singularity analysis investigated in this article also opens a new way to construct an elliptic version of the Landau bootstrap program \cite{Hannesdottir:2024hke,Bargiela:2026lje} for the massive WQFT integrals. The primary step, which is missing in this article, is to solve the Landau equation for the complete uncut integrals. The computed global monodromy structure, Picard–Lefschetz transformations, and the developed function space may be used as bootstrap data. However, it is understood that integrals with an iterative $d\log$ kernel can be completely localized to the Feynman maximal cut integral. On the other hand, the Feynman integrals that evaluates to more general class of integrals having higher order curve like Elliptic (or Calabi-Yau) like our case, it is not possible to do a complete localization, rather as we see, one left out with period integrals over non trivial manifold. Some progress has been made in bootstrapping the elliptic integrals  using Schubert analysis \cite{Morales:2022csr}. Nonetheless, together with the analyticity conditions, asymptotic power countings, physical discontinuities, and boundary values, they may constrain an ansatz from the eMPLs. One of the advantages of this method is that one may only focus on the integrals that are coming up to the particular observable without caring about the full family of integrals. It would be particularly interesting to investigate whether the invariant bilinear forms and quadratic relations can be promoted to bootstrap constraints on the coefficients of the ansatz.
\item A second interesting direction is to evaluate the function space for the full Fourier transformed integral rather than studying its maximal cuts or Taylor expansion around some physical limits. The occurrence of $I_0^2$, the relevant object combines the elliptic period with local Bessel function insertion. One
should therefore derive a coupled holonomic system in the kinematic modulus and the impact
parameter, determine its boundary conditions, and study its analytic continuation in both
variables. In particular, one may try to find a suitable connection to the Bessel spectral curve.
\begin{align}
    x_{B}(\lambda)=\frac{\lambda^2}{2}\,,\quad y_{B}=\frac{1}{\lambda}\,,\quad 2x_By_B^2=1\,.
\end{align}
whose quantization gives the modified Bessel equation. In the Fourier cut, the corresponding
Bessel wavefunction is pulled back through the composite argument $\lambda=mb/z$ and inserted
into a period of the elliptic covering curve. It remains to understand whether this composition
defines a useful “Bessel-twisted elliptic curve”, a product or twisted quantum geometry, or more
naturally a holonomic D-module rather than an ordinary algebraic curve. Such a formulation
could relate the large-$b$ expansion to the WKB expansion \cite{DoNorbury2018} of the Bessel quantum curve and
explain how its Stokes data interact with elliptic monodromy. It should also clarify why the  exact Bessel series projects onto the rational quotient, whereas the large-impact-parameter
expansion probes both rational and elliptic quotient sectors.
\end{itemize}
We hope to report on some of these questions in future work(s).
\section*{Acknowledgements}
 We thank Sounak Pal, Ankit Mishra, and Anisha Sharma for useful discussions and collaboration in related projects. S.G. would like to thank Aninda Sinha, Arnab P. Saha, P.V. Athira, Arhum Ansari, Diptarka Das, and Apratim Kaviraj for discussion on aspects of scattering amplitudes, including the bootstrap program and related collaboration(s). AB is supported by the Core Research Grant (CRG/2023/005112) by ANRF (erstwhile SERB) of the Indian Government. S.G. is supported by ANRF grant ANRF/ARG/2025/001338/PS. We acknowledge the use of ChatGPT-Codex (5.6 Sol-6.0 ASTRA) (OpenAI) and Claude Opus 5.5 (Anthropic) in writing the Mathematica code for numerical cross-checks, language editing, and formatting the draft, as well as many useful discussions. All the logical flow behind the AI coding has been directed and verified by the authors.
\appendix

\section{Intersection-matrix entries}
\label{app:intersection-entries}

The parity selection rule gives
\begin{equation}
c_{ij}=f_{ij}=0
\qquad\text{for}\qquad
i+j\ \text{odd}.
\end{equation}
The nonvanishing entries used in \eqref{eq:intersection-matrices} are listed below.


\begingroup
\small
\allowdisplaybreaks[4]
\setlength{\abovedisplayskip}{3pt}
\setlength{\belowdisplayskip}{3pt}
\setlength{\jot}{0pt}

\begin{flalign*}
c_{11}={}&
\frac{2\tau+4\tau\epsilon+\epsilon-2}
     {8\epsilon(\epsilon+2)+6},\quad 
f_{11}=
\frac{-\tau+\epsilon^2-2(\tau+1)\epsilon}
     {2\tau\bigl(4\epsilon(\epsilon+2)+3\bigr)},
&&\\[0.2ex]
c_{13}={}&
-\frac{1}
{8(2\epsilon+1)(2\epsilon+3)(2\epsilon+5)}
\Bigl\{
16(\tau^2+\tau+1)-16(\tau+1)\epsilon^2
+\bigl(8\tau(4\tau+3)+3\bigr)\epsilon
\Bigr\},
&&\\[-0.2ex]
f_{13}={}&
\frac{1}
{8\tau(2\epsilon+1)(2\epsilon+3)(2\epsilon+5)}
\Bigl\{
4\tau(2\tau+1)+8(\tau+2)\epsilon^3
-(16\tau+3)\epsilon^2
+2\bigl(\tau(8\tau-1)-8\bigr)\epsilon
\Bigr\},
&&\\[0.2ex]
c_{15}={}&
\frac{1}
{32(2\epsilon+1)(2\epsilon+3)(2\epsilon+5)(2\epsilon+7)}
\Bigl\{
32(\tau-1)\bigl(\tau(8\tau+13)+8\bigr)+16(\tau+1)(4\tau+13)\epsilon^3
&&\\[-0.4ex]
&\quad+8\bigl(\tau(17-32\tau)+47\bigr)\epsilon^2
+\bigl(16\tau(\tau(32\tau+11)-20)-133\bigr)\epsilon
\Bigr\},
&&\\[-0.2ex]
f_{15}={}&
\frac{1}
{32\tau(2\epsilon+1)(2\epsilon+3)(2\epsilon+5)(2\epsilon+7)}
\Bigl\{
64\tau-128\tau^3(2\epsilon+1)
-8\tau^2(\epsilon-2)(2\epsilon+1)(10\epsilon-1)
&&\\[-0.4ex]
&\quad
+2\tau\epsilon
 \Bigl(4\epsilon\bigl(2\epsilon(9\epsilon-5)-23\bigr)+37\Bigr)
+\epsilon\Bigl(\epsilon\bigl(8\epsilon(26\epsilon+47)-133\bigr)-256\Bigr)
\Bigr\},
&&\\[0.2ex]
c_{22}={}&
-\frac{1}
{32\epsilon(\epsilon+1)(\epsilon+2)}
\Bigl\{
4(\tau+1)^2-8(\tau+1)\epsilon^2
+\bigl(4\tau(2\tau+3)+5\bigr)\epsilon
\Bigr\},
&&\\[-0.2ex]
f_{22}={}&
\frac{\epsilon\bigl(-6\tau+4(\tau+2)\epsilon-5\bigr)-4(\tau+1)}
{32\tau(\epsilon+1)(\epsilon+2)},
&&\\[0.2ex]
c_{24}={}&
\frac{1}
{64\epsilon(\epsilon+1)(\epsilon+2)(\epsilon+3)}
\Bigl\{
12(\tau-1)(\tau+1)^2
+2(\tau+1)(4\tau+13)\epsilon^3
&&\\[-0.4ex]
&\quad
+\bigl(3\tau(1-8\tau)+25\bigr)\epsilon^2
+\bigl(\tau(2\tau(12\tau+5)-31)-22\bigr)\epsilon
\Bigr\},
&&\\[-0.2ex]
f_{24}={}&
\frac{1}
{64\tau(\epsilon+1)(\epsilon+2)(\epsilon+3)}
\Bigl\{
-6\tau^2(\epsilon-2)(2\epsilon+1)
+9\tau(\epsilon-1)\epsilon(2\epsilon+1)
+\epsilon\bigl(\epsilon(26\epsilon+25)-22\bigr)-12
\Bigr\},
&&\\[0.2ex]
c_{31}={}&
-\frac{
16\tau-16(\tau+1)\epsilon^2+(24\tau+17)\epsilon}
{8(8\epsilon^3+12\epsilon^2-2\epsilon-3)},\quad
f_{31}=
\frac{(16\epsilon-17)\epsilon^2
+2\tau(\epsilon-2)(2\epsilon+1)^2}
{8\tau(2\epsilon+3)(4\epsilon^2-1)},
&&\\[0.2ex]
c_{33}={}&
\frac{1}
{32(2\epsilon-1)(2\epsilon+1)(2\epsilon+3)(2\epsilon+5)}
\Bigl\{
32(\tau-1)\tau
+16(\tau+1)(4\tau+13)\epsilon^3
-8\bigl(\tau(16\tau+11)-3\bigr)\epsilon^2&&\\[-0.4ex]
&\quad
-\bigl(16\tau(\tau+11)+145\bigr)\epsilon
\Bigr\},
&&\\[-0.2ex]
f_{33}={}&
\frac{1}
{32\tau(2\epsilon-1)(2\epsilon+1)(2\epsilon+3)(2\epsilon+5)}
\Bigl\{
-16\tau(\tau+2)+16(9\tau+13)\epsilon^4
-8\bigl(4\tau(\tau+2)-3\bigr)\epsilon^3
+\bigl(16\tau(4\tau-1)-145\bigr)\epsilon^2&&\\[-0.4ex]
&\quad
+2\tau(4\tau-19)\epsilon
\Bigr\},
&&\\[0.2ex]
c_{35}={}&
\frac{1}
{128(2\epsilon-1)(2\epsilon+1)(2\epsilon+3)(2\epsilon+5)(2\epsilon+7)}
\Bigl\{
-256\tau^3(\epsilon-2)(2\epsilon-1)(2\epsilon+1)
&&\\[-0.4ex]
&\quad
+16\tau^2(2\epsilon-1)(2\epsilon+1)
 \bigl(\epsilon(20\epsilon-33)+16\bigr)
+8\tau(2\epsilon+1)
\bigl(\epsilon(10\epsilon(24\epsilon+13)-159)-64\bigr)&&\\[-0.4ex]
&\quad
+5\epsilon
 \bigl(32\epsilon(\epsilon(16\epsilon+31)-4)-491\bigr)
\Bigr\},
&&\\[-0.2ex]
f_{35}={}&
\frac{1}
{128\tau(2\epsilon-1)(2\epsilon+1)(2\epsilon+3)(2\epsilon+5)(2\epsilon+7)}
\Bigl\{
8\tau^2(2\epsilon-1)(2\epsilon+1)
&&\\[-0.4ex]
&\quad
\times\bigl(\epsilon(4\epsilon^2-62\epsilon+59)+8\bigr)
+5\epsilon^2
 \bigl(32\epsilon(\epsilon(16\epsilon+31)-4)-491\bigr)
&&\\[-0.4ex]
&\quad
+128\tau^3(\epsilon-2)(2\epsilon-1)(2\epsilon+1)
+2\tau(2\epsilon+1)
 \Bigl(
 \epsilon\bigl(2\epsilon(4\epsilon(68\epsilon-3)+3)+123\bigr)-256
 \Bigr)
\Bigr\},
&&\\[0.2ex]
c_{42}={}&
\frac{1}
{64(\epsilon-1)\epsilon(\epsilon+1)(\epsilon+2)}
\Bigl\{
26\epsilon^3-19\epsilon^2-4\tau^3(2\epsilon+1)
+2(\epsilon-2)(\tau+2\tau\epsilon)^2
+\tau(\epsilon-1)(2\epsilon+1)(17\epsilon-4)-5\epsilon+4
\Bigr\},
&&\\[-0.2ex]
f_{42}={}&
\frac{1}
{64\tau(\epsilon-1)\epsilon(\epsilon+1)(\epsilon+2)}
\Bigl\{
-4\tau(\tau+1)^2+2(9\tau+13)\epsilon^4
+\bigl(\tau(4\tau-7)-19\bigr)\epsilon^3
+\bigl(2(\tau-1)\tau-5\bigr)\epsilon^2
&&\\[-0.4ex]
&\quad
-\bigl(\tau(8\tau(\tau+2)+5)-4\bigr)\epsilon
\Bigr\},
&&\\[0.2ex]
c_{44}={}&
\frac{1}
{128(\epsilon-1)\epsilon(\epsilon+1)(\epsilon+2)(\epsilon+3)}
\Bigl\{
12(\tau^2-1)^2+40(\tau+1)(\tau+2)\epsilon^4
+\bigl(56-2\tau(\tau(8\tau+23)-15)\bigr)\epsilon^3
&&\\[-0.4ex]
&\quad
+\bigl(\tau(\tau(24\tau+41)-57)-80\bigr)\epsilon^2
+(\tau-1)\bigl(\tau(8\tau(3\tau+5)+29)+8\bigr)\epsilon
\Bigr\},
&&\\[-0.2ex]
f_{44}={}&
\frac{1}
{128\tau(\epsilon-1)\epsilon(\epsilon+1)(\epsilon+2)(\epsilon+3)}
\Bigl\{
12\tau^4(2\epsilon+1)
-2\tau^3(\epsilon-2)(\epsilon+3)(2\epsilon+1)
&&\\[-0.4ex]
&\quad
+\tau^2(\epsilon-1)(2\epsilon+1)
 \bigl(\epsilon(2(\epsilon-6)\epsilon+31)+12\bigr)
+2\tau(2\epsilon+1)
 \bigl(\epsilon(\epsilon(\epsilon(17\epsilon-7)+13)-5)-6\bigr)
&&\\[-0.4ex]
&\quad
+4\epsilon(2\epsilon-1)
 \bigl(2\epsilon(\epsilon(5\epsilon+6)-2)-3\bigr)
\Bigr\},
&&\\[0.2ex]
c_{51}={}&
\frac{1}
{32(16\epsilon^4-40\epsilon^2+9)}
\Bigl\{
-96(\tau-1)\tau
+16(\tau+1)(4\tau+13)\epsilon^3
-8(39\tau+41)\epsilon^2
+\bigl(139-16\tau(13\tau+2)\bigr)\epsilon
\Bigr\},
&&\\[-0.2ex]
f_{51}={}&
\frac{1}
{32\tau(16\epsilon^4-40\epsilon^2+9)}
\Bigl\{
24\tau^2(4\epsilon^3-13\epsilon-6)
+\bigl(8\epsilon(26\epsilon-41)+139\bigr)\epsilon^2
+6\tau(2\epsilon+1)
 \bigl(2\epsilon(6\epsilon-5)-1\bigr)\epsilon
\Bigr\},
&&\\[0.2ex]
c_{53}={}&
\frac{1}
{128(2\epsilon-3)(2\epsilon-1)(2\epsilon+1)(2\epsilon+3)(2\epsilon+5)}
\Bigl\{
-768\tau^2+1280(\tau+1)(\tau+2)\epsilon^4
&&\\[-0.4ex]
&\quad
-32\bigl(13\tau(2\tau+5)+43\bigr)\epsilon^3
+32(32\tau^2-30\tau-59)\epsilon^2
+\bigl(8\tau(65-82\tau)+1211\bigr)\epsilon
\Bigr\},
&&\\[-0.2ex]
f_{53}={}&
\frac{1}
{128\tau(2\epsilon-3)(2\epsilon-1)(2\epsilon+1)(2\epsilon+3)(2\epsilon+5)}
\Bigl\{
\bigl(32\epsilon(\epsilon(80\epsilon-43)-59)+1211\bigr)\epsilon^2
&&\\[-0.4ex]
&\quad
+8\tau^2(2\epsilon+1)
\Bigl(\epsilon\bigl(4\epsilon(2\epsilon(\epsilon+2)+35)-93\bigr)-72\Bigr)
+2\tau(2\epsilon+1)
 \bigl(2\epsilon(4\epsilon(68\epsilon-53)+311)-435\bigr)\epsilon
\Bigr\},
&&\\[0.2ex]
c_{55}={}&
\frac{1}
{512(2\epsilon-3)(2\epsilon-1)(2\epsilon+1)
 (2\epsilon+3)(2\epsilon+5)(2\epsilon+7)}
\Bigl\{
16\tau^2(2\epsilon+1)\Bigl(
\epsilon\bigl(2\epsilon(340\epsilon^2-406\epsilon+989)-943\bigr)-288
\Bigr)
&&\\[-0.4ex]
&\quad
+128\tau^3(2\epsilon-3)(2\epsilon-1)(2\epsilon+1)
 \bigl((\epsilon-11)\epsilon+12\bigr)
+8\tau\epsilon(2\epsilon+1)
 \bigl(\epsilon(16\epsilon(202\epsilon+43)-1951)+278\bigr)
&&\\[-0.4ex]
&\quad
+\epsilon
 \Bigl(
 8\epsilon\bigl(2\epsilon(16\epsilon(121\epsilon+149)-3087)-2893\bigr)
 +20861
 \Bigr)
\Bigr\},
&&\\[-0.2ex]
f_{55}={}&
\frac{1}
{512\tau(2\epsilon-3)(2\epsilon-1)(2\epsilon+1)
 (2\epsilon+3)(2\epsilon+5)(2\epsilon+7)}
\Bigl\{
\epsilon^2\Bigl(
8\epsilon\bigl(2\epsilon(16\epsilon(121\epsilon+149)-3087)-2893\bigr)
+20861
\Bigr)
&&\\[-0.4ex]
&\quad
-64\tau^3(2\epsilon-3)(2\epsilon-1)(2\epsilon+1)
 \bigl((\epsilon-11)\epsilon+12\bigr)
+8\tau^2(2\epsilon+1)
 \Bigl(
 \epsilon\bigl(
 2\epsilon(4\epsilon(\epsilon(34\epsilon-113)+599)-349)-393
 \bigr)&&\\[-0.4ex]
&\quad-1152
 \Bigr)
+2\tau\epsilon(2\epsilon+1)
 \Bigl(
 2\epsilon\bigl(
 2\epsilon(32\epsilon(61\epsilon+4)+2089)+2051
 \bigr)-7365
 \Bigr)
\Bigr\}.
&&
\end{flalign*}

\endgroup




\section{Reduction of the asymmetric two-loop integral}
\label{app:twoloop}
We now derive the compact two-parameter representation of the asymmetric integral,
\begin{align}
I^{\mu}(|b|)
=
\int_{\ell_1,\ell_2,q}
\frac{
\ell_1^\mu\,
\hat\delta(\ell_1\!\cdot\! v_1)\,
\hat\delta(\ell_2\!\cdot\! v_2)\,
\hat\delta(q\!\cdot\! v_1)\,
\hat\delta(q\!\cdot\! v_2)\,
\ee^{\ii q\cdot b}
}{
(\ell_1^2-m^2)(\ell_2^2-m^2)\big[(\ell_1+\ell_2)^2-m^2\big]\big[(q-\ell_1)^2-m^2\big]
}\,.
\label{eq:S2-start}
\end{align}
\textcolor{black}{Throughout this appendix, we track only the functional form of the integrand. This is sufficient for our purpose, which is to exhibit the algebraic structure entering Sec.~(\ref{sec9}). The elliptic curve, its modulus, and the Bessel weight are all unaffected by the overall constant. Whenever a normalization is required, it is fixed directly from the Baikov representation of Sec.~(\ref{sec:twoloop}) rather than from the chain below.} 

Next, we introduce the spacelike unit vector, 
\begin{align}
u_1^\mu=\frac{\gamma v_1^\mu-v_2^\mu}{\zeta}\,,
\qquad
\zeta=\sqrt{\gamma^2-1}\,,
\qquad
u_1^2=-1\,,
\qquad
u_1\!\cdot\!v_1=0\,,
\label{eq:S2-u1}
\end{align}
\textcolor{black}{(in this appendix and in Appendix~(\ref{app:maximal-cut}), $\zeta$ always denotes
$\sqrt{\gamma^2-1}$, not the monodromy variable $e^{i\pi d}$ of
Sec.~(\ref{sec:global-monodromy}))}, which in the chosen frame ($v_2^\mu=(1,0,0,0)$) is $u_1^\mu=(\zeta,\gamma,0,0)$. Because $\hat\delta(\ell_1\!\cdot\! v_1)$ constrains $\ell_1^\mu$ to the hyperplane orthogonal to $v_1^\mu$, it is natural to decompose
\begin{align}
\ell_1^\mu=a\,u_1^\mu+\ell^\mu\,,
\qquad
\ell\!\cdot\!v_1=\ell\!\cdot\!v_2=0\,.
\label{eq:S2-kdecomp}
\end{align}
The pair of delta functions $\hat\delta(q\!\cdot\! v_1)\hat\delta(q\!\cdot\! v_2)$ forces $q^\mu$ to be purely transverse as well. Thus, both $\ell^\mu$ and $q^\mu$ live in the two-dimensional plane orthogonal to the scattering plane, which is why the remaining angular integrals reduce to ordinary Fourier--Bessel transforms. Writing
\begin{align}
\ell^\mu=(0,0,\bm{\ell}_{\perp})\,,
\qquad
q^\mu=(0,0,\bm{q}_{\perp})\,,
\label{eq:S2-transverse}
\end{align}
the relevant propagators become
\begin{align}
&\ell_1^2-m^2=-(a^2+\bm{\ell}_\perp^2+m^2)\,,\quad
(q-\ell_1)^2-m^2=-\big[(\bm q_\perp-\bm\ell_\perp)^2+a^2+m^2\big]\,.
\label{eq:S2-props}
\end{align}
The Fourier transform over $q$ is then immediate:
\begin{align}
\int \ee^{\ii q\cdot b}\,
\frac{\hat\delta(q\!\cdot\! v_1)\hat\delta(q\!\cdot\! v_2)}
{(q-\ell_1)^2-m^2}
\propto
-\frac{1}{\sqrt{\gamma^2-1}}\,
\ee^{-\ii\bm\ell_\perp\cdot\bm b}\,
K_0\!\left(|b|\sqrt{a^2+m^2}\right)\,.
\label{eq:S2-qint}
\end{align}

The remaining nontrivial subintegral is the $\ell_2$ bubble,
\begin{align}
\mathcal B(a,\ell)
=
\int
\frac{\hat\delta(\ell_2\!\cdot\! v_2)}
{(\ell_2^2-m^2)\big[(\ell_1+\ell_2)^2-m^2\big]}\,.
\label{eq:S2-bubble-def}
\end{align}
In the $v_2^\mu=(1,0,0,0)$ frame we parametrize $\ell_2^\mu=(0,c,\bm k)$, so
\begin{align}
\mathcal B(a,\ell)
=
\int
\frac{\dd c\,\dd^2\bm k}{(2\pi)^3}
\frac{1}{
(c^2+\bm k^2+m^2)
\left[(a\gamma+c)^2+(\bm\ell_\perp+\bm k)^2+m^2-a^2(\gamma^2-1)\right]
}\,.
\label{eq:S2-bubble-int}
\end{align}
Combining denominators with a Feynman parameter $s$ (not the Mandelstam $s=-q^2$) and performing the Gaussian integrations yields
\begin{align}
\mathcal B(a,\ell)
=\frac{1}{8\pi}\int_0^1 \dd s\,
\frac{1}{
\sqrt{
m^2+s(1-s)\bm\ell_\perp^2+(1-s)\bigl(s\gamma^2-\zeta^2\bigr)a^2+\ii 0
}
}\,.
\label{eq:S2-bubble-fp}
\end{align}
The $s$-integral can be evaluated explicitly,
\begin{align}
\mathcal B(a,\ell)
=\textcolor{black}{\frac{1}{4\pi}}\,
\frac{1}{P_\ell(a)}
\arctan\!\left(\frac{P_\ell(a)}{M_a}\right),
\label{eq:S2-bubble-arctan}
\end{align}
where
\begin{align}
M_a=m+\sqrt{m^2-a^2\zeta^2}\,,
\qquad
P_\ell(a)=\sqrt{a^2\gamma^2+\bm\ell_\perp^2}\,.
\label{eq:S2-MP}
\end{align}
For the subsequent Fourier reduction, it is more convenient to rewrite the same result as
\begin{align}
\mathcal B(a,\ell)
=\textcolor{black}{\frac{1}{4\pi}}\,
M_a
\int_0^1 \frac{\dd s}{s^2}\,
\frac{1}{M_a^2/s^2+P_\ell(a)^2}\,,
\label{eq:S2-bubble-compact}
\end{align}
which is equivalent to eq.~\eqref{eq:S2-bubble-arctan} after the elementary $s$-integration.
Substituting eqs.~\eqref{eq:S2-qint} and \eqref{eq:S2-bubble-compact} back into eq.~\eqref{eq:S2-start}, one finds
\begin{align}
I^\mu
\propto
\int_{-\infty}^{\infty}\dd a
\int \dd^2\bm\ell_\perp\,
\ee^{-\ii\bm\ell_\perp\cdot\bm b}
\frac{a\,u_1^\mu+\ell^\mu}{\bm\ell_\perp^2+a^2+m^2}
\mathcal B(a,\ell)
K_0\!\left(|b|\sqrt{a^2+m^2}\right)\,.
\label{eq:S2-afterbubble}
\end{align}
The term proportional to $a\,u_1^\mu$ is odd under $a\to-a$ and therefore drops out. Rotational invariance in the transverse plane then forces the surviving tensor structure to be proportional to $b^\mu$, so
\begin{align}
I^\mu
=-\frac{\ii\,b^\mu}{\zeta\,|b|}
\partial_{|b|}\,\mathcal I(|b|)\,.
\label{eq:S2-tensor}
\end{align}
Using eq.~\eqref{eq:S2-bubble-compact}, this scalar kernel can be written as
\begin{align}
\begin{split}
\mathcal I(|b|)
\propto
\int_0^1 \frac{\dd s}{s^2}
\int_{-\infty}^{\infty}\dd a\,
M_a\,
K_0\!\left(|b|\sqrt{a^2+m^2}\right)
\int \dd^2\bm\ell_\perp\,
\ee^{-\ii\bm\ell_\perp\cdot\bm b}
\frac{1}{
(\bm\ell_\perp^2+\Lambda_a^2)
(\bm\ell_\perp^2+\Lambda_{a,s}^2)
}
\end{split}
\label{eq:S2-scalar-pre-radial}
\end{align}
with
\begin{align}
\Lambda_a=\sqrt{a^2+m^2}\,,
\qquad
\Lambda_{a,s}=\sqrt{a^2\gamma^2+\frac{M_a^2}{s^2}}\,.
\label{eq:S2-lambdas}
\end{align}
Passing to polar coordinates in the transverse plane produces a single radial integral,
\begin{align}
\begin{split}
\mathcal I(|b|)
\propto
\int_0^1 \frac{\dd s}{s^2}
\int_{-\infty}^{\infty}\dd a\,
M_a\,
K_0(\Lambda_a|b|)
\int_0^\infty \dd \ell\,\ell\,
\frac{J_0(\ell|b|)}
{(\ell^2+\Lambda_a^2)(\ell^2+\Lambda_{a,s}^2)}\,.
\end{split}
\label{eq:S2-radial}
\end{align}
Using the standard Fourier--Bessel identity for the radial integral then gives
\begin{align}
\begin{split}
\mathcal I(|b|)
\propto
-2\int_0^1 \frac{\dd s}{s^2}
\int_0^\infty \dd a\,
M_a\,
\frac{
K_0^2(\Lambda_a|b|)
-K_0(\Lambda_{a,s}|b|)\,K_0(\Lambda_a|b|)
}{
\Lambda_{a,s}^2-\Lambda_a^2
}\,.
\end{split}
\label{eq:S2-pre-u}
\end{align}

Finally, map the semi-infinite $a$-integration range to the unit interval with
$
a=m\,\frac{\sqrt{1-u^2}}{u}\,,
$
for which $\Lambda_a=m/u$. The remaining scalar kernel becomes
\begin{align}
\begin{split}
\mathcal I(|b|)
&\propto
\int_0^1 \frac{\dd u}{u^2\sqrt{1-u^2}}
\int_0^1 \dd s\,
\\
&\quad
\times
\frac{1+\rho(u)}{(1+\rho(u))^2-s^2\rho(u)^2}
\Bigg[
K_0^2\!\left(\frac{m|b|}{u}\right)
-K_0\!\big(m|b|L(u,s)\big)
K_0\!\left(\frac{m|b|}{u}\right)
\Bigg]\,.
\end{split}
\label{eq:S2-final}
\end{align}
with
\begin{align}
\rho(u)=\frac{\sqrt{\gamma^2u^2-\zeta^2+\ii 0}}{u}\,,
\qquad
L(u,s)=
\sqrt{
\frac{\gamma^2(1-u^2)}{u^2}
+\frac{(1+\rho(u))^2}{s^2}
}\,.
\label{eq:S2-rhoL}
\end{align}
\textcolor{black}{This is the representation quoted in Sec.~(\ref{newsec4}) and in the next Appendix~(\ref{app:maximal-cut})}. Its main practical advantage is that both integrations now run over finite ranges, making the real and imaginary parts straightforward to study numerically.

\section{Elliptic structure from the maximal cut in momentum space}
\label{app:maximal-cut}

Equation~\eqref{eq:S2-final} already makes the algebraic structure visible. Introduce
\begin{align}
k=\frac{1}{\gamma}\,,
\qquad
k'=\frac{\zeta}{\gamma}=\sqrt{1-k^2}\,,
\label{eq:S5-moduli}
\end{align}
so that
\begin{align}
\rho(z)=\frac{1}{kz}\sqrt{z^2-k'^2+\ii 0}\,.
\label{eq:S5-rho}
\end{align}
The non-rational dependence is then generated by the two square roots
$
\sqrt{1-z^2}\,,\sqrt{z^2-k'^2}
$
which define the genus-one curve
\begin{align}
w^2=(1-z^2)(z^2-k'^2)\,.
\label{eq:S5-curve}
\end{align}
Under the following change of variables,
\begin{align}
x=\frac{1-z^2}{k^2}\,,
\qquad
Y=\frac{zw}{k^2}\,,
\label{eq:S5-legendre-map}
\end{align}
this becomes the standard Legendre form
\begin{align}
Y^2=x(1-x)(1-k^2x)\,,
\label{eq:S5-legendre}
\end{align}
which makes the genus-one character explicit. The same curve appears on the maximal cut. Starting from
\begin{align}
\MaxCut_4(I)
\propto
\int_{\Gamma\subset \mathbb{C}^{3d}} d^4 \ell_1\,d^4 \ell_2\,d^4 q\;
e^{i q\cdot b}\,
\hat\delta(\ell_1\!\cdot\! v_1)\,
\hat\delta(\ell_2\!\cdot\! v_2)\,
\hat\delta(q\!\cdot\! v_1)\,
\hat\delta(q\!\cdot\! v_2)\,
\prod_{i=1}^4 \delta(D_i)\,,
\label{eq:S5-cut-start}
\end{align}
with
\begin{align}
&D_1=\ell_1^2-m^2\,,\quad
D_2=\ell_2^2-m^2\,,\quad
D_3=(q-\ell_1)^2-m^2\,,\quad 
D_4=(\ell_1+\ell_2)^2-m^2\,,\quad
\label{eq:S5-Ds}
\end{align}
The parametrization of Appendix~(\ref{app:twoloop}) gives 
\textcolor{black}{\begin{align}
&D_1=-(a^2+\bm\ell_\perp^2+m^2)\,,\quad
D_2=-(c^2+\bm k^2+m^2)\,,\nonumber\\&
D_3=-\Big[(\bm q_\perp-\bm\ell_\perp)^2+a^2+m^2\Big]\,,\quad
D_4=-\Big[(a\gamma+c)^2+(\bm\ell_\perp+\bm k)^2+m^2-a^2\zeta^2\Big]\,.
\label{eq:S5-Ds-param}
\end{align}}
Shifting $\bm r_\perp=\bm q_\perp-\bm\ell_\perp$ and performing the angular integrals over the phases of $\bm\ell_\perp$ and $\bm r_\perp$ yields a pair of Bessel factors. On the cut the corresponding radial variables localize to the analytically continued roots
\begin{align}
l_*=\pm \ii A(a)\,,
\qquad
r_*=\pm \ii S(c)\,,
\label{eq:S5-LRQ}
\end{align}
where
\begin{align}
A(a)=\sqrt{a^2+m^2}\,,
\qquad
S(c)=\sqrt{c^2+m^2}\,.
\label{eq:S5-AS}
\end{align}
Using $J_0(\ii x)=I_0(x)\,,$ the cut therefore reduces to
\begin{align}
\MaxCut_4(I)
\propto
\int \dd a\,\dd c\,
\frac{I_0\!\big(A(a)|b|\big)^2}{\sqrt{\Delta(a,c)}}\,,
\label{eq:S5-cut-dadc}
\end{align}
with
\begin{align}
\Delta(a,c)
=4A(a)^2S(c)^2-(2a\gamma c-m^2)^2
=4(m^2-\zeta^2a^2)c^2+4a\gamma m^2 c+m^2(4a^2+3m^2)\,.
\label{eq:S5-delta}
\end{align}
For fixed $a$, the quadratic dependence on $c$ can be integrated out by residues in the analytically continued maximal-unitarity sense, leading to the projected differential
\begin{align}
\widetilde\Omega^{(4)}_{\rm cut}(a;|b|)
\propto
I_0\!\big(|b|\sqrt{a^2+m^2}\big)^2\,
\frac{\dd a}{\sqrt{m^2-\zeta^2a^2}}\,.
\label{eq:S5-cut-a}
\end{align}

\textcolor{black}{Two families of algebraic periods occur on this cut,
\begin{align}
\oint_{\mathcal C_B}\frac{\dd u}{u^{2n}w}\,,
\qquad
\oint_{\mathcal C_B}\frac{\dd u}{u^{2n+1}w}\,,
\label{eq:S5-families}
\end{align}
with $w=\sqrt{(1-u^2)(u^2-k'^2)}$, and they are selected by the two involutions of Sec.~(\ref{newsec4}) rather than by one and the same expansion. The even family
is $e_2$-even and pulls back to the elliptic quotient $\mathcal E_x$; it yields
complete elliptic integrals,
\begin{align}
&\oint_{\mathcal C_B}\frac{\dd u}{w}
=\frac{2}{k'}K\!\left(1-\frac{1}{k'^2}\right)\,,\quad
\oint_{\mathcal C_B}\frac{\dd u}{u^2w}
=\frac{2}{k'}E\!\left(1-\frac{1}{k'^2}\right)\,,\\&
\oint_{\mathcal C_B}\frac{\dd u}{u^4w}
=-2\frac{K(1-1/k'^2)-2(1+k'^2)E(1-1/k'^2)}{3k'^3}\,,
\label{eq:S5-even3}
\end{align}
where the arguments are elliptic parameters, not moduli. The odd family is
$e_1$-even, pulls back to the rational quotient $\mathcal R_x$, and collapses to
rational functions of the modulus:
\begin{align}
&\oint_{\mathcal C_B}\frac{\dd u}{uw}
=\frac{\pi}{k'}\,,\quad \oint_{\mathcal C_B}\frac{\dd u}{u^3w}
=\frac{\pi(1+k'^2)}{2k'^3}\,,\quad
\oint_{\mathcal C_B}\frac{\dd u}{u^5w}
=\frac{\pi(3k'^4+2k'^2+3)}{8k'^5}\,.
\end{align}
\textcolor{black}{Finally, with $a=m\sqrt{1-z^2}/z$, one has
\begin{align}
\widetilde\Omega^{(4)}_{\rm cut}(z;|b|)
\propto
I_0\!\left(\frac{m|b|}{z}\right)^2
\frac{\dd z}{z\sqrt{1-z^2}\sqrt{z^2-k'^2}}\,.
\label{eq:S5-cut}
\end{align}
Because of the transcendental factor $I_0^2$, this is not a pure algebraic differential; it is better viewed as a twisted differential on the elliptic curve.
The maximal cut itself is the contour integral
\begin{align}
\MaxCut_4(I)\propto \oint_{\mathcal C_B}\widetilde\Omega^{(4)}_{\rm cut}(z;|b|)\,,
\label{eq:S5-cut-contour}
\end{align}
where $\mathcal C_B$ circles the points $[-k',k']$ as shown in
Fig.~(\ref{fig:bessel-cut}). Since the square root changes sign across the two sides
of the cut, this contour is twice the line integral on a single sheet.
Two families of algebraic periods occur on this cut,
\begin{align}
\oint_{\mathcal C_B}\frac{\dd z}{z^{2n}w}\,,
\qquad
\oint_{\mathcal C_B}\frac{\dd z}{z^{2n+1}w}\,,
\label{eq:S5-families}
\end{align}
with $w=\sqrt{(1-z^2)(z^2-k'^2)}$, and they are selected by the two involutions of
Sec.~(\ref{newsec4}) rather than by one and the same expansion. The even family
is $e_2$-even and pulls back to the elliptic quotient $\mathcal E_x$; it yields
complete elliptic integrals,
\begin{align}
&\oint_{\mathcal C_B}\frac{\dd z}{w}
=\frac{2}{k'}K\!\left(1-\frac{1}{k'^2}\right)\,,\quad
\oint_{\mathcal C_B}\frac{\dd z}{z^2w}
=\frac{2}{k'}E\!\left(1-\frac{1}{k'^2}\right)\,,\\&
\oint_{\mathcal C_B}\frac{\dd z}{z^4w}
=-2\frac{K(1-1/k'^2)-2(1+k'^2)E(1-1/k'^2)}{3k'^3}\,,
\label{eq:S5-even3}
\end{align}
where the arguments are elliptic parameters, not moduli. The odd family is
$e_1$-even, pulls back to the rational quotient $\mathcal R_x$, and collapses to
rational functions of the modulus times $\pi$:
\begin{align}
&\oint_{\mathcal C_B}\frac{\dd z}{zw}
=\frac{\pi}{k'}\,,\quad \oint_{\mathcal C_B}\frac{\dd z}{z^3w}
=\frac{\pi(1+k'^2)}{2k'^3}\,,\quad
\oint_{\mathcal C_B}\frac{\dd z}{z^5w}
=\frac{\pi(3k'^4+2k'^2+3)}{8k'^5}\,.
\end{align}
Which family the Bessel weight selects depends on the regime. The convergent Taylor
series of $I_0^{2}(m|b|/z)$ contains only even powers of $1/z$, so together with the
measure $\dd z/(z\,w)$ it produces the odd family alone. Every term of
\eqref{9.21kkkk} is therefore elementary: this is the complete internal genus drop
of Sec.~(\ref{newsec3}). The large-$|b|$ asymptotics is not of this type at all.
There $I_0^{2}(\beta/z)$ is not a power series in $1/z$ but carries the exponential
weight $e^{2\beta/z}$, which is odd under $z\to-z\,.$ The individual terms of
\eqref{eq:new10} are therefore not periods of either quotient, each one being
dominated by the endpoint $z=k'$ and carrying the common factor $e^{2\beta/k'}$. The
alternation with $r$ in Sec.~(\ref{newsec3}) is an alternation of the quotient from
which the algebraic measure is pulled back, not of the function space of the answer}.}


\nocite{Bhattacharyya:2024aeq,Mogull:2020sak,Zhou:2018tva,DoNorbury2018,EynardOrantin2007}
\bibliographystyle{jhep}
\bibliography{refs_local}

@article{DoNorbury2018,
  author = {Do, Norman and Norbury, Paul},
  title = {Topological recursion on the Bessel curve},
  journal = {Communications in Number Theory and Physics},
  volume = {12},
  pages = {53--73},
  year = {2018},
  eprint = {1608.02781},
  archivePrefix = {arXiv},
  primaryClass = {math-ph},
  doi = {10.4310/CNTP.2018.v12.n1.a2}
}

@article{Duhr:2024rxe,
    author = "Duhr, Claude and Porkert, Franziska and Semper, Cathrin and Stawinski, Sven F.",
    title = "{Twisted Riemann bilinear relations and Feynman integrals}",
    eprint = "2407.17175",
    archivePrefix = "arXiv",
    primaryClass = "hep-th",
    reportNumber = "BONN-TH-2024-10",
    doi = "10.1007/JHEP03(2025)019",
    journal = "JHEP",
    volume = "03",
    pages = "019",
    year = "2025"
}

@article{Morales:2022csr,
    author = "Morales, Roger and Spiering, Anne and Wilhelm, Matthias and Yang, Qinglin and Zhang, Chi",
    title = "{Bootstrapping Elliptic Feynman Integrals Using Schubert Analysis}",
    eprint = "2212.09762",
    archivePrefix = "arXiv",
    primaryClass = "hep-th",
    doi = "10.1103/PhysRevLett.131.041601",
    journal = "Phys. Rev. Lett.",
    volume = "131",
    number = "4",
    pages = "041601",
    year = "2023"
}

@article{EynardOrantin2007,
  author = {Eynard, Bertrand and Orantin, Nicolas},
  title = {Invariants of algebraic curves and topological expansion},
  journal = {Communications in Number Theory and Physics},
  volume = {1},
  number = {2},
  pages = {347--452},
  year = {2007},
  eprint = {math-ph/0702045},
  archivePrefix = {arXiv},
  doi = {10.4310/CNTP.2007.v1.n2.a4}
}

@article{Mastrolia:2018uzb,
    author = "Mastrolia, Pierpaolo and Mizera, Sebastian",
    title = "{Feynman Integrals and Intersection Theory}",
    eprint = "1810.03818",
    archivePrefix = "arXiv",
    primaryClass = "hep-th",
    doi = "10.1007/JHEP02(2019)139",
    journal = "JHEP",
    volume = "02",
    pages = "139",
    year = "2019"
}

@article{Broedel:2017kkb,
    author = "Broedel, Johannes and Duhr, Claude and Dulat, Falko and Tancredi, Lorenzo",
    title = "{Elliptic polylogarithms and iterated integrals on elliptic curves. Part I: general formalism}",
    eprint = "1712.07089",
    archivePrefix = "arXiv",
    primaryClass = "hep-th",
    reportNumber = "CERN-TH-2017-273, CP3-17-57, HU-EP-17-29, HU-Mathematik-2017-09, SLAC-PUB-17194",
    doi = "10.1007/JHEP05(2018)093",
    journal = "JHEP",
    volume = "05",
    pages = "093",
    year = "2018"
}

@article{Dlapa:2022wdu,
    author = "Dlapa, Christoph and Henn, Johannes M. and Wagner, Fabian J.",
    title = "{An algorithmic approach to finding canonical differential equations for elliptic Feynman integrals}",
    eprint = "2211.16357",
    archivePrefix = "arXiv",
    primaryClass = "hep-ph",
    reportNumber = "DESY 22-189, MPP-2022-138, TUM-HEP-1427/22",
    doi = "10.1007/JHEP08(2023)120",
    journal = "JHEP",
    volume = "08",
    pages = "120",
    year = "2023"
}

@article{Frellesvig:2019kgj,
    author = "Frellesvig, Hjalte and Gasparotto, Federico and Laporta, Stefano and Mandal, Manoj K. and Mastrolia, Pierpaolo and Mattiazzi, Luca and Mizera, Sebastian",
    title = "{Decomposition of Feynman Integrals on the Maximal Cut by Intersection Numbers}",
    eprint = "1901.11510",
    archivePrefix = "arXiv",
    primaryClass = "hep-ph",
    doi = "10.1007/JHEP05(2019)153",
    journal = "JHEP",
    volume = "05",
    pages = "153",
    year = "2019"
}

@article{Mogull:2020sak,
    author = "Mogull, Gustav and Plefka, Jan and Steinhoff, Jan",
    title = "{Classical black hole scattering from a worldline quantum field theory}",
    eprint = "2010.02865",
    archivePrefix = "arXiv",
    primaryClass = "hep-th",
    reportNumber = "UUITP-37/20, HU-EP-20/22-RTG",
    doi = "10.1007/JHEP02(2021)048",
    journal = "JHEP",
    volume = "02",
    pages = "048",
    year = "2021"
}

@article{Bhattacharyya:2024aeq,
    author = "Bhattacharyya, Arpan and Ghosh, Debodirna and Ghosh, Saptaswa and Pal, Sounak",
    title = "{Observables from classical black hole scattering in Scalar-Tensor theory of gravity from worldline quantum field theory}",
    eprint = "2401.05492",
    archivePrefix = "arXiv",
    primaryClass = "hep-th",
    doi = "10.1007/JHEP04(2024)015",
    journal = "JHEP",
    volume = "04",
    pages = "015",
    year = "2024"
}

@article{Baikov:1996iu,
    author = "Baikov, P. A.",
    editor = "Werlen, M. and Perret-Gallix, D.",
    title = "{Explicit solutions of the multiloop integral recurrence relations and its application}",
    eprint = "hep-ph/9611449",
    archivePrefix = "arXiv",
    reportNumber = "INP-96-42-449",
    doi = "10.1016/S0168-9002(97)00126-5",
    journal = "Nucl. Instrum. Meth. A",
    volume = "389",
    pages = "347--349",
    year = "1997"
}

@article{Lee:2025rik,
    author = "Lee, Roman N. and Pomeransky, Andrei A.",
    title = "{Monodromy of multiloop integrals in d dimensions}",
    eprint = "2506.18452",
    archivePrefix = "arXiv",
    primaryClass = "hep-th",
    doi = "10.1007/JHEP10(2025)009",
    journal = "JHEP",
    volume = "10",
    pages = "009",
    year = "2025"
}

@article{Henn:2013pwa,
    author = "Henn, Johannes M.",
    title = "{Multiloop integrals in dimensional regularization made simple}",
    eprint = "1304.1806",
    archivePrefix = "arXiv",
    primaryClass = "hep-th",
    doi = "10.1103/PhysRevLett.110.251601",
    journal = "Phys. Rev. Lett.",
    volume = "110",
    pages = "251601",
    year = "2013"
}

@article{Primo:2016ebd,
    author = "Primo, Amedeo and Tancredi, Lorenzo",
    title = "{On the maximal cut of Feynman integrals and the solution of their differential equations}",
    eprint = "1610.08397",
    archivePrefix = "arXiv",
    primaryClass = "hep-ph",
    reportNumber = "TTP16-046",
    doi = "10.1016/j.nuclphysb.2016.12.021",
    journal = "Nucl. Phys. B",
    volume = "916",
    pages = "94--116",
    year = "2017"
}

@article{Bosma:2017ens,
    author = "Bosma, Jorrit and Sogaard, Mads and Zhang, Yang",
    title = "{Maximal Cuts in Arbitrary Dimension}",
    eprint = "1704.04255",
    archivePrefix = "arXiv",
    primaryClass = "hep-th",
    doi = "10.1007/JHEP08(2017)051",
    journal = "JHEP",
    volume = "08",
    pages = "051",
    year = "2017"
}

@article{Frellesvig:2017aai,
    author = "Frellesvig, Hjalte and Papadopoulos, Costas G.",
    title = "{Cuts of Feynman Integrals in Baikov representation}",
    eprint = "1701.07356",
    archivePrefix = "arXiv",
    primaryClass = "hep-ph",
    doi = "10.1007/JHEP04(2017)083",
    journal = "JHEP",
    volume = "04",
    pages = "083",
    year = "2017"
}

@article{Bloch:2013tra,
    author = "Bloch, Spencer and Vanhove, Pierre",
    title = "{The elliptic dilogarithm for the sunset graph}",
    eprint = "1309.5865",
    archivePrefix = "arXiv",
    primaryClass = "hep-th",
    doi = "10.1016/j.jnt.2014.09.032",
    journal = "J. Number Theor.",
    volume = "148",
    pages = "328--364",
    year = "2015"
}

@article{Kita1994IntersectionTF,
  title={Intersection Theory for Twisted Cycles},
  author={Michitake Kita and Masaaki Yoshida},
  journal={Mathematische Nachrichten},
  year={1994},
  volume={166},
  pages={287-304},
  url={https://api.semanticscholar.org/CorpusID:121947629}
}

@inproceedings{Zhou:2018tva,
    author = "Zhou, Yajun",
    title = "{Some algebraic and arithmetic properties of Feynman diagrams}",
    booktitle = "{KMPB Conference}: {Elliptic Integrals, Elliptic Functions and Modular Forms in Quantum Field Theory}",
    eprint = "1801.05555",
    archivePrefix = "arXiv",
    primaryClass = "math.NT",
    doi = "10.1007/978-3-030-04480-0_19",
    pages = "485--509",
    year = "2019"
}

@article{Remiddi:1999ew,
    author = "Remiddi, E. and Vermaseren, J. A. M.",
    title = "{Harmonic polylogarithms}",
    eprint = "hep-ph/9905237",
    archivePrefix = "arXiv",
    reportNumber = "NIKHEF-99-005, TTP-99-08",
    doi = "10.1142/S0217751X00000367",
    journal = "Int. J. Mod. Phys. A",
    volume = "15",
    pages = "725--754",
    year = "2000"
}

@article{Lee:2020zfb,
    author = "Lee, Roman N.",
    title = "{Libra: A package for transformation of differential systems for multiloop integrals}",
    eprint = "2012.00279",
    archivePrefix = "arXiv",
    primaryClass = "hep-ph",
    doi = "10.1016/j.cpc.2021.108058",
    journal = "Comput. Phys. Commun.",
    volume = "267",
    pages = "108058",
    year = "2021"
}

@article{Remiddi:1997ny,
    author = "Remiddi, Ettore",
    title = "{Differential equations for Feynman graph amplitudes}",
    eprint = "hep-th/9711188",
    archivePrefix = "arXiv",
    reportNumber = "DFUB-97-15, DFUB 97-15",
    doi = "10.1007/BF03185566",
    journal = "Nuovo Cim. A",
    volume = "110",
    pages = "1435--1452",
    year = "1997"
}

@article{Kotikov:1991pm,
    author = "Kotikov, A. V.",
    title = "{Differential equation method: The Calculation of N point Feynman diagrams}",
    doi = "10.1016/0370-2693(91)90536-Y",
    journal = "Phys. Lett. B",
    volume = "267",
    pages = "123--127",
    year = "1991",
    note = "[Erratum: Phys.Lett.B 295, 409--409 (1992)]"
}

@article{Alessio:2024wmz,
    author = "Alessio, Francesco and Di Vecchia, Paolo",
    title = "{2PM waveform from loop corrected soft theorems}",
    eprint = "2402.06533",
    archivePrefix = "arXiv",
    primaryClass = "hep-th",
    doi = "10.1088/1751-8121/ad8b02",
    journal = "J. Phys. A",
    volume = "57",
    number = "47",
    pages = "475402",
    year = "2024"
}

@article{AccettulliHuber:2019jqo,
    author = "Accettulli Huber, Manuel and Brandhuber, Andreas and De Angelis, Stefano and Travaglini, Gabriele",
    title = "{Note on the absence of $R^2$ corrections to Newton\textquoteright{}s potential}",
    eprint = "1911.10108",
    archivePrefix = "arXiv",
    primaryClass = "hep-th",
    reportNumber = "QMUL-PH-19-31, SAGEX-19-29-E",
    doi = "10.1103/PhysRevD.101.046011",
    journal = "Phys. Rev. D",
    volume = "101",
    number = "4",
    pages = "046011",
    year = "2020"
}

@article{AccettulliHuber:2020oou,
    author = "Accettulli Huber, Manuel and Brandhuber, Andreas and De Angelis, Stefano and Travaglini, Gabriele",
    title = "{Eikonal phase matrix, deflection angle and time delay in effective field theories of gravity}",
    eprint = "2006.02375",
    archivePrefix = "arXiv",
    primaryClass = "hep-th",
    reportNumber = "QMUL-PH-20-05, SAGEX-20-04-E",
    doi = "10.1103/PhysRevD.102.046014",
    journal = "Phys. Rev. D",
    volume = "102",
    number = "4",
    pages = "046014",
    year = "2020"
}

@article{Brandhuber:2019qpg,
    author = "Brandhuber, Andreas and Travaglini, Gabriele",
    title = "{On higher-derivative effects on the gravitational potential and particle bending}",
    eprint = "1905.05657",
    archivePrefix = "arXiv",
    primaryClass = "hep-th",
    reportNumber = "QMUL-PH-19-09, SAGEX-19-05",
    doi = "10.1007/JHEP01(2020)010",
    journal = "JHEP",
    volume = "01",
    pages = "010",
    year = "2020"
}

@article{Bern:2020gjj,
    author = "Bern, Zvi and Ita, Harald and Parra-Martinez, Julio and Ruf, Michael S.",
    title = "{Universality in the classical limit of massless gravitational scattering}",
    eprint = "2002.02459",
    archivePrefix = "arXiv",
    primaryClass = "hep-th",
    reportNumber = "UCLA/TEP/2020/101, FR-PHENO-2020-001",
    doi = "10.1103/PhysRevLett.125.031601",
    journal = "Phys. Rev. Lett.",
    volume = "125",
    number = "3",
    pages = "031601",
    year = "2020"
}

@article{Bern:2019nnu,
    author = "Bern, Zvi and Cheung, Clifford and Roiban, Radu and Shen, Chia-Hsien and Solon, Mikhail P. and Zeng, Mao",
    title = "{Scattering Amplitudes and the Conservative Hamiltonian for Binary Systems at Third Post-Minkowskian Order}",
    eprint = "1901.04424",
    archivePrefix = "arXiv",
    primaryClass = "hep-th",
    reportNumber = "CALT-TH 2019-002, UCLA/TEP/2019/101",
    doi = "10.1103/PhysRevLett.122.201603",
    journal = "Phys. Rev. Lett.",
    volume = "122",
    number = "20",
    pages = "201603",
    year = "2019"
}

@article{Bern:2021dqo,
    author = "Bern, Zvi and Parra-Martinez, Julio and Roiban, Radu and Ruf, Michael S. and Shen, Chia-Hsien and Solon, Mikhail P. and Zeng, Mao",
    title = "{Scattering Amplitudes and Conservative Binary Dynamics at $\mathcal{O}(G^4)$}",
    eprint = "2101.07254",
    archivePrefix = "arXiv",
    primaryClass = "hep-th",
    reportNumber = "CALT-TH-2021-004, FR-PHENO-2021-03, OUTP-21-03P",
    doi = "10.1103/PhysRevLett.126.171601",
    journal = "Phys. Rev. Lett.",
    volume = "126",
    number = "17",
    pages = "171601",
    year = "2021"
}

@article{Bern:2021yeh,
    author = "Bern, Zvi and Parra-Martinez, Julio and Roiban, Radu and Ruf, Michael S. and Shen, Chia-Hsien and Solon, Mikhail P. and Zeng, Mao",
    title = "{Scattering Amplitudes, the Tail Effect, and Conservative Binary Dynamics at $\mathcal{O}(G^4)$}",
    eprint = "2112.10750",
    archivePrefix = "arXiv",
    primaryClass = "hep-th",
    doi = "10.1103/PhysRevLett.128.161103",
    journal = "Phys. Rev. Lett.",
    volume = "128",
    number = "16",
    pages = "161103",
    year = "2022"
}

@article{Bern:2022kto,
    author = "Bern, Zvi and Kosmopoulos, Dimitrios and Luna, Andr\'es and Roiban, Radu and Teng, Fei",
    title = "{Binary Dynamics through the Fifth Power of Spin at $\mathcal{O}(G^2)$}",
    eprint = "2203.06202",
    archivePrefix = "arXiv",
    primaryClass = "hep-th",
    doi = "10.1103/PhysRevLett.130.201402",
    journal = "Phys. Rev. Lett.",
    volume = "130",
    number = "20",
    pages = "201402",
    year = "2023"
}

@article{Bern:2023ity,
    author = "Bern, Zvi and Kosmopoulos, Dimitrios and Luna, Andres and Roiban, Radu and Scheopner, Trevor and Teng, Fei and Vines, Justin",
    title = "{Quantum field theory, worldline theory, and spin magnitude change in orbital evolution}",
    eprint = "2308.14176",
    archivePrefix = "arXiv",
    primaryClass = "hep-th",
    doi = "10.1103/PhysRevD.109.045011",
    journal = "Phys. Rev. D",
    volume = "109",
    number = "4",
    pages = "045011",
    year = "2024"
}

@article{Bern:2024adl,
    author = "Bern, Zvi and Herrmann, Enrico and Roiban, Radu and Ruf, Michael S. and Smirnov, Alexander V. and Smirnov, Vladimir A. and Zeng, Mao",
    title = "{Amplitudes, supersymmetric black hole scattering at $ \mathcal{O}\left({G}^5\right) $, and loop integration}",
    eprint = "2406.01554",
    archivePrefix = "arXiv",
    primaryClass = "hep-th",
    doi = "10.1007/JHEP10(2024)023",
    journal = "JHEP",
    volume = "10",
    pages = "023",
    year = "2024"
}

@article{Frellesvig:2023bbf,
    author = "Frellesvig, Hjalte and Morales, Roger and Wilhelm, Matthias",
    title = "{Calabi-Yau Meets Gravity: A Calabi-Yau Threefold at Fifth Post-Minkowskian Order}",
    eprint = "2312.11371",
    archivePrefix = "arXiv",
    primaryClass = "hep-th",
    doi = "10.1103/PhysRevLett.132.201602",
    journal = "Phys. Rev. Lett.",
    volume = "132",
    number = "20",
    pages = "201602",
    year = "2024"
}

@article{Klemm:2024wtd,
    author = "Klemm, Albrecht and Nega, Christoph and Sauer, Benjamin and Plefka, Jan",
    title = "{Calabi-Yau periods for black hole scattering in classical general relativity}",
    eprint = "2401.07899",
    archivePrefix = "arXiv",
    primaryClass = "hep-th",
    reportNumber = "HU-EP-24/02-RTG, TUM-HEP-1492/24",
    doi = "10.1103/PhysRevD.109.124046",
    journal = "Phys. Rev. D",
    volume = "109",
    number = "12",
    pages = "124046",
    year = "2024"
}

@article{Driesse:2024xad,
    author = "Driesse, Mathias and Jakobsen, Gustav Uhre and Mogull, Gustav and Plefka, Jan and Sauer, Benjamin and Usovitsch, Johann",
    title = "{Conservative Black Hole Scattering at Fifth Post-Minkowskian and First Self-Force Order}",
    eprint = "2403.07781",
    archivePrefix = "arXiv",
    primaryClass = "hep-th",
    reportNumber = "HU-EP-24/08-RTG, CERN-TH-2024-032",
    doi = "10.1103/PhysRevLett.132.241402",
    journal = "Phys. Rev. Lett.",
    volume = "132",
    number = "24",
    pages = "241402",
    year = "2024"
}

@article{Wang:2022ntx,
    author = "Wang, Tianheng",
    title = "{Binary dynamics from worldline QFT for scalar QED}",
    eprint = "2205.15753",
    archivePrefix = "arXiv",
    primaryClass = "hep-th",
    reportNumber = "HU-EP-22/20",
    doi = "10.1103/PhysRevD.107.085011",
    journal = "Phys. Rev. D",
    volume = "107",
    number = "8",
    pages = "085011",
    year = "2023"
}

@article{Jakobsen:2022psy,
    author = "Jakobsen, Gustav Uhre and Mogull, Gustav and Plefka, Jan and Sauer, Benjamin",
    title = "{All things retarded: radiation-reaction in worldline quantum field theory}",
    eprint = "2207.00569",
    archivePrefix = "arXiv",
    primaryClass = "hep-th",
    reportNumber = "HU-EP-22/24-RTG",
    doi = "10.1007/JHEP10(2022)128",
    journal = "JHEP",
    volume = "10",
    pages = "128",
    year = "2022"
}

@article{Gehrmann:1999as,
    author = "Gehrmann, T. and Remiddi, E.",
    title = "{Differential equations for two loop four point functions}",
    eprint = "hep-ph/9912329",
    archivePrefix = "arXiv",
    reportNumber = "TTP-99-49",
    doi = "10.1016/S0550-3213(00)00223-6",
    journal = "Nucl. Phys. B",
    volume = "580",
    pages = "485--518",
    year = "2000"
}

@article{Kotikov:1990kg,
    author = "Kotikov, A. V.",
    title = "{Differential equations method: New technique for massive Feynman diagrams calculation}",
    reportNumber = "ITF-90-31E",
    doi = "10.1016/0370-2693(91)90413-K",
    journal = "Phys. Lett. B",
    volume = "254",
    pages = "158--164",
    year = "1991"
}

@article{Brandhuber:2022qbk,
    author = "Brandhuber, Andreas and Plefka, Jan and Travaglini, Gabriele",
    title = "{Chapter 1: Modern fundamentals of amplitudes}",
    eprint = "2203.13012",
    archivePrefix = "arXiv",
    primaryClass = "hep-th",
    reportNumber = "SAGEX-22-02, HU-EP-22/06, QMUL-PH-22-01",
    doi = "10.1088/1751-8121/ac8254",
    journal = "J. Phys. A",
    volume = "55",
    number = "44",
    pages = "443002",
    year = "2022"
}

@article{Kuntz:2019zef,
    author = "Kuntz, Adrien and Piazza, Federico and Vernizzi, Filippo",
    title = "{Effective field theory for gravitational radiation in scalar-tensor gravity}",
    eprint = "1902.04941",
    archivePrefix = "arXiv",
    primaryClass = "gr-qc",
    doi = "10.1088/1475-7516/2019/05/052",
    journal = "JCAP",
    volume = "05",
    pages = "052",
    year = "2019"
}

@book{Weinzierl:2022eaz,
    author = "Weinzierl, Stefan",
    title = "{Feynman Integrals}",
    eprint = "2201.03593",
    archivePrefix = "arXiv",
    primaryClass = "hep-th",
    reportNumber = "MITP/22-001",
    doi = "10.1007/978-3-030-99558-4",
    month = "1",
    year = "2022"
}

@article{LIGOScientific:2016aoc,
    author = "Abbott, B. P. and others",
    collaboration = "LIGO Scientific, Virgo",
    title = "{Observation of Gravitational Waves from a Binary Black Hole Merger}",
    eprint = "1602.03837",
    archivePrefix = "arXiv",
    primaryClass = "gr-qc",
    reportNumber = "LIGO-P150914",
    doi = "10.1103/PhysRevLett.116.061102",
    journal = "Phys. Rev. Lett.",
    volume = "116",
    number = "6",
    pages = "061102",
    year = "2016"
}

@article{LIGOScientific:2016vlm,
    author = "Abbott, B. P. and others",
    collaboration = "LIGO Scientific, Virgo",
    title = "{Properties of the Binary Black Hole Merger GW150914}",
    eprint = "1602.03840",
    archivePrefix = "arXiv",
    primaryClass = "gr-qc",
    reportNumber = "LIGO-P1500218",
    doi = "10.1103/PhysRevLett.116.241102",
    journal = "Phys. Rev. Lett.",
    volume = "116",
    number = "24",
    pages = "241102",
    year = "2016"
}

@article{LIGOScientific:2016sjg,
    author = "Abbott, B. P. and others",
    collaboration = "LIGO Scientific, Virgo",
    title = "{GW151226: Observation of Gravitational Waves from a 22-Solar-Mass Binary Black Hole Coalescence}",
    eprint = "1606.04855",
    archivePrefix = "arXiv",
    primaryClass = "gr-qc",
    reportNumber = "LIGO-P151226",
    doi = "10.1103/PhysRevLett.116.241103",
    journal = "Phys. Rev. Lett.",
    volume = "116",
    number = "24",
    pages = "241103",
    year = "2016"
}

@article{Huang:2018pbu,
    author = "Huang, Junwu and Johnson, Matthew C. and Sagunski, Laura and Sakellariadou, Mairi and Zhang, Jun",
    title = "{Prospects for axion searches with Advanced LIGO through binary mergers}",
    eprint = "1807.02133",
    archivePrefix = "arXiv",
    primaryClass = "hep-ph",
    doi = "10.1103/PhysRevD.99.063013",
    journal = "Phys. Rev. D",
    volume = "99",
    number = "6",
    pages = "063013",
    year = "2019"
}

@article{Zhang:2021mks,
    author = "Zhang, Jun and Lyu, Zhenwei and Huang, Junwu and Johnson, Matthew C. and Sagunski, Laura and Sakellariadou, Mairi and Yang, Huan",
    title = "{First Constraints on Nuclear Coupling of Axionlike Particles from the Binary Neutron Star Gravitational Wave Event GW170817}",
    eprint = "2105.13963",
    archivePrefix = "arXiv",
    primaryClass = "hep-ph",
    reportNumber = "Imperial/TP/2021/JZ/01, KCL-PH-TH-2021-25, CERN-TH-2021-061,
  LIGO-P2100161, CERN-TH-2021-061, LIGO-P2100161",
    doi = "10.1103/PhysRevLett.127.161101",
    journal = "Phys. Rev. Lett.",
    volume = "127",
    number = "16",
    pages = "161101",
    year = "2021"
}

@article{Adamo:2024oxy,
    author = "Adamo, Tim and Gonzo, Riccardo and Ilderton, Anton",
    title = "{Gravitational bound waveforms from amplitudes}",
    eprint = "2402.00124",
    archivePrefix = "arXiv",
    primaryClass = "hep-th",
    doi = "10.1007/JHEP05(2024)034",
    journal = "JHEP",
    volume = "05",
    pages = "034",
    year = "2024"
}

@article{Laporta:2000dsw,
    author = "Laporta, S.",
    title = "{High-precision calculation of multiloop Feynman integrals by difference equations}",
    eprint = "hep-ph/0102033",
    archivePrefix = "arXiv",
    doi = "10.1142/S0217751X00002159",
    journal = "Int. J. Mod. Phys. A",
    volume = "15",
    pages = "5087--5159",
    year = "2000"
}

@article{Jakobsen:2021lvp,
    author = "Jakobsen, Gustav Uhre and Mogull, Gustav and Plefka, Jan and Steinhoff, Jan",
    title = "{Gravitational Bremsstrahlung and Hidden Supersymmetry of Spinning Bodies}",
    eprint = "2106.10256",
    archivePrefix = "arXiv",
    primaryClass = "hep-th",
    reportNumber = "HU-EP-21/15-RTG",
    doi = "10.1103/PhysRevLett.128.011101",
    journal = "Phys. Rev. Lett.",
    volume = "128",
    number = "1",
    pages = "011101",
    year = "2022"
}

@article{Dlapa:2023hsl,
    author = {Dlapa, Christoph and K{\"a}lin, Gregor and Liu, Zhengwen and Porto, Rafael A.},
    title = "{Bootstrapping the relativistic two-body problem}",
    eprint = "2304.01275",
    archivePrefix = "arXiv",
    primaryClass = "hep-th",
    reportNumber = "DESY 23-041",
    doi = "10.1007/JHEP08(2023)109",
    journal = "JHEP",
    volume = "08",
    pages = "109",
    year = "2023"
}

@article{Hannesdottir:2024hke,
    author = "Hannesdottir, Holmfridur S. and McLeod, Andrew J. and Schwartz, Matthew D. and Vergu, Cristian",
    title = "{Applications of the Landau bootstrap}",
    eprint = "2410.02424",
    archivePrefix = "arXiv",
    primaryClass = "hep-ph",
    doi = "10.1103/PhysRevD.111.085003",
    journal = "Phys. Rev. D",
    volume = "111",
    number = "8",
    pages = "085003",
    year = "2025"
}

@inproceedings{Bargiela:2026lje,
    author = "Bargie{\l}a, Piotr and Crisanti, Giulio and Larkin, Craig and Lippstreu, Luke and McLeod, Andrew J. and Polackova, Maria and Walsh, Laura",
    title = "{New Tools in the Landau Bootstrap}",
    booktitle = "{Loops and Legs in Quantum Field Theories}",
    eprint = "2607.16034",
    archivePrefix = "arXiv",
    primaryClass = "hep-th",
    month = "7",
    year = "2026"
}

@article{Pham:1965,
    author  = {Pham, Fr{\'e}d{\'e}ric},
    title   = {Formules de Picard--Lefschetz g{\'e}n{\'e}ralis{\'e}es
               et ramification des int{\'e}grales},
    journal = {Bull. Soc. Math. France},
    volume  = {93},
    pages   = {333--367},
    year    = {1965}
}

@article{Mizera:2017,
    author        = {Mizera, Sebastian},
    title         = {Scattering Amplitudes from Intersection Theory},
    journal       = {Phys. Rev. Lett.},
    volume        = {120},
    number        = {14},
    pages         = {141602},
    year          = {2018},
    eprint        = {1711.00469},
    archivePrefix = {arXiv},
    primaryClass  = {hep-th},
    doi           = {10.1103/PhysRevLett.120.141602}
}

@article{Atiyah:2017erd,
    author = "Atiyah, Michael and Dunajski, Maciej and Mason, Lionel",
    title = "{Twistor theory at fifty: from contour integrals to twistor strings}",
    eprint = "1704.07464",
    archivePrefix = "arXiv",
    primaryClass = "hep-th",
    reportNumber = "DAMTP-2017-17",
    doi = "10.1098/rspa.2017.0530",
    journal = "Proc. Roy. Soc. Lond. A",
    volume = "473",
    number = "2206",
    pages = "20170530",
    year = "2017"
}

@article{Arkani-Hamed:2013jha,
    author = "Arkani-Hamed, Nima and Trnka, Jaroslav",
    title = "{The Amplituhedron}",
    eprint = "1312.2007",
    archivePrefix = "arXiv",
    primaryClass = "hep-th",
    doi = "10.1007/JHEP10(2014)030",
    journal = "JHEP",
    volume = "10",
    pages = "030",
    year = "2014"
}

@book{Arkani-Hamed:2012zlh,
    author = "Arkani-Hamed, Nima and Bourjaily, Jacob L. and Cachazo, Freddy and Goncharov, Alexander B. and Postnikov, Alexander and Trnka, Jaroslav",
    title = "{Grassmannian Geometry of Scattering Amplitudes}",
    eprint = "1212.5605",
    archivePrefix = "arXiv",
    primaryClass = "hep-th",
    reportNumber = "PUPT-2435",
    doi = "10.1017/CBO9781316091548",
    isbn = "978-1-107-08658-6, 978-1-316-57296-2",
    publisher = "Cambridge University Press",
    month = "4",
    year = "2016"
}

@article{Witten:2003nn,
    author = "Witten, Edward",
    title = "{Perturbative gauge theory as a string theory in twistor space}",
    eprint = "hep-th/0312171",
    archivePrefix = "arXiv",
    doi = "10.1007/s00220-004-1187-3",
    journal = "Commun. Math. Phys.",
    volume = "252",
    pages = "189--258",
    year = "2004"
}

@article{Bauer:2000cp,
    author = "Bauer, Christian W. and Frink, Alexander and Kreckel, Richard",
    title = "{Introduction to the GiNaC framework for symbolic computation within the C++ programming language}",
    eprint = "cs/0004015",
    archivePrefix = "arXiv",
    reportNumber = "MZ-TH-00-17",
    doi = "10.1006/jsco.2001.0494",
    journal = "J. Symb. Comput.",
    volume = "33",
    pages = "1--12",
    year = "2002"
}

@book{Lefschetz1924,
  author    = {Lefschetz, Solomon},
  title     = {L'Analysis Situs et la G{\'e}om{\'e}trie Alg{\'e}brique},
  year      = {1924},
  publisher = {Gauthier-Villars et Cie},
  address   = {Paris},
  series    = {Collection de monographies sur la th{\'e}orie des fonctions}
}

@article{Kosower:2011ty,
    author = "Kosower, David A. and Larsen, Kasper J.",
    title = "{Maximal Unitarity at Two Loops}",
    eprint = "1108.1180",
    archivePrefix = "arXiv",
    primaryClass = "hep-th",
    reportNumber = "UUITP-21-11, IPHT-T11-208",
    doi = "10.1103/PhysRevD.85.045017",
    journal = "Phys. Rev. D",
    volume = "85",
    pages = "045017",
    year = "2012"
}

@book{PicardSimart1897,
  author    = {Picard, {\'E}mile and Simart, Georges},
  title     = {Th{\'e}orie des fonctions alg{\'e}briques de deux variables ind{\'e}pendantes},
  volume    = {1},
  year      = {1897},
  publisher = {Gauthier-Villars et fils},
  address   = {Paris}
}

@phdthesis{Ghosh:2026gev,
    author = "Ghosh, Saptaswa",
    title = "{Classical Black Hole Scattering to Celestial Amplitudes in Effective Theories of Gravity}",
    eprint = "2608.00784",
    archivePrefix = "arXiv",
    primaryClass = "hep-th",
    school = "Indian Institute of Technology Gandhinagar",
    year = "2026"
}

@article{Duhr:2023eld,
    author = "Duhr, Claude and Klemm, Albrecht and Loebbert, Florian and Nega, Christoph and Porkert, Franziska",
    title = "{The Basso-Dixon formula and Calabi-Yau geometry}",
    eprint = "2310.08625",
    archivePrefix = "arXiv",
    primaryClass = "hep-th",
    reportNumber = "BONN-TH-2023-08, TUM-HEP-1470-23",
    doi = "10.1007/JHEP03(2024)177",
    journal = "JHEP",
    volume = "03",
    pages = "177",
    year = "2024"
}

@article{Duhr:2025ppd,
    author = "Duhr, Claude and Maggio, Sara",
    title = "{Feynman integrals, elliptic integrals and two-parameter K3 surfaces}",
    eprint = "2502.15326",
    archivePrefix = "arXiv",
    primaryClass = "hep-th",
    reportNumber = "BONN-TH-2025-05",
    doi = "10.1007/JHEP06(2025)250",
    journal = "JHEP",
    volume = "06",
    pages = "250",
    year = "2025"
}

@article{Duhr:2025lbz,
    author = "Duhr, Claude and Maggio, Sara and Nega, Christoph and Sauer, Benjamin and Tancredi, Lorenzo and Wagner, Fabian J.",
    title = "{Aspects of canonical differential equations for Calabi-Yau geometries and beyond}",
    eprint = "2503.20655",
    archivePrefix = "arXiv",
    primaryClass = "hep-th",
    reportNumber = "BONN-TH-2025-11, TUM-HEP 1559/25, HU-EP-25/13-RTG",
    doi = "10.1007/JHEP06(2025)128",
    journal = "JHEP",
    volume = "06",
    pages = "128",
    year = "2025"
}

@article{Frellesvig:2019uqt,
    author        = {Frellesvig, Hjalte and Gasparotto, Federico
                     and Mandal, Manoj K. and Mastrolia, Pierpaolo
                     and Mattiazzi, Luca and Mizera, Sebastian},
    title         = {Vector Space of Feynman Integrals and
                     Multivariate Intersection Numbers},
    journal       = {Phys. Rev. Lett.},
    volume        = {123},
    number        = {20},
    pages         = {201602},
    year          = {2019},
    eprint        = {1907.02000},
    archivePrefix = {arXiv},
    primaryClass  = {hep-th},
    doi           = {10.1103/PhysRevLett.123.201602}
}

@article{Mizera:2019vvs,
    author        = {Mizera, Sebastian and Pokraka, Andrzej},
    title         = {From Infinity to Four Dimensions:
                     Higher Residue Pairings and Feynman Integrals},
    journal       = {JHEP},
    volume        = {02},
    pages         = {159},
    year          = {2020},
    eprint        = {1910.11852},
    archivePrefix = {arXiv},
    primaryClass  = {hep-th},
    doi           = {10.1007/JHEP02(2020)159}
}

@article{CaronHuot:2021,
    author        = {Caron-Huot, Simon and Pokraka, Andrzej},
    title         = {Duals of Feynman Integrals. Part I.
                     Differential Equations},
    journal       = {JHEP},
    volume        = {12},
    pages         = {045},
    year          = {2021},
    eprint        = {2104.06898},
    archivePrefix = {arXiv},
    primaryClass  = {hep-th},
    doi           = {10.1007/JHEP12(2021)045}
}

@article{CaronHuot:2021xqj,
    author        = {Caron-Huot, Simon and Pokraka, Andrzej},
    title         = {Duals of Feynman Integrals. Part II.
                     Generalized Unitarity},
    journal       = {JHEP},
    volume        = {04},
    pages         = {078},
    year          = {2022},
    eprint        = {2112.00055},
    archivePrefix = {arXiv},
    primaryClass  = {hep-th},
    doi           = {10.1007/JHEP04(2022)078}
}

@article{Kosower:2018adc,
    author = "Kosower, David A. and Maybee, Ben and O'Connell, Donal",
    title = "{Amplitudes, Observables, and Classical Scattering}",
    eprint = "1811.10950",
    archivePrefix = "arXiv",
    primaryClass = "hep-th",
    doi = "10.1007/JHEP02(2019)137",
    journal = "JHEP",
    volume = "02",
    pages = "137",
    year = "2019"
}

@article{Bhattacharyya:2023kbh,
    author = "Bhattacharyya, Arpan and Ghosh, Saptaswa and Pal, Sounak",
    title = "{Worldline effective field theory of inspiralling black hole binaries in presence of dark photon and axionic dark matter}",
    eprint = "2305.15473",
    archivePrefix = "arXiv",
    primaryClass = "hep-th",
    doi = "10.1007/JHEP08(2023)207",
    journal = "JHEP",
    volume = "08",
    pages = "207",
    year = "2023"
}

@article{Kalin:2020mvi,
    author = {K\"alin, Gregor and Porto, Rafael A.},
    title = "{Post-Minkowskian Effective Field Theory for Conservative Binary Dynamics}",
    eprint = "2006.01184",
    archivePrefix = "arXiv",
    primaryClass = "hep-th",
    reportNumber = "DESY20-077, SLAC-PUB-17529",
    doi = "10.1007/JHEP11(2020)106",
    journal = "JHEP",
    volume = "11",
    pages = "106",
    year = "2020"
}

@article{Kalin:2020lmz,
    author = {K\"alin, Gregor and Liu, Zhengwen and Porto, Rafael A.},
    title = "{Conservative Tidal Effects in Compact Binary Systems to Next-to-Leading Post-Minkowskian Order}",
    eprint = "2008.06047",
    archivePrefix = "arXiv",
    primaryClass = "hep-th",
    reportNumber = "DESY-20-131, DESY 20-131, SLAC-PUB-17555",
    doi = "10.1103/PhysRevD.102.124025",
    journal = "Phys. Rev. D",
    volume = "102",
    pages = "124025",
    year = "2020"
}

@article{Bjerrum-Bohr:2013bxa,
    author = "Bjerrum-Bohr, N. E. J. and Donoghue, John F. and Vanhove, Pierre",
    title = "{On-shell Techniques and Universal Results in Quantum Gravity}",
    eprint = "1309.0804",
    archivePrefix = "arXiv",
    primaryClass = "hep-th",
    reportNumber = "IHES-P-13-23, IPHT-T13-019",
    doi = "10.1007/JHEP02(2014)111",
    journal = "JHEP",
    volume = "02",
    pages = "111",
    year = "2014"
}

@article{Bjerrum-Bohr:2018xdl,
    author = "Bjerrum-Bohr, N. E. J. and Damgaard, Poul H. and Festuccia, Guido and Plant\'e, Ludovic and Vanhove, Pierre",
    title = "{General Relativity from Scattering Amplitudes}",
    eprint = "1806.04920",
    archivePrefix = "arXiv",
    primaryClass = "hep-th",
    doi = "10.1103/PhysRevLett.121.171601",
    journal = "Phys. Rev. Lett.",
    volume = "121",
    number = "17",
    pages = "171601",
    year = "2018"
}

@article{Aoude:2023vdk,
    author = "Aoude, Rafael and Haddad, Kays and Helset, Andreas",
    title = "{Classical gravitational scattering amplitude at $O(G^2S_1^\infty S_2^\infty)$}",
    eprint = "2304.13740",
    archivePrefix = "arXiv",
    primaryClass = "hep-th",
    reportNumber = "CP3-23-19, UUITP-10/23, CALT-TH-2023-010",
    doi = "10.1103/PhysRevD.108.024050",
    journal = "Phys. Rev. D",
    volume = "108",
    number = "2",
    pages = "024050",
    year = "2023"
}

@article{Aoude:2023dui,
    author = "Aoude, Rafael and Haddad, Kays and Heissenberg, Carlo and Helset, Andreas",
    title = "{Leading-order gravitational radiation to all spin orders}",
    eprint = "2310.05832",
    archivePrefix = "arXiv",
    primaryClass = "hep-th",
    doi = "10.1103/PhysRevD.109.036007",
    journal = "Phys. Rev. D",
    volume = "109",
    number = "3",
    pages = "036007",
    year = "2024"
}

@article{Cristofoli:2019neg,
    author = "Cristofoli, Andrea and Bjerrum-Bohr, N. E. J. and Damgaard, Poul H. and Vanhove, Pierre",
    title = "{Post-Minkowskian Hamiltonians in general relativity}",
    eprint = "1906.01579",
    archivePrefix = "arXiv",
    primaryClass = "hep-th",
    doi = "10.1103/PhysRevD.100.084040",
    journal = "Phys. Rev. D",
    volume = "100",
    number = "8",
    pages = "084040",
    year = "2019"
}

@article{Mougiakakos:2022sic,
    author = "Mougiakakos, Stavros and Riva, Massimiliano Maria and Vernizzi, Filippo",
    title = "{Gravitational Bremsstrahlung with Tidal Effects in the Post-Minkowskian Expansion}",
    eprint = "2204.06556",
    archivePrefix = "arXiv",
    primaryClass = "hep-th",
    doi = "10.1103/PhysRevLett.129.121101",
    journal = "Phys. Rev. Lett.",
    volume = "129",
    number = "12",
    pages = "121101",
    year = "2022"
}

@article{Chen:2024mmm,
    author = "Chen, Gang and Wang, Tianheng",
    title = "{Dynamics of spinning binary at 2PM}",
    eprint = "2406.09086",
    archivePrefix = "arXiv",
    primaryClass = "hep-th",
    reportNumber = "SNUTP24-002",
    doi = "10.1007/JHEP12(2024)213",
    journal = "JHEP",
    volume = "12",
    pages = "213",
    year = "2025"
}

@article{Bjerrum-Bohr:2016hpa,
    author = "Bjerrum-Bohr, N. E. J. and Donoghue, John F. and Holstein, Barry R. and Plante, Ludovic and Vanhove, Pierre",
    title = "{Light-like Scattering in Quantum Gravity}",
    eprint = "1609.07477",
    archivePrefix = "arXiv",
    primaryClass = "hep-th",
    reportNumber = "IPHT-T16-082, ACFI-T16-23",
    doi = "10.1007/JHEP11(2016)117",
    journal = "JHEP",
    volume = "11",
    pages = "117",
    year = "2016"
}

@article{Bjerrum-Bohr:2019kec,
    author = "Bjerrum-Bohr, N. E. J. and Cristofoli, Andrea and Damgaard, Poul H.",
    title = "{Post-Minkowskian Scattering Angle in Einstein Gravity}",
    eprint = "1910.09366",
    archivePrefix = "arXiv",
    primaryClass = "hep-th",
    reportNumber = "SAGEX-19-24-E",
    doi = "10.1007/JHEP08(2020)038",
    journal = "JHEP",
    volume = "08",
    pages = "038",
    year = "2020"
}

@article{Bjerrum-Bohr:2020syg,
    author = "Bjerrum-Bohr, N. Emil J. and Brown, Taro V. and Gomez, Humberto",
    title = "{Scattering of Gravitons and Spinning Massive States from Compact Numerators}",
    eprint = "2011.10556",
    archivePrefix = "arXiv",
    primaryClass = "hep-th",
    doi = "10.1007/JHEP04(2021)234",
    journal = "JHEP",
    volume = "04",
    pages = "234",
    year = "2021"
}

@article{Bjerrum-Bohr:2021vuf,
    author = "Bjerrum-Bohr, N. Emil J. and Damgaard, Poul H. and Plant\'e, Ludovic and Vanhove, Pierre",
    title = "{Classical gravity from loop amplitudes}",
    eprint = "2104.04510",
    archivePrefix = "arXiv",
    primaryClass = "hep-th",
    reportNumber = "IPhT-t21/015, CERN-TH-2021-052",
    doi = "10.1103/PhysRevD.104.026009",
    journal = "Phys. Rev. D",
    volume = "104",
    number = "2",
    pages = "026009",
    year = "2021"
}

@article{Bjerrum-Bohr:2021wwt,
    author = "Bjerrum-Bohr, N. Emil J. and Plant\'e, Ludovic and Vanhove, Pierre",
    title = "{Post-Minkowskian radial action from soft limits and velocity cuts}",
    eprint = "2111.02976",
    archivePrefix = "arXiv",
    primaryClass = "hep-th",
    reportNumber = "IPhT-t21/072",
    doi = "10.1007/JHEP03(2022)071",
    journal = "JHEP",
    volume = "03",
    pages = "071",
    year = "2022"
}

@article{Bern:2019crd,
    author = "Bern, Zvi and Cheung, Clifford and Roiban, Radu and Shen, Chia-Hsien and Solon, Mikhail P. and Zeng, Mao",
    title = "{Black Hole Binary Dynamics from the Double Copy and Effective Theory}",
    eprint = "1908.01493",
    archivePrefix = "arXiv",
    primaryClass = "hep-th",
    reportNumber = "CERN-TH-2019-128, CALT-TH 2019-026, UCLA/TEP/2019/103",
    doi = "10.1007/JHEP10(2019)206",
    journal = "JHEP",
    volume = "10",
    pages = "206",
    year = "2019"
}

@article{Jakobsen:2023ndj,
    author = "Jakobsen, Gustav Uhre and Mogull, Gustav and Plefka, Jan and Sauer, Benjamin and Xu, Yingxuan",
    title = "{Conservative Scattering of Spinning Black Holes at Fourth Post-Minkowskian Order}",
    eprint = "2306.01714",
    archivePrefix = "arXiv",
    primaryClass = "hep-th",
    reportNumber = "HU-EP-23/16-RTG",
    doi = "10.1103/PhysRevLett.131.151401",
    journal = "Phys. Rev. Lett.",
    volume = "131",
    number = "15",
    pages = "151401",
    year = "2023"
}

@article{Jakobsen:2023hig,
    author = "Jakobsen, Gustav Uhre and Mogull, Gustav and Plefka, Jan and Sauer, Benjamin",
    title = "{Dissipative Scattering of Spinning Black Holes at Fourth Post-Minkowskian Order}",
    eprint = "2308.11514",
    archivePrefix = "arXiv",
    primaryClass = "hep-th",
    reportNumber = "HU-EP-23/47-RTG",
    doi = "10.1103/PhysRevLett.131.241402",
    journal = "Phys. Rev. Lett.",
    volume = "131",
    number = "24",
    pages = "241402",
    year = "2023"
}

@article{Beneke:1997zp,
    author = "Beneke, M. and Smirnov, Vladimir A.",
    title = "{Asymptotic expansion of Feynman integrals near threshold}",
    eprint = "hep-ph/9711391",
    archivePrefix = "arXiv",
    reportNumber = "CERN-TH-97-315",
    doi = "10.1016/S0550-3213(98)00138-2",
    journal = "Nucl. Phys. B",
    volume = "522",
    pages = "321--344",
    year = "1998"
}

@article{Bern:2020buy,
    author = "Bern, Zvi and Luna, Andres and Roiban, Radu and Shen, Chia-Hsien and Zeng, Mao",
    title = "{Spinning black hole binary dynamics, scattering amplitudes, and effective field theory}",
    eprint = "2005.03071",
    archivePrefix = "arXiv",
    primaryClass = "hep-th",
    doi = "10.1103/PhysRevD.104.065014",
    journal = "Phys. Rev. D",
    volume = "104",
    number = "6",
    pages = "065014",
    year = "2021"
}

@article{Broedel:2017siw,
    author = "Broedel, Johannes and Duhr, Claude and Dulat, Falko and Tancredi, Lorenzo",
    title = "{Elliptic polylogarithms and iterated integrals on elliptic curves II: an application to the sunrise integral}",
    eprint = "1712.07095",
    archivePrefix = "arXiv",
    primaryClass = "hep-ph",
    reportNumber = "CERN-TH-2017-274, CP3-17-58, HU-EP-17-30, HU-Mathematik-2017-10, SLAC-PUB-17195",
    doi = "10.1103/PhysRevD.97.116009",
    journal = "Phys. Rev. D",
    volume = "97",
    number = "11",
    pages = "116009",
    year = "2018"
}

@article{Marzucca:2023gto,
    author = {Marzucca, Robin and McLeod, Andrew J. and Page, Ben and P{\"o}gel, Sebastian and Weinzierl, Stefan},
    title = "{Genus drop in hyperelliptic Feynman integrals}",
    eprint = "2307.11497",
    archivePrefix = "arXiv",
    primaryClass = "hep-th",
    reportNumber = "CERN-TH-2023-133, MITP-23-033, ZU-TH 33/23",
    doi = "10.1103/PhysRevD.109.L031901",
    journal = "Phys. Rev. D",
    volume = "109",
    number = "3",
    pages = "L031901",
    year = "2024"
}

@article{Cho_Matsumoto_1995, title={Intersection theory for twisted cohomologies and twisted Riemann’s period relations I}, volume={139}, DOI={10.1017/S0027763000005304}, journal={Nagoya Mathematical Journal}, author={Cho, Koji and Matsumoto, Keiji}, year={1995}, pages={67–86}}

@article{ashoke1,
    author = "Laddha, Alok and Sen, Ashoke",
    title = "{Gravity Waves from Soft Theorem in General Dimensions}",
    eprint = "1801.07719",
    archivePrefix = "arXiv",
    primaryClass = "hep-th",
    doi = "10.1007/JHEP09(2018)105",
    journal = "JHEP",
    volume = "09",
    pages = "105",
    year = "2018"
}

@article{Bini:2024rsy,
    author = "Bini, Donato and Damour, Thibault and De Angelis, Stefano and Geralico, Andrea and Herderschee, Aidan and Roiban, Radu and Teng, Fei",
    title = "{Gravitational waveforms: A tale of two formalisms}",
    eprint = "2402.06604",
    archivePrefix = "arXiv",
    primaryClass = "hep-th",
    doi = "10.1103/PhysRevD.109.125008",
    journal = "Phys. Rev. D",
    volume = "109",
    number = "12",
    pages = "125008",
    year = "2024"
}

@article{Georgoudis:2024pdz,
    author = "Georgoudis, Alessandro and Heissenberg, Carlo and Russo, Rodolfo",
    title = "{Post-Newtonian multipoles from the next-to-leading post-Minkowskian gravitational waveform}",
    eprint = "2402.06361",
    archivePrefix = "arXiv",
    primaryClass = "hep-th",
    reportNumber = "QMUL-PH-24-03",
    doi = "10.1103/PhysRevD.109.106020",
    journal = "Phys. Rev. D",
    volume = "109",
    number = "10",
    pages = "106020",
    year = "2024"
}

@article{Chetyrkin:1981qh,
    author = "Chetyrkin, K. G. and Tkachov, F. V.",
    title = "{Integration by parts: The algorithm to calculate $\beta$-functions in 4 loops}",
    doi = "10.1016/0550-3213(81)90199-1",
    journal = "Nucl. Phys. B",
    volume = "192",
    pages = "159--204",
    year = "1981"
}

@article{Dlapa:2024cje,
    author = {Dlapa, Christoph and K\"alin, Gregor and Liu, Zhengwen and Porto, Rafael A.},
    title = "{Local in Time Conservative Binary Dynamics at Fourth Post-Minkowskian Order}",
    eprint = "2403.04853",
    archivePrefix = "arXiv",
    primaryClass = "hep-th",
    reportNumber = "DESY 24-029",
    doi = "10.1103/PhysRevLett.132.221401",
    journal = "Phys. Rev. Lett.",
    volume = "132",
    number = "22",
    pages = "221401",
    year = "2024"
}

@article{DeAngelis:2023lvf,
    author = "De Angelis, Stefano and Novichkov, Pavel P. and Gonzo, Riccardo",
    title = "{Spinning waveforms from the Kosower-Maybee-O{\textquoteright}Connell formalism at leading order}",
    eprint = "2309.17429",
    archivePrefix = "arXiv",
    primaryClass = "hep-th",
    doi = "10.1103/PhysRevD.110.L041502",
    journal = "Phys. Rev. D",
    volume = "110",
    number = "4",
    pages = "L041502",
    year = "2024"
}

@article{Adamo:2021rfq,
    author = "Adamo, Tim and Cristofoli, Andrea and Tourkine, Piotr",
    title = "{Eikonal amplitudes from curved backgrounds}",
    eprint = "2112.09113",
    archivePrefix = "arXiv",
    primaryClass = "hep-th",
    doi = "10.21468/SciPostPhys.13.2.032",
    journal = "SciPost Phys.",
    volume = "13",
    number = "2",
    pages = "032",
    year = "2022"
}

@article{Gonzo:2024zxo,
    author = "Gonzo, Riccardo and Shi, Canxin",
    title = "{Scattering and Bound Observables for Spinning Particles in Kerr Spacetime with Generic Spin Orientations}",
    eprint = "2405.09687",
    archivePrefix = "arXiv",
    primaryClass = "hep-th",
    doi = "10.1103/PhysRevLett.133.221401",
    journal = "Phys. Rev. Lett.",
    volume = "133",
    number = "22",
    pages = "221401",
    year = "2024"
}

@article{Alessio:2023kgf,
    author = "Alessio, Francesco",
    title = "{Kerr binary dynamics from minimal coupling and double copy}",
    eprint = "2303.12784",
    archivePrefix = "arXiv",
    primaryClass = "hep-th",
    doi = "10.1007/JHEP04(2024)058",
    journal = "JHEP",
    volume = "04",
    pages = "058",
    year = "2024"
}

@article{Alessio:2022kwv,
    author = "Alessio, Francesco and Di Vecchia, Paolo",
    title = "{Radiation reaction for spinning black-hole scattering}",
    eprint = "2203.13272",
    archivePrefix = "arXiv",
    primaryClass = "hep-th",
    doi = "10.1016/j.physletb.2022.137258",
    journal = "Phys. Lett. B",
    volume = "832",
    pages = "137258",
    year = "2022"
}

@article{Cristofoli:2021vyo,
    author = "Cristofoli, Andrea and Gonzo, Riccardo and Kosower, David A. and O'Connell, Donal",
    title = "{Waveforms from amplitudes}",
    eprint = "2107.10193",
    archivePrefix = "arXiv",
    primaryClass = "hep-th",
    doi = "10.1103/PhysRevD.106.056007",
    journal = "Phys. Rev. D",
    volume = "106",
    number = "5",
    pages = "056007",
    year = "2022"
}

@article{Brandhuber:2023hhy,
    author = "Brandhuber, Andreas and Brown, Graham R. and Chen, Gang and De Angelis, Stefano and Gowdy, Joshua and Travaglini, Gabriele",
    title = "{One-loop gravitational bremsstrahlung and waveforms from a heavy-mass effective field theory}",
    eprint = "2303.06111",
    archivePrefix = "arXiv",
    primaryClass = "hep-th",
    reportNumber = "QMUL-22-28,SAGEX-22-32-E",
    doi = "10.1007/JHEP06(2023)048",
    journal = "JHEP",
    volume = "06",
    pages = "048",
    year = "2023"
}

@article{Georgoudis:2023eke,
    author = "Georgoudis, Alessandro and Heissenberg, Carlo and Russo, Rodolfo",
    title = "{An eikonal-inspired approach to the gravitational scattering waveform}",
    eprint = "2312.07452",
    archivePrefix = "arXiv",
    primaryClass = "hep-th",
    reportNumber = "QMUL-PH-23-34",
    doi = "10.1007/JHEP03(2024)089",
    journal = "JHEP",
    volume = "03",
    pages = "089",
    year = "2024"
}

@article{Bhattacharyya:2025ghl,
    author = "Bhattacharyya, Arpan and Ghosh, Saptaswa and Mishra, Ankit and Pal, Sounak",
    title = "{Heterotic footprints in classical gravity: PM dynamics from on-shell soft amplitudes at one loop}",
    eprint = "2510.07390",
    archivePrefix = "arXiv",
    primaryClass = "hep-th",
    doi = "10.1007/JHEP06(2026)008",
    journal = "JHEP",
    volume = "06",
    pages = "008",
    year = "2026"
}

@article{Bhattacharyya:2024kxj,
    author = "Bhattacharyya, Arpan and Ghosh, Debodirna and Ghosh, Saptaswa and Pal, Sounak",
    title = "{Bootstrapping the spinning two body problem in dynamical Chern-Simons gravity using worldline QFT}",
    eprint = "2407.07195",
    archivePrefix = "arXiv",
    primaryClass = "hep-th",
    doi = "10.1007/JHEP04(2025)175",
    journal = "JHEP",
    volume = "04",
    pages = "175",
    year = "2025"
}

@article{Dlapa:2025biy,
    author = {Dlapa, Christoph and K{\"a}lin, Gregor and Liu, Zhengwen and Porto, Rafael A.},
    title = "{Local-in-Time Conservative Binary Dynamics at Fifth Post-Minkowskian and First Self-Force Orders}",
    eprint = "2506.20665",
    archivePrefix = "arXiv",
    primaryClass = "hep-th",
    reportNumber = "DESY 25-089",
    month = "6",
    year = "2025"
}

@article{Bern:2025zno,
    author = "Bern, Zvi and Herrmann, Enrico and Roiban, Radu and Ruf, Michael S. and Smirnov, Alexander V. and Smirnov, Vladimir A. and Zeng, Mao",
    title = "{Second-order self-force potential-region binary dynamics at $O(G^5)$ in supergravity}",
    eprint = "2509.17412",
    archivePrefix = "arXiv",
    primaryClass = "hep-th",
    month = "9",
    year = "2025"
}

@article{Driesse:2024feo,
    author = "Driesse, Mathias and Jakobsen, Gustav Uhre and Klemm, Albrecht and Mogull, Gustav and Nega, Christoph and Plefka, Jan and Sauer, Benjamin and Usovitsch, Johann",
    title = "{Emergence of Calabi{\textendash}Yau manifolds in high-precision black-hole scattering}",
    eprint = "2411.11846",
    archivePrefix = "arXiv",
    primaryClass = "hep-th",
    reportNumber = "HU-EP-24/32-RTG, QMUL-PH-24-26, BONN-TH-2024-15, TUM-HEP-1532/24",
    doi = "10.1038/s41586-025-08984-2",
    journal = "Nature",
    volume = "641",
    number = "8063",
    pages = "603--607",
    year = "2025"
}

@article{Akpinar:2024meg,
    author = "Akpinar, Dogan and Febres Cordero, Fernando and Kraus, Manfred and Ruf, Michael S. and Zeng, Mao",
    title = "{Spinning black hole scattering at $ \mathcal{O} $(G$^{3}$S$^{2}$): Casimir terms, radial action and hidden symmetry}",
    eprint = "2407.19005",
    archivePrefix = "arXiv",
    primaryClass = "hep-th",
    doi = "10.1007/JHEP03(2025)126",
    journal = "JHEP",
    volume = "03",
    pages = "126",
    year = "2025"
}

@article{Driesse:2026qiz,
    author = "Driesse, Mathias and Jakobsen, Gustav Uhre and Mogull, Gustav and Nega, Christoph and Plefka, Jan and Sauer, Benjamin and Usovitsch, Johann",
    title = "{Conservative Black Hole Scattering at Fifth Post-Minkowskian and Second Self-Force Order}",
    eprint = "2601.16256",
    archivePrefix = "arXiv",
    primaryClass = "hep-th",
    reportNumber = "HU-EP-26/04-RTG",
    doi = "10.1103/4dvk-nglx",
    journal = "Phys. Rev. Lett.",
    volume = "137",
    number = "8",
    pages = "081402",
    year = "2026"
}

@article{Biendarra:2026nig,
    author = "Biendarra, Birgitta and Haddad, Kays and Plefka, Jan",
    title = "{Classical gravitational scattering with a massive scalar mediator}",
    eprint = "2608.14267",
    archivePrefix = "arXiv",
    primaryClass = "hep-th",
    reportNumber = "HU-EP-26/26",
    month = "8",
    year = "2026"
}

@article{Sabry:1962,
    author  = "Sabry, Afaf",
    title   = "{Fourth order spectral functions for the electron propagator}",
    journal = "Nucl. Phys.",
    volume  = "33",
    pages   = "401--430",
    year    = "1962",
    doi     = "10.1016/0029-5582(62)90535-7"
}

@article{Broadhurst:1990,
    author  = "Broadhurst, David J.",
    title   = "{The Master Two Loop Diagram With Masses}",
    journal = "Z. Phys. C",
    volume  = "47",
    pages   = "115--124",
    year    = "1990",
    doi     = "10.1007/BF01551921"
}

@article{Bauberger:1994,
    author        = "Bauberger, S. and Berends, F. A. and B{\"o}hm, M. and Buza, M.",
    title         = "{Analytical and numerical methods for massive two-loop self-energy diagrams}",
    journal       = "Nucl. Phys. B",
    volume        = "434",
    pages         = "383--407",
    year          = "1995",
    eprint        = "hep-ph/9409388",
    archivePrefix = "arXiv",
    doi           = "10.1016/0550-3213(94)00475-T"
}

@article{Caffo:1998du,
    author        = "Caffo, Michele and Czyz, H. and Laporta, S. and Remiddi, E.",
    title         = "{The Master differential equations for the two loop sunrise selfmass amplitudes}",
    journal       = "Nuovo Cim. A",
    volume        = "111",
    pages         = "365--389",
    year          = "1998",
    eprint        = "hep-th/9805118",
    archivePrefix = "arXiv"
}

@article{Laporta:2004rb,
    author        = "Laporta, S. and Remiddi, E.",
    title         = "{Analytic treatment of the two loop equal mass sunrise graph}",
    journal       = "Nucl. Phys. B",
    volume        = "704",
    pages         = "349--386",
    year          = "2005",
    eprint        = "hep-ph/0406160",
    archivePrefix = "arXiv",
    doi           = "10.1016/j.nuclphysb.2004.10.044"
}

@article{MullerStach:2011ru,
    author        = "M{\"u}ller-Stach, Stefan and Weinzierl, Stefan and Zayadeh, Raphael",
    title         = "{A Second-Order Differential Equation for the Two-Loop Sunrise Graph with Arbitrary Masses}",
    journal       = "Commun. Num. Theor. Phys.",
    volume        = "6",
    pages         = "203--222",
    year          = "2012",
    eprint        = "1112.4360",
    archivePrefix = "arXiv",
    primaryClass  = "hep-ph",
    doi           = "10.4310/CNTP.2012.v6.n1.a5"
}

@article{Adams:2013kgc,
    author        = "Adams, Luise and Bogner, Christian and Weinzierl, Stefan",
    title         = "{The two-loop sunrise graph with arbitrary masses}",
    journal       = "J. Math. Phys.",
    volume        = "54",
    pages         = "052303",
    year          = "2013",
    eprint        = "1302.7004",
    archivePrefix = "arXiv",
    primaryClass  = "hep-ph",
    doi           = "10.1063/1.4804996"
}

@article{Remiddi:2013joa,
    author        = "Remiddi, Ettore and Tancredi, Lorenzo",
    title         = "{Schouten identities for Feynman graph amplitudes; The Master Integrals for the two-loop massive sunrise graph}",
    journal       = "Nucl. Phys. B",
    volume        = "880",
    pages         = "343--377",
    year          = "2014",
    eprint        = "1311.3342",
    archivePrefix = "arXiv",
    primaryClass  = "hep-ph",
    doi           = "10.1016/j.nuclphysb.2014.01.009"
}

@article{Adams:2014vja,
    author        = "Adams, Luise and Bogner, Christian and Weinzierl, Stefan",
    title         = "{The two-loop sunrise graph in two space-time dimensions with arbitrary masses in terms of elliptic dilogarithms}",
    journal       = "J. Math. Phys.",
    volume        = "55",
    pages         = "102301",
    year          = "2014",
    eprint        = "1405.5640",
    archivePrefix = "arXiv",
    primaryClass  = "hep-ph",
    doi           = "10.1063/1.4896563"
}

@article{Adams:2015gva,
    author        = "Adams, Luise and Bogner, Christian and Weinzierl, Stefan",
    title         = "{The two-loop sunrise integral around four space-time dimensions and generalisations of the Clausen and Glaisher functions towards the elliptic case}",
    journal       = "J. Math. Phys.",
    volume        = "56",
    pages         = "072303",
    year          = "2015",
    eprint        = "1504.03255",
    archivePrefix = "arXiv",
    primaryClass  = "hep-ph",
    doi           = "10.1063/1.4926985"
}

@article{Adams:2015ydq,
    author        = "Adams, Luise and Bogner, Christian and Weinzierl, Stefan",
    title         = "{The iterated structure of the all-order result for the two-loop sunrise integral}",
    journal       = "J. Math. Phys.",
    volume        = "57",
    pages         = "032304",
    year          = "2016",
    eprint        = "1512.05630",
    archivePrefix = "arXiv",
    primaryClass  = "hep-ph",
    doi           = "10.1063/1.4944722"
}

@article{Remiddi:2016gno,
    author        = "Remiddi, Ettore and Tancredi, Lorenzo",
    title         = "{Differential equations and dispersion relations for Feynman amplitudes. The two-loop massive sunrise and the kite integral}",
    journal       = "Nucl. Phys. B",
    volume        = "907",
    pages         = "400--444",
    year          = "2016",
    eprint        = "1602.01481",
    archivePrefix = "arXiv",
    primaryClass  = "hep-ph",
    doi           = "10.1016/j.nuclphysb.2016.04.013"
}

@article{Adams:2016xah,
    author        = "Adams, Luise and Bogner, Christian and Schweitzer, Armin and Weinzierl, Stefan",
    title         = "{The kite integral to all orders in terms of elliptic polylogarithms}",
    journal       = "J. Math. Phys.",
    volume        = "57",
    pages         = "122302",
    year          = "2016",
    eprint        = "1607.01571",
    archivePrefix = "arXiv",
    primaryClass  = "hep-ph",
    doi           = "10.1063/1.4969060"
}

@article{Primo:2017ipr,
    author        = "Primo, Amedeo and Tancredi, Lorenzo",
    title         = "{Maximal cuts and differential equations for Feynman integrals. An application to the three-loop massive banana graph}",
    journal       = "Nucl. Phys. B",
    volume        = "921",
    pages         = "316--356",
    year          = "2017",
    eprint        = "1704.05465",
    archivePrefix = "arXiv",
    primaryClass  = "hep-ph",
    doi           = "10.1016/j.nuclphysb.2017.05.018"
}

@article{vonManteuffel:2017hms,
    author        = "von Manteuffel, Andreas and Tancredi, Lorenzo",
    title         = "{A non-planar two-loop three-point function beyond multiple polylogarithms}",
    journal       = "JHEP",
    volume        = "06",
    pages         = "127",
    year          = "2017",
    eprint        = "1701.05905",
    archivePrefix = "arXiv",
    primaryClass  = "hep-ph",
    doi           = "10.1007/JHEP06(2017)127"
}

@article{Adams:2017ejb,
    author        = "Adams, Luise and Weinzierl, Stefan",
    title         = "{Feynman integrals and iterated integrals of modular forms}",
    journal       = "Commun. Num. Theor. Phys.",
    volume        = "12",
    pages         = "193--251",
    year          = "2018",
    eprint        = "1704.08895",
    archivePrefix = "arXiv",
    primaryClass  = "hep-ph",
    doi           = "10.4310/CNTP.2018.v12.n2.a1"
}

@article{Broedel:2018iwv,
    author        = "Broedel, Johannes and Duhr, Claude and Dulat, Falko and Penante, Brenda and Tancredi, Lorenzo",
    title         = "{Elliptic Feynman integrals and pure functions}",
    journal       = "JHEP",
    volume        = "01",
    pages         = "023",
    year          = "2019",
    eprint        = "1809.10698",
    archivePrefix = "arXiv",
    primaryClass  = "hep-th",
    doi           = "10.1007/JHEP01(2019)023"
}

@article{Eichten:1989zv,
    author = "Eichten, Estia and Hill, Brian Russell",
    title = "{An Effective Field Theory for the Calculation of Matrix Elements Involving Heavy Quarks}",
    reportNumber = "FERMILAB-PUB-89-184-T",
    doi = "10.1016/0370-2693(90)92049-O",
    journal = "Phys. Lett. B",
    volume = "234",
    pages = "511--516",
    year = "1990"
}

@article{Georgi:1990um,
    author = "Georgi, Howard",
    title = "{An Effective Field Theory for Heavy Quarks at Low-energies}",
    reportNumber = "HUTP-90/A007",
    doi = "10.1016/0370-2693(90)91128-X",
    journal = "Phys. Lett. B",
    volume = "240",
    pages = "447--450",
    year = "1990"
}

@article{Broadhurst:1994se,
    author = "Broadhurst, David J. and Grozin, A. G.",
    title = "{Matching QCD and heavy-quark effective theory heavy-light currents at two loops and beyond}",
    eprint = "hep-ph/9410240",
    archivePrefix = "arXiv",
    reportNumber = "OUT-4102-52",
    doi = "10.1103/PhysRevD.52.4082",
    journal = "Phys. Rev. D",
    volume = "52",
    pages = "4082--4098",
    year = "1995"
}

@article{Grozin:2000jv,
    author = "Grozin, A. G.",
    title = "{Calculating three loop diagrams in heavy quark effective theory with integration by parts recurrence relations}",
    eprint = "hep-ph/0002266",
    archivePrefix = "arXiv",
    reportNumber = "TTP-00-01",
    doi = "10.1088/1126-6708/2000/03/013",
    journal = "JHEP",
    volume = "03",
    pages = "013",
    year = "2000"
}

@article{Grozin:2006xm,
    author = "Grozin, Andrey G. and Smirnov, Alexander V. and Smirnov, Vladimir A.",
    title = "{Decoupling of heavy quarks in HQET}",
    eprint = "hep-ph/0609280",
    archivePrefix = "arXiv",
    reportNumber = "TTP06-25, SFB-CPP-06-46",
    doi = "10.1088/1126-6708/2006/11/022",
    journal = "JHEP",
    volume = "11",
    pages = "022",
    year = "2006"
}

@article{Grozin:2007ap,
    author = "Grozin, A. G. and Huber, T. and Maitre, D.",
    title = "{On one master integral for three-loop on-shell HQET propagator diagrams with mass}",
    eprint = "0705.2609",
    archivePrefix = "arXiv",
    primaryClass = "hep-ph",
    reportNumber = "SLAC-PUB-12567",
    doi = "10.1088/1126-6708/2007/07/033",
    journal = "JHEP",
    volume = "07",
    pages = "033",
    year = "2007"
}

@article{LIGOScientific:2026oim,
    author = "Abac, A. G. and others",
    collaboration = "LIGO Scientific, VIRGO, KAGRA",
    title = "{GWTC-5.0: Tests of General Relativity}",
    eprint = "2607.19293",
    archivePrefix = "arXiv",
    primaryClass = "gr-qc",
    reportNumber = "LIGO-P2500781",
    month = "7",
    year = "2026"
}

@article{LIGOScientific:2026qni,
    author = "Abac, A. G. and others",
    collaboration = "LIGO Scientific, VIRGO, KAGRA",
    title = "{GWTC-4.0: Tests of General Relativity. I. Overview and General Tests}",
    eprint = "2603.19019",
    archivePrefix = "arXiv",
    primaryClass = "gr-qc",
    reportNumber = "LIGO-P2500065",
    month = "3",
    year = "2026"
}

@article{LIGOScientific:2021sio,
    author = "Abbott, R. and others",
    collaboration = "LIGO Scientific, VIRGO, KAGRA",
    title = "{Tests of General Relativity with GWTC-3}",
    eprint = "2112.06861",
    archivePrefix = "arXiv",
    primaryClass = "gr-qc",
    reportNumber = "LIGO-P2100275",
    doi = "10.1103/PhysRevD.112.084080",
    journal = "Phys. Rev. D",
    volume = "112",
    number = "8",
    pages = "084080",
    year = "2025"
}

@article{Damour:2025oys,
    author = "Damour, Thibault and Jain, Tamanna and Sperhake, Ulrich",
    title = "{Gravitational Scattering of Solitonic Boson Stars: Analytics vs Numerics}",
    eprint = "2512.00945",
    archivePrefix = "arXiv",
    primaryClass = "gr-qc",
    doi = "10.1103/24g1-k5p2",
    journal = "Phys. Rev. Lett.",
    volume = "137",
    number = "2",
    pages = "021401",
    year = "2026"
}

@article{Pogel:2022vat,
    author = {P{\"o}gel, Sebastian and Wang, Xing and Weinzierl, Stefan},
    title = "{Bananas of equal mass: any loop, any order in the dimensional regularisation parameter}",
    eprint = "2212.08908",
    archivePrefix = "arXiv",
    primaryClass = "hep-th",
    doi = "10.1007/JHEP04(2023)117",
    journal = "JHEP",
    volume = "04",
    pages = "117",
    year = "2023"
}

@article{Pogel:2022yat,
    author = {P{\"o}gel, Sebastian and Wang, Xing and Weinzierl, Stefan},
    title = "{The three-loop equal-mass banana integral in {\ensuremath{\varepsilon}}-factorised form with meromorphic modular forms}",
    eprint = "2207.12893",
    archivePrefix = "arXiv",
    primaryClass = "hep-th",
    doi = "10.1007/JHEP09(2022)062",
    journal = "JHEP",
    volume = "09",
    pages = "062",
    year = "2022"
}

@article{Duhr:2022dxb,
    author = "Duhr, Claude and Klemm, Albrecht and Nega, Christoph and Tancredi, Lorenzo",
    title = "{The ice cone family and iterated integrals for Calabi-Yau varieties}",
    eprint = "2212.09550",
    archivePrefix = "arXiv",
    primaryClass = "hep-th",
    reportNumber = "BONN-TH-2022-24, TUM-HEP-1444/22",
    doi = "10.1007/JHEP02(2023)228",
    journal = "JHEP",
    volume = "02",
    pages = "228",
    year = "2023"
}

@article{Forner:2024ojj,
    author = "Forner, Felix and Nega, Christoph and Tancredi, Lorenzo",
    title = "{On the photon self-energy to three loops in QED}",
    eprint = "2411.19042",
    archivePrefix = "arXiv",
    primaryClass = "hep-th",
    reportNumber = "TUM-HEP-1539/24",
    doi = "10.1007/JHEP03(2025)148",
    journal = "JHEP",
    volume = "03",
    pages = "148",
    year = "2025"
}

@article{Duhr:2024uid,
    author = "Duhr, Claude and Porkert, Franziska and Stawinski, Sven F.",
    title = "{Canonical differential equations beyond genus one}",
    eprint = "2412.02300",
    archivePrefix = "arXiv",
    primaryClass = "hep-th",
    reportNumber = "BONN-TH-2024-17",
    doi = "10.1007/JHEP02(2025)014",
    journal = "JHEP",
    volume = "02",
    pages = "014",
    year = "2025"
}

@article{Maggio:2025jel,
    author = "Maggio, Sara and Sohnle, Yoann",
    title = "{On canonical differential equations for Calabi-Yau multi-scale Feynman integrals}",
    eprint = "2504.17757",
    archivePrefix = "arXiv",
    primaryClass = "hep-th",
    reportNumber = "BONN-TH-2025-16, UUITP--14/25",
    doi = "10.1007/JHEP10(2025)202",
    journal = "JHEP",
    volume = "10",
    pages = "202",
    year = "2025"
}

@article{Duhr:2025kkq,
    author = "Duhr, Claude and Maggio, Sara and Porkert, Franziska and Semper, Cathrin and Stawinski, Sven F.",
    title = "{Three-loop banana integrals with four unequal masses}",
    eprint = "2507.23061",
    archivePrefix = "arXiv",
    primaryClass = "hep-th",
    reportNumber = "BONN-TH-2025-24",
    doi = "10.1007/JHEP12(2025)034",
    journal = "JHEP",
    volume = "12",
    pages = "034",
    year = "2025"
}

@article{Pogel:2025bca,
    author = {P{\"o}gel, Sebastian and Teschke, Toni and Wang, Xing and Weinzierl, Stefan},
    title = "{The unequal-mass three-loop banana integral}",
    eprint = "2507.23594",
    archivePrefix = "arXiv",
    primaryClass = "hep-th",
    doi = "10.1007/JHEP01(2026)021",
    journal = "JHEP",
    volume = "01",
    pages = "021",
    year = "2026"
}

@article{Pogel:2022ken,
    author = {P{\"o}gel, Sebastian and Wang, Xing and Weinzierl, Stefan},
    title = "{Taming Calabi-Yau Feynman Integrals: The Four-Loop Equal-Mass Banana Integral}",
    eprint = "2211.04292",
    archivePrefix = "arXiv",
    primaryClass = "hep-th",
    reportNumber = "MITP/22-094, TUM-HEP-1431/22",
    doi = "10.1103/PhysRevLett.130.101601",
    journal = "Phys. Rev. Lett.",
    volume = "130",
    number = "10",
    pages = "101601",
    year = "2023"
}

\end{document}